\documentclass{article}
\usepackage{iclr2024_conference,times}

\usepackage{graphicx}
\usepackage{booktabs}

\usepackage{amsmath,amsfonts,bm}

\def\eqref#1{equation~\ref{#1}}
\def\1{\bm{1}}

\DeclareMathAlphabet{\mathsfit}{\encodingdefault}{\sfdefault}{m}{sl}
\SetMathAlphabet{\mathsfit}{bold}{\encodingdefault}{\sfdefault}{bx}{n}

\usepackage{url}

\usepackage{breakurl}
\usepackage[breaklinks]{hyperref}

\usepackage[capitalize,noabbrev]{cleveref}

\title{On the Shape of Brainscores for Large Language Models (LLMs)}

\author{Jingkai Li \\
OpenSci.World \\
Montréal, Québec H4R 2R9, Canada \\
\texttt{jingkai.li@opensci.world} \enspace  OR\\ 
\texttt{jkli898@126.com}
}

\iclrfinalcopy 
\begin{document}

\maketitle

\begin{abstract}

With the rise of Large Language Models (LLMs), the novel metric "Brainscore" emerged as a means to evaluate the functional similarity between LLMs and human brain/neural systems. Our efforts were dedicated to mining the meaning of the novel score by constructing topological features derived from both human fMRI data involving 190 subjects, and 39 LLMs plus their untrained counterparts. Subsequently, we trained 36 Linear Regression Models and conducted thorough statistical analyses to discern reliable and valid features from our constructed ones. Our findings reveal distinctive feature combinations conducive to interpreting existing brainscores across various brain regions of interest (ROIs) and hemispheres, thereby significantly contributing to advancing interpretable machine learning (iML) studies. The study is enriched by our further discussions and analyses concerning existing brainscores. To our knowledge, this study represents the first attempt to comprehend the novel metric brainscore within this interdisciplinary domain.
\end{abstract}

\section{Introduction}

There has been a notable surge in the proliferation of Large Language Models (LLMs) \citep{open-llm-leaderboard} in recent times. Within this burgeoning landscape, certain models could be reasonably regarded as early, albeit incomplete, versions of artificial general intelligence (AGI) systems \citep{bubeck2023sparks}. Preceding the realization of AGI, a crucial imperative arises: we want to, and \textit{need} to know the extent of human-likeness inherent in the LLMs under development.

Inspired by biological neural networks, Artificial Neural Networks (ANNs) were created in the 1940s by Warren McCullough and Walter Pitts \citep{russell2010artificial}, marking the inception of one of the three principal schools of the machine learning community: \textit{Connectionism} \citep{bognar2022prospects}. Subsequently, deep and very deep neural networks emerged, and LLMs nowadays.  Nevertheless, \textit{the extent to which ANNs have diverged from biological neural networks} remains largely unknown and vastly underexplored, thereby motivating our study.

In addressing the aforementioned inquiries, scientists have commenced employing state-of-the-art LMs to investigate neural activity in the human brain during language processing. Simultaneously, NLP researchers have been motivated to incorporate neuroimaging data to improve their model, as delineated in \citet{karamolegkou2023mapping}. Within this context, the novel metrics "brainscores" \citep{schrimpf2018brain} and "brain hierarchy score" \citep{nonaka2021brain}, have been created to assess the functional similarity between AI models and human brain/neural systems.

The Pearson Correlation/Brainscore metric, as surveyed in \citet{karamolegkou2023mapping}, stands as the predominant method in mapping brains with LMs. In addition to those efforts to calculate "brainscores," our study seeks to address the fundamental questions: \textit{What is the meaning of the score? Can we derive features to interpret it?} Building on prior efforts to quantify similarities between LMs and human neural responses, \citet{li2023structural}'s groundbreaking study  employed Procrustes analysis, a form of statistical shape analysis, to align brain fMRI representations with those of LMs. This approach is substantiated by their discussions on the philosophical foundations regarding the non-trivial nature of structural similarities. Our study extends the concept of "Shape" by systematically comparing the distinctions in the "shapes" between collected human brain fMRI data and the internal embeddings of various LLMs \footnote{A total of 39 models plus their untrained configurations, spanning from albert-base to Llama2-70B (quantized).}. We employ the term "shape" deliberately, as our observation and quantification of the data are from a topological perspective, focusing solely on their "shapes" rather than their geometric properties such as lengths, areas, volumes, angles, etc. Leveraging the Topological Data Analysis (TDA) tool Persistent Homology (PH), we characterize data representations in both realms and construct features by computing $q$-Wasserstein Distances between pairs of their persistence diagrams. Subsequently, we learn Linear Regression Models to fit the existing "brainscores," followed by rigorous statistical analyses to identify reliable and valid features among our constructed ones. Our primary contributions are outlined below:

\begin{itemize}

\item We present detailed results concerning reliable and valid features derived from our constructed ones. It exhibits distinct feature combinations that facilitate the interpretation of existing brainscores for each ROI and each hemisphere. The identified features hold significant value for advancing studies in interpretable machine learning (iML). Our original research findings are augmented with further discussions and analyses on existing brainscores.

\item We demonstrate the efficacy of the TDA tool Persistent Homology (PH), and the $q$-Wasserstein Distance metric in constructing features that are rarely seen in current methods, analytical frameworks, and metrics within the domain of brainscores research. This showcases the potential for learning from data apart from existing conventions.
\item To the best of our knowledge, this study represents the inaugural endeavor in interpreting the novel metric brainscore within this interdisciplinary domain. Overall, we take small steps in trying to address the fundamental inquiries: \textit{"In what sense are the LLMs we create human-like?"}

\end{itemize}

\section{Methods and Data Processing Pipelines}

\subsection{Overview of Methods and Data Processing Pipelines}

\begin{figure}[!t]
\vskip -0.2in
\begin{center}
\centerline{\includegraphics[width=\columnwidth]{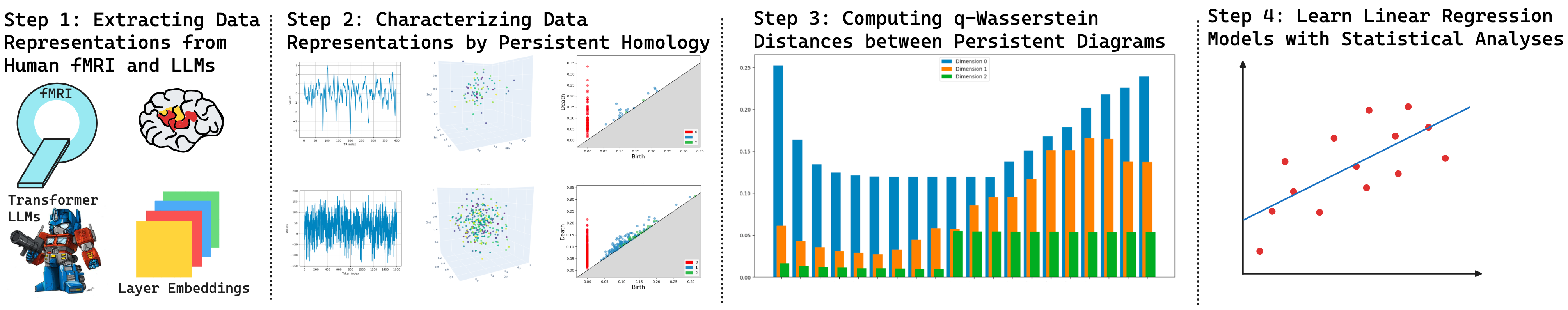}}
\caption{This figure delineates the four steps of our proposed approach: (1) Extracting data representations from human fMRI and diverse LLMs, (2) Characterising those data representations by Persistent Homology, (3) Constructing features by computing $q$-Wasserstein Distances between pairs of Persistence Diagrams, and (4) Systematically filter in reliable and valid features by learning Linear Regression Models with statistical analyses.}
\label{methods_overview}
\end{center}
\vskip -0.2in
\end{figure}

The brainscores that we try to interpret are based on \citet{caucheteux2023evidence}. \cref{methods_overview} illustrates the four steps of our proposed approach.

We firstly extracted data representations from existing human fMRI data and embeddings from specific layers of various Transformer \citep{vaswani2017attention} based LLMs (\cref{Extracting Data Representations from human fMRI and LLMs}). Subsequently, we characterized these data representations using PH (\cref{Characterizing the Data Representations by Persistent Homology (PH)}). Simultaneously, we reproduced the results from \citet{caucheteux2023evidence} and computed corresponding brainscores for our selected LLMs against our specified ROIs as the learning target. Following this, we constructed a set of features by computing $q$-Wasserstein Distances between pairs of persistence diagrams (\cref{Computing q-Wasserstein Distances between Persistence Diagrams}). Finally, we systematically identified reliable and valid features from our constructed ones by learning arrays of Linear Regression Models, accompanied by subsequent statistical analyses (\cref{Learning Linear Regression Models from q-Wasserstein Distances and Existing Brainscores}).

\subsection{Intuitions behind the Methods}
\label{Intuitions behind the Methods}

This section proffers an intuitive explication of the foundational techniques underpinning our methodological approaches and data processing pipelines, namely, the neuroimaging technique of functional Magnetic Resonance Imaging (fMRI) (\cref{An Intuitive to fMRI}), the TDA tool of Persistent Homology (PH) (\cref{An Intuitive Introduction to Persistent Homology (PH)}), and the $q$-Wasserstein Distance metric (\cref{An Intuitive Introduction to Persistence Diagram and q-Wasserstein Distance}). These expository overviews aim to facilitate a comprehensive understanding of the integral components constituting our investigative framework.

\subsubsection{An Intuitive Introduction to fMRI}
\label{An Intuitive to fMRI}

Functional Magnetic Resonance Imaging (fMRI) is a powerful technique used in neuroscience and medicine to study the brain's activity and functionality. It is a non-invasive imaging method that measures changes in blood oxygenation levels within the brain, which are related to neural activity \citep{martini_oermann_opie_panov_oxley_yaeger_2019}.

\textbf{Brain activity}: When a particular region of the brain is active, it requires more oxygen and nutrients. This increased demand for oxygen is met by an increase in blood flow to that region. \textbf{Magnetic properties}: Oxygenated blood (blood carrying oxygen) and deoxygenated blood (blood without oxygen) have different magnetic properties. This difference is known as the "Blood Oxygenation Level Dependent" (BOLD) signal. \textbf{fMRI scanning}: During an fMRI scan, powerful magnetic fields and radio waves are used to detect and measure the BOLD signal in different parts of the brain. This allows researchers and doctors to infer which brain regions are more active or less active during a particular task or condition. \textbf{Spatial and temporal resolution}: fMRI provides good spatial resolution, meaning it can pinpoint the location of brain activity with reasonable accuracy. It also has a relatively good temporal resolution, allowing researchers to track changes in brain activity over time, although not as precise as some other techniques like electroencephalography (EEG). \textbf{Applications}: fMRI is widely used in cognitive neuroscience research to study brain function during various mental processes, such as perception, decision-making, memory, and language processing. It is also used in clinical settings to diagnose and study brain disorders, such as strokes, tumors, and neurodegenerative diseases \citep{ulmer_olav_jansen_2010, ulmer_jansen_2013}.

\begin{figure}[!t]
\vskip -0.2in
\begin{center}
\centerline{\includegraphics[width=\columnwidth * 3/4]{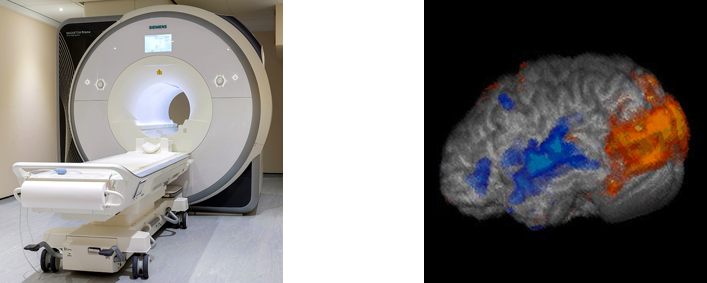}}
\caption{The original images are from \href{https://www.ndcn.ox.ac.uk/divisions/fmrib/what-is-fmri/introduction-to-fmri}{Introduction to fMRI}. The image on the left is a typical research scanner that has a field strength of 3 teslas (T), about 50,000 times greater than the Earth’s field. The right one is the result of an fMRI experiment, and the processed image is the brain's "activation map" during the experiment.}
\label{scanner_and_fmri_result}
\end{center}
\vskip -0.2in
\end{figure}

An MRI scanner and a corresponding brain "activation map", generated from an fMRI experiment (as illustrated in \href{https://www.ndcn.ox.ac.uk/divisions/fmrib/what-is-fmri/introduction-to-fmri}{Introduction to fMRI}), are depicted in \cref{scanner_and_fmri_result}. The processed fMRI data for our current investigation are presented in \cref{averaged-fMRI-for-each-task-roi-pair}.

By measuring changes in blood oxygenation levels, fMRI provides a window into the brain's functional activity, enabling researchers and clinicians to better understand how the brain works and to investigate potential abnormalities or dysfunctions in different brain regions \citep{ulmer_olav_jansen_2010, ulmer_jansen_2013}.

In accordance with prior studies such as \citet{caucheteux2023evidence} and \citet{oota2023does}, the fMRI data utilized in our investigation originate from the Narratives dataset \citep{nastase2021narratives}, which is publicly accessible on \href{https://openneuro.org/datasets/ds002345/versions/1.1.4}{OpenNeuro}. The dataset is a collection which aggregates a variety of fMRI datasets collected while human subjects listened to real spoken stories. 

Utilizing the same Narratives dataset \citep{nastase2021narratives}, \citet{ye2024query} investigated 8 participants, \citet{hahamy2023human} studied one task (21st year) with 25 participants, and \citet{liu2023coupling} examined two tasks: "Pie Man" with 75 participants and "Shapes" with 59 subjects, whereas, \citet{oota2023does} analyzed data from 82 subjects listening to the single story titled "Pie Man".

Considering the scale of our study, we extracted fMRI data from 190 out of 345 participants (subjects) listening to 3 out of 27 stories (tasks) from the dataset's pool. The selected tasks, namely "Pie man," "Shapes," and "It’s Not the Fall That Gets You," were chosen due to their inclusion of the top 3 most subjects (refer to \cref{Summaries for the 3 Selected Tasks from Narratives Dataset} and \cref{narratives-3-tasks} for a concise overview).

\citet{schrimpf2021neural} conducted a large-scale investigation into the relationships between brain activity and 43 contemporary neural network models (encompassing embedding, recurrent, and transformer architectures). Their analysis leveraged three datasets: two fMRI and one electrocardiography (ECG) signals. To assess model-brain alignment, they employed a novel metric termed the Brainscore, which incorporates noise ceiling normalization of the Pearson correlation coefficient. 

As a large-scale study, we examined and analyzed 39 LLMs, spanning from albert-base to Llama-70B (quantized), plus their untrained counterparts \footnote{Untrained versions were unavailable for LLMs: Llama-70B (quantized), Llama-13B (quantized), and Llama-7B (quantized), thus yielding a total of 36 untrained LLMs.}. To be in line with \citet{schrimpf2021neural}, we didn't make a distinction between masked language models and causal language models, but limit the architecture of our interest as Transformer \citep{vaswani2017attention} based one.

\subsubsection{An Intuitive Introduction to Persistent Homology (PH)}
\label{An Intuitive Introduction to Persistent Homology (PH)}

Persistent homology (PH) is a powerful tool in topological data analysis (TDA) that allows us to study the shape and structure of data at different scales or resolutions. It provides a way to quantify and understand the topological features of data, such as connected components, holes, and higher-dimensional voids, and how these features persist or change as the scale or resolution changes \citep{fugacci2016persistent, carter2020data, koplik_2022}.

Imagine the data as a point cloud in a high-dimensional space. For example, if the data consists of images, each image can be represented as a point in a high-dimensional space, where each dimension corresponds to a pixel value. Then, imagine placing small balls or spheres around each data point. The size of these balls can be thought of as a scale parameter or resolution at which we are looking at the data. As the size of the balls increases, they start to overlap and merge, forming connected components, loops, and higher-dimensional voids or cavities. Persistent homology tracks the birth and death of these topological features (connected components, loops, voids) as the scale or resolution changes (i.e., as the ball size increases or decreases). The features that persist over a wide range of scales are considered more significant or robust, while features that appear and disappear quickly are considered noise or less important. Persistent homology represents this information in a concise way, typically using persistence diagrams or barcodes, which show the birth and death times (or scales) of the topological features.

\begin{figure}[!t]
\vskip -0.2in
\begin{center}
\centerline{\includegraphics[width=\columnwidth * 2/3]{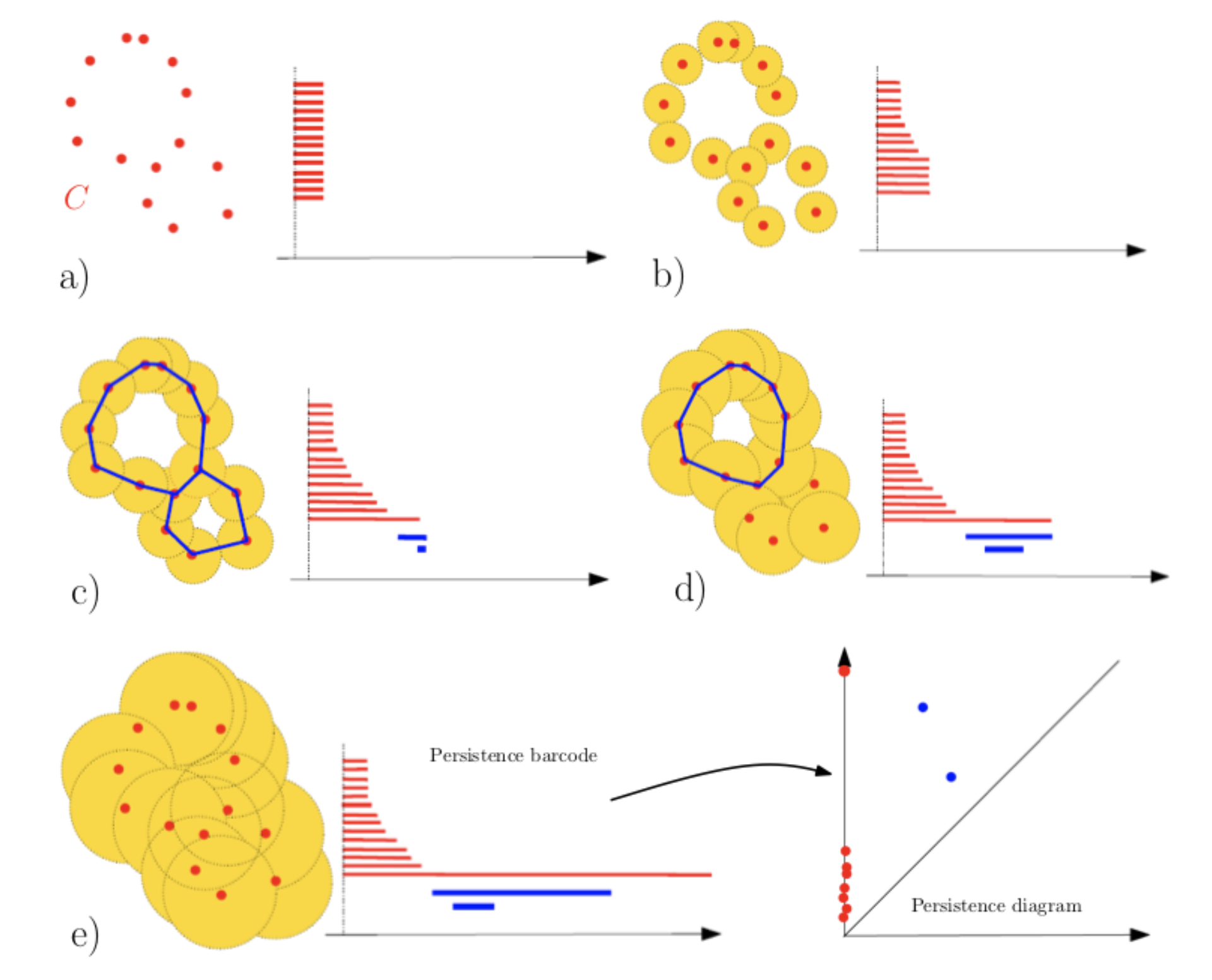}}
\caption{The original figures are from \href{https://github.com/GUDHI/TDA-tutorial/blob/master/Tuto-GUDHI-persistence-diagrams.ipynb}{TDA with Python using the Gudhi Library - Persistent homology and persistence diagrams}, and also appear in \citet{chazal2021introduction} and \citet{hashemi2024deep}. a) For the radius $r = 0$, the union of balls is reduced to the initial finite set of point, each of them corresponding to a $0$-dimensional feature, i.e. a connected component; an interval is created for the birth for each of these features at $r = 0$. b) Some of the balls started to overlap resulting in the death of some connected components that get merged together; the persistence diagram keeps track of these deaths, putting an end point to the corresponding intervals as they disappear. c) New components have merged giving rise to a single connected component and, so, all the intervals associated to a $0$-dimensional feature have been ended, except the one corresponding to the remaining components; two new $1$-dimensional features, have appeared resulting in two new intervals (in blue) starting at their birth scale. d) One of the two $1$-dimensional cycles has been filled, resulting in its death in the filtration and the end of the corresponding blue interval. e) all the $1$-dimensional features have died, it only remains the long (and never dying) red interval. The final barcode can also be equivalently represented as a persistence diagram where every interval $(a,b)$ is represented by the the point of coordinate $(a,b)$ in $\mathbb{R}$. Intuitively the longer is an interval in the barcode or, equivalently the farther from the diagonal is the corresponding point in the diagram, the more persistent, and thus relevant, is the corresponding homological feature across the filtration.}
\label{pers}
\end{center}
\vskip -0.2in
\end{figure}

In the example shown in \cref{pers} (from the original notebook: \href{https://github.com/GUDHI/TDA-tutorial/blob/master/Tuto-GUDHI-persistence-diagrams.ipynb}{TDA with Python using the Gudhi Library - Persistent homology and persistence diagrams}, and also appear in \citet{chazal2021introduction} and \citet{hashemi2024deep}) we witness the filtration given by a union of growing balls centered on the finite set of points $C$. We can see how the topological features emerge, change, and disappear at different scales or resolutions.

The power of persistent homology lies in its ability to capture and quantify the topological structure of data at different scales, without relying on specific assumptions about the data distribution or shape. It can reveal hidden patterns, clusters, and higher-dimensional structures that may not be apparent from traditional geometric or statistical methods \citep{carlsson2009topology, edelsbrunner2010computational, carlsson2021topological, virk_2022_introduction}.

\citet{koplik_2022} provides a non-technical introduction to PH, supplemented by illustrative examples. Another accessible one catering to newcomers is available in \citet{fugacci2016persistent}. Readers seeking more formal and in-depth treatments may refer to \citet{carlsson2009topology, edelsbrunner2010computational, boissonnat2018geometric, carter2020data, carlsson2021topological, dey2022computational, virk_2022_introduction}.

In our investigation, we utilized the potent tool PH to characterize the extracted data representations obtained from fMRI and LLMs embeddings, as elaborated in \cref{Characterizing the Data Representations by Persistent Homology (PH)}.

\subsubsection{An Intuitive Introduction to Persistence Diagram and $q$-Wasserstein Distance}
\label{An Intuitive Introduction to Persistence Diagram and q-Wasserstein Distance}

Imagine the PH is applied to the data, identifying all the important holes and voids at different scales. Now, a persistence diagram is a way to represent this information visually.

Think of it as a scatter plot. On the x-axis, there is the birth time of each hole, and on the y-axis, presented is the death time. Each point on the plot represents a hole in the data, with its coordinates indicating when the hole was born and when it "died" as zoomed in or out \citep{carlsson2009topology, edelsbrunner2010computational, carter2020data, carlsson2021topological, virk_2022_introduction}.

The beauty of persistence diagrams is that they capture the essential topological features of the data in a concise and informative way. By looking at the distribution of points and their positions relative to each other, one can gain insights into the structure and complexity of the data. The intuitive examples are shown in \cref{pers}, whereas examples of persistence diagrams for our current study can be found in \cref{persistence_diagram_for_397_TRs_and_75_voxels_from_54_participants_y_periodic_embedded_ts_by_TR_pca} and \cref{persistence_diagram_layer_21_ouput_time_series_by_tokens_pca}.

The $q$-Wasserstein distance is a way to measure the dissimilarity between two distributions. It's based on a concept from optimal transport theory \citep{mukherjee2021outlier, nietert2022outlier}, which is all about finding the most efficient way to "transport" one distribution of points to another.

In the context of persistence diagrams, the $q$-Wasserstein distance tells how much "work" is required to transform one set of holes into another. Here, "work" refers to the amount of "mass" (i.e., the persistence of each hole) that needs to be moved from one point to another.

\begin{figure}[!t]
\vskip -0.2in
\begin{center}
\centerline{\includegraphics[width=\columnwidth * 3/4]{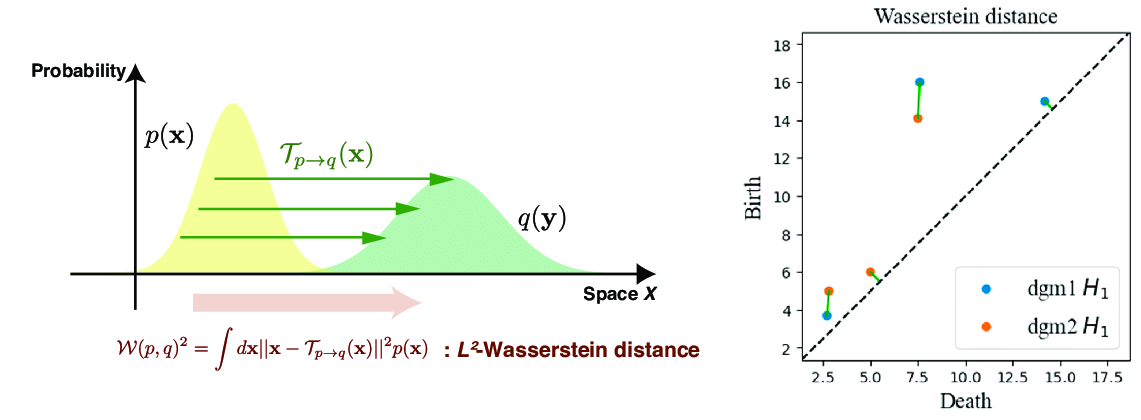}}
\caption{The figure on the left is originally from \citet{nakazato2021geometrical}, which is a schematic of the $L^2$-Wasserstein distance. They here consider optimal transport from the probability distribution $p(\mathbf{x})$ to the probability distribution $p(\mathbf{y})$. The length of the green arrow shows the optimal transportation distance $||\mathbf{x} - \mathcal{T}_{p \to q}(\mathbf{x})||$, and the square of the $L^2$-Wasserstein distance is given by the expected value of the square of its optimal transportation distance. The plot on the right, originally from \citet{zhang2021time}, shows that the Wasserstein distance is the sum of $p$-th power of the distance between all points matching two persistence diagrams. If no corresponding matching point is found, it will match to diagonal.}
\label{wasserstein_distance}
\end{center}
\vskip -0.2in
\end{figure}

\cref{wasserstein_distance} illustrates the $q$-Wasserstein distances for two distributions (left) \citep{nakazato2021geometrical} and two persistence diagrams (right) \citep{zhang2021time}.

By calculating the $q$-Wasserstein distance between two persistence diagrams, one can quantitatively measure how similar or dissimilar their topological structures are. This is useful for tasks like comparing datasets, identifying outliers, or tracking changes in data over time \citep{carlsson2009topology, edelsbrunner2010computational, carter2020data, carlsson2021topological, virk_2022_introduction}.

A concise overview of the $q$-Wasserstein Distance is provided in \citet{rahman2023}, while more comprehensive and formal introductions can be found in \citet{edelsbrunner2010computational, boissonnat2018geometric, carlsson2021topological, dey2022computational}.

In our study, series of $q$-Wasserstein Distances are calculated with the aim of deriving features that subsequently aid in the interpretation of the corresponding brainscores.

\subsection{Implementation Details and Experiments}
\label{Implementation Details and Experiments}

Expanding upon the introductory context, we further present the implementation details and experimental procedures, as outlined in \cref{Implementation Details and Experiments (Appendix)}.

\section{Results}
\label{Results}

We present our results in terms of the filtered-in reliable and valid features w.r.t $q$-Wasserstein Distances with specific $q$ values for each ROI, and each hemisphere. Given the division of our training data into "Only Trained" and "Trained plus Untrained" LLMs, the aforementioned results are presented separately for each group.

\cref{max_q_best_test_R2_scores} in \cref{Filter in Reliable and Valid Features: First Pass} illustrates the maximum $q$ for each ROI and each hemisphere that achieves the highest averaged test $R^2$ score as our result for the First Pass (detailed in \cref{First Pass}).

The detailed results of the Second Pass are elucidated in \cref{Filter in Reliable and Valid Features: Second Pass}. In this phase, we refined features by scrutinizing whether each averaged $p$ value met the criterion $p < 5\%$, as elaborated in \cref{Second Pass}. Each table within \cref{Filter in Reliable and Valid Features: Second Pass} provides a summary of the ultimately filtered-in features \footnote{The Intercept is consistently presented, although it is not utilized for interpretations, irrespective of its $p$ value.} \footnote{Only the features meeting the criterion $p < 5\%$ are presented.} for each ROI (including the whole brain mask), and each hemisphere, along with their respective weights, including both lower and upper bounds of $95\%$ confidence intervals (CI), standard error (SE) averaged across all 25 runs, feature importance $t$-statistic value $|t|$, and the $p$ value \footnote{All values are averaged across the 25 runs, except for SE.}.

Each table is then further accompanied with two figures in its following. The first figure depicts each feature's weight along with both the lower and upper bounds of the $95\%$ CIs. While the second one illustrates the coefficient importance and its variability.

\section{Discussions and Limitations}
\label{Discussions and Limitations}

The brainscores, derived through the re-implementation of \citet{caucheteux2023evidence}, served a dual purpose in our study. Not only did we employ them for learning Linear Regression Models, but they also played a crucial role in addressing our research questions. Simultaneously, we conducted various analyses to augment our investigation. The discussions and limitations outlined in \cref{Notes on Existing Brainscores}, \cref{Comparisons of the Brainscores between LLMs}, \cref{Does the Training Help Increase the LLM's Brainscore?}, \cref{Future Direction in Discussions and Limitations}, and \cref{Analogy of Brain ROIs for Different Layers of LLMs} are directly rooted in the computed brainscores. Subsequently, a follow-up discussion ensues, offering interpretations for these brainscores based on our original research findings, as detailed in \cref{Interpretations for the Learned Linear Regression Models}.

\subsection{Notes on Existing Brainscores}
\label{Notes on Existing Brainscores}

The novel meric brainscore \citep{schrimpf2018brain} has emerged as the primary one for mapping brains with LLMs \citep{karamolegkou2023mapping}. However, being a metric under ongoing development, divergent scores and even conflicting results have been observed across different studies \citep{schrimpf2018brain, nonaka2021brain, macpherson2021natural}, thereby introducing significant uncertainty into interpretation studies, including our current one.

The absolute values of the calculated brainscores remain notably low, suggesting that our developed LLMs still exhibit considerable disparity from human brain functionality. Meanwhile, advancements in Brain–Computer Interface (BCI) technology \footnote{BCI implementations encompass a spectrum from non-invasive (EEG, MEG, MRI) and partially invasive (ECoG and endovascular) to invasive (microelectrode array), based on how close electrodes get to brain tissue \citep{martini_oermann_opie_panov_oxley_yaeger_2019}.} promise to yield higher-quality data, fostering further interdisciplinary investigations.

\subsection{Comparisons of the Brainscores between LLMs: \# Parameters Matter}
\label{Comparisons of the Brainscores between LLMs}

We are interested in the brainscores for diverse LLMs incorporating innovations in architectures, representations, optimizations, generalizations, and the size i.e. \# of parameters of the model, in light of the evolving landscape i.e. recent super giant LLMs \citep{open-llm-leaderboard}.

We pose a direct inquiry: \textit{"Does the brainscore exhibit an increase with a larger number of parameters in the LLM?} The response aligns with our intuitive expectation, leaning towards an affirmative answer, but it is crucial to note that this correlation is not universal. While most instances in \cref{notthefallintact-two_thirds_layer-L-PCC} and \cref{pieman-linspace_layer_8-L-Evidence} within \cref{Illustrations for Brainscores in Discussions and Limitations} support the affirmative correlation, counterexamples also exist. Specifically, the trained LLMs, namely Llama-2-7B-GPTQ, Lama-2-13B-GPTQ, and Lama-2-70B-GPTQ, as illustrated in \cref{notthefallintact-two_thirds_layer-L-PCC}, challenge the notion that an increase in parameters consistently leads to higher brainscores.

These brainscores enable further statistical analyses, such as investigating the correlation between the number of parameters in each LLM and its associated brainscore. The outcomes of these analyses may vary based on the specific groups sampled, i.e. different ROIs, hemispheres, LLM layers, and task conditions, among others.

\subsection{Does the Training Help Increase the LLM's Brainscore?}
\label{Does the Training Help Increase the LLM's Brainscore?}

Researchers and practitioners have devoted extensive efforts to the training of the developed LLMs. The training process is instrumental in endowing the model with intelligence. The pivotal question arises: \textit{Does the training contribute to an elevation in the LLM's brainscore?} Conversely, is there an inverse correlation, \textit{does the training result in a decrease in the brainscore?}

Once again, the response to the aforementioned inquiries is predominantly \textit{Yes}, though, \textit{not always}. Notably, the instances in \cref{notthefallintact-two_thirds_layer-L-PCC} and \cref{pieman-linspace_layer_8-L-Evidence} within \cref{Illustrations for Brainscores in Discussions and Limitations} exemplify both scenarios. In most examples, the brainscores of trained LLMs surpass those of their untrained counterparts. However, negative cases, such as bert-large-uncased and bert-base-cased in \cref{notthefallintact-two_thirds_layer-L-PCC}, and bert-base-uncased and bert-large-uncased in \cref{pieman-linspace_layer_8-L-Evidence}, deviate from this trend.

We, again, assert further analyses can unveil more patterns based on our computed brainscores, contributing to a deeper understanding of our created \textit{intelligence} i.e. LLMs.

\subsection{How do the Compression Techniques i.e. Quantization, Pruning, Distillation, Low-Rank Factorization etc. Affect the Brainscores?}
\label{Future Direction in Discussions and Limitations}

Efforts have been undertaken to explore various compression techniques, i.e. Quantization, Pruning, Distillation, and Low-Rank Factorization \citep{zhu2023survey}, as a response to the escalating training costs associated with the growing number of parameters in LLMs. Notably, some compression methods, such as quantization, impact the performance of LLMs \citep{DBLP:conf/iclr/FrantarAHA23, kawrakow_2023}. In light of this, we are interested in examining \textit{How is the brainscore influenced when we compress the LLM compared to its full-resolution configuration?}

Regrettably, our sample size is insufficient to provide a conclusive answer, as our calculated brainscores only include Llama2-7b and its quantized counterpart relevant to this question. However, the situation is intriguingly nuanced and \textit{depends} on specific instances. The comparison between the original and quantized LLMs, exemplified by Llama-2-7b-hf vs. Llama-2-7B-GPTQ in \cref{notthefallintact-two_thirds_layer-L-PCC} and \cref{pieman-linspace_layer_8-L-Evidence} in \cref{Illustrations for Brainscores in Discussions and Limitations}, aligns with the aforementioned performance decline. While, an intriguing observation surfaces in \cref{notthefallintact-linspace_layer_8-L-Evidence}, where the quantized version in the pair exhibits an even higher brainscore. We posit this question as another avenue for future research.

\subsection{Analogy of Brain ROIs for Different LLM Layers}
\label{Analogy of Brain ROIs for Different Layers of LLMs}

Brain Regions of Interest (ROIs) are often used as the nodes of functional brain networks. ROIs consist of several fMRI measurement voxels that are assumed to exhibit functional homogeneity \citep{poldrack2007region, ryyppo2018regions}. Concurrently, Transformer \citep{vaswani2017attention} based LLMs consist of varying numbers of Transformer-block layers. Drawing an analogy, our inquiry revolves around \textit{Do different layers of LLMs play different functional roles compared to those delineated as Brain ROIs?}

Researches \citep{jawahar2019does, manning2020emergent} in the field have indicated that the deep layers of some LLM encode higher-level and more contextualized representations than their initial layers. Additionally, interdisciplinary investigations involving Brain/Neural Science have sought to determine which LLM layers exhibit the most brain-like behavior \citep{tucker2023increasing, li2023structural}. However, previous studies such as \citet{tucker2023increasing} only examined two LLMs, namely BERT-large-cased and GPT2-XL, for selected ROIs, while \citet{li2023structural} explored 14 LLMs without explicitly specifying any particular ROI. Consequently, this area remains significantly underexplored.

We present several heatmaps depicting the averaged brainscores across uniformly sampled eight layers \footnote{The sampling procedures are detailed in \cref{Characterizing the LLMs Embeddings by PH}} of the LLM: Llama-2-7b-hf, computed for all eight language-related brain ROIs. These visualizations are provided in \cref{notthefallintact-Llama-2-7b-hf-L-and-R}, \cref{shapessocial-Llama-2-7b-hf-L-and-R}, and \cref{pieman-Llama-2-7b-hf-L-and-R} within \cref{Illustrations for Brainscores in Discussions and Limitations}. Notably, the ROI labeled as PTL consistently exhibits elevated scores compared to others. The PTL is known for its significant involvement in lexicalized syntactic processing \citep{https://doi.org/10.1002/hbm.24403, matchin2020syntax}. This raises the question: \textit{Do our developed LLMs primarily serve the functions associated with this specific ROI?}

\subsection{Interpretations for the Learned Linear Regression Models}
\label{Interpretations for the Learned Linear Regression Models}

Finally, it comes to interpret our learned Linear Regression Models. This process unfolds in \cref{Results}, with additional granularity provided in \cref{Filter in Reliable and Valid Features: First Pass} and \cref{Filter in Reliable and Valid Features: Second Pass}. Within each table, the $t$-statistic $|t|$ values serve as indicators of the significance of individual features. These distinct feature combinations facilitate the interpretation of the current brainscores associated with different ROIs and hemispheres.

It is notable that while the brainscore serves as a metric denoting the \textit{similarity} between fMRI and LLM data representations, the $q$-Wasserstein Distance quantifies the \textit{dissimilarities} between two distributions. Therefore, particular attention should be directed towards features bearing negative weights.

Our learned Linear Regression Models and the subsequently filtered-in features are not yet in their definitive optimal state, as evidenced by \cref{max_q_best_test_R2_scores}, where all the highest averaged test $R^2$ scores remain below the ideal threshold to varying degrees. Potential avenues for improvement includes considerations on removal of outliers on persistence diagrams (\cref{Characterizing the Data Representations by Persistent Homology (PH)}), i.e. the $q$-Wasserstein Distance is reported to be sensitive to outliers \citep{mukherjee2021outlier, nietert2022outlier}, the exploration of different $q$ and $p$ values for constructing features through $q$-Wasserstein Distances (\cref{Computing q-Wasserstein Distances between Persistence Diagrams}), and the utilization of diverse Linear Regression Models (\cref{Learning Linear Regression Models from q-Wasserstein Distances and Existing Brainscores}), i.e. Ridge vs. OLS. It is evident that these considerations will lead to algorithm optimization and increased complexity.

Moreover, there has been limited comparison of the structural properties between human brain/neural systems and LLMs \footnote{For instance, with the exception of studies such as the Brain Hierarchy (BH) score, which quantifies the degree of hierarchical correspondence between DNN and human brain activity in image recognition \citep{nonaka2021brain}}. We thus propose it as another promising avenue for research.

\section{Conclusions}

We devoted our efforts in mining the meaning of the novel metric brainscores through the construction of topological features derived from both human fMRI data, encompassing 190 subjects, and 39 LLMs along with their untrained counterparts. Subsequently, we trained a total of 36 Linear Regression Models and conducted thorough statistical analyses to discern reliable and valid features from our constructed ones. Our findings reveal distinctive feature combinations conducive to interpreting existing brainscores across various ROIs and hemispheres, thereby contributing significantly to advancing iML studies. The study is enhanced by our further discussions and analyses regarding existing brainscores. To our knowledge, this study represents the first attempt to comprehend the novel metric brainscore within this interdisciplinary domain.  Overall, we take incremental steps toward addressing fundamental inquiries: \textit{"In what sense are the LLMs we create human-like?"}.

\section*{Impact Statements}

Our study does not present direct negative societal impacts, as we did not generate any new fMRI data nor develop any new LLMs. All the data and tools utilized in our research are publicly accessible.

It is crucial to recognize that the biological brain and neural system do not represent the exclusive or ultimate objective for AI development and research. This perspective arises from a fundamental philosophical inquiry: \textit{"Is the biological brain and neural system the epitome of purity and perfection? Can intelligence be created in forms distinct from our own?"} Consequently, future advancements in AI must factor in this perspective. However, comprehending the disparities between human intelligence and artificially created one is paramount. This understanding is propelled by our curiosity about the unknown and our societal responsibilities.

\bibliography{iclr2024_conference}
\bibliographystyle{iclr2024_conference}

\newpage
\appendix

\section{Implementation Details and Experiments}
\label{Implementation Details and Experiments (Appendix)}

\subsection{Extracting Data Representations from human fMRI and LLMs}
\label{Extracting Data Representations from human fMRI and LLMs}

\subsubsection{Extracting Data Representations from human fMRI}
\label{Extracting Data Representations from human fMRI}

Consistent with the approach in \citet{caucheteux2023evidence}, we used the preprocessed fMRI signals from the Narratives dataset \citep{nastase2021narratives}, without spatial smoothing (referred to as "afni-nosmooth" in the repository) and sampled a repetition time (TR) of 1.5s. The preprocessing steps were performed using fMRIPrep \citep{esteban2019fmriprep}; no temporal filtering was applied. The resulting preprocessing led to the analysis of cortical voxels projected onto the surface and morphed onto an "fsaverage" template brain; hereafter, they are referred to as voxels for simplicity. 

We systematically derived data representations for \textit{each task, subject, hemisphere, and Region of Interest (ROI)} from the aforementioned fMRI dataset. The term "hemisphere" denotes either the \textit{L}(eft) or \textit{R}(ight) half of the brain. We referred to the Glasser Atlas \citep{glasser2016multi} (consist of 180 ROIs in each hemisphere) to mask the Brain ROI, since the Narratives dataset \citep{nastase2021narratives} contains annotations tied to this atlas. Consistent with the approach in \citet{oota2023does}, eight language-related brain ROIs, subcategorized according to \citet{fedorenko2010new, fedorenko2012lexical, fedorenko2016neural, baker2018connectomic, milton2021parcellation, schrimpf2021neural, desai2023proper}, were selected: (i) angular gyrus (AG: PFm, PGs, PGi, TPOJ2, and TPOJ3); (ii) anterior temporal lobe (ATL: STSda, STSva, STGa, TE1a, TE2a, TGv, and TGd); (iii) posterior temporal lobe (PTL: A5, STSdp, STSvp, PSL, STV, TPOJ1); (iv) inferior frontal gyrus (IFG: 44, 45, IFJa, IFSp); (v) middle frontal gyrus (MFG: 55b); (vi) inferior frontal gyrus orbital (IFGOrb: a47r, p47r, a9-46v), (vii) posterior cingulate cortex (PCC: 31pv, 31pd, PCV, 7m, 23, RSC); and (viii) dorsal medial pre-frontal cortex (dmPFC: 9m, 10d, d32). Additionally, the whole brain mask was incorporated into the data processing pipeline \footnote{See figures for the fMRI collected and averaged for each ROI and task in \cref{averaged-fMRI-for-each-task-roi-pair}.}. We derived 3,366 matrices in total, each shape of those being like \# TR $\times$ 40,962 voxels \footnote{See \cref{fMRI Data Representation Stats} and \cref{fMRI-data-representations} for details.}. The noise, as suggested in the original paper \citep{nastase2021narratives}, led to the exclusion of specific individual–story pairs, resulting in reduced TRs and a diminished number of subjects for each of the selected 3 tasks \footnote{See \cref{Summaries for the 3 Selected Tasks from Narratives Dataset} and \cref{narratives-3-tasks} for details.}. To simplify the analysis, a subdivision of the Destrieux atlas \citep{destrieux2010automatic} was employed, where regions with more than 500 vertices were subdivided, resulting in 142 regions per hemisphere, each containing fewer than 500 vertices, which reduced the number of voxels from 40,962 to 75 for each of the 3,366 matrices generated.

Similar to \citet{caucheteux2023evidence}, formally, we denote:

\begin{itemize}
\item $Y$ as the fMRI recordings elicited by a subject listening to a task, of size $T \times V$, with $T$ as the number of fMRI time samples (TRs) and $V$ as the number of voxels;
\end{itemize}

\subsubsection{Extracting Data Representations from LLMs}
\label{Extracting Data Representations from LLMs}

We examined and analyzed 39 LLMs, spanning from albert-base to Llama-70B (quantized), plus their untrained counterparts \footnote{Untrained versions were unavailable for LLMs: Llama-70B (quantized), Llama-13B (quantized), and Llama-7B (quantized), thus yielding a total of 36 untrained LLMs.}. All of these LLMs are constructed with Transformer \citep{vaswani2017attention} blocks as their principal components, signifying their composition of varying numbers of Transformer-block layers.

Upon inputting sequences of words into each of the aforementioned LLMs, we sampled and extracted embeddings for \textit{each LLM across each task}. This process yielded a total of 225 tensors, encompassing 39 trained LLMs and 36 untrained ones, each multiplied by 3 tasks. The tensor is three-dimensional, representing \# LLM layers, \# tokens in the task, and the LLM embedding dimension. \footnote{Refer to \cref{LLMs Data Representation Stats} and \cref{LLMs-data-representations} for details.} All the LLMs analysed are publicly accessible via \href{https://huggingface.co/}{Hugging Face}.

Similar to \citet{caucheteux2023evidence}, we denote:

\begin{itemize}
\item $w$ as a sequence of $M$ words (that is, our selected 3 tasks)
\item $X$ as the embeddings of a LLM model input with $w$, of size $M \times U$, with $U$ as the dimensionality of the embeddings (for a layer of i.e. GPT-2, U = 768). We explicitly denote $X_k$ as the embeddings extracted from layer $k$.
\end{itemize}

Diverging from \citet{caucheteux2023evidence}, we refrained from summing the LLM embeddings of words presented within the same TR to align with the sampling frequency of the fMRI and LLMs. This deviation is attributed to the nature of the LLMs under investigation, which process text (transcript) directly rather than audio input \footnote{They in fact lack the capability to process audio signals and are exclusively designed for text processing.}. Our objective is to examine the intrinsic nature of the internal embeddings without introducing any supplementary interference.

\subsection{Characterizing the Data Representations by Persistent Homology (PH)}
\label{Characterizing the Data Representations by Persistent Homology (PH)}

\subsubsection{Characterizing the fMRI Data Representations by PH}
\label{Characterizing the fMRI Data Representations by PH}

\begin{figure}[H]
\vskip 0.2in
\begin{center}
\centerline{\includegraphics[width=\columnwidth]{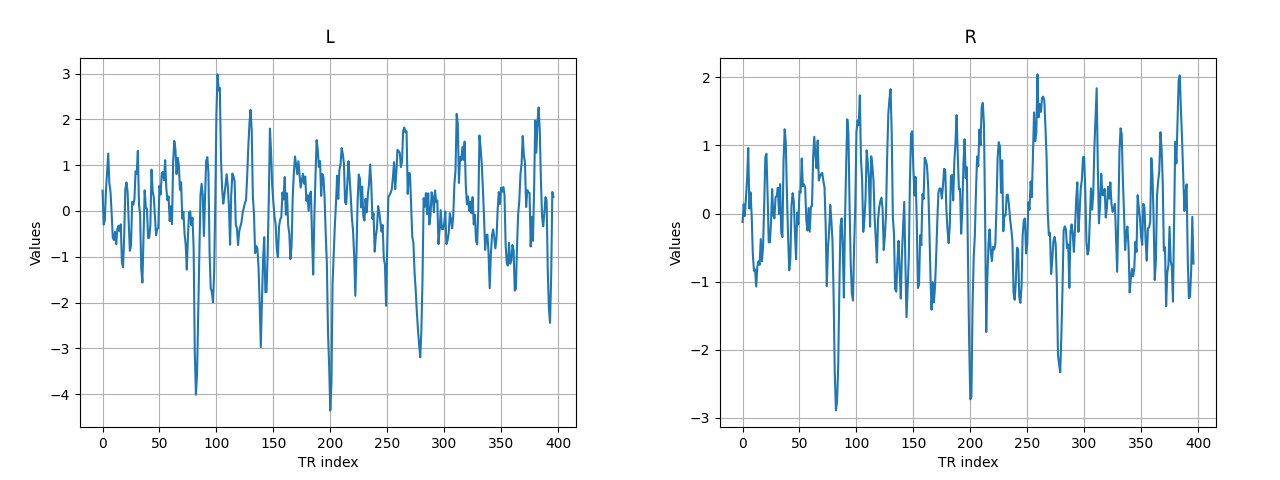}}
\caption{This figure depicts the time-series of the fMRI BOLD values for both the (L)eft and (R)ight hemispheres and the region of interest (ROI): PCC. These values are averaged across 54 out of 56 subjects while they listened to the task "It's Not the Fall that Gets You," and subsequently aggregated along the TR dimension.}
\label{ts_for_397_TRs_and_75_voxels_from_54_participants_time_series_by_TR}
\end{center}
\vskip -0.2in
\end{figure}

Continuing from \cref{Extracting Data Representations from human fMRI} and considering the temporal dimension (TR), we observe an array of time-series signals. We then averaged across all valid subjects for \textit{each task, hemisphere, and ROI}, resulting in a diminished number of matrices while preserving consistent dimensions. Subsequently, we aggregated each of these matrices along the TR dimension, transforming each matrix into a 1-D time-series vector. The length of each vector corresponds to the number of TRs for the respective task (refer to \cref{ts_for_397_TRs_and_75_voxels_from_54_participants_time_series_by_TR} in \cref{Illustrations for Characterizing the fMRI Data Representations by PH} for an illustrative example).

The TDA Toolkit \href{https://giotto-ai.github.io/gtda-docs/latest/index.html}{Giotto-TDA} \citep{tauzin2021giotto} was employed to project the aforementioned time-series signals into a 3-D space as a point cloud. Following the tutorial provided in \href{https://giotto-ai.github.io/gtda-docs/latest/notebooks/topology_time_series.html}{Topology of time series}, we utilized the algorithm embedded in the \href{https://giotto-ai.github.io/gtda-docs/latest/index.html}{Giotto-TDA} toolkit to determine as optimal choices for the embedding dimension ($d$) and time delay ($\tau$). The search started from relatively large values of $(d, \tau)$ to prevent convergence to sub-optimal minimum. However, we restricted the optimal embedding dimension $d$ not to exceed 3. This restriction stems from the fact that surpassing $d=3$ would necessitate truncating $d$ to 3 through dimensionality reduction, thereby incurring information loss (refer to our employed \cref{Transform 1-D Time-Series into 3-D Point Clouds} in practice).

\begin{algorithm}[H]
   \caption{Transform 1-D Time-Series into 3-D Point Cloud}
   \label{Transform 1-D Time-Series into 3-D Point Clouds}
\begin{algorithmic}
   \STATE {\bfseries Input:} max embedding dimension $d$, optimal time delay $\tau$
   \STATE Initialize $d = 3$, $\tau = \text{\# TR}$.
   \FOR{$i=1$ {\bfseries to} \# TR}
   \STATE pointCloud = searchAlgo($\tau // i)$, $d$)
   \IF{pointCloud.shape[1] $< 3$}
   \STATE continue
   \ENDIF
   \STATE $\tau = \tau // i$
   \STATE break
   \ENDFOR
\end{algorithmic}
\end{algorithm}

\begin{figure}[H]
\vskip 0.2in
\begin{center}
\centerline{\includegraphics[width=\columnwidth]{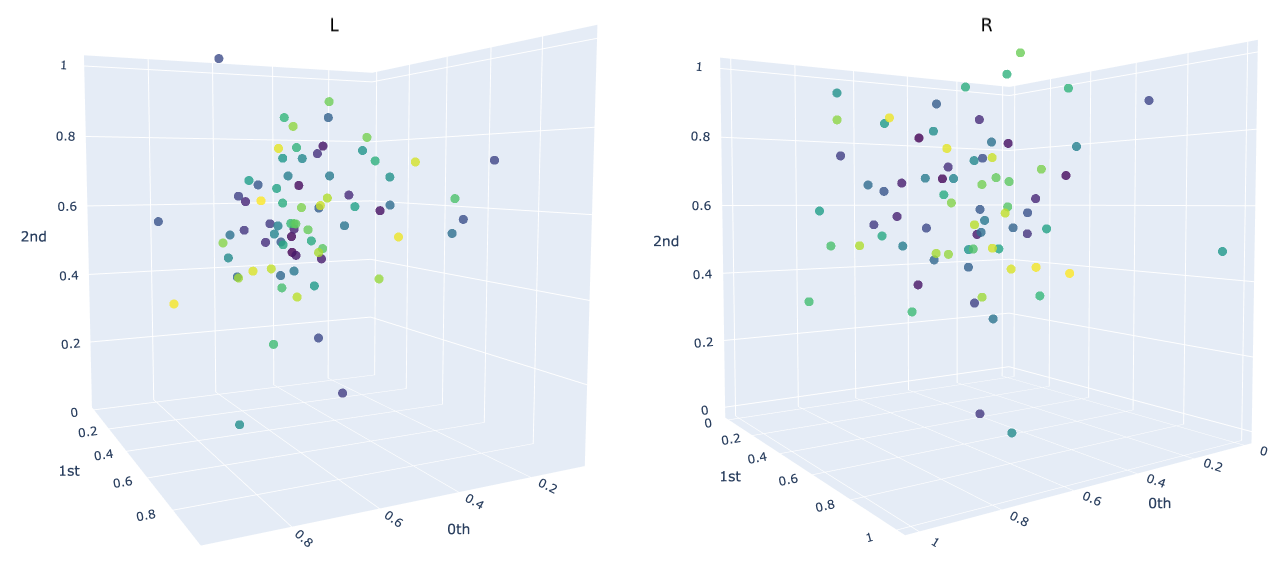}}
\caption{This figure presents the projected point cloud in 3-D space for the time-series signals illustrated in \cref{ts_for_397_TRs_and_75_voxels_from_54_participants_time_series_by_TR}. In this specific instance, the embedding dimension is constrained to 3, and the optimal time delay is determined to be $\tau = 78$ for the (L)eft hemisphere and $\tau = 76$ for the (R)ight one. This result pertains to the task "It's Not the Fall that Gets You," which comprised a total of 397 valid TRs.}
\label{397_TRs_and_75_voxels_from_54_participants_y_periodic_embedded_ts_by_TR_pca}
\end{center}
\vskip -0.2in
\end{figure}

The shape of our projected point cloud is interpreted as optimal time delay $\tau \times$ restricted embedding dimension $d=3$. All values of the point cloud were normalized to fall within the range of $[0,1]$. This normalization ensures that the threshold "birth-death" time points, delineated on each persistence diagram \footnote{We utilized the \href{https://gudhi.inria.fr/index.html}{GUDHI} library \citep{maria2014gudhi} to generate summary representations, i.e. persistence diagrams and persistence barcodes, from our computed PH. This library was also employed in computing $q$-Wasserstein Distances as outlined in \cref{Computing q-Wasserstein Distances between Persistence Diagrams}.}, maintain controlled scales, since Our emphasis lies in the topological "shape" of these point clouds, and we deliberately mitigate the influence of their geometrical sizes (refer to \cref{397_TRs_and_75_voxels_from_54_participants_y_periodic_embedded_ts_by_TR_pca} in \cref{Illustrations for Characterizing the fMRI Data Representations by PH} for an illustrative example of a point cloud in 3-D space).

\begin{figure}[H]
\vskip 0.2in
\begin{center}
\centerline{\includegraphics[width=\columnwidth]{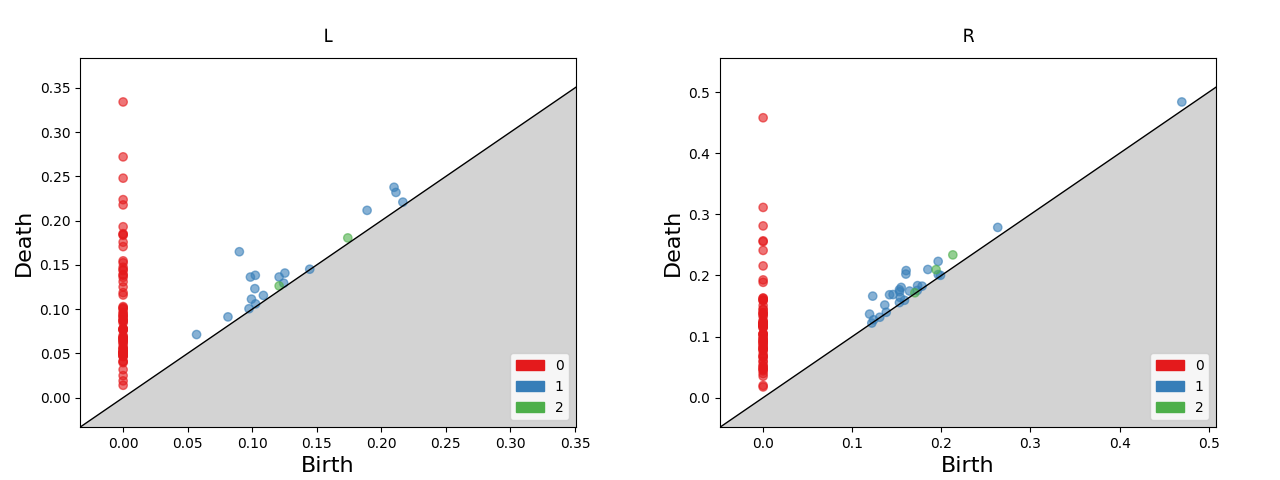}}
\caption{This figure depicts the persistence diagram derived from the Persistent Homology analysis conducted on the point cloud detailed in \cref{397_TRs_and_75_voxels_from_54_participants_y_periodic_embedded_ts_by_TR_pca}.}
\label{persistence_diagram_for_397_TRs_and_75_voxels_from_54_participants_y_periodic_embedded_ts_by_TR_pca}
\end{center}
\vskip -0.2in
\end{figure}

\begin{figure}[H]
\vskip 0.2in
\begin{center}
\centerline{\includegraphics[width=\columnwidth]{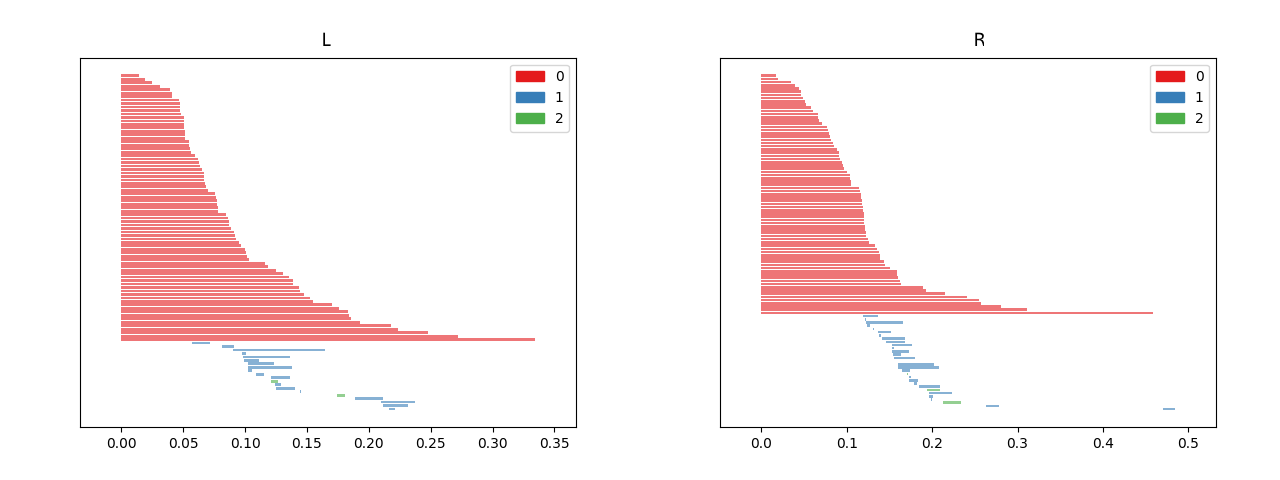}}
\caption{This figure depicts the persistence barcode derived from the Persistent Homology analysis conducted on the point cloud detailed in \cref{397_TRs_and_75_voxels_from_54_participants_y_periodic_embedded_ts_by_TR_pca}.}
\label{persistence_barcode_for_397_TRs_and_75_voxels_from_54_participants_y_periodic_embedded_ts_by_TR_pca}
\end{center}
\vskip -0.2in
\end{figure}

Finally, we employed the "VietorisRipsPersistence" transformer from \href{https://giotto-ai.github.io/gtda-docs/latest/index.html}{Giotto-TDA} to execute the computation for persistence. Considering computation complexity and cost, we restricted the homology dimensions to 0,1,2 (refer to \cref{persistence_diagram_for_397_TRs_and_75_voxels_from_54_participants_y_periodic_embedded_ts_by_TR_pca} and \cref{persistence_barcode_for_397_TRs_and_75_voxels_from_54_participants_y_periodic_embedded_ts_by_TR_pca} in \cref{Illustrations for Characterizing the fMRI Data Representations by PH} for the summary representations i.e. persistence diagram and persistence barcode respectively).

\subsubsection{Characterizing the LLMs Embeddings by PH}
\label{Characterizing the LLMs Embeddings by PH}

\begin{figure}[H]
\vskip 0.2in
\begin{center}
\centerline{\includegraphics[width=\columnwidth]{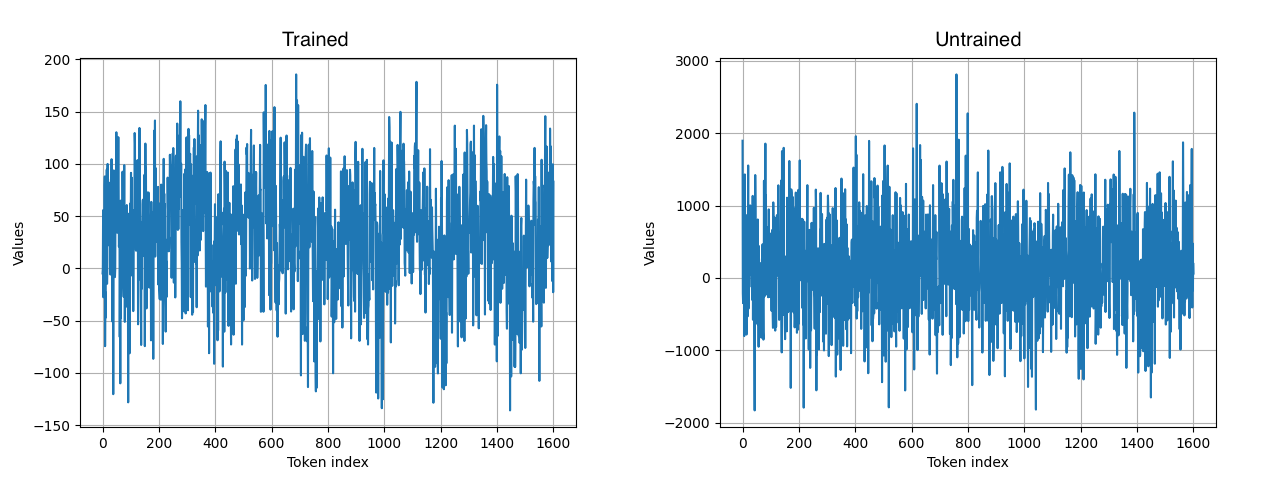}}
\caption{This figure presents the time-series for the embedding from Layer 21 of the LLM \href{https://huggingface.co/meta-llama/Llama-2-7b-hf}{meta-llama/Llama-2-7b-hf}, both trained and untrained, when provided with the task "It's not the Fall that Gets You". The embeddings are then aggregated along the token dimension. Layer 21 from \href{https://huggingface.co/meta-llama/Llama-2-7b-hf}{meta-llama/Llama-2-7b-hf} corresponds to the intermediate-to-deep layer of the LLM ($l = \frac{2}{3} n_{\text{layers}}$), as defined by \citet{caucheteux2023evidence}.}
\label{ts_by_tokens_for_layer_21_ouput_time_series_by_tokens}
\end{center}
\vskip -0.2in
\end{figure}

Following the preceding step (\cref{Extracting Data Representations from LLMs}) and focusing on the token dimension for each LLM layer, we obtained again a set of time-series signals. For each specific LLM layer, we firstly aggregated the matrices (\# tokens of the task $\times$ LLM embedding dimension) along the token dimension. This transformation converted each matrix into a 1-D time-series vector, with its length corresponding to \# tokens for the respective task (refer to \cref{ts_by_tokens_for_layer_21_ouput_time_series_by_tokens} in \cref{Illustrations for Characterizing the LLMs Embeddings by PH}). While we analyzed all 39 LLMs \footnote{Including their 36 untrained counterparts.}, we did not process every individual layer of each LLM. Instead, we uniformly sampled eight layers, evenly spaced from the first one to the last (inclusive at both ends), from each LLM. These selected layers are referred to as "Used Layers" in \cref{LLMs Data Representation Stats} and \cref{LLMs-data-representations}, where the layer numbers marked with an underline indicate the intermediate-to-deep layers of each LLM ($l = \frac{2}{3} n_{\text{layers}}$) according to \citet{caucheteux2023evidence}, warranting our special attention due to their superior predictive capabilities for brain activity, as reported in \citet{schrimpf2021neural, caucheteux2022brains, caucheteux2023evidence}.

In a manner akin to \cref{Characterizing the fMRI Data Representations by PH}, we projected the aforementioned time-series signals into 3-D space as point clouds. Employing the same \cref{Transform 1-D Time-Series into 3-D Point Clouds}, we pursued the determination of the optimal time delay $\tau$, constraining the optimal embedding dimension $d = 3$, where the value initialized for $\tau = \text{\# tokens}$ in this instance.

\begin{figure}[H]
\vskip 0.2in
\begin{center}
\centerline{\includegraphics[width=\columnwidth]{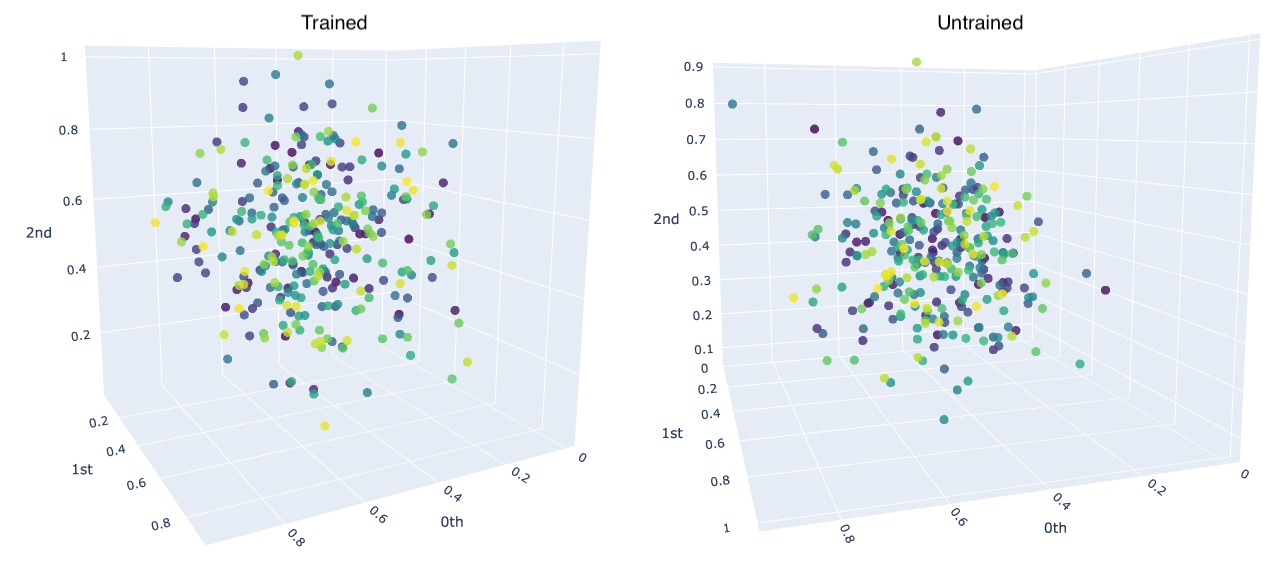}}
\caption{This figure depicts the projected point cloud in 3-D space for the time-series signals specified in \cref{ts_by_tokens_for_layer_21_ouput_time_series_by_tokens}. In this specific instance, the embedding dimension is constrained to $3$, with the optimal time delay searched being $\tau = 320$ for the trained model and $\tau = 318$ for the untrained one. The total number of tokens for the task "It's not the Fall that Gets You" is 1601.}
\label{y_periodic_embedded_ts_by_tokens_pca_for_layer_21_ouput_time_series_by_tokens_pca}
\end{center}
\vskip -0.2in
\end{figure}

\begin{figure}[H]
\vskip 0.2in
\begin{center}
\centerline{\includegraphics[width=\columnwidth]{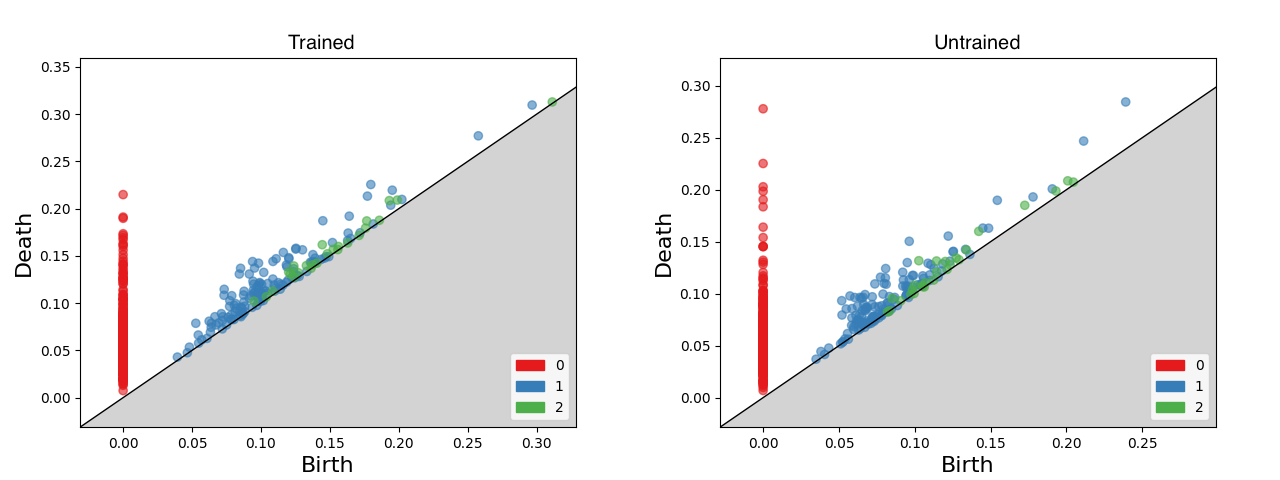}}
\caption{This figure presents the persistence diagram summarized from the Persistent Homology computed from the point cloud specified in \cref{y_periodic_embedded_ts_by_tokens_pca_for_layer_21_ouput_time_series_by_tokens_pca}.}
\label{persistence_diagram_layer_21_ouput_time_series_by_tokens_pca}
\end{center}
\vskip -0.2in
\end{figure}

\begin{figure}[H]
\vskip 0.2in
\begin{center}
\centerline{\includegraphics[width=\columnwidth]{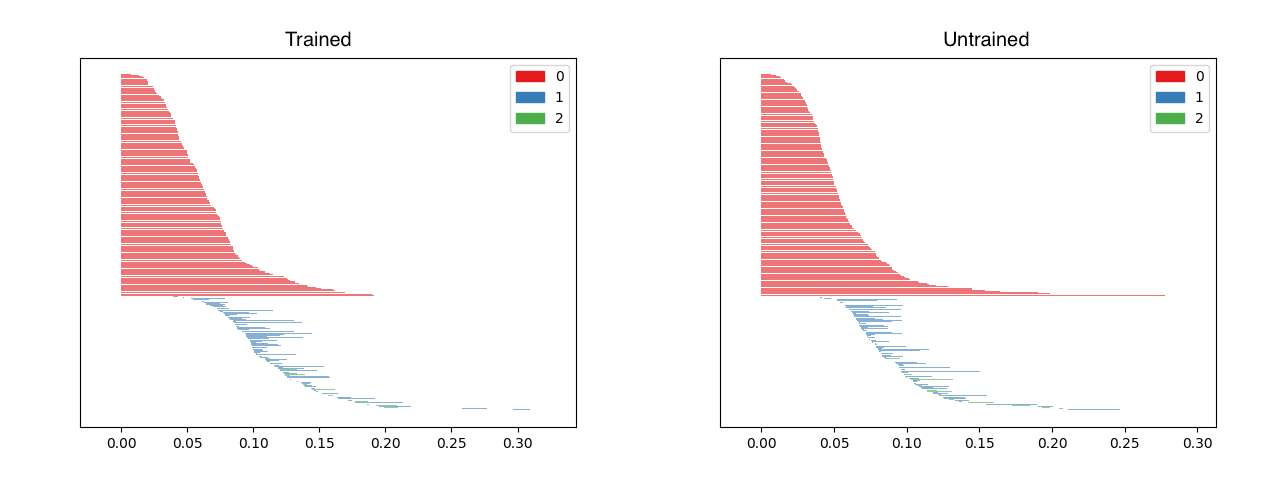}}
\caption{This figure presents the persistence barcode summarized from the Persistent Homology computed from the point cloud specified in \cref{y_periodic_embedded_ts_by_tokens_pca_for_layer_21_ouput_time_series_by_tokens_pca}.}
\label{persistence_barcode_21_ouput_time_series_by_tokens_pca}
\end{center}
\vskip -0.2in
\end{figure}

Subsequently, we obtained the projected point cloud, where the size is interpreted as optimal time delay $\tau \times$ restricted embedding dimension $d=3$. In alignment with the rationale elucidated in \cref{Characterizing the fMRI Data Representations by PH}, we normalized all its values within the range of $[0,1]$ (refer to \cref{y_periodic_embedded_ts_by_tokens_pca_for_layer_21_ouput_time_series_by_tokens_pca} in \cref{Illustrations for Characterizing the LLMs Embeddings by PH} for a representative point cloud projected into the 3-D space). Homologously, the computation for homology dimensions was constrained to 0,1,2 for considerations of computational complexity and cost (see \cref{persistence_diagram_layer_21_ouput_time_series_by_tokens_pca} for corresponding summary representations as the persistence diagram, and \cref{persistence_barcode_21_ouput_time_series_by_tokens_pca} in \cref{Illustrations for Characterizing the LLMs Embeddings by PH} for the persistence barcode, respectively).

\subsection{Computing $q$-Wasserstein Distances between Persistence Diagrams}
\label{Computing q-Wasserstein Distances between Persistence Diagrams}

A total of 34,416 combination pairs were generated encompassing both fMRI data and LLMs Embeddings. Each pair was associated with a specific task listened to by the human subject and processed by the LLM. More specifically, the identification of each pair involved a combination of a \textit{particular ROI, hemisphere, LLM layer, and training status}, consuming the same \textit{task}. The ensuing step required the computation of $q$-Wasserstein Distances \footnote{As advocated by \citet{turner_2019_katharine}, we constrained the two parameters $q=p$ when computing the $q$-Wasserstein Distances.} for each unique combination. It is noteworthy that the computation of averaged layer embeddings was restricted to the whole brain mask exclusively, as the averaged layer represents the entire LLM embedding, which is inherently incomparable to a specific ROI, while a distinct layer embedding better serves this purpose.

\subsection{Learning Linear Regression Models from $q$-Wasserstein Distances and Existing Brainscores}
\label{Learning Linear Regression Models from q-Wasserstein Distances and Existing Brainscores}

The aim of our study is to comprehend the meaning of brainscores through constructing interpretative features. To achieve this, we utilized the Linear Regression Model, chosen for its simplicity in estimation and interpretability of weights \citep{molnar2020interpretable}.

The training data was partitioned to learn Linear Regression Models for \textit{each ROI across each hemisphere} in two: "Only Trained" LLMs and "Trained plus Untrained" LLMs. The latter group involved the brainscores for each untrained LLM. Consequently, a total of 36 models were learned, covering (8+1) ROIs $\times$ 2 hemispheres $\times$ 2 training groups. We employed the \href{https://scikit-learn.org/}{scikit-learn} toolkit \citep{scikit-learn} to construct the models with cross validations.

\begin{figure}[H]
\vskip 0.2in
\begin{center}
\centerline{\includegraphics[width=\columnwidth/13*10]{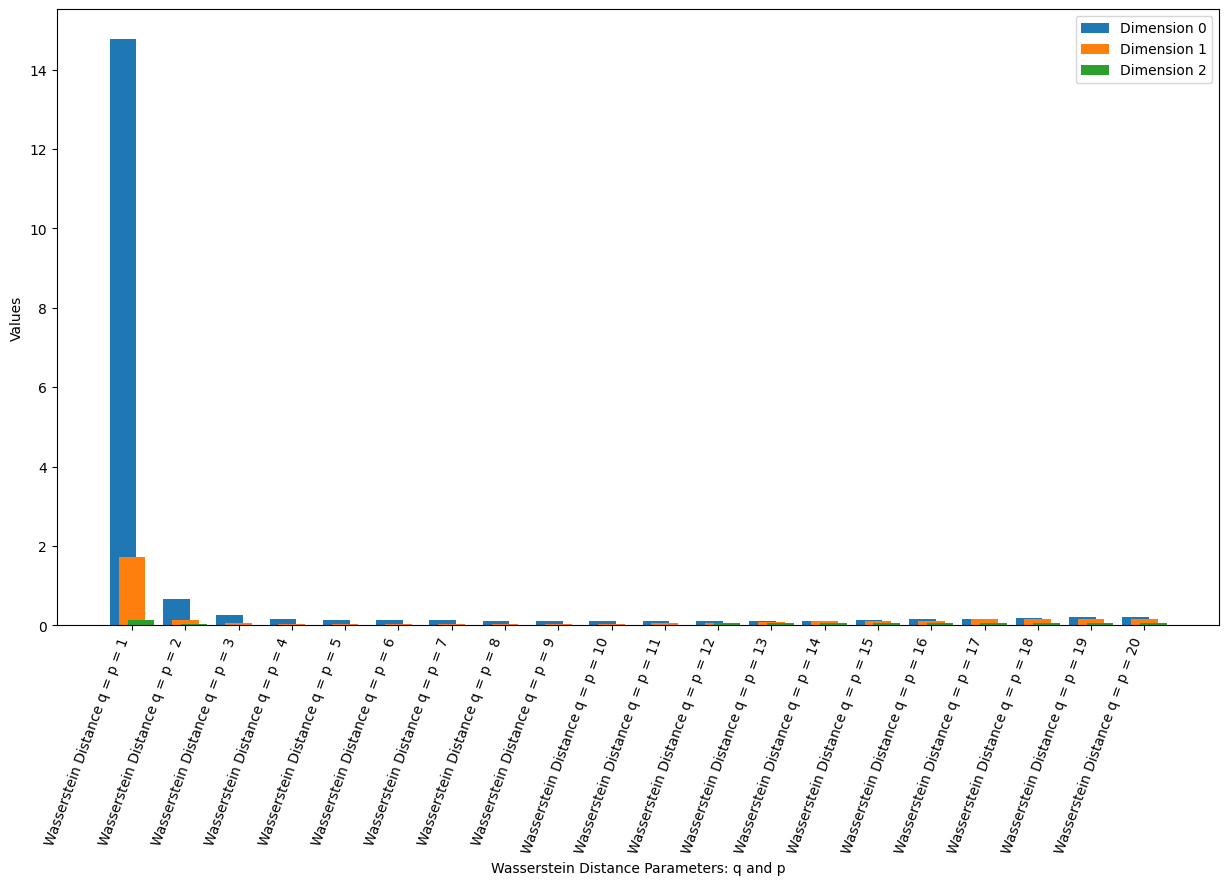}}
\caption{The figure depicts the $q$-Wasserstein Distances with parameters $q$ and $p$ spanning from $q=p=1$ to $q=p=20$.}
\label{notthefallintact-Llama-2-7b-hf-trained-layer_21-L-PCC-1-20}
\end{center}
\vskip -0.2in
\end{figure}

\begin{figure}[H]
\vskip 0.2in
\begin{center}
\centerline{\includegraphics[width=\columnwidth/13*10]{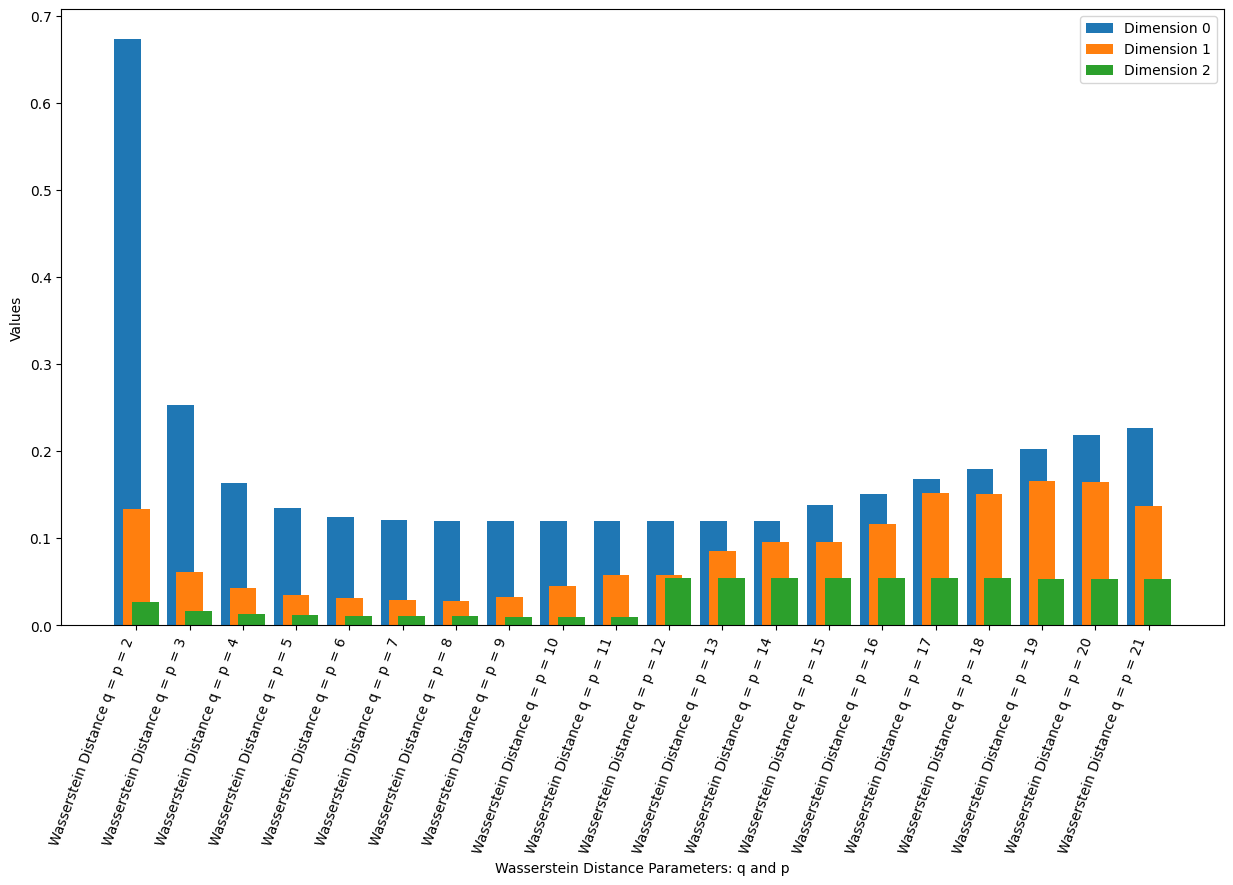}}
\caption{The figure depicts the $q$-Wasserstein Distances with parameters $q$ and $p$ spanning from $q=p=2$ to $q=p=21$.}
\label{notthefallintact-Llama-2-7b-hf-trained-layer_21-L-PCC-2-21}
\end{center}
\vskip -0.2in
\end{figure}

\begin{figure}[H]
\vskip 0.2in
\begin{center}
\centerline{\includegraphics[width=\columnwidth/13*10]{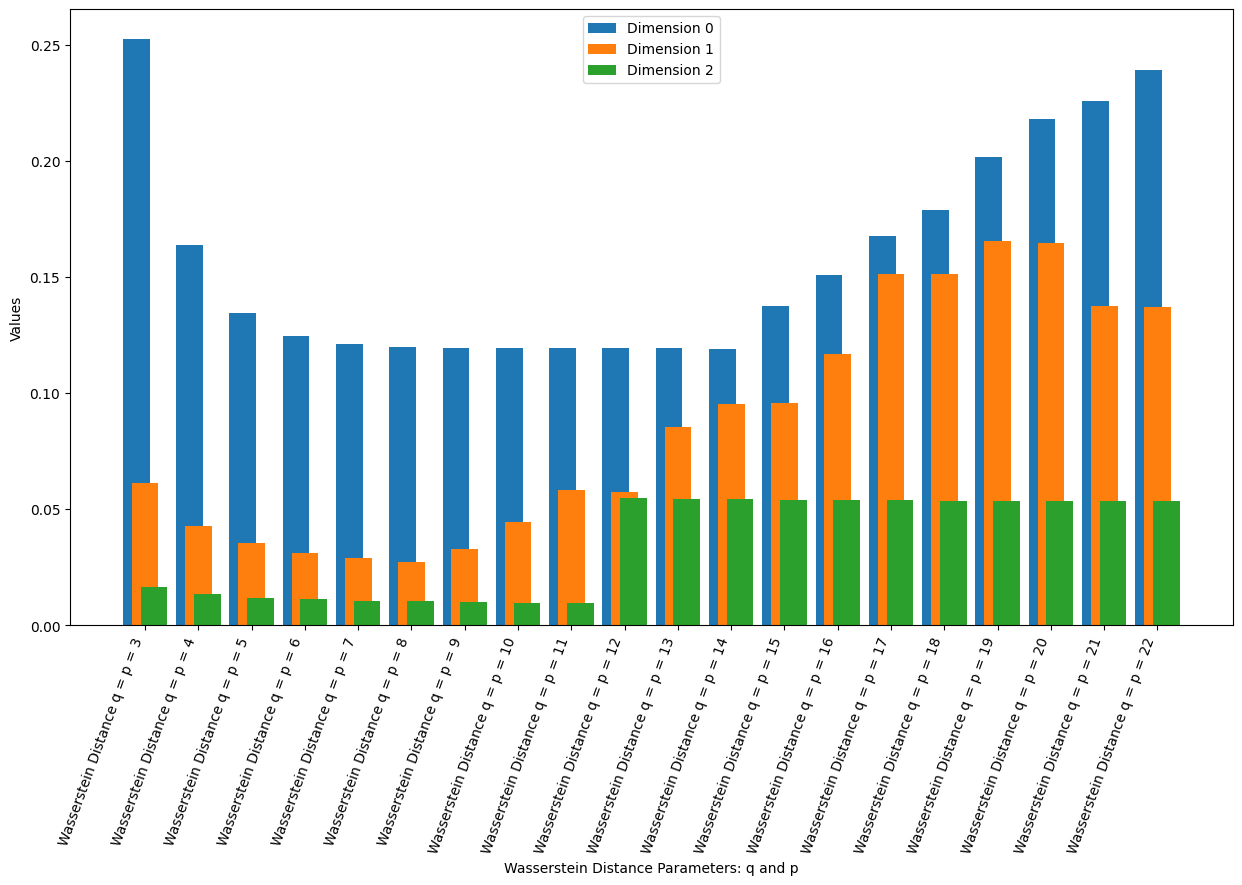}}
\caption{The figure depicts the $q$-Wasserstein Distances with parameters $q$ and $p$ spanning from $q=p=3$ to $q=p=22$.}
\label{notthefallintact-Llama-2-7b-hf-trained-layer_21-L-PCC-3-22}
\end{center}
\vskip -0.2in
\end{figure}

\begin{figure}[H]
\vskip 0.2in
\begin{center}
\centerline{\includegraphics[width=\columnwidth/13*10]{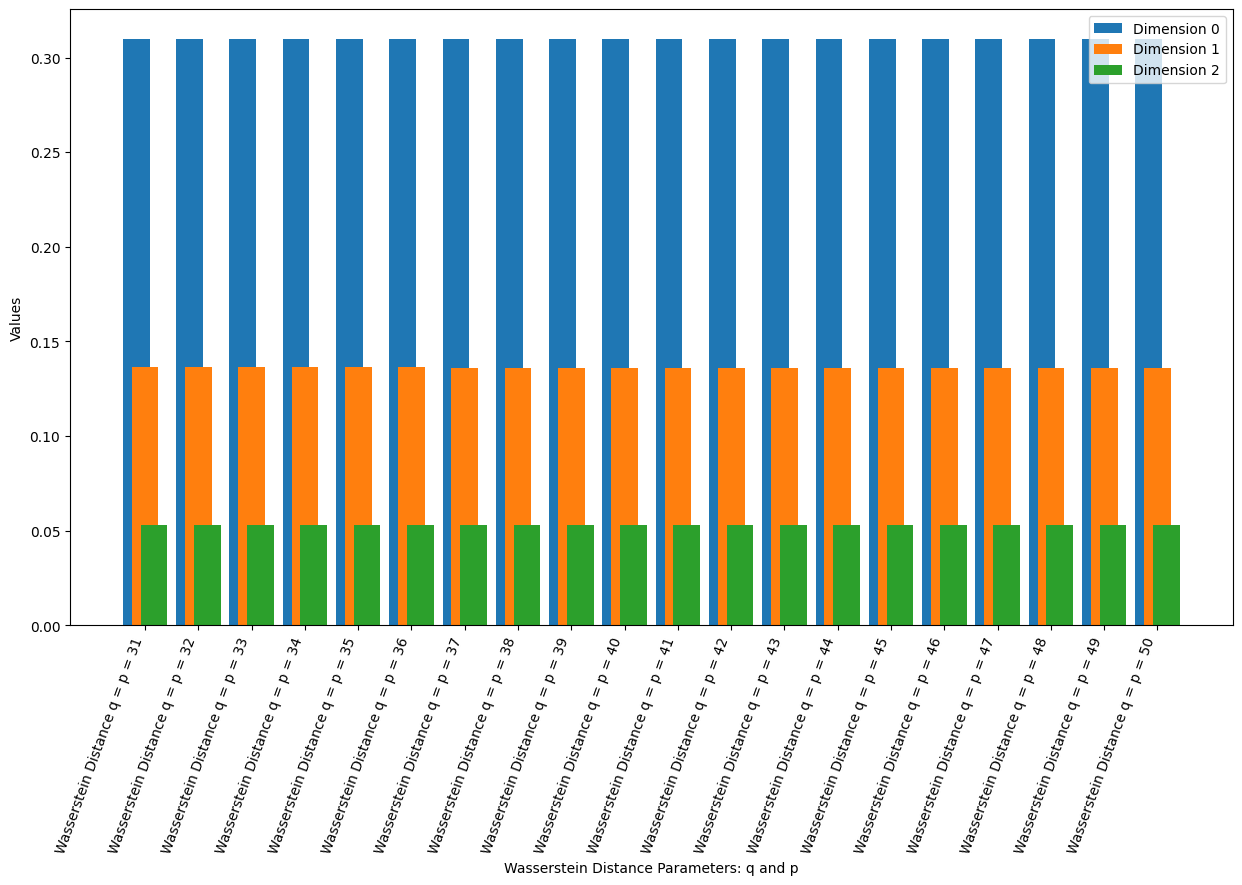}}
\caption{The figure depicts the $q$-Wasserstein Distances with parameters $q$ and $p$ spanning from $q=p=31$ to $q=p=50$.}
\label{notthefallintact-Llama-2-7b-hf-trained-layer_21-L-PCC-31-50}
\end{center}
\vskip -0.2in
\end{figure}

\begin{figure}[H]
\vskip 0.2in
\begin{center}
\centerline{\includegraphics[width=\columnwidth/13*10]{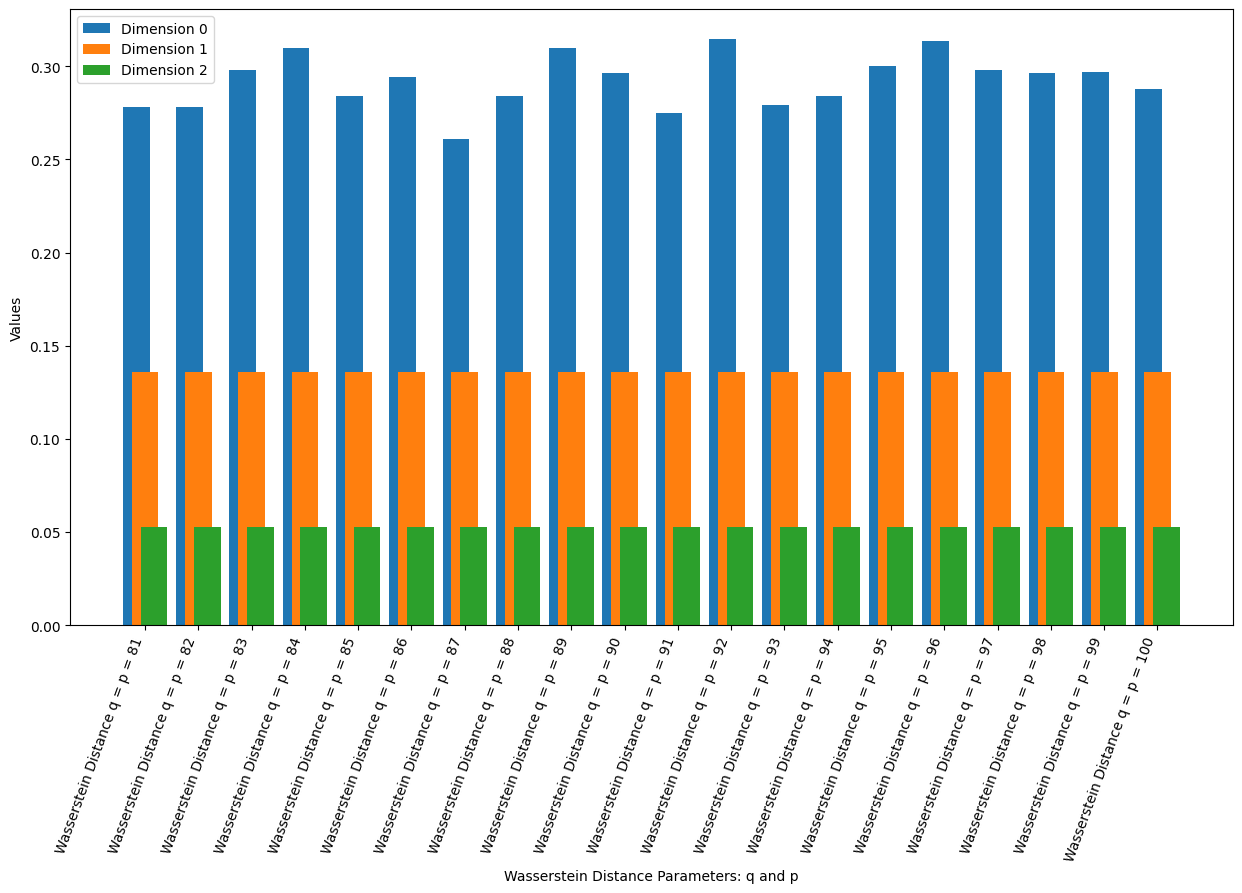}}
\caption{The figure depicts the $q$-Wasserstein Distances with parameters $q$ and $p$ spanning from $q=p=81$ to $q=p=100$.}
\label{notthefallintact-Llama-2-7b-hf-trained-layer_21-L-PCC-81-100}
\end{center}
\vskip -0.2in
\end{figure}

\begin{figure}[H]
\vskip 0.2in
\begin{center}
\centerline{\includegraphics[width=\columnwidth/13*10]{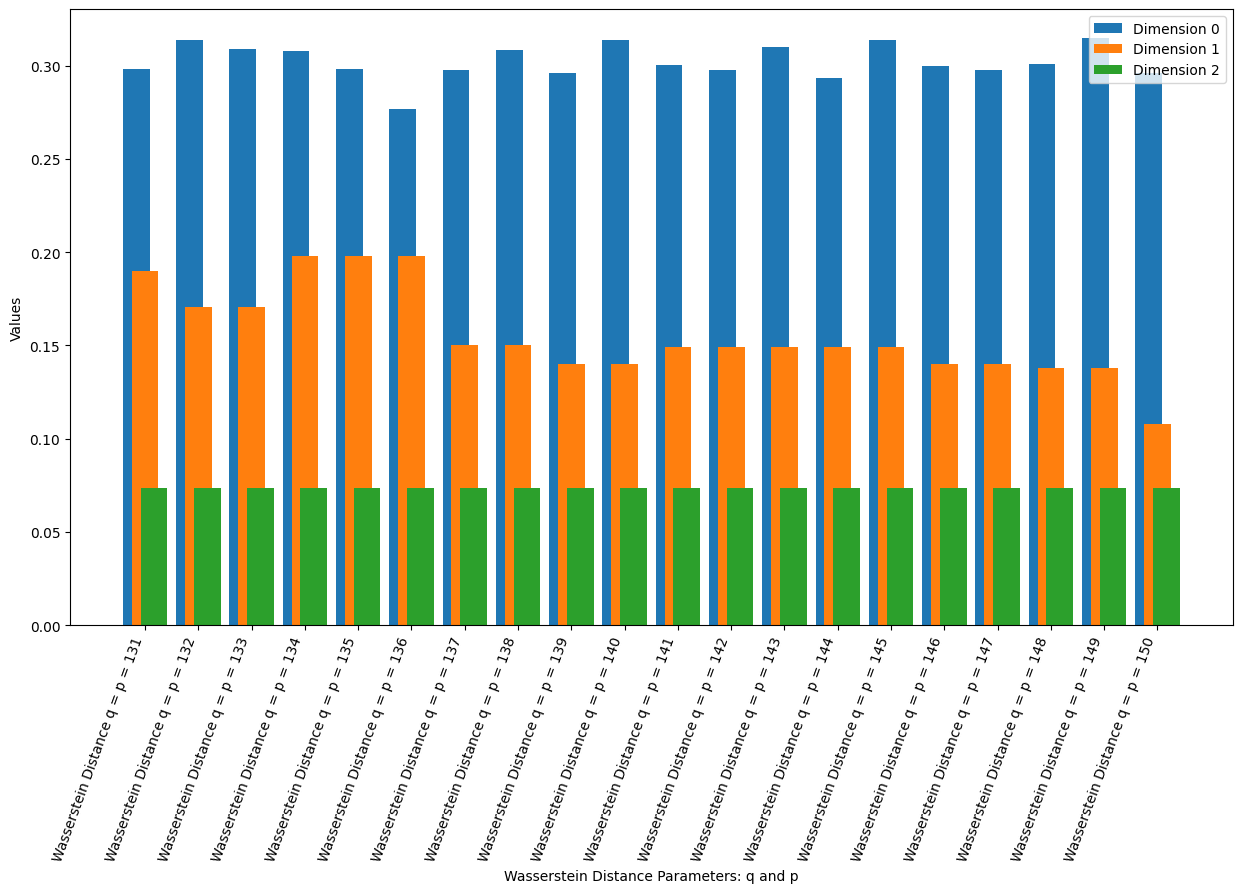}}
\caption{The figure depicts the $q$-Wasserstein Distances with parameters $q$ and $p$ spanning from $q=p=131$ to $q=p=150$.}
\label{notthefallintact-Llama-2-7b-hf-trained-layer_21-L-PCC-131-150}
\end{center}
\vskip -0.2in
\end{figure}

\begin{figure}[H]
\vskip 0.2in
\begin{center}
\centerline{\includegraphics[width=\columnwidth/13*10]{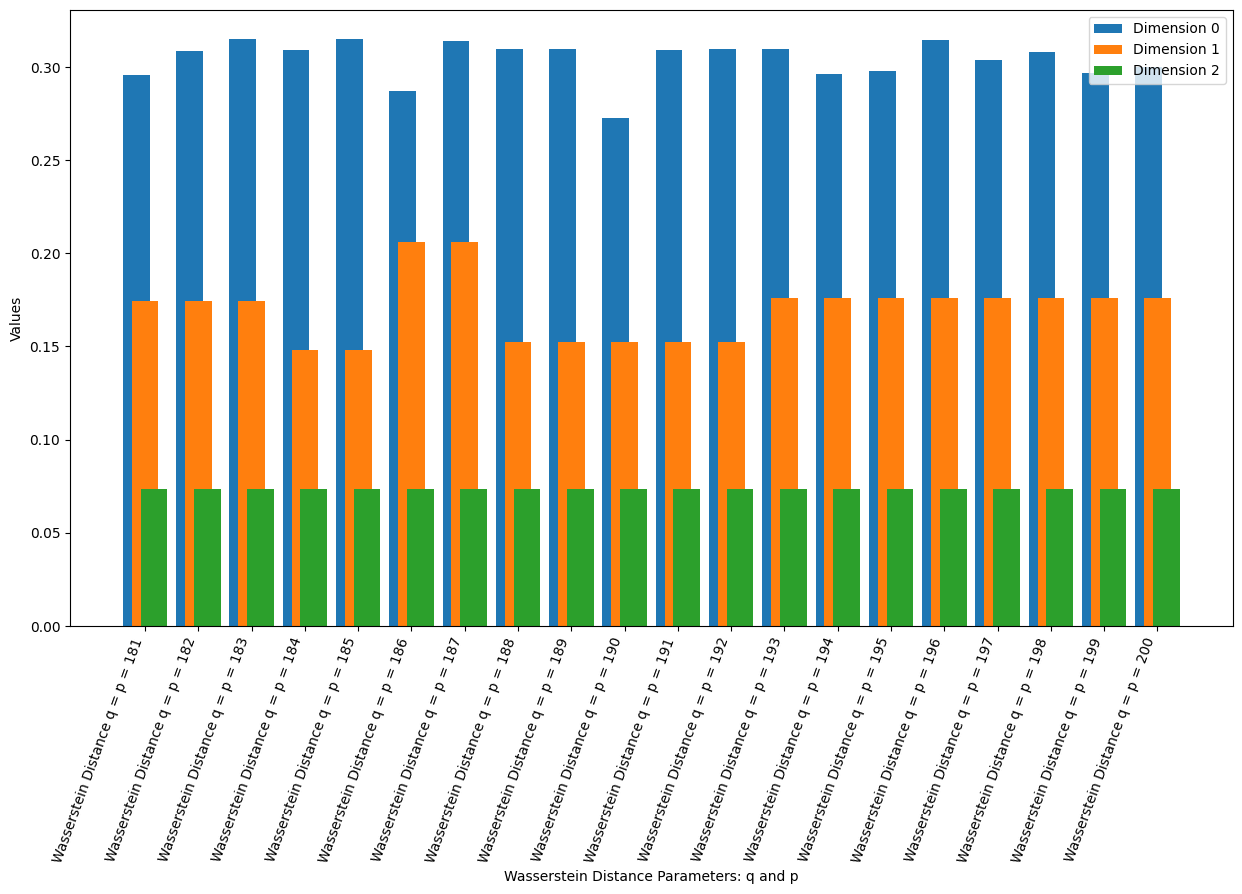}}
\caption{The figure depicts the $q$-Wasserstein Distances with parameters $q$ and $p$ spanning from $q=p=181$ to $q=p=200$.}
\label{notthefallintact-Llama-2-7b-hf-trained-layer_21-L-PCC-181-200}
\end{center}
\vskip -0.2in
\end{figure}

\begin{figure}[H]
\vskip 0.2in
\begin{center}
\centerline{\includegraphics[width=\columnwidth/13*10]{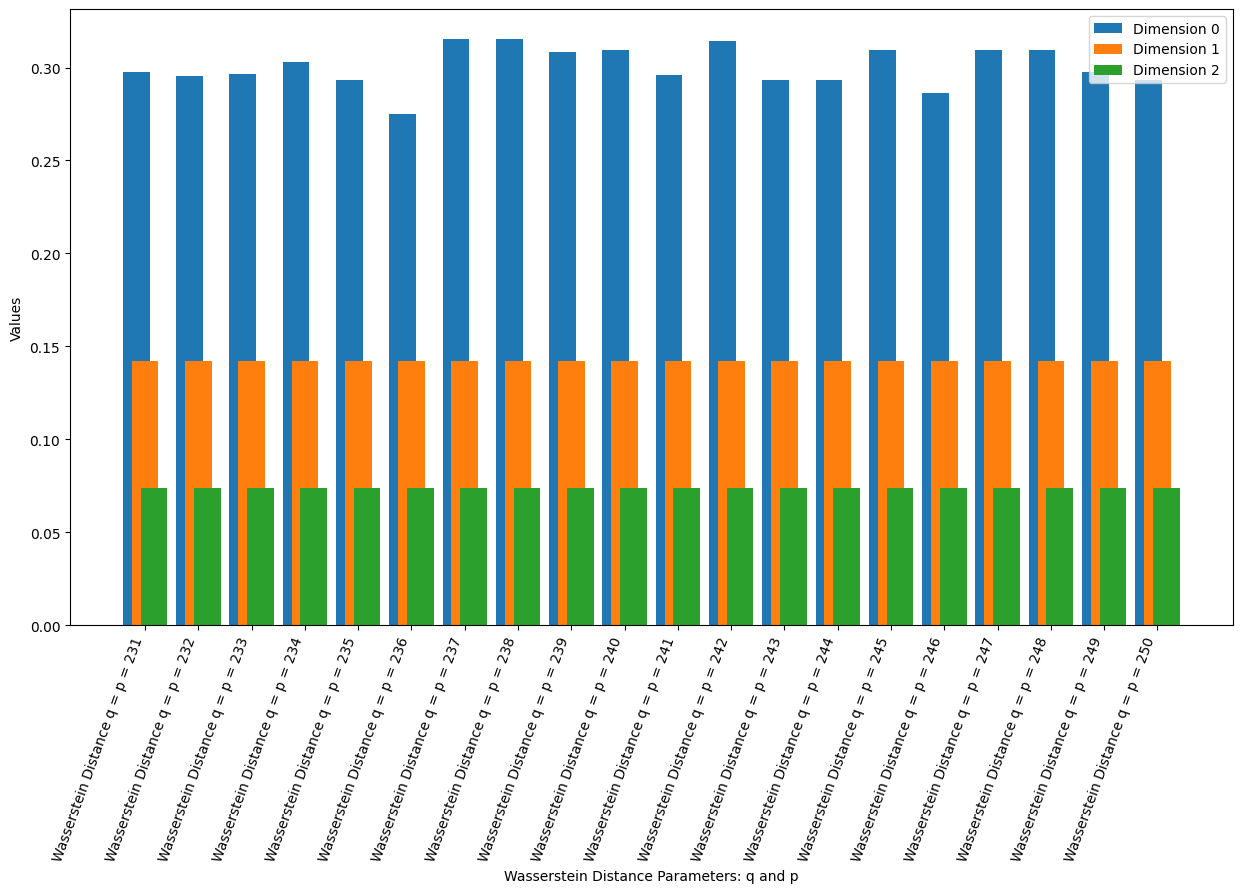}}
\caption{The figure depicts the $q$-Wasserstein Distances with parameters $q$ and $p$ spanning from $q=p=231$ to $q=p=250$.}
\label{notthefallintact-Llama-2-7b-hf-trained-layer_21-L-PCC-231-50}
\end{center}
\vskip -0.2in
\end{figure}

\begin{figure}[H]
\vskip 0.2in
\begin{center}
\centerline{\includegraphics[width=\columnwidth/13*10]{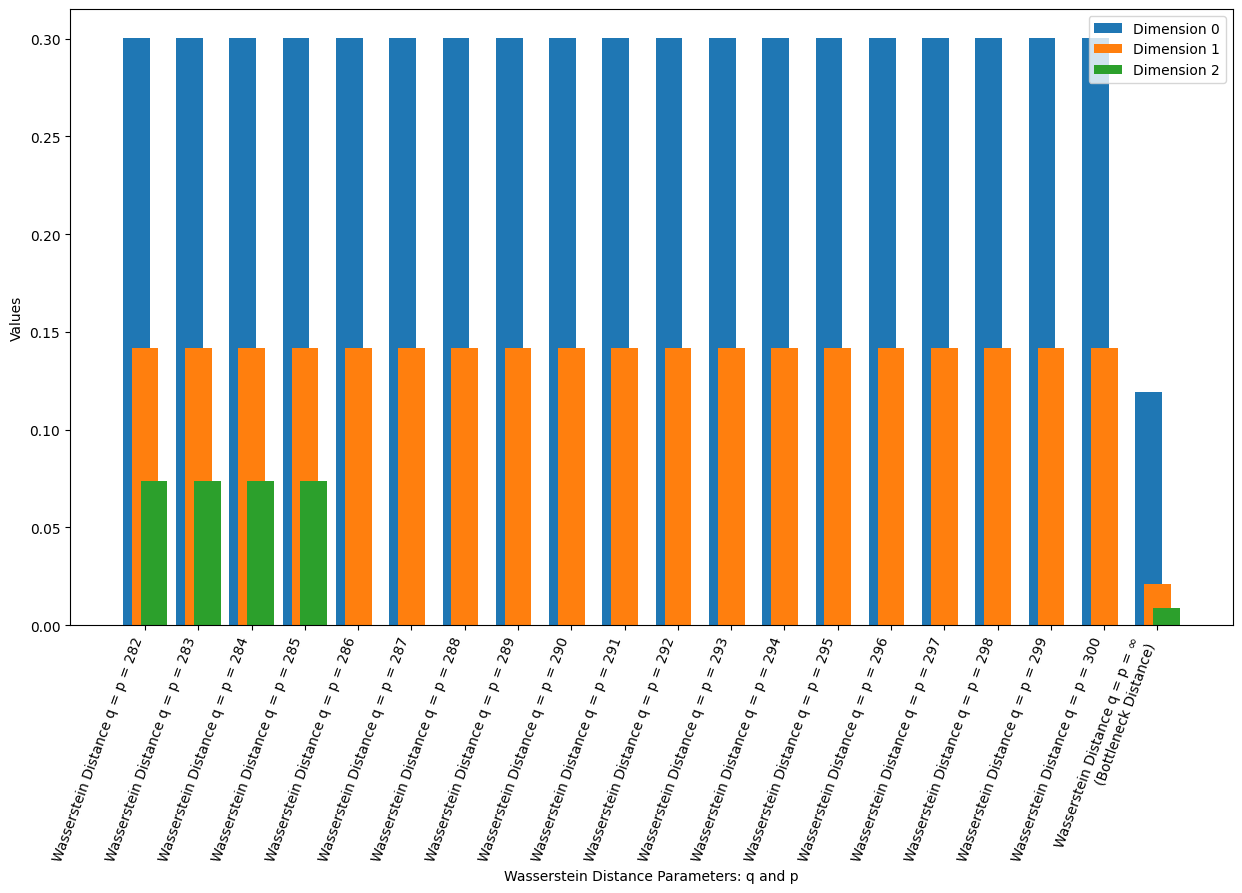}}
\caption{The figure presents the $q$-Wasserstein Distances with parameters $q$ and $p$ varying from $q=p=282$ to $q=p=\infty$. In this specific instance, the distance is a.k.a Bottleneck Distance.}
\label{notthefallintact-Llama-2-7b-hf-trained-layer_21-L-PCC-282-301}
\end{center}
\vskip -0.2in
\end{figure}

Each of the 34,416 combination pairs is characterized by 903 features, representing the $q$-Wasserstein Distances, where $q = p$, ranging from 1 to 300, along with the special case $q = p = \infty$, across 3 PH dimensions. However, not all features are deemed reliable and valid; hence, a systematic filtration process is essential. Therefore firstly, we conducted an Exploratory Data Analysis (EDA) on the constructed features. The observed trend indicated a consistent "long tail" distribution followed by the $q$-Wasserstein Distances spanning from $q=p=1$ to $q=p=300$ and $q=p=\infty$ for each pair in the 34,416 combinations, with minor fluctuations in the tail (refer to \cref{notthefallintact-Llama-2-7b-hf-trained-layer_21-L-PCC-1-20} to \cref{notthefallintact-Llama-2-7b-hf-trained-layer_21-L-PCC-282-301} in \cref{Exploratory Data Analysis on Constructed Features: q-Wasserstein Distances}). Formally, the $q$-Wasserstein Distance is bounded by the Wasserstein Stability Theorem \citep{edelsbrunner2010computational, cohen_steiner_edelsbrunner_harer_yuriy_mileyko_2010}.

We devised a two-pass process to systematically identify reliable and valid features capable of elucidating a substantial portion of the variability present in the brainscores while demonstrating statistical significance as well.

\subsubsection{First Pass}
\label{First Pass}

\begin{figure}[H]
\vskip 0.2in
\begin{center}
\centerline{\includegraphics[width=\columnwidth]{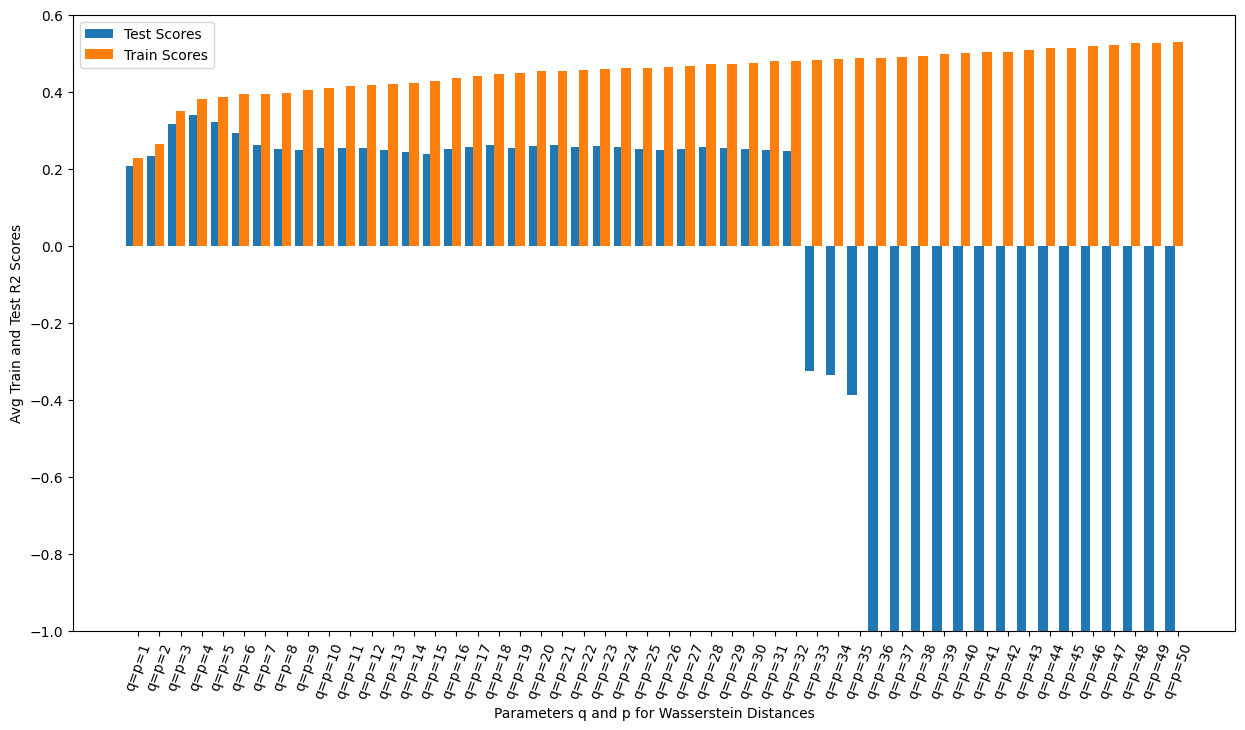}}
\caption{The figure depicts the averaged train and test $R^2$ scores acquired during the learning process of the Linear Regression model to correlate with the respective brainscores using "Only Trained" LLMs. The train scores exhibit a gradual increase as additional $q$-Wasserstein Distances are incorporated progressively. Conversely, the test scores deviate from the pattern, reaching a peak at $q=p=4$ before sharply declining, with a significant downturn observed beyond $q=p=35$ (we just hide those very large negative test scores below $-1$, refer to \cref{train test R2} and \cref{only test R2} for details). It is noteworthy that the trend observed in the test scores is not exclusive to "Only Trained" LLMs; it is pervasive across "Trained plus Untrained" LLMs, as well as both the L(eft) and R(ight) hemispheres and all ROIs. The critical distinction lies in the specific $q$ values where the test score peaks and subsequently descends into negative territory.}
\label{qs_train_and_test_R2_scores}
\end{center}
\vskip -0.2in
\end{figure}

In the First Pass, we incrementally introduced in features from $q=p=1$ to $q=p=300$ \footnote{The special case where $q=p=\infty$ is consistently included in each iteration, as it lies outside the $[1,300]$ range.} given the observed distribution of our constructed ones. At each iteration, we implemented a 5-time repeated 5-fold cross-validation (CV) pipeline to train the Linear Regression Model. We assessed the model's performance using the averaged test $R^2$ score across the 25 outcomes to gauge the extent to which our learned model could elucidate variation \citep{chicco2021coefficient}. The regular $R^2$ score falls within the [0,1] range (higher is better in our context), while negative values typically indicates overfitting \citep{nau_2019} (refer to \cref{qs_train_and_test_R2_scores}, \cref{train test R2}, and \cref{only test R2} in \cref{Filter in Reliable and Valid Features: First Pass} for an illustrative example). Therefore, we employed a stopping criterion triggered by encountering a negative averaged test $R^2$ score. Prior to that, we recorded all $q$ values along with their corresponding averaged test $R^2$ scores, and ultimately the $q$ value yielding the highest averaged test $R^2$ score was determined.

Formally, we established a mapping $\phi: q \to s$ where $q \in [1, 300] \cup \{+\infty\}$ denotes the search space, and $s \in [0, 1]$ represents the corresponding averaged test $R^2$ score for each iteration. The $q$ value is determined as $q_{\max} = \arg\max \phi$. (See the pseudocode for the First Pass in \cref{first pass for q}).

\begin{algorithm}[H]
   \caption{$q_{\max}$ on Highest Averaged Test $R^2$ Score}
   \label{first pass for q}
\begin{algorithmic}
   \STATE Initialize $\phi$-list $= []$
   \FOR{$q=1$ {\bfseries to} $300$}
   \STATE X = featuresContainingMax$q$WassersteinDistances \\ PlusBottleneckDistance
   \STATE Y = brainscores
   \STATE regressor = linearRegressor()
   \STATE cvModel = crossValidate(regressor, X, Y, scoring = ("r2"))
   \IF{mean(cvModel["test-score"]) $\le 0$}
   \STATE break
   \ENDIF
   \STATE $\phi$-list.append((q, mean(cvModel["test-score"])))
   
   \ENDFOR
   \STATE $q_{\max} = \arg\max$ ($\phi$-list)
\end{algorithmic}
\end{algorithm}

\subsubsection{Second Pass}
\label{Second Pass}

Determining the optimal $q$ alone does not suffice to identify truly reliable and valid features from our constructed multicollinear ones. Subsequently, we designated the $p$-value \citep{thiese2016p, bzovsky2022clinician} with a threshold of $p < 5\%$ to selectively retain desired features given the determined $q$ value. Specifically, we computed the $p$-value for each iteration, averaged the 25 $p$-values corresponding to each feature, and ultimately retained those with $p < 5\%$.

\section{Summaries for the Three Selected Tasks from Narratives Dataset}
\label{Summaries for the 3 Selected Tasks from Narratives Dataset}

This section provides a concise overview of our three tasks selected from the Narratives dataset \citep{nastase2021narratives}.

It is important to acknowledge certain discrepancies observed in the tables between \cref{narratives-3-tasks} and those presented in \citet{nastase2021narratives}. Here, we present the precise values for "TRs," "Valid TRs", "Subjects", and "Valid Subjects" based on the repository from OpenNeuro (\url{https://openneuro.org/datasets/ds002345/versions/1.1.4}).

\begin{table}[H]
\caption{Summary for our three selected tasks from Narratives dataset \citep{nastase2021narratives}.}
\label{narratives-3-tasks}
\vskip 0.15in
\begin{center}
\begin{small}
\begin{sc}
\begin{tabular}{lccccccc}
\toprule
Story & Duration & TRs & Words & Subjects & Onset & Valid & Valid \\
 &  &  &  &  &  & TRs & Subjects \\
\midrule
"Pie Man"    & 07:02 &  300 & 957 & 82 & 0 & 300 & 75 \\
"Shapes"    & 06:45 &  313 & 910 & 59 & 3 & 310 & 58 \\
"It’s Not the Fall     & 09:07 &  400 & 1,601 & 56 & 3 & 397 & 54 \\
That Gets You"    &  &   &  &  &  &  &  \\
\bottomrule
\end{tabular}
\end{sc}
\end{small}
\end{center}
\vskip -0.1in
\end{table}

\section{Averaged fMRI for each Task-ROI Pair}
\label{averaged-fMRI-for-each-task-roi-pair}

This section presents the averaged fMRI data for each Task-ROI Pair, derived from varying numbers of subjects engaged in the three tasks: "Pie Man," "Shapes," and "It’s Not the Fall That Gets You," as documented in \citet{nastase2021narratives}. Summaries are available in \cref{narratives-3-tasks}. Here, we display all eight language-related ROIs along with the whole brain mask (refer to \cref{Extracting Data Representations from human fMRI} for specifics). Initially, for each subject's fMRI measurements across all valid TRs, we computed the average across TRs, resulting in 40,962 voxel-level blood-oxygen-level-dependent (BOLD) signal values. Subsequently, we calculated the mean across subjects participating in each task to obtain the final set of 40,962 values. Each figure comprises 12 sub-figures representing six viewing perspectives—lateral, medial, dorsal, ventral, anterior, posterior—for each hemisphere: (L)eft and (R)ight. The colorbar on the right side of each figure illustrates the BOLD value ranges, while the task-ROI information is denoted in the title of each one.

Please note that the averaging across TRs for each figure is solely for visualization purposes; we did not employ this step when calculating the "brainscores" based on \citet{caucheteux2023evidence} or processing fMRI data representations in the subsequent data processing pipelines.

\begin{figure}[H]
\vskip 0.2in
\begin{center}
\centerline{\includegraphics[width=\columnwidth]{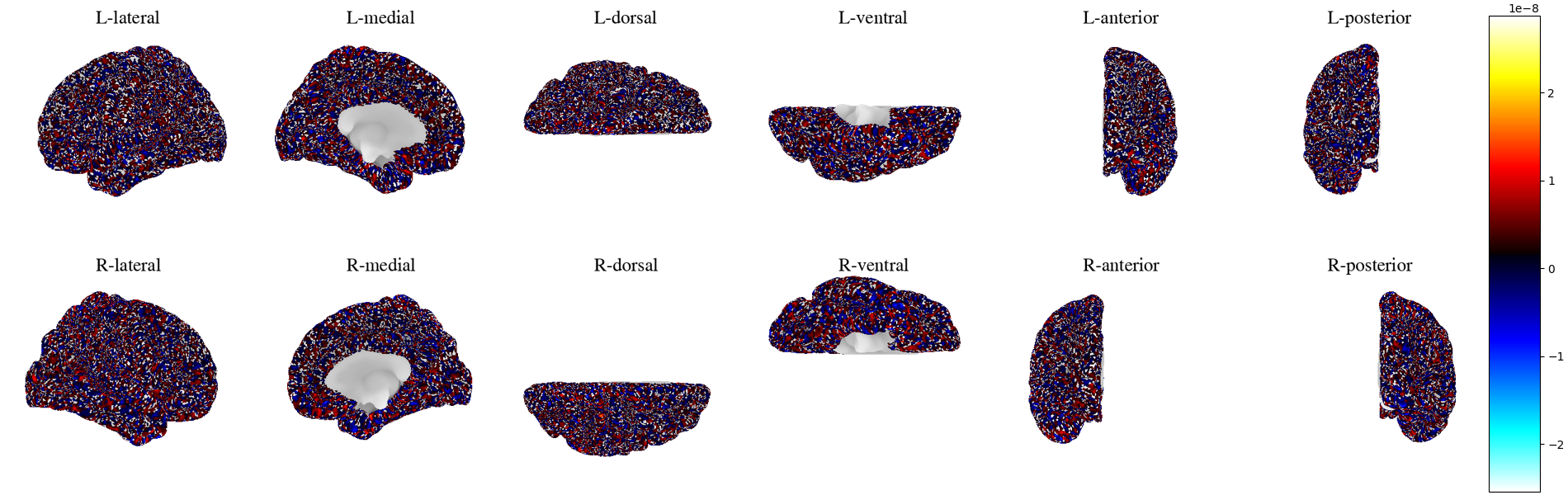}}
\caption{The averaged fMRI for the whole brain mask for the task: "Pie Man"}
\label{appendix-pieman-Evidence-pial}
\end{center}
\vskip -0.2in
\end{figure}

\begin{figure}[H]
\vskip 0.2in
\begin{center}
\centerline{\includegraphics[width=\columnwidth]{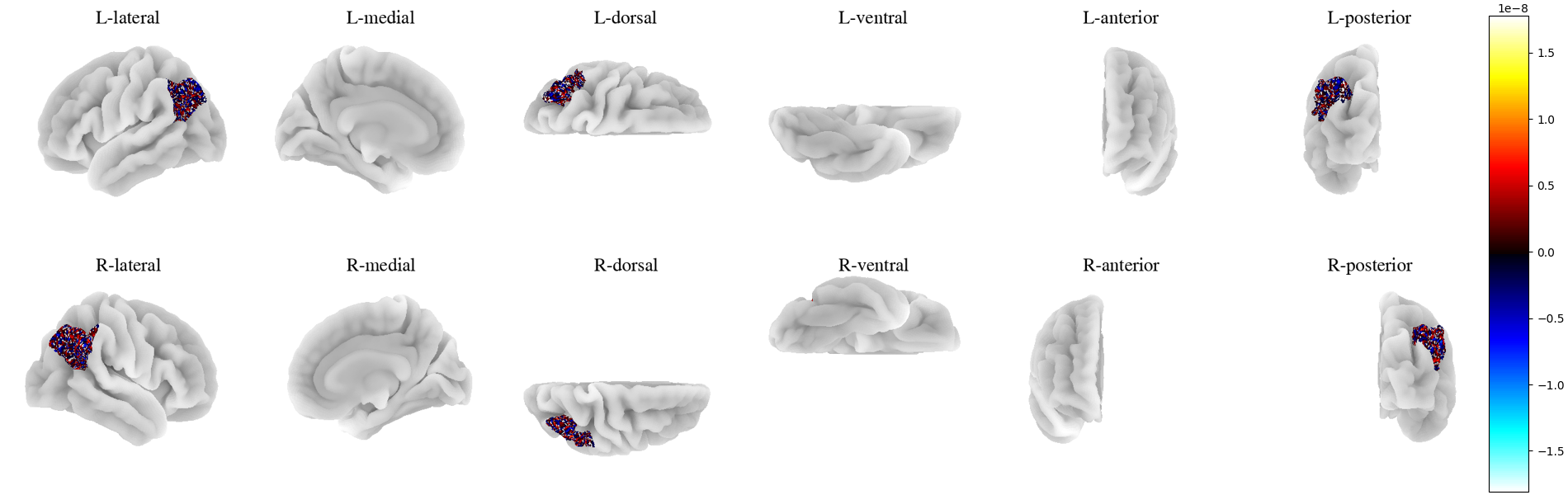}}
\caption{The averaged fMRI for the ROI: AG for the task: "Pie Man"}
\label{appendix-pieman-AG-pial}
\end{center}
\vskip -0.2in
\end{figure}

\begin{figure}[H]
\vskip 0.2in
\begin{center}
\centerline{\includegraphics[width=\columnwidth]{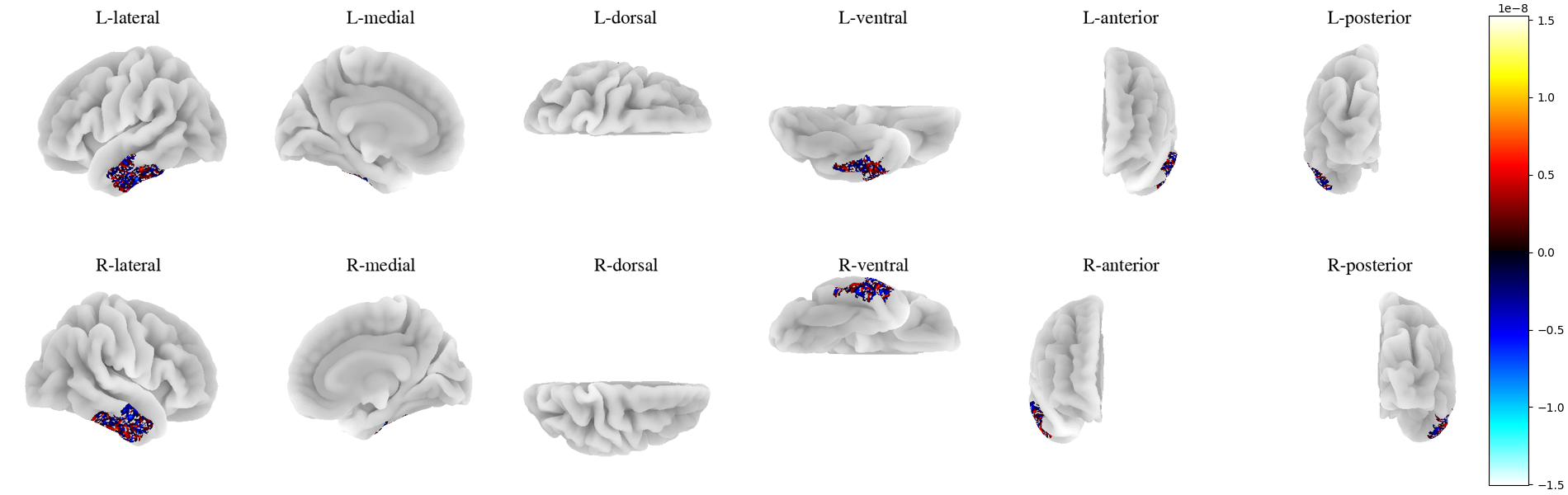}}
\caption{The averaged fMRI for the ROI: ATL for the task: "Pie Man"}
\label{appendix-pieman-ATL-pial}
\end{center}
\vskip -0.2in
\end{figure}

\begin{figure}[H]
\vskip 0.2in
\begin{center}
\centerline{\includegraphics[width=\columnwidth]{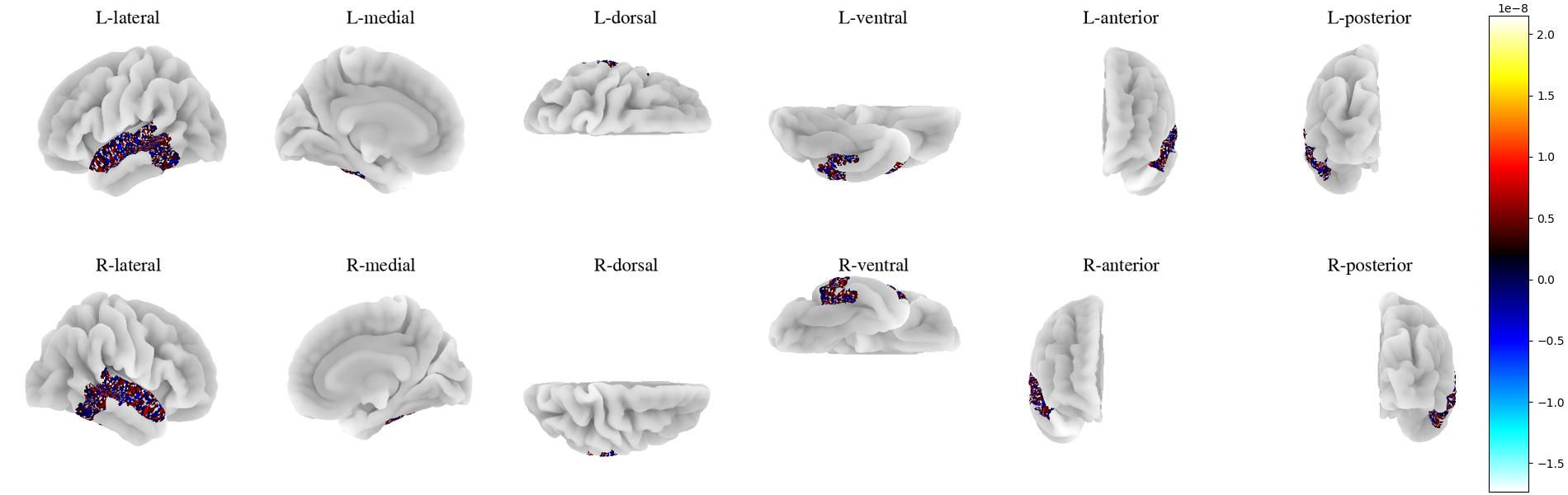}}
\caption{The averaged fMRI for the ROI: PTL for the task: "Pie Man"}
\label{appendix-pieman-PTL-pial}
\end{center}
\vskip -0.2in
\end{figure}

\begin{figure}[H]
\vskip 0.2in
\begin{center}
\centerline{\includegraphics[width=\columnwidth]{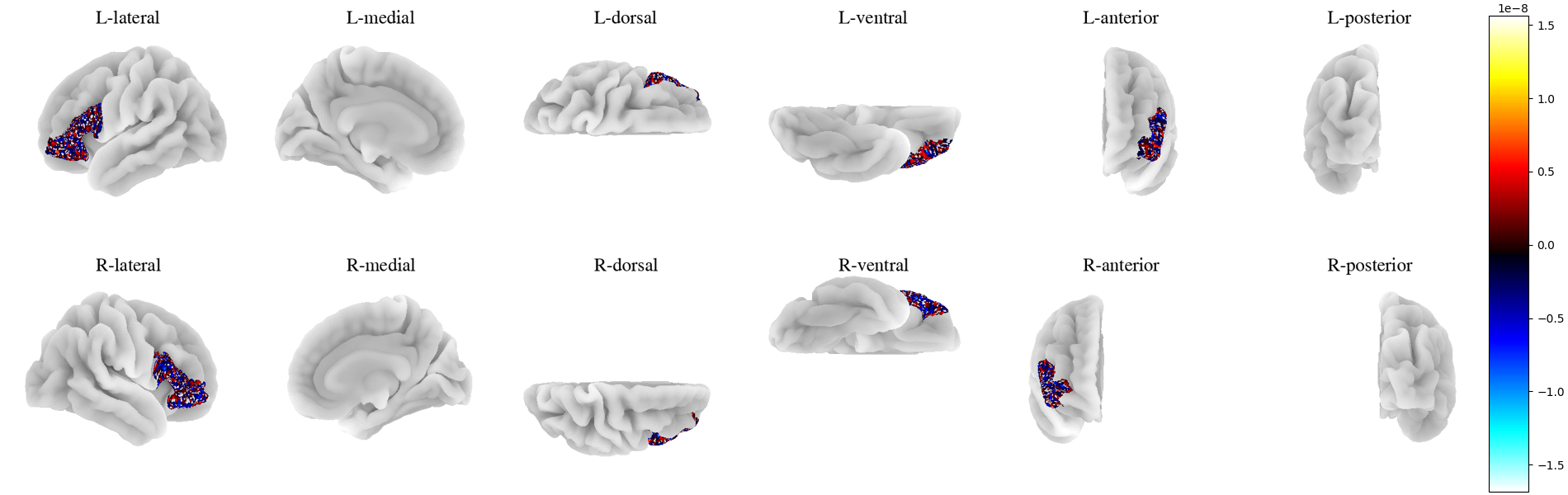}}
\caption{The averaged fMRI for the ROI: IFG for the task: "Pie Man"}
\label{appendix-pieman-IFG-pial}
\end{center}
\vskip -0.2in
\end{figure}

\begin{figure}[H]
\vskip 0.2in
\begin{center}
\centerline{\includegraphics[width=\columnwidth]{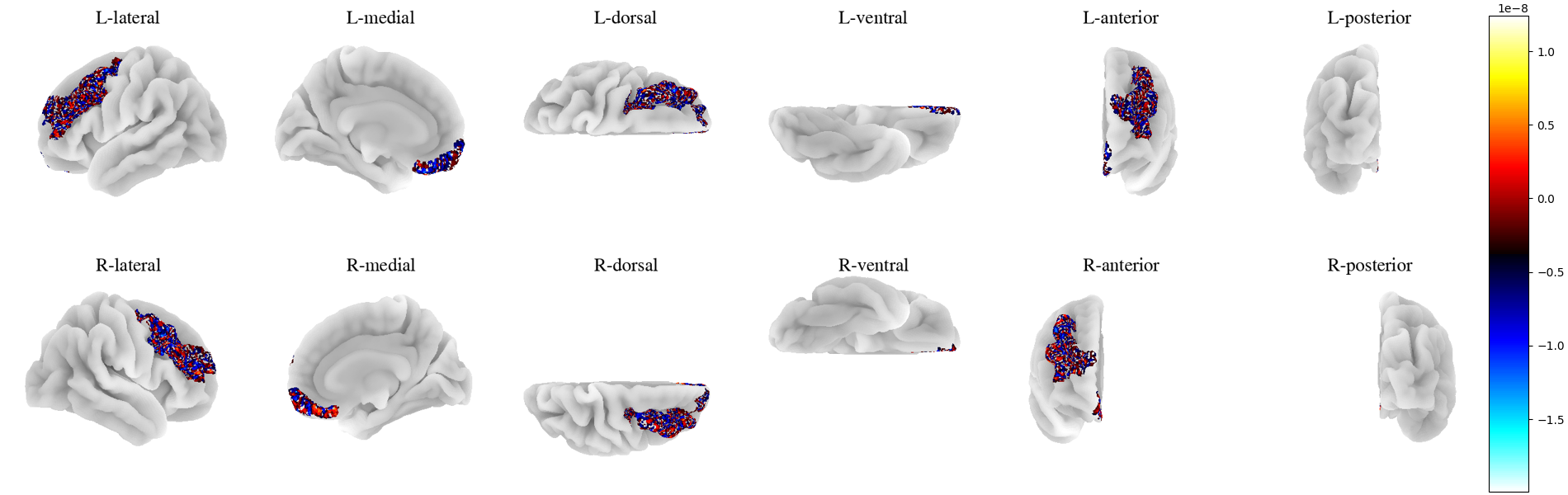}}
\caption{The averaged fMRI for the ROI: MFG for the task: "Pie Man"}
\label{appendix-pieman-MFG-pial}
\end{center}
\vskip -0.2in
\end{figure}

\begin{figure}[H]
\vskip 0.2in
\begin{center}
\centerline{\includegraphics[width=\columnwidth]{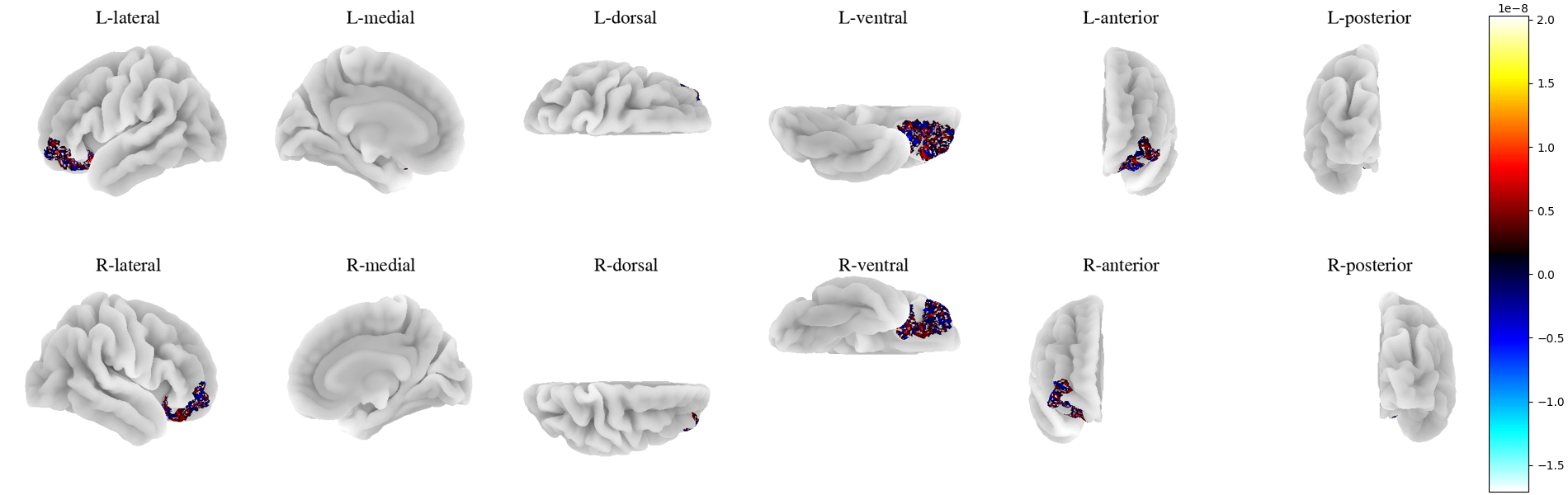}}
\caption{The averaged fMRI for the ROI: IFGorb for the task: "Pie Man"}
\label{appendix-pieman-IFGorb-pial}
\end{center}
\vskip -0.2in
\end{figure}

\begin{figure}[H]
\vskip 0.2in
\begin{center}
\centerline{\includegraphics[width=\columnwidth]{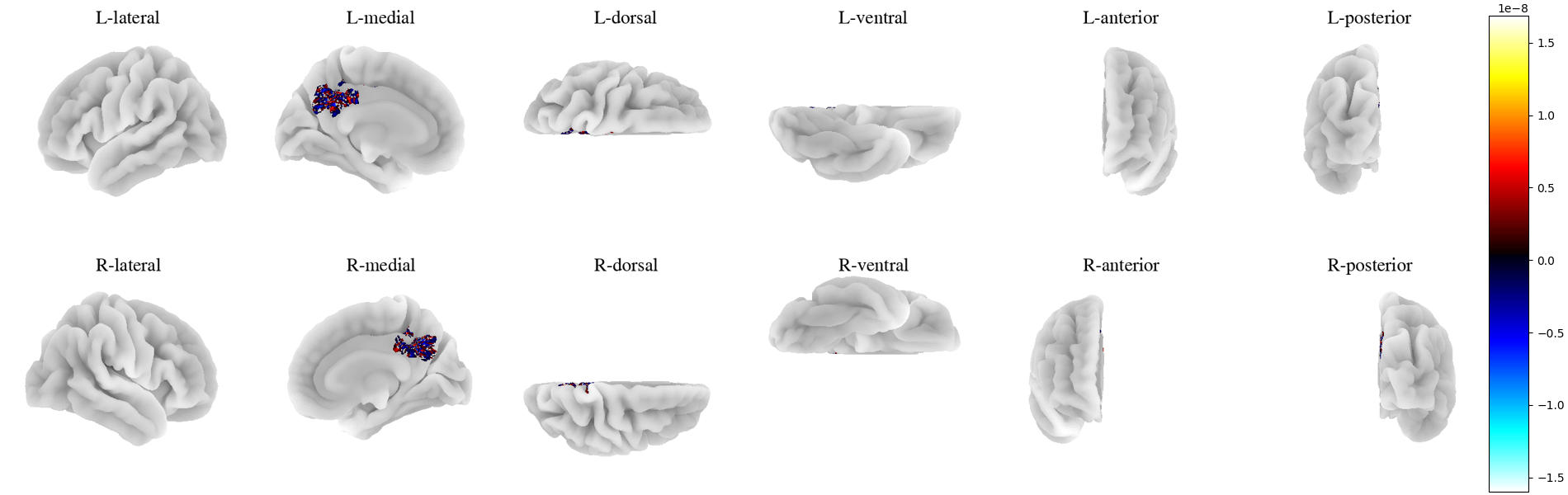}}
\caption{The averaged fMRI for the ROI: PCC for the task: "Pie Man"}
\label{appendix-pieman-PCC-pial}
\end{center}
\vskip -0.2in
\end{figure}

\begin{figure}[H]
\vskip 0.2in
\begin{center}
\centerline{\includegraphics[width=\columnwidth]{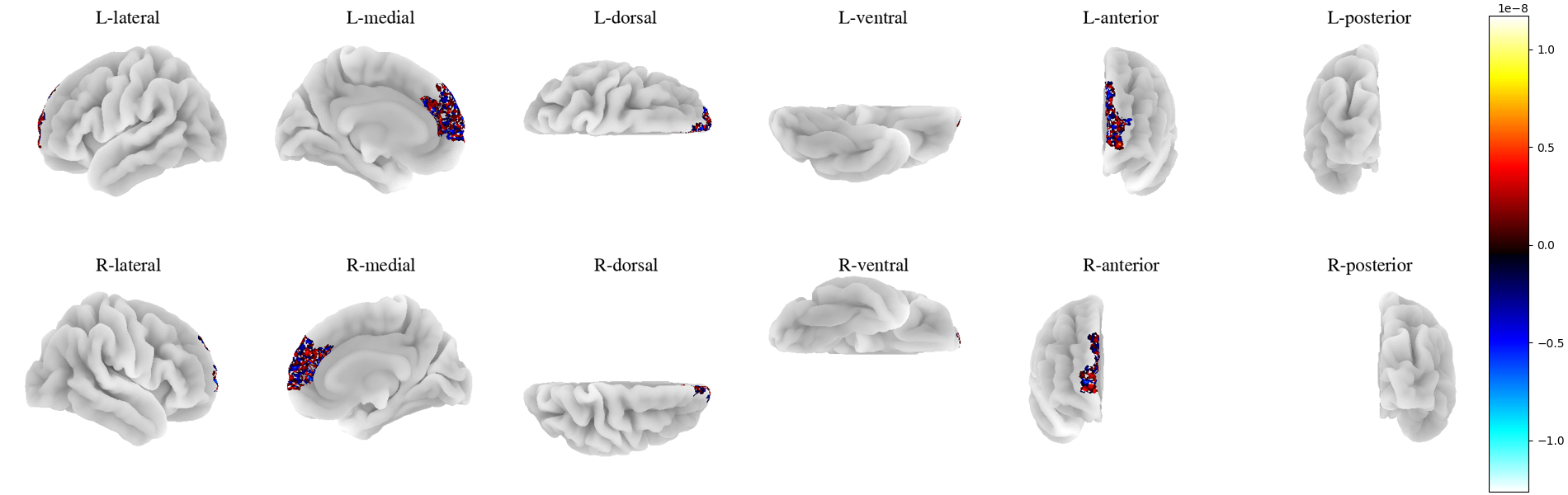}}
\caption{The averaged fMRI for the ROI: dmPFC for the task: "Pie Man"}
\label{appendix-pieman-dmPFC-pial}
\end{center}
\vskip -0.2in
\end{figure}

\begin{figure}[H]
\vskip 0.2in
\begin{center}
\centerline{\includegraphics[width=\columnwidth]{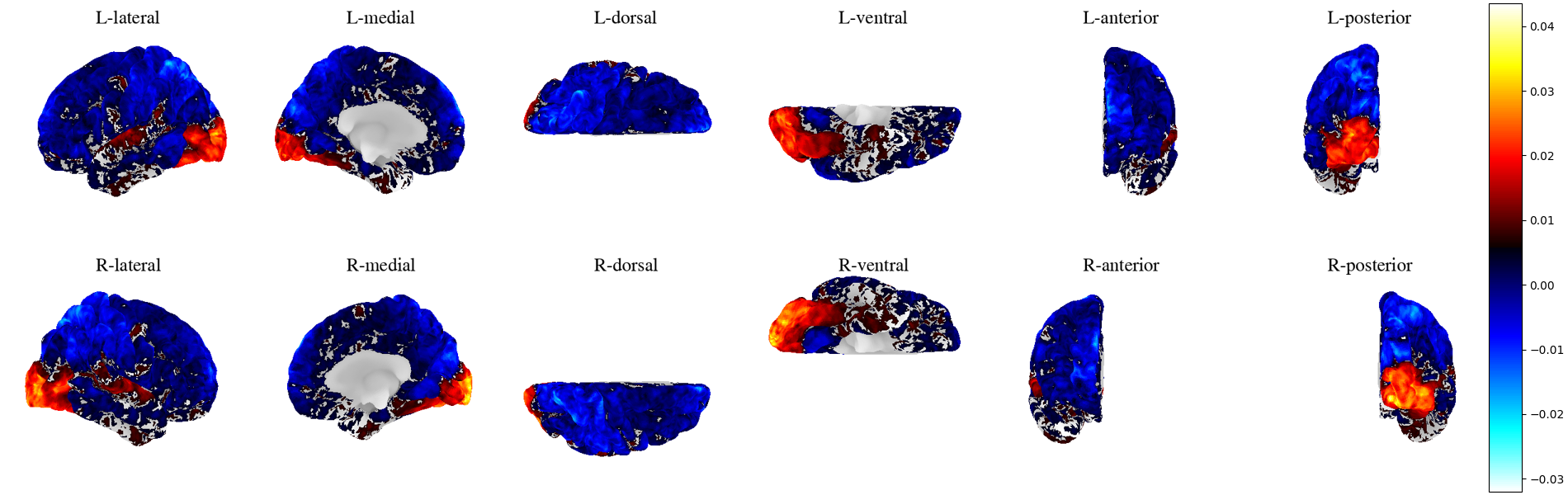}}
\caption{The averaged fMRI for the whole brain mask for the task: "Shapes"}
\label{appendix-shapessocial-Evidence-pial}
\end{center}
\vskip -0.2in
\end{figure}

\begin{figure}[H]
\vskip 0.2in
\begin{center}
\centerline{\includegraphics[width=\columnwidth]{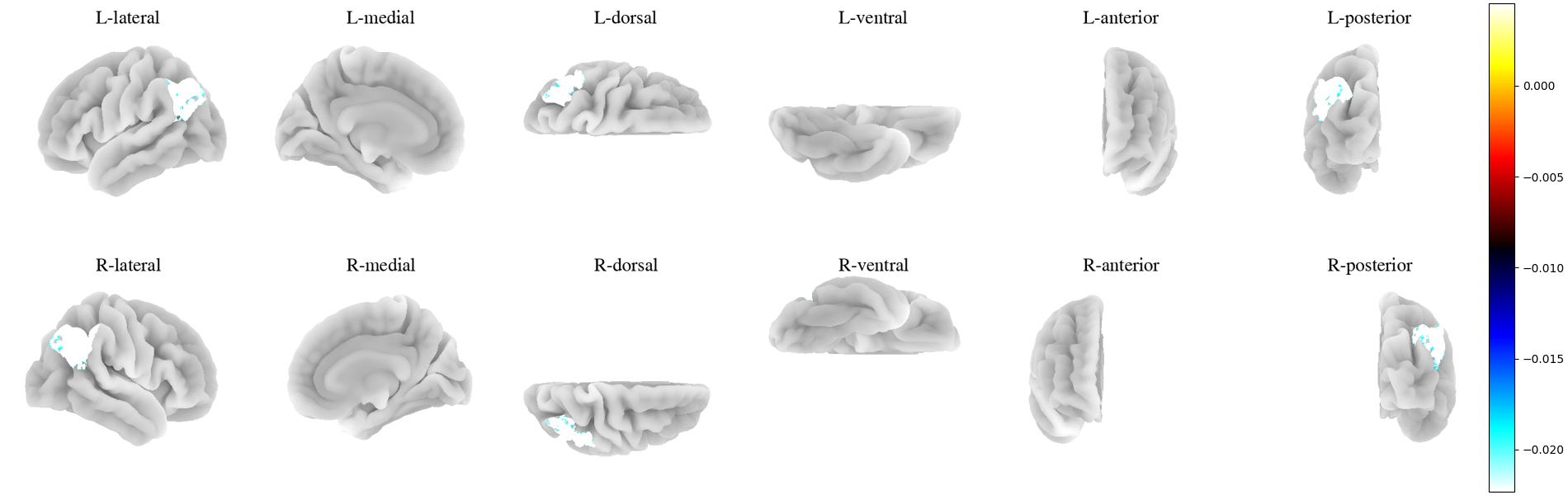}}
\caption{The averaged fMRI for the ROI: AG for the task: "Shapes"}
\label{appendix-shapessocial-AG-pial}
\end{center}
\vskip -0.2in
\end{figure}

\begin{figure}[H]
\vskip 0.2in
\begin{center}
\centerline{\includegraphics[width=\columnwidth]{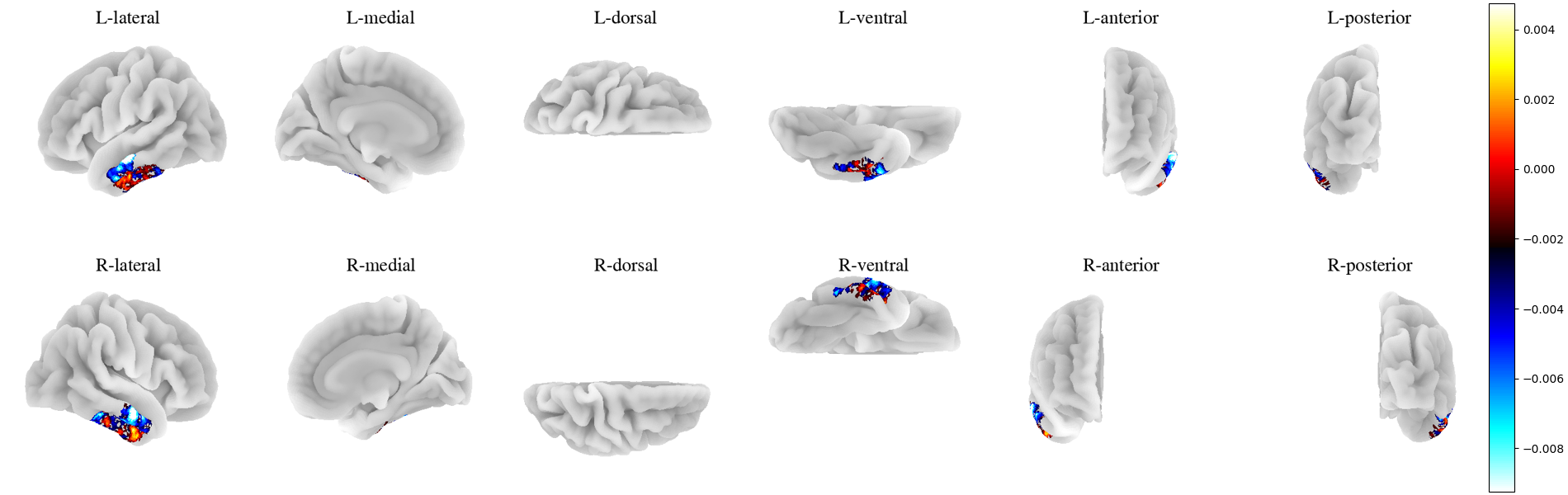}}
\caption{The averaged fMRI for the ROI: ATL for the task: "Shapes"}
\label{appendix-shapessocial-ATL-pial}
\end{center}
\vskip -0.2in
\end{figure}

\begin{figure}[H]
\vskip 0.2in
\begin{center}
\centerline{\includegraphics[width=\columnwidth]{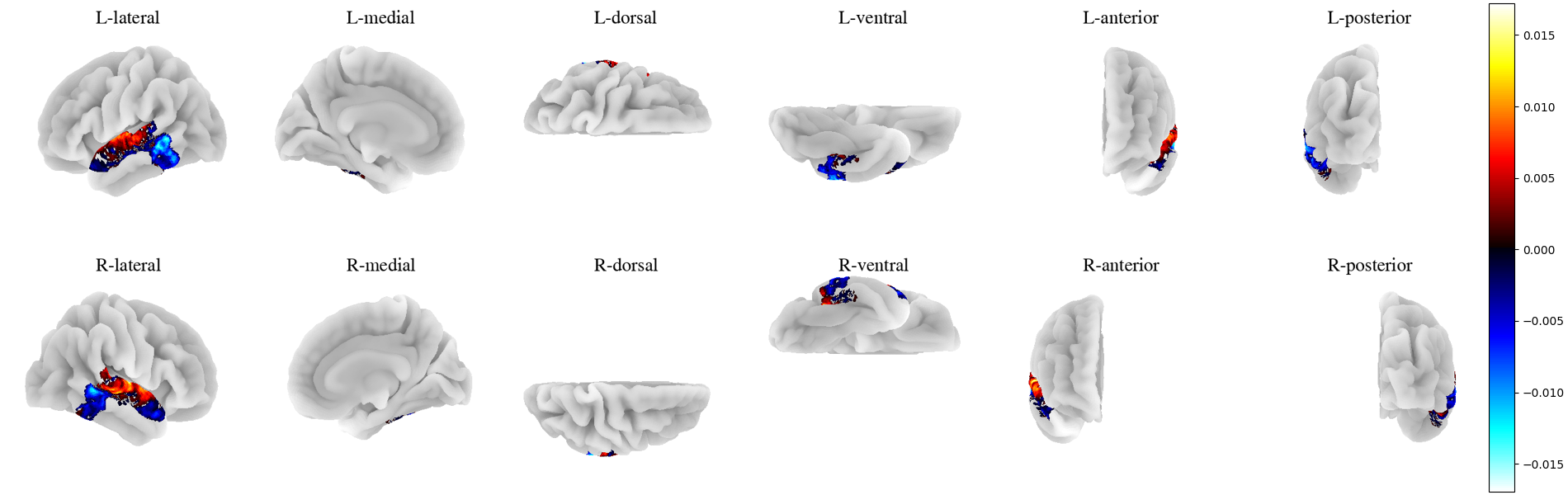}}
\caption{The averaged fMRI for the ROI: PTL for the task: "Shapes"}
\label{appendix-shapessocial-PTL-pial}
\end{center}
\vskip -0.2in
\end{figure}

\begin{figure}[H]
\vskip 0.2in
\begin{center}
\centerline{\includegraphics[width=\columnwidth]{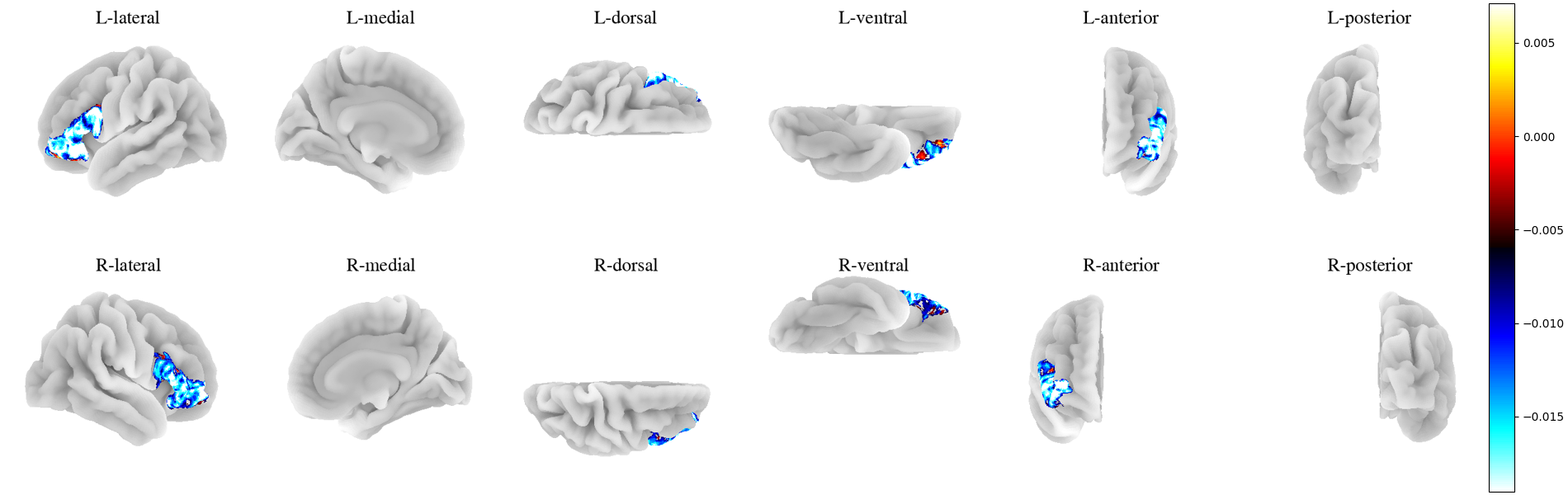}}
\caption{The averaged fMRI for the ROI: IFG for the task: "Shapes"}
\label{appendix-shapessocial-IFG-pial}
\end{center}
\vskip -0.2in
\end{figure}

\begin{figure}[H]
\vskip 0.2in
\begin{center}
\centerline{\includegraphics[width=\columnwidth]{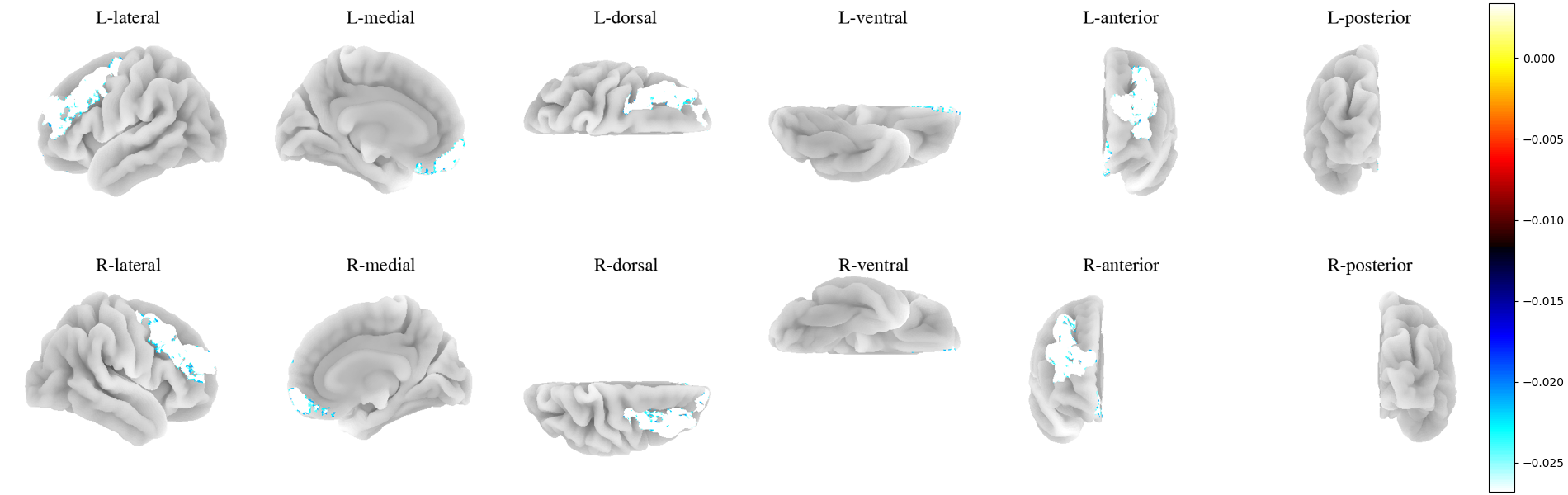}}
\caption{The averaged fMRI for the ROI: MFG for the task: "Shapes"}
\label{appendix-shapessocial-MFG-pial}
\end{center}
\vskip -0.2in
\end{figure}

\begin{figure}[H]
\vskip 0.2in
\begin{center}
\centerline{\includegraphics[width=\columnwidth]{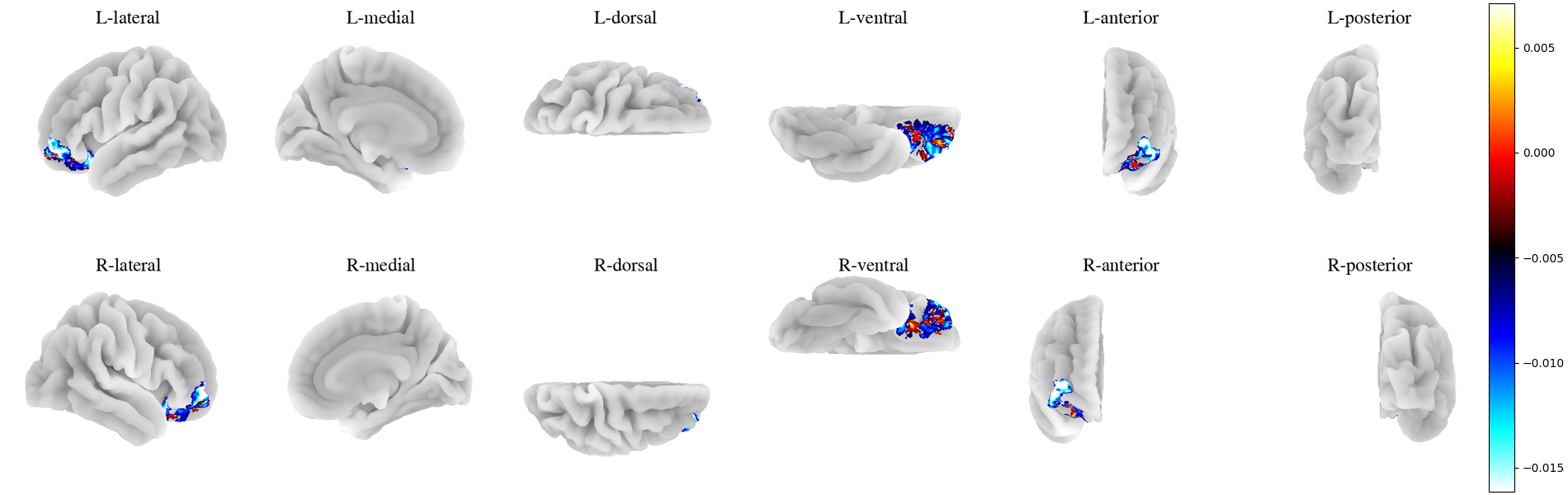}}
\caption{The averaged fMRI for the ROI: IFGorb for the task: "Shapes"}
\label{appendix-shapessocial-IFGorb-pial}
\end{center}
\vskip -0.2in
\end{figure}

\begin{figure}[H]
\vskip 0.2in
\begin{center}
\centerline{\includegraphics[width=\columnwidth]{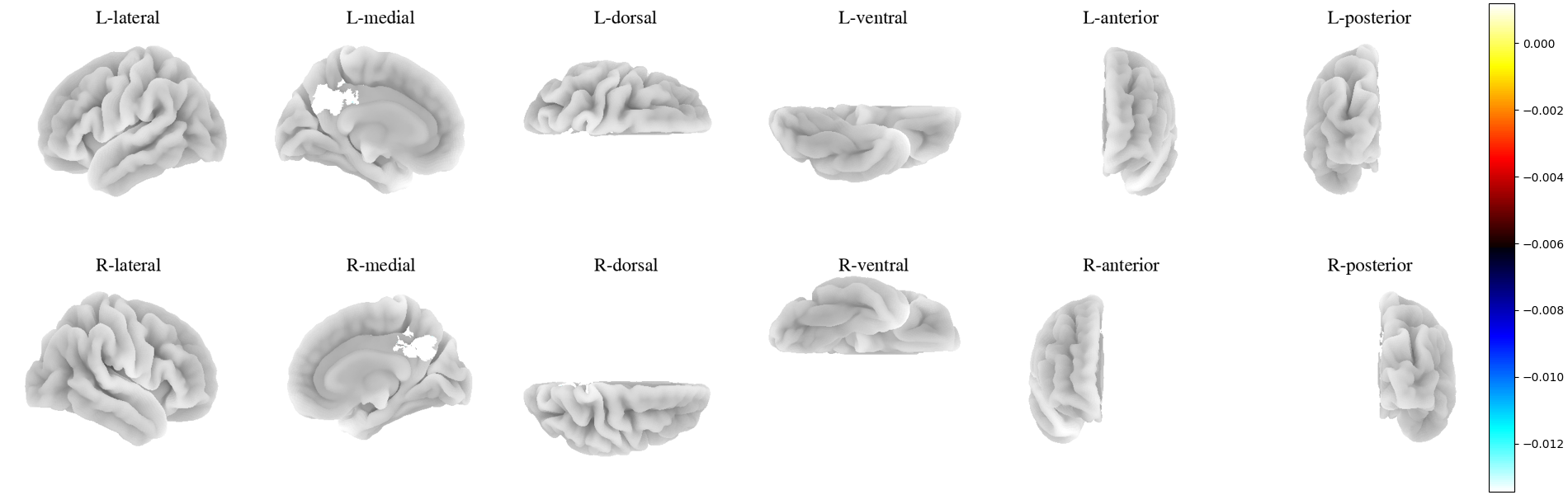}}
\caption{The averaged fMRI for the ROI: PCC for the task: "Shapes"}
\label{appendix-shapessocial-PCC-pial}
\end{center}
\vskip -0.2in
\end{figure}

\begin{figure}[H]
\vskip 0.2in
\begin{center}
\centerline{\includegraphics[width=\columnwidth]{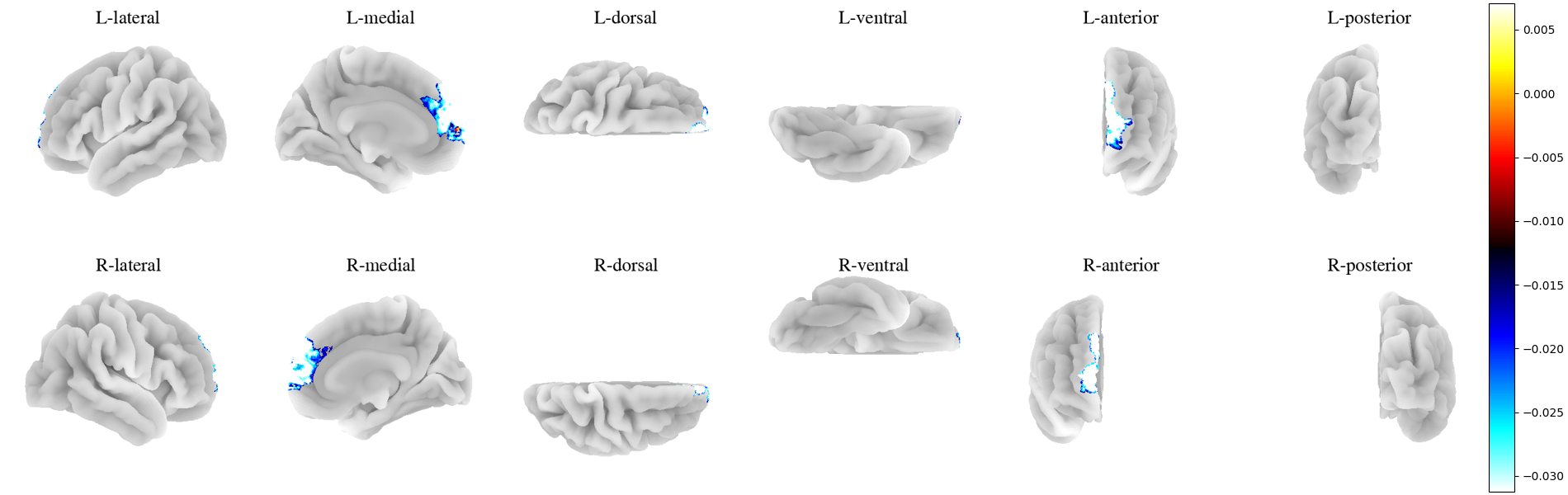}}
\caption{The averaged fMRI for the ROI: dmPFC for the task: "Shapes"}
\label{appendix-shapessocial-dmPFC-pial}
\end{center}
\vskip -0.2in
\end{figure}

\begin{figure}[H]
\vskip 0.2in
\begin{center}
\centerline{\includegraphics[width=\columnwidth]{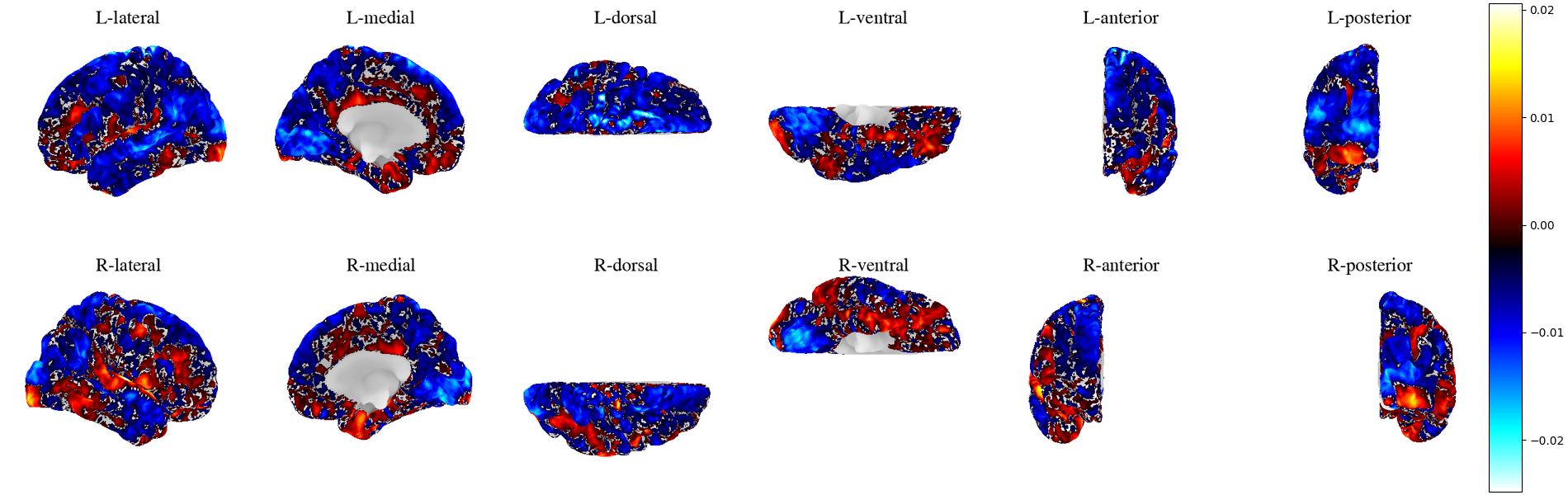}}
\caption{The averaged fMRI for the whole brain mask for the task: "It’s Not the Fall That Gets You"}
\label{appendix-notthefallintact-Evidence-pial}
\end{center}
\vskip -0.2in
\end{figure}

\begin{figure}[H]
\vskip 0.2in
\begin{center}
\centerline{\includegraphics[width=\columnwidth]{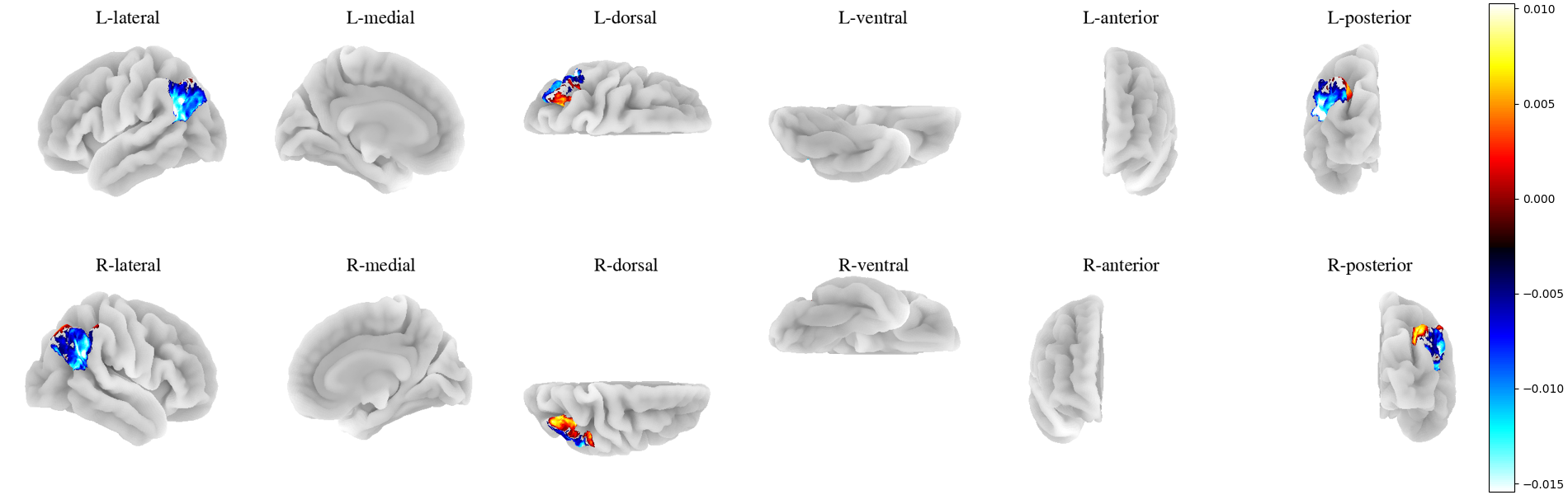}}
\caption{The averaged fMRI for the ROI: AG for the task: "It’s Not the Fall That Gets You"}
\label{appendix-notthefallintact-AG-pial}
\end{center}
\vskip -0.2in
\end{figure}

\begin{figure}[H]
\vskip 0.2in
\begin{center}
\centerline{\includegraphics[width=\columnwidth]{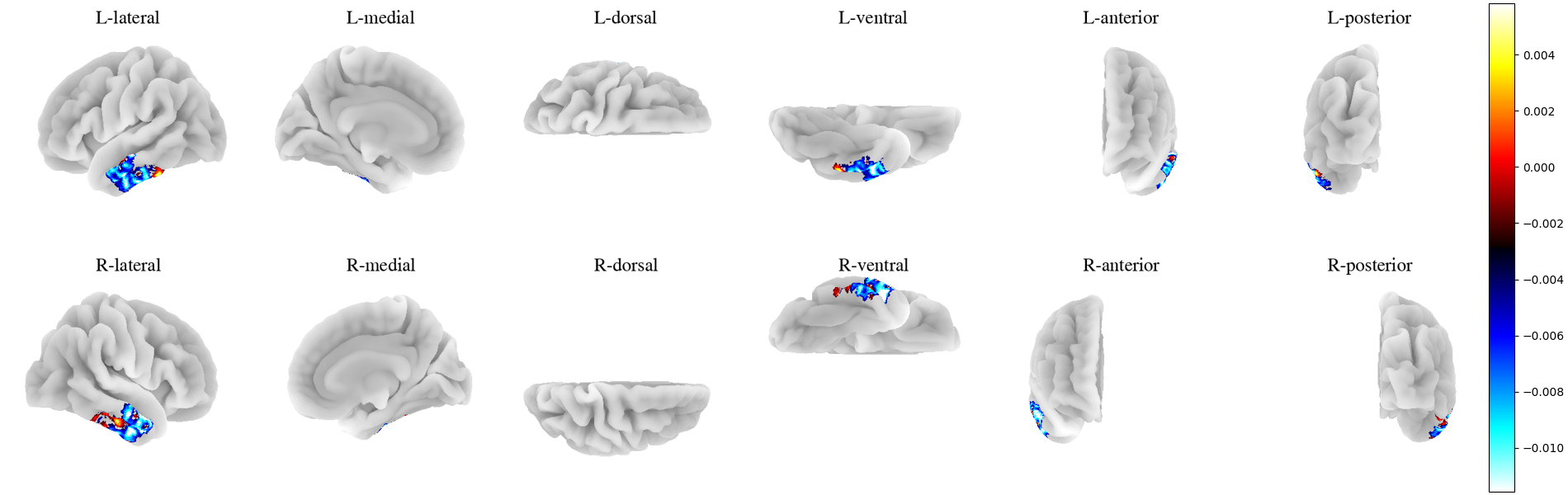}}
\caption{The averaged fMRI for the ROI: ATL for the task: "It’s Not the Fall That Gets You"}
\label{appendix-notthefallintact-ATL-pial}
\end{center}
\vskip -0.2in
\end{figure}

\begin{figure}[H]
\vskip 0.2in
\begin{center}
\centerline{\includegraphics[width=\columnwidth]{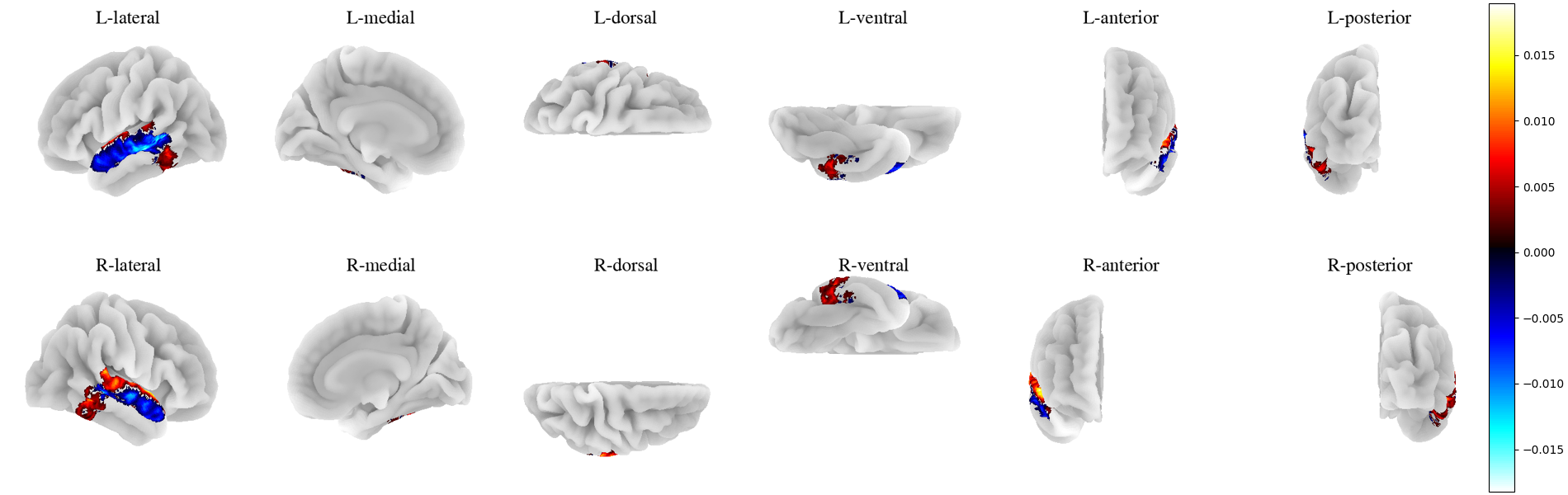}}
\caption{The averaged fMRI for the ROI: PTL for the task: "It’s Not the Fall That Gets You"}
\label{appendix-notthefallintact-PTL-pial}
\end{center}
\vskip -0.2in
\end{figure}

\begin{figure}[H]
\vskip 0.2in
\begin{center}
\centerline{\includegraphics[width=\columnwidth]{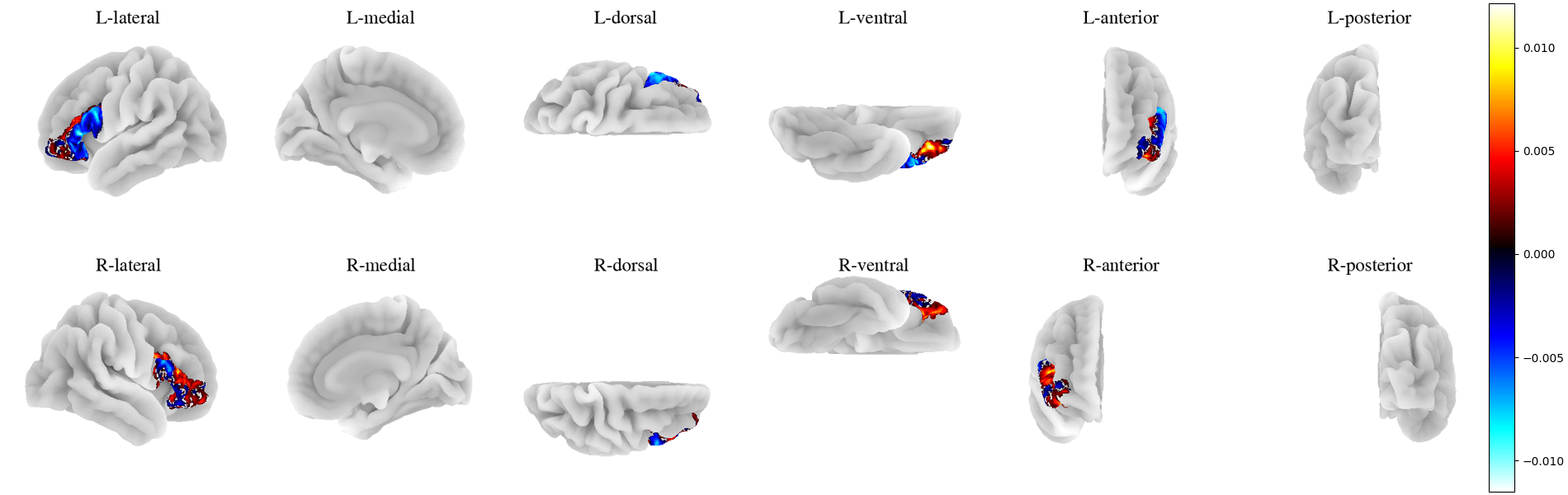}}
\caption{The averaged fMRI for the ROI: IFG for the task: "It’s Not the Fall That Gets You"}
\label{appendix-notthefallintact-IFG-pial}
\end{center}
\vskip -0.2in
\end{figure}

\begin{figure}[H]
\vskip 0.2in
\begin{center}
\centerline{\includegraphics[width=\columnwidth]{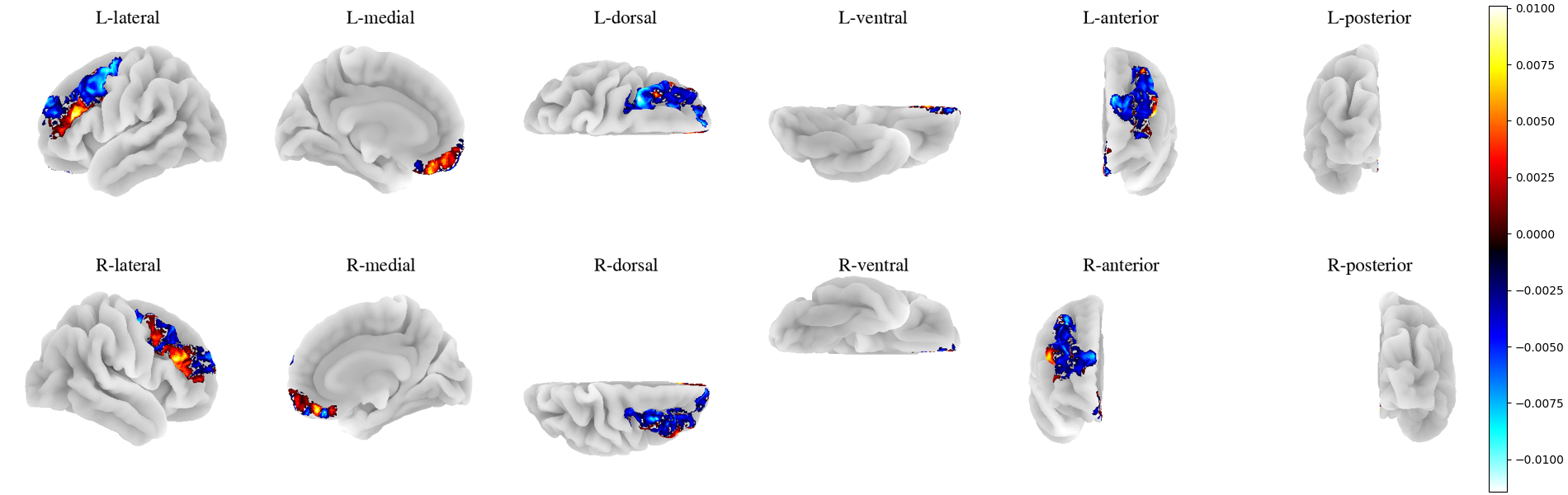}}
\caption{The averaged fMRI for the ROI: MFG for the task: "It’s Not the Fall That Gets You"}
\label{appendix-notthefallintact-MFG-pial}
\end{center}
\vskip -0.2in
\end{figure}

\begin{figure}[H]
\vskip 0.2in
\begin{center}
\centerline{\includegraphics[width=\columnwidth]{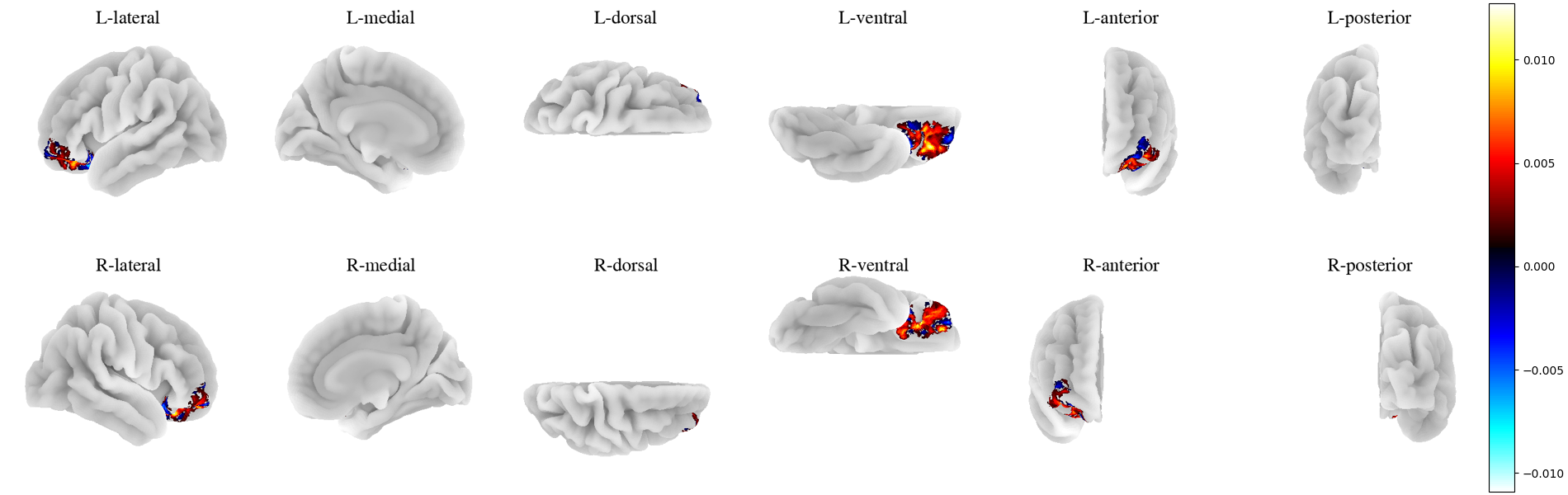}}
\caption{The averaged fMRI for the ROI: IFGorb for the task: "It’s Not the Fall That Gets You"}
\label{appendix-notthefallintact-IFGorb-pial}
\end{center}
\vskip -0.2in
\end{figure}

\begin{figure}[H]
\vskip 0.2in
\begin{center}
\centerline{\includegraphics[width=\columnwidth]{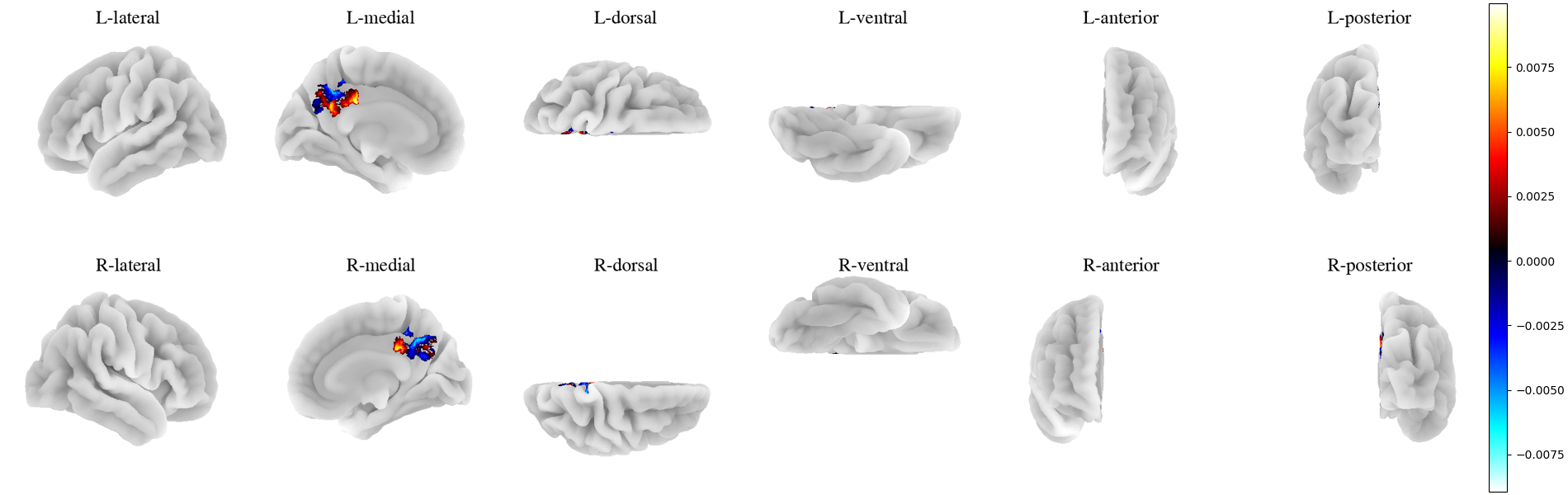}}
\caption{The averaged fMRI for the ROI: PCC for the task: "It’s Not the Fall That Gets You"}
\label{appendix-notthefallintact-PCC-pial}
\end{center}
\vskip -0.2in
\end{figure}

\begin{figure}[H]
\vskip 0.2in
\begin{center}
\centerline{\includegraphics[width=\columnwidth]{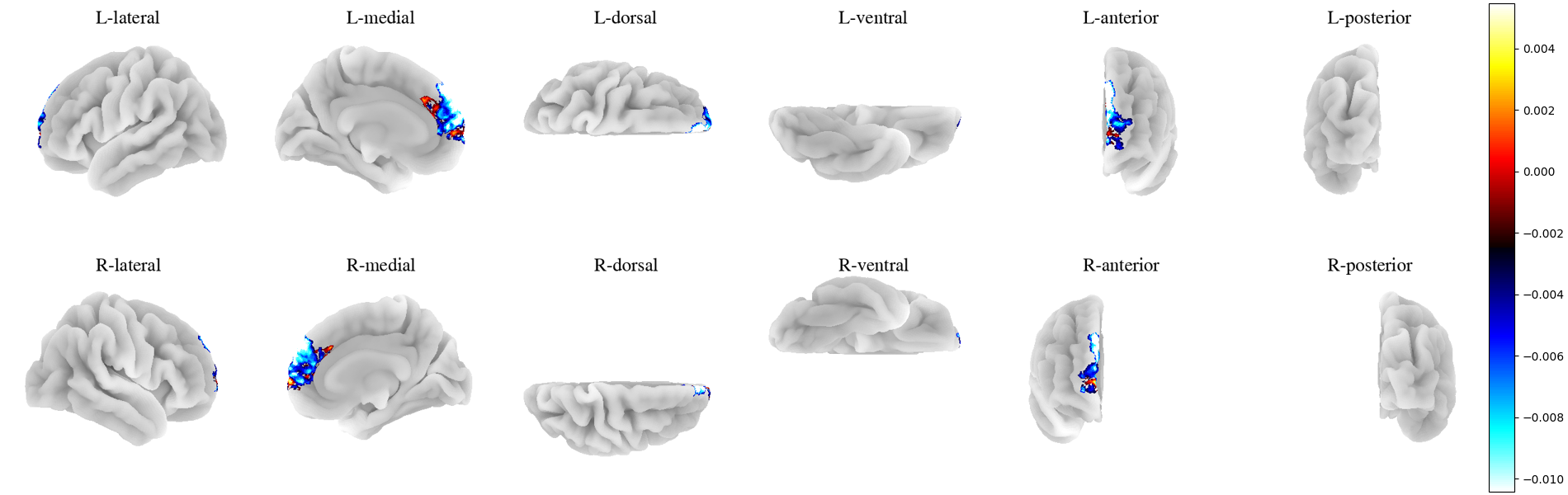}}
\caption{The averaged fMRI for the ROI: dmPFC for the task: "It’s Not the Fall That Gets You"}
\label{appendix-notthefallintact-dmPFC-pial}
\end{center}
\vskip -0.2in
\end{figure}

\section{fMRI Data Representation Summaries}
\label{fMRI Data Representation Stats}

This section provides a concise overview of the fMRI data representations pertaining to our selected three tasks from the Narratives dataset \citep{nastase2021narratives}.

\begin{table}[H]
\caption{Summary for fMRI data representations for selected three tasks from Narratives dataset \citep{nastase2021narratives}.}
\label{fMRI-data-representations}
\vskip 0.15in
\begin{center}
\begin{small}
\begin{sc}
\begin{tabular}{lcccr}
\toprule
Story & \# Subjects & \# Hemispheres & \# ROIs & Total  \\
\midrule
"Pie Man"    & 75 &  2 & 8+1 & $75 \times 2 \times (8+1) = 1350$ \\
"Shapes"    & 58 &  2 & 8+1 & $58 \times 2 \times (8+1) = 1044$  \\
"It’s Not the Fall & 54 &  2 & 8+1 & $54 \times 2 \times (8+1) = 972$  \\
That Gets You" &  &   &  &  \\
\toprule
Total    &  &   &  & 3,366  \\
\bottomrule
\end{tabular}
\end{sc}
\end{small}
\end{center}
\vskip -0.1in
\end{table}

\section{LLMs Data Representation Summaries}
\label{LLMs Data Representation Stats}

\Cref{LLMs-data-representations} provides an overview of all the LLMs sampled and investigated in this study. The layer numbers underlined signify the intermediate-to-deep layers of each LLM ($l = \frac{2}{3} n_{\text{layers}}$), a detail we scrutinized closely due to its reported significance in predicting brain activity \citep{schrimpf2021neural, caucheteux2022brains, caucheteux2023evidence}.

We employed the quantized versions of Llama-2-13B and Llama-2-70B to accommodate the limitations of our NVIDIA A100 card with 40 GB GPU memory. However, the randomly initialized weights for the quantized versions of Llama-2-7B, Llama-2-13B, and Llama-2-70B were unavailable during our study. Consequently, we didn't have these three untrained LLMs for data processing.

The Hugging Face webpage URLs for each of the 39 LLMs are integrated into the model names presented in both tables: \cref{LLMs-refs} \footnote{The number of parameters for each LLM is computed directly from the respective model configurations available on the \href{https://huggingface.co/}{Hugging Face} platform.} and \cref{LLMs-data-representations}. The references for these models can be accessed through the column "Reference" in \cref{LLMs-refs}.

\begin{table}[H]
\caption{Number of Parameters, Embedding Dimensions and References for our selected 39 LLMs.}
\label{LLMs-refs}
\vskip 0.15in
\begin{center}
\begin{small}
\begin{sc}
\begin{tabular}{lccc}
\toprule
LLM & \# Parameters & Embedding & Reference  \\
 &  & Dimension &  \\
\midrule
\href{https://huggingface.co/distilbert-base-uncased}{distilbert-base-uncased}    & 67M & 768 & \citep{sanh2019distilbert} \\
\href{https://huggingface.co/bert-base-uncased}{bert-base-uncased}    & 110M & 768 & \citep{devlin2018bert} \\
\href{https://huggingface.co/bert-large-uncased}{bert-large-uncased}   & 336M & 1024 & \citep{devlin2018bert} \\
\href{https://huggingface.co/bert-large-uncased-whole-word-masking}{bert-large-uncased-} & 336M & 1024 & \citep{devlin2018bert} \\
\href{https://huggingface.co/bert-large-uncased-whole-word-masking}{whole-word-masking} &  & \\

\href{https://huggingface.co/distilbert-base-cased}{distilbert-base-cased}    & 65.8M & 768 & \citep{sanh2019distilbert} \\
\href{https://huggingface.co/bert-base-cased}{bert-base-cased} & 109M & 768   & \citep{devlin2018bert} \\
\href{https://huggingface.co/bert-large-cased}{bert-large-cased}    & 335M & 1024 &  \citep{devlin2018bert} \\
\href{https://huggingface.co/bert-large-cased-whole-word-masking}{bert-large-cased-} & 335M & 1024  & \citep{devlin2018bert} \\
\href{https://huggingface.co/bert-large-cased-whole-word-masking}{whole-word-masking} & & \\

\href{https://huggingface.co/distilroberta-base}{distilroberta-base}    & 82.8M & 768 & \citep{sanh2019distilbert} \\
\href{https://huggingface.co/roberta-base}{roberta-base}    & 125M & 768 & \citep{liu2019roberta} \\
\href{https://huggingface.co/roberta-large}{roberta-large}    & 355M & 1024 & \citep{liu2019roberta} \\

\href{https://huggingface.co/xlm-mlm-en-2048}{xlm-mlm-en-2048}    & 668M & 2048 & \citep{conneau2019cross} \\
\href{https://huggingface.co/xlm-roberta-base}{xlm-roberta-base}    & 279M & 768 & \citep{conneau2019unsupervised} \\
\href{https://huggingface.co/xlm-roberta-large}{xlm-roberta-large}    & 561M & 1024 & \citep{conneau2019unsupervised} \\

\href{https://huggingface.co/xlnet-base-cased}{xlnet-base-cased}    & 117M & 768 & \citep{yang2019xlnet} \\
\href{https://huggingface.co/xlnet-large-cased}{xlnet-large-cased}    & 361M & 1024 & \citep{yang2019xlnet} \\

\href{https://huggingface.co/ctrl}{ctrl}    &1.64B & 1280 & \citep{keskar2019ctrl} \\

\href{https://huggingface.co/albert-base-v1}{albert-base-v1}    & 11.8M & 768 & \citep{lan2019albert} \\
\href{https://huggingface.co/albert-base-v2}{albert-base-v2}    & 11.8M & 768 & \citep{lan2019albert} \\
\href{https://huggingface.co/albert-large-v1}{albert-large-v1}    & 18.7M & 1024 & \citep{lan2019albert} \\
\href{https://huggingface.co/albert-large-v2}{albert-large-v2}    & 18.7M & 1024 & \citep{lan2019albert} \\
\href{https://huggingface.co/albert-xlarge-v1}{albert-xlarge-v1}    & 59M & 2048 & \citep{lan2019albert} \\
\href{https://huggingface.co/albert-xlarge-v2}{albert-xlarge-v2}    & 59M & 2048 & \citep{lan2019albert} \\
\href{https://huggingface.co/albert-xxlarge-v1}{albert-xxlarge-v1}    & 223M & 4096 & \citep{lan2019albert} \\
\href{https://huggingface.co/albert-xxlarge-v2}{albert-xxlarge-v2}    & 223M & 4096 & \citep{lan2019albert} \\

\href{https://huggingface.co/distilgpt2}{distilgpt2}    & 88.2M & 768 & \citep{sanh2019distilbert} \\
\href{https://huggingface.co/gpt2}{gpt2}    & 137M & 768 & \citep{radford2019language} \\
\href{https://huggingface.co/gpt2-medium}{gpt2-medium}    & 380M & 1024 & \citep{radford2019language} \\
\href{https://huggingface.co/gpt2-large}{gpt2-large}    & 812M & 1280 &  \citep{radford2019language} \\
\href{https://huggingface.co/gpt2-xl}{gpt2-xl}    & 1.61B & 1600 & \citep{radford2019language} \\

\href{https://huggingface.co/microsoft/phi-1}{microsoft/phi-1}    & 1.42B & 2048 & \citep{gunasekar2023textbooks} \\
\href{https://huggingface.co/microsoft/phi-1_5}{microsoft/phi-1\_5}    & 1.42B & 2048 & \citep{li2023textbooks} \\
\href{https://huggingface.co/microsoft/phi-2}{microsoft/phi-2}    & 2.78B & 2560 & \citep{hughes_2023_phi2} \\

\href{https://huggingface.co/mosaicml/mpt-7b}{mosaicml/mpt-7b}    & 6.65B & 4096 & \citep{MosaicML2023Introducing} \\

\href{https://huggingface.co/tiiuae/falcon-7b}{tiiuae/falcon-7b}    & 6.93B & 4544 & \citep{penedo2023refinedweb} \\

\href{https://huggingface.co/meta-llama/Llama-2-7b-hf}{meta-llama/Llama-2-7b-hf}    & 6.74B & 4096 & \citep{touvron2023llama} \\

\href{https://huggingface.co/TheBloke/Llama-2-7B-GPTQ}{TheBloke/Llama-2-7B-GPTQ}    & 263M & 4096 & \citep{frantar2022gptq} \\
 & & & \citep{touvron2023llama} \\

\href{https://huggingface.co/TheBloke/Llama-2-13B-GPTQ}{TheBloke/Llama-2-13B-GPTQ}    & 329M & 5120 & \citep{frantar2022gptq} \\
 & & &  \citep{touvron2023llama} \\

\href{https://huggingface.co/TheBloke/Llama-2-70B-GPTQ}{TheBloke/Llama-2-70B-GPTQ}    & 526M & 8192 & \citep{frantar2022gptq} \\
 & & & \citep{touvron2023llama} \\

\bottomrule
\end{tabular}
\end{sc}
\end{small}
\end{center}
\vskip -0.1in
\end{table}

\begin{table}[H]
\caption{Summary for our selected 39 LLMs.}
\label{LLMs-data-representations}
\vskip 0.15in
\begin{center}
\begin{small}
\begin{sc}
\begin{tabular}{lccc}
\toprule
LLM & \# Layers & Used & Has  \\
 & & Layers & Untrained?  \\
\midrule
\href{https://huggingface.co/distilbert-base-uncased}{distilbert-base-uncased}    & 6 & 1,2,3,\underline{4},5,6,avg & Yes \\
\href{https://huggingface.co/bert-base-uncased}{bert-base-uncased}    & 12 & 1, 2, 4, 5, 7, \underline{8}, 10, 12,avg & Yes \\
\href{https://huggingface.co/bert-large-uncased}{bert-large-uncased}    & 24  & 1, 4, 7, 10, 14, \underline{16}, 17, 20, 24,avg & Yes \\
\href{https://huggingface.co/bert-large-uncased-whole-word-masking}{bert-large-uncased-} & 24 & 1, 4, 7, 10, 14, \underline{16}, 17, 20, 24,avg & Yes \\
\href{https://huggingface.co/bert-large-uncased-whole-word-masking}{whole-word-masking} &  &  &  \\

\href{https://huggingface.co/distilbert-base-cased}{distilbert-base-cased}    & 6 & 1,2,3,\underline{4},5,6,avg & Yes \\
\href{https://huggingface.co/bert-base-cased}{bert-base-cased}    & 12 & 1, 2, 4, 5, 7, \underline{8}, 10, 12,avg & Yes \\
\href{https://huggingface.co/bert-large-cased}{bert-large-cased}    & 24 & 1, 4, 7, 10, 14, \underline{16}, 17, 20, 24,avg & Yes \\
\href{https://huggingface.co/bert-large-cased-whole-word-masking}{bert-large-cased-} & 24 & 1, 4, 7, 10, 14, \underline{16}, 17, 20, 24,avg & Yes \\
\href{https://huggingface.co/bert-large-cased-whole-word-masking}{whole-word-masking} & & & \\

\href{https://huggingface.co/distilroberta-base}{distilroberta-base}    & 6  & 1,2,3,\underline{4},5,6,avg & Yes \\
\href{https://huggingface.co/roberta-base}{roberta-base}    & 12 & 1, 2, 4, 5, 7, \underline{8}, 10, 12,avg & Yes \\
\href{https://huggingface.co/roberta-large}{roberta-large}    & 24 & 1, 4, 7, 10, 14, \underline{16}, 17, 20, 24,avg & Yes \\

\href{https://huggingface.co/xlm-mlm-en-2048}{xlm-mlm-en-2048}    & 12 & 1, 2, 4, 5, 7, \underline{8}, 10, 12,avg & Yes \\
\href{https://huggingface.co/xlm-roberta-base}{xlm-roberta-base}    & 12 & 1, 2, 4, 5, 7, \underline{8}, 10, 12,avg & Yes \\
\href{https://huggingface.co/xlm-roberta-large}{xlm-roberta-large}    & 24 & 1, 4, 7, 10, 14, \underline{16}, 17, 20, 24,avg & Yes \\

\href{https://huggingface.co/xlnet-base-cased}{xlnet-base-cased}    & 24 & 1, 2, 4, 5, 7, \underline{8}, 10, 12,avg & Yes \\
\href{https://huggingface.co/xlnet-large-cased}{xlnet-large-cased}    & 24 & 1, 4, 7, 10, 14, \underline{16}, 17, 20, 24,avg & Yes \\

\href{https://huggingface.co/ctrl}{ctrl}    & 48 & 1, 7, 14, 21, 27, \underline{32}, 34, 41, 48,avg & Yes \\

\href{https://huggingface.co/albert-base-v1}{albert-base-v1}    & 12 & 1, 2, 4, 5, 7, \underline{8}, 10, 12,avg & Yes \\
\href{https://huggingface.co/albert-base-v2}{albert-base-v2}    & 12 & 1, 2, 4, 5, 7, \underline{8}, 10, 12,avg & Yes \\
\href{https://huggingface.co/albert-large-v1}{albert-large-v1}    & 24 & 1, 4, 7, 10, 14, \underline{16}, 17, 20, 24,avg & Yes \\
\href{https://huggingface.co/albert-large-v2}{albert-large-v2}    & 24 & 1, 4, 7, 10, 14, \underline{16}, 17, 20, 24,avg & Yes \\
\href{https://huggingface.co/albert-xlarge-v1}{albert-xlarge-v1}    & 24 & 1, 4, 7, 10, 14, \underline{16}, 17, 20, 24,avg & Yes \\
\href{https://huggingface.co/albert-xlarge-v2}{albert-xlarge-v2}    & 24 & 1, 4, 7, 10, 14, \underline{16}, 17, 20, 24,avg & Yes \\
\href{https://huggingface.co/albert-xxlarge-v1}{albert-xxlarge-v1}    & 12 & 1, 2, 4, 5, 7, \underline{8}, 10, 12,avg & Yes \\
\href{https://huggingface.co/albert-xxlarge-v2}{albert-xxlarge-v2}    & 12 & 1, 2, 4, 5, 7, \underline{8}, 10, 12,avg & Yes \\

\href{https://huggingface.co/distilgpt2}{distilgpt2}    & 6 & 1,2,3,\underline{4},5,6,avg & Yes \\
\href{https://huggingface.co/gpt2}{gpt2}    & 12 & 1, 2, 4, 5, 7, \underline{8}, 10, 12,avg & Yes \\
\href{https://huggingface.co/gpt2-medium}{gpt2-medium}    & 24 & 1, 4, 7, 10, 14, \underline{16}, 17, 20, 24,avg & Yes \\
\href{https://huggingface.co/gpt2-large}{gpt2-large}    & 36 & 1, 6, 11, 16, 21, \underline{24}, 26, 31, 36,avg & Yes \\
\href{https://huggingface.co/gpt2-xl}{gpt2-xl}    & 48 & 1, 7, 14, 21, 27, \underline{32}, 34, 41, 48,avg & Yes \\

\href{https://huggingface.co/microsoft/phi-1}{microsoft/phi-1}    & 24 & 1, 4, 7, 10, 14, \underline{16}, 17, 20, 24,avg & Yes \\
\href{https://huggingface.co/microsoft/phi-1_5}{microsoft/phi-1\_5}    & 24 & 1, 4, 7, 10, 14, \underline{16}, 17, 20, 24,avg & Yes \\
\href{https://huggingface.co/microsoft/phi-2}{microsoft/phi-2}    & 32 & 1, 5, 9, 14, 18, \underline{21}, 23, 27, 32,avg & Yes \\

\href{https://huggingface.co/mosaicml/mpt-7b}{mosaicml/mpt-7b}    & 32 & 1, 5, 9, 14, 18, \underline{21}, 23, 27, 32,avg & Yes \\

\href{https://huggingface.co/tiiuae/falcon-7b}{tiiuae/falcon-7b}    & 32 & 1, 5, 9, 14, 18, \underline{21}, 23, 27, 32,avg & Yes \\

\href{https://huggingface.co/meta-llama/Llama-2-7b-hf}{meta-llama/Llama-2-7b-hf}    & 32 & 1, 5, 9, 14, 18, \underline{21}, 23, 27, 32,avg & Yes \\

\href{https://huggingface.co/TheBloke/Llama-2-7B-GPTQ}{TheBloke/Llama-2-7B-GPTQ}    & 32 & 1, 5, 9, 14, 18, \underline{21}, 23, 27, 32,avg & No \\

\href{https://huggingface.co/TheBloke/Llama-2-13B-GPTQ}{TheBloke/Llama-2-13B-GPTQ}    & 40 & 1, 6, 12, 17, 23, \underline{26}, 28, 34, 40,avg & No \\

\href{https://huggingface.co/TheBloke/Llama-2-70B-GPTQ}{TheBloke/Llama-2-70B-GPTQ}    & 80 & 1, 12, 23, 34, 46, \underline{53}, 57, 68, 80,avg & No \\

\bottomrule
\end{tabular}
\end{sc}
\end{small}
\end{center}
\vskip -0.1in
\end{table}

\section{Illustrations for Characterizing the fMRI Data Representations by PH}
\label{Illustrations for Characterizing the fMRI Data Representations by PH}

This section elucidates the process by which we converted the fMRI Data Representations into 1-D time-series vectors, followed by projecting them into 3-D spaces as point clouds, and subsequently computed the Persistent Homology (PH) for these point clouds, depicted through summary representations such as persistence diagrams and persistence barcodes. Further elaboration on this process is provided in \cref{Characterizing the fMRI Data Representations by PH}.

\section{Illustrations for Characterizing the LLMs Embeddings by PH}
\label{Illustrations for Characterizing the LLMs Embeddings by PH}

This section elucidates the process by which we converted the LLM Embeddings into 1-D time-series vectors, then projected them into 3-D spaces as point clouds, and subsequently computed the Persistent Homology for these point clouds. The resultant summary representations, in the form of persistence diagrams and persistence barcodes, are illustrated herein. Further details regarding this process can be found in \cref{Characterizing the LLMs Embeddings by PH}.

\section{Exploratory Data Analysis on Constructed Features: $q$-Wasserstein Distances}
\label{Exploratory Data Analysis on Constructed Features: q-Wasserstein Distances}

This section illustrates the $q$-Wasserstein Distances computed between the persistence diagrams for fMRI data from ROI: PCC, hemisphere: L, and embeddings from LLM: Llama-2-7b-hf, training status: trained, layer: 21, under the same task: "It's not the Fall That Gets You". Each figure displays the results for all three persistent homology dimensions: 0, 1, and 2, respectively, with the parameter ranges for $q$ and $p$ marked in each figure to denote the differences.

Across the $q$-Wasserstein Distances, we observe "long tail" distributions, where the values converge to approximately 0.30 for persistent homology dimension 0, and approximately 0.14 and 0.07 for persistent homology dimensions 1 and 2, respectively, in this specific example among all the 34,416 pairs in \cref{Computing q-Wasserstein Distances between Persistence Diagrams}. It is noteworthy that this distribution pattern is not unique to this particular example; rather, it is prevalent across all 34,416 pairs. The variations lie only in the specific values to which the $q$-Wasserstein Distances ultimately converge for each of the three persistent homology dimensions. Formally, the $q$-Wasserstein Distance is bounded by the Wasserstein Stability Theorem \citep{edelsbrunner2010computational, cohen_steiner_edelsbrunner_harer_yuriy_mileyko_2010}.

\section{Filter in Reliable and Valid Features: First Pass}
\label{Filter in Reliable and Valid Features: First Pass}

This section provides an illustrative example of our filtering process "First Pass" outlined in \cref{Learning Linear Regression Models from q-Wasserstein Distances and Existing Brainscores}. Specifically, the example focuses on the Region of Interest (ROI): PCC, considering both "Only Trained" LLMs and a combination of "Trained plus Untrained" ones.

At the end of this section, we present the maximum value of $q$ for each Region of Interest (ROI) across each hemisphere, which corresponds to the highest averaged test $R^2$ score achieved during the "First Pass", as detailed in \cref{First Pass}.

\begin{table}[H]
\caption{The table compiles the averaged \textbf{train} $R^2$ scores across four scenarios: the L(eft) and R(ight) Hemispheres, considering whether the training dataset comprises "Only Trained" or "Trained plus Untrained" LLMs. The $q$-Wasserstein Distances are progressively included as features for training the Linear Regression Model, ranging from $q=p=1$ to $q=p=50$.}
\label{train test R2}
\vskip 0.15in
\begin{center}
\begin{small}
\begin{sc}
\begin{tabular}{lcccc}
\toprule
$q(p)$ & L Only & R Only & L Trained & R Trained  \\
value & Trained & Trained & and Untrained & and Untrained  \\
\midrule
$q=p=1$ & $2.2676 e{-1}$ & $2.8019 e{-1}$ & $1.9626 e{-1}$ & $2.1948 e{-1}$ \\
$q=p=2$ & $2.6528 e{-1}$ & $3.0415 e{-1}$ & $2.6011 e{-1}$ & $2.8302 e{-1}$ \\
$q=p=3$ & $3.5044 e{-1}$ & $3.7017 e{-1}$ & $3.3508 e{-1}$ & $3.4951 e{-1}$ \\
$q=p=4$ & $3.8208 e{-1}$ & $3.8145 e{-1}$ & $3.5319 e{-1}$ & $3.5761 e{-1}$ \\
$q=p=5$ & $3.8609 e{-1}$ & $3.8510 e{-1}$ & $3.5758 e{-1}$ & $3.6216 e{-1}$ \\
$q=p=6$ & $3.9301 e{-1}$ & $3.8735 e{-1}$ & $3.6052 e{-1}$ & $3.6619 e{-1}$ \\
$q=p=7$ & $3.9482 e{-1}$ & $3.9077 e{-1}$ & $3.6230 e{-1}$ & $3.6840 e{-1}$ \\
$q=p=8$ & $3.9646 e{-1}$ & $3.9167 e{-1}$ & $3.7142 e{-1}$ & $3.6907 e{-1}$ \\
$q=p=9$ & $4.0346 e{-1}$ & $3.9569 e{-1}$ & $3.8007 e{-1}$ & $3.7048 e{-1}$ \\
$q=p=10$ & $4.0933 e{-1}$ & $3.9800 e{-1}$ & $3.8601 e{-1}$ & $3.7439 e{-1}$ \\
$q=p=11$ & $4.1479 e{-1}$ & $4.0077 e{-1}$ & $3.9041 e{-1}$ & $3.7749 e{-1}$ \\
$q=p=12$ & $4.1723 e{-1}$ & $4.0173 e{-1}$ & $3.9141 e{-1}$ & $3.7866 e{-1}$ \\
$q=p=13$ & $4.1881 e{-1}$ & $4.0354 e{-1}$ & $3.9192 e{-1}$ & $3.8125 e{-1}$ \\
$q=p=14$ & $4.2224 e{-1}$ & $4.0605 e{-1}$ & $3.9326 e{-1}$ & $3.8477 e{-1}$ \\
$q=p=15$ & $4.2834 e{-1}$ & $4.0894 e{-1}$ & $3.9729 e{-1}$ & $3.8780 e{-1}$ \\
$q=p=16$ & $4.3450 e{-1}$ & $4.0994 e{-1}$ & $4.0117 e{-1}$ & $3.9079 e{-1}$ \\
$q=p=17$ & $4.4185 e{-1}$ & $4.1091 e{-1}$ & $4.0410 e{-1}$ & $3.9106 e{-1}$ \\
$q=p=18$ & $4.4598 e{-1}$ & $4.1226 e{-1}$ & $4.0634 e{-1}$ & $3.9196 e{-1}$ \\
$q=p=19$ & $4.4742 e{-1}$ & $4.1395 e{-1}$ & $4.0721 e{-1}$ & $3.9242 e{-1}$ \\
$q=p=20$ & $4.5304 e{-1}$ & $4.2334 e{-1}$ & $4.0868 e{-1}$ & $3.9586 e{-1}$ \\
$q=p=21$ & $4.5470 e{-1}$ & $4.2466 e{-1}$ & $4.0905 e{-1}$ & $3.9676 e{-1}$ \\
$q=p=22$ & $4.5631 e{-1}$ & $4.2585 e{-1}$ & $4.0950 e{-1}$ & $3.9750 e{-1}$ \\
$q=p=23$ & $4.5956 e{-1}$ & $4.2699 e{-1}$ & $4.1044 e{-1}$ & $3.9854 e{-1}$ \\
$q=p=24$ & $4.6116 e{-1}$ & $4.3547 e{-1}$ & $4.1130 e{-1}$ & $4.0235 e{-1}$ \\
$q=p=25$ & $4.6171 e{-1}$ & $4.4021 e{-1}$ & $4.1180 e{-1}$ & $4.0539 e{-1}$ \\
$q=p=26$ & $4.6372 e{-1}$ & $4.4153 e{-1}$ & $4.1297 e{-1}$ & $4.0669 e{-1}$ \\
$q=p=27$ & $4.6627 e{-1}$ & $4.4698 e{-1}$ & $4.1423 e{-1}$ & $4.0975 e{-1}$ \\
$q=p=28$ & $4.7133 e{-1}$ & $4.4992 e{-1}$ & $4.1742 e{-1}$ & $4.1135 e{-1}$ \\
$q=p=29$ & $4.7193 e{-1}$ & $4.5163 e{-1}$ & $4.1820 e{-1}$ & $4.1225 e{-1}$ \\
$q=p=30$ & $4.7406 e{-1}$ & $4.5370 e{-1}$ & $4.1925 e{-1}$ & $4.1628 e{-1}$ \\
$q=p=31$ & $4.7962 e{-1}$ & $4.6614 e{-1}$ & $4.2364 e{-1}$ & $4.2017 e{-1}$ \\
$q=p=32$ & $4.8062 e{-1}$ & $4.6891 e{-1}$ & $4.2486 e{-1}$ & $4.2158 e{-1}$ \\
$q=p=33$ & $4.8186 e{-1}$ & $4.7294 e{-1}$ & $4.2567 e{-1}$ & $4.2331 e{-1}$ \\
$q=p=34$ & $4.8566 e{-1}$ & $4.7470 e{-1}$ & $4.2822 e{-1}$ & $4.2413 e{-1}$ \\
$q=p=35$ & $4.8657 e{-1}$ & $4.7610 e{-1}$ & $4.2857 e{-1}$ & $4.2449 e{-1}$ \\
$q=p=36$ & $4.8875 e{-1}$ & $4.7986 e{-1}$ & $4.2935 e{-1}$ & $4.2619 e{-1}$ \\
$q=p=37$ & $4.9102 e{-1}$ & $4.8264 e{-1}$ & $4.3021 e{-1}$ & $4.2736 e{-1}$ \\
$q=p=38$ & $4.9382 e{-1}$ & $4.9000 e{-1}$ & $4.3099 e{-1}$ & $4.3422 e{-1}$ \\
$q=p=39$ & $4.9731 e{-1}$ & $4.9332 e{-1}$ & $4.3311 e{-1}$ & $4.3571 e{-1}$ \\
$q=p=40$ & $5.0121 e{-1}$ & $5.0561 e{-1}$ & $4.3550 e{-1}$ & $4.4134 e{-1}$ \\
$q=p=41$ & $5.0265 e{-1}$ & $5.0855 e{-1}$ & $4.3625 e{-1}$ & $4.4330 e{-1}$ \\
$q=p=42$ & $5.0424 e{-1}$ & $5.0902 e{-1}$ & $4.3733 e{-1}$ & $4.4418 e{-1}$ \\
$q=p=43$ & $5.0766 e{-1}$ & $5.1294 e{-1}$ & $4.3981 e{-1}$ & $4.4527 e{-1}$ \\
$q=p=44$ & $5.1244 e{-1}$ & $5.1654 e{-1}$ & $4.4163 e{-1}$ & $4.4744 e{-1}$ \\
$q=p=45$ & $5.1410 e{-1}$ & $5.1992 e{-1}$ & $4.4265 e{-1}$ & $4.4914 e{-1}$ \\
$q=p=46$ & $5.1848 e{-1}$ & $5.2137 e{-1}$ & $4.4355 e{-1}$ & $4.5063 e{-1}$ \\
$q=p=47$ & $5.2083 e{-1}$ & $5.2489 e{-1}$ & $4.4484 e{-1}$ & $4.5307 e{-1}$ \\
$q=p=48$ & $5.2561 e{-1}$ & $5.2628 e{-1}$ & $4.4699 e{-1}$ & $4.5348 e{-1}$ \\
$q=p=49$ & $5.2772 e{-1}$ & $5.2814 e{-1}$ & $4.4831 e{-1}$ & $4.5483 e{-1}$ \\
$q=p=50$ & $5.2951 e{-1}$ & $5.3148 e{-1}$ & $4.4872 e{-1}$ & $4.5604 e{-1}$ \\

\bottomrule
\end{tabular}
\end{sc}
\end{small}
\end{center}
\vskip -0.1in
\end{table}

\begin{table}[H]
\caption{The table compiles the averaged \textbf{test} $R^2$ scores across four scenarios: the L(eft) and R(ight) Hemispheres, considering whether the training dataset comprises "Only Trained" or "Trained plus Untrained" LLMs. The $q$-Wasserstein Distances are progressively included as features for training the Linear Regression Model, ranging from $q=p=1$ to $q=p=50$.}
\label{only test R2}
\vskip 0.15in
\begin{center}
\begin{small}
\begin{sc}
\begin{tabular}{lcccc}
\toprule
$q(p)$ & L Only & R Only & L Trained & R Trained  \\
value & Trained & Trained & and Untrained & and Untrained  \\
\midrule
$q=p=1$ & $2.0603 e{-1}$ & $2.5393 e{-1}$ & $1.8616 e{-1}$ & $2.0459 e{-1}$ \\
$q=p=2$ & $2.3421 e{-1}$ & $2.6107 e{-1}$ & $2.4498 e{-1}$ & $2.5628 e{-1}$ \\
$q=p=3$ & $3.1531 e{-1}$ & $3.3301 e{-1}$ & $3.1849 e{-1}$ & $3.2631 e{-1}$ \\
$q=p=4$ & $3.3954 e{-1}$ & $3.4015 e{-1}$ & $3.3480 e{-1}$ & $3.3212 e{-1}$ \\
$q=p=5$ & $3.2199 e{-1}$ & $3.3588 e{-1}$ & $3.3036 e{-1}$ & $3.3411 e{-1}$ \\
$q=p=6$ & $2.9304 e{-1}$ & $3.3314 e{-1}$ & $3.1853 e{-1}$ & $3.3560 e{-1}$ \\
$q=p=7$ & $2.6157 e{-1}$ & $3.2523 e{-1}$ & $3.0384 e{-1}$ & $3.3400 e{-1}$ \\
$q=p=8$ & $2.5110 e{-1}$ & $3.1828 e{-1}$ & $3.0736 e{-1}$ & $3.3102 e{-1}$ \\
$q=p=9$ & $2.4856 e{-1}$ & $3.1503 e{-1}$ & $3.1189 e{-1}$ & $3.2760 e{-1}$ \\
$q=p=10$ & $2.5271 e{-1}$ & $3.0938 e{-1}$ & $3.1651 e{-1}$ & $3.2852 e{-1}$ \\
$q=p=11$ & $2.5458 e{-1}$ & $3.0568 e{-1}$ & $3.1946 e{-1}$ & $3.2645 e{-1}$ \\
$q=p=12$ & $2.5511 e{-1}$ & $2.9922 e{-1}$ & $3.1697 e{-1}$ & $3.2432 e{-1}$ \\
$q=p=13$ & $2.4993 e{-1}$ & $2.9481 e{-1}$ & $3.1485 e{-1}$ & $3.2252 e{-1}$ \\
$q=p=14$ & $2.4366 e{-1}$ & $2.9236 e{-1}$ & $3.1423 e{-1}$ & $3.2412 e{-1}$ \\
$q=p=15$ & $2.3893 e{-1}$ & $2.8742 e{-1}$ & $3.1676 e{-1}$ & $3.2639 e{-1}$ \\
$q=p=16$ & $2.5084 e{-1}$ & $2.8472 e{-1}$ & $3.2079 e{-1}$ & $3.2775 e{-1}$ \\
$q=p=17$ & $2.5561 e{-1}$ & $2.8285 e{-1}$ & $3.2194 e{-1}$ & $3.2608 e{-1}$ \\
$q=p=18$ & $2.6034 e{-1}$ & $2.7908 e{-1}$ & $3.2266 e{-1}$ & $3.2417 e{-1}$ \\
$q=p=19$ & $2.5477 e{-1}$ & $2.7730 e{-1}$ & $3.2078 e{-1}$ & $3.2158 e{-1}$ \\
$q=p=20$ & $2.6030 e{-1}$ & $2.8381 e{-1}$ & $3.2033 e{-1}$ & $3.2274 e{-1}$ \\
$q=p=21$ & $2.6076 e{-1}$ & $2.8038 e{-1}$ & $3.1821 e{-1}$ & $3.2089 e{-1}$ \\
$q=p=22$ & $2.5601 e{-1}$ & $2.7905 e{-1}$ & $3.1649 e{-1}$ & $3.1885 e{-1}$ \\
$q=p=23$ & $2.5837 e{-1}$ & $2.7532 e{-1}$ & $3.1515 e{-1}$ & $3.1664 e{-1}$ \\
$q=p=24$ & $2.5548 e{-1}$ & $2.8413 e{-1}$ & $3.1456 e{-1}$ & $3.1879 e{-1}$ \\
$q=p=25$ & $2.5206 e{-1}$ & $2.8468 e{-1}$ & $3.1260 e{-1}$ & $3.2021 e{-1}$ \\
$q=p=26$ & $2.4848 e{-1}$ & $2.7970 e{-1}$ & $3.1212 e{-1}$ & $3.1858 e{-1}$ \\
$q=p=27$ & $2.5065 e{-1}$ & $2.8108 e{-1}$ & $3.1118 e{-1}$ & $3.1892 e{-1}$ \\
$q=p=28$ & $2.5681 e{-1}$ & $2.7902 e{-1}$ & $3.1319 e{-1}$ & $3.1824 e{-1}$ \\
$q=p=29$ & $2.5490 e{-1}$ & $2.7208 e{-1}$ & $3.1301 e{-1}$ & $3.1640 e{-1}$ \\
$q=p=30$ & $2.5025 e{-1}$ & $2.6217 e{-1}$ & $3.1275 e{-1}$ & $3.1849 e{-1}$ \\
$q=p=31$ & $2.4757 e{-1}$ & $2.7185 e{-1}$ & $3.1357 e{-1}$ & $3.1888 e{-1}$ \\
$q=p=32$ & $2.4519 e{-1}$ & $2.6816 e{-1}$ & $3.1397 e{-1}$ & $3.1772 e{-1}$ \\
$q=p=33$ & $-3.2415 e{-1}$ & $2.6636 e{-1}$ & $2.5329 e{-1}$ & $3.1610 e{-1}$ \\
$q=p=34$ & $-3.3517 e{-1}$ & $-2.2066 e{1}$ & $2.5427 e{-1}$ & $-1.0702 e{1}$ \\
$q=p=35$ & $-3.8890 e{-1}$ & $-1.9224 e{1}$ & $2.2239 e{-1}$ & $-1.0613 e{1}$ \\
$q=p=36$ & $-1.5627 e{1}$ & $-2.9217 e{1}$ & $-7.6096 e{-1}$ & $-1.1058 e{1}$ \\
$q=p=37$ & $-1.3481 e{1}$ & $-8.9582 e{3}$ & $-9.2275 e{1}$ & $-5.4787 e{1}$ \\
$q=p=38$ & $-1.5631 e{1}$ & $-6.8164 e{4}$ & $-9.2302 e{1}$ & $-7.1104 e{1}$ \\
$q=p=39$ & $-3.6913 e{1}$ & $-1.0058 e{5}$ & $-1.5284 e{2}$ & $-7.2362 e{1}$ \\
$q=p=40$ & $-3.4904 e{1}$ & $-1.1385 e{6}$ & $-1.3840 e{2}$ & $-8.0620 e{3}$ \\
$q=p=41$ & $-5.7584 e{3}$ & $-1.5004 e{6}$ & $-1.4184 e{2}$ & $-1.0231 e{4}$ \\
$q=p=42$ & $-3.6511 e{8}$ & $-1.4485 e{6}$ & $-2.1095 e{3}$ & $-6.3590 e{4}$ \\
$q=p=43$ & $-2.0746 e{8}$ & $-4.7278 e{8}$ & $-4.9774 e{3}$ & $-4.7846 e{7}$ \\
$q=p=44$ & $-2.5970 e{8}$ & $-5.8001 e{10}$ & $-3.7748 e{3}$ & $-1.5573 e{10}$ \\
$q=p=45$ & $-1.6952 e{8}$ & $-5.4524 e{10}$ & $-3.3020 e{4}$ & $-1.3105 e{10}$ \\
$q=p=46$ & $-1.3855 e{8}$ & $-5.9515 e{10}$ & $-1.2126 e{5}$ & $-1.5097 e{10}$ \\
$q=p=47$ & $-1.3019 e{12}$ & $-7.1320 e{10}$ & $-6.4879 e{11}$ & $-1.9357 e{10}$ \\
$q=p=48$ & $-9.4337 e{15}$ & $-4.9336 e{12}$ & $-3.6670 e{15}$ & $-7.1362 e{10}$ \\
$q=p=49$ & $-9.5105 e{15}$ & $-1.9473 e{16}$ & $-3.8326 e{15}$ & $-6.7940 e{10}$ \\
$q=p=50$ & $-2.8495 e{15}$ & $-1.0416 e{18}$ & $-8.7427 e{14}$ & $-1.1248 e{13}$ \\

\bottomrule
\end{tabular}
\end{sc}
\end{small}
\end{center}
\vskip -0.1in
\end{table}

\begin{figure}[H]
\vskip 0.2in
\begin{center}
\centerline{\includegraphics[width=\columnwidth]{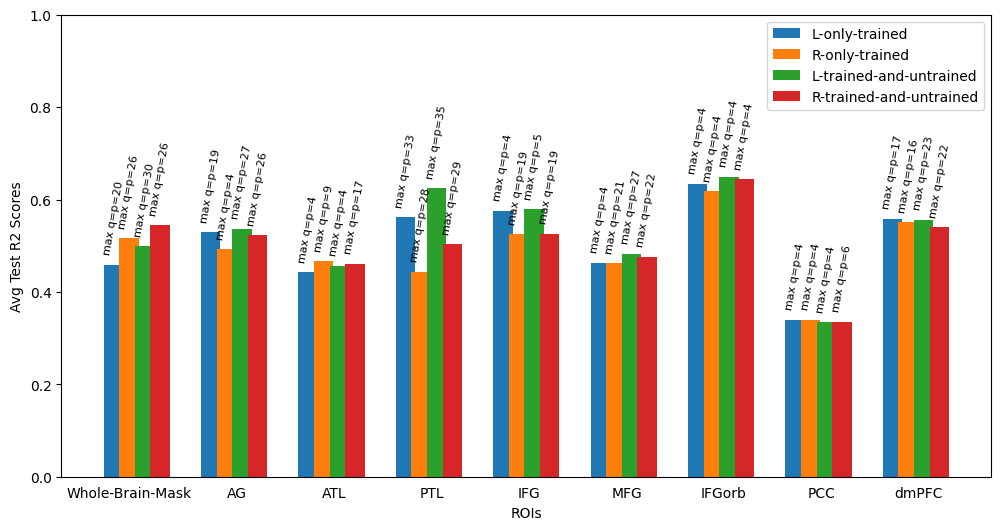}}
\caption{The figure depicts the maximum $q$ value corresponding to each ROI and hemisphere, which yields the highest averaged test $R^2$ score, as elucidated in our results for the First Pass detailed in \cref{First Pass}. Additionally, it employs distinct colors to denote whether the training dataset encompasses untrained LLMs.}
\label{max_q_best_test_R2_scores}
\end{center}
\vskip -0.2in
\end{figure}

\section{Filter in Reliable and Valid Features: Second Pass}
\label{Filter in Reliable and Valid Features: Second Pass}

This section provides the detailed results stemming from the Second Pass, as delineated in \cref{Results}. Specifically, we conducted further filtration to eliminate insignificant features by scrutinizing whether each averaged $p$ value satisfies $p < 5\%$, as detailed in \cref{Second Pass}. Within this section, each table encapsulates the features that ultimately passed through the filtering process \footnote{The feature "Intercept" is consistently included, albeit not utilized for interpretations, and is listed regardless of its $p$ value.} \footnote{We exclusively present features derived from the Second Pass in \cref{Second Pass}, ensuring their compliance with $p < 5\%$.} for each ROI (including the whole brain mask) and hemisphere (L and R) with each feature's Weight (plus both Lower and Upper Bound of $95\%$ Confidence Intervals), the Standard Error (SE) across all 25 runs, the Feature Importance $t$-statistic value $|t|$, and the corresponding $p$ value \footnote{All the above values are averaged across all the 25 runs except for the Standard Error (SE).}.

Each table is then further accompanied with two figures in its following, illustrating each feature's Weight across both Lower and Upper Bound of $95\%$ Confidence Intervals, and its Coefficient Importance and Variability respectively.

\begin{table}[H]
\caption{This table summarizes the filtered-in features generated from the Second Pass (\cref{Second Pass}) for the ROI: Whole Brain Mask and the hemisphere: L(eft) where the training data contains "Only Trained" LLMs.}
\label{only_trained_L_Whole_Brain_Mask}
\vskip 0.15in
\begin{center}
\begin{small}
\begin{sc}
\begin{tabular}{ccccccc}
\toprule
Feature & Weight & $95\%$ CI & $95\%$ CI & SE & $|t|$ & $p$  \\
 &  & Lower & Upper &  & &  \\
\midrule
(Intercept) & $1.9346 e{-2}$ & $1.8800 e{-4}$ & $3.8504 e{-2}$ & $5.8683 e{-3}$ & $3.4248 e{0}$ & $8.1760 e{-2}$ \\
Dim 0 $q = p = 1$ & $9.4292 e{-3}$ & $6.6046 e{-3}$ & $1.2254 e{-2}$ & $1.0329 e{-3}$ & $8.0441 e{0}$ & $0.0000 e{0}$ \\
Dim 0 $q = p = 2$ & $-3.5147 e{-1}$ & $-4.8224 e{-1}$ & $-2.2070 e{-1}$ & $3.4591 e{-2}$ & $9.3502 e{0}$ & $0.0000 e{0}$ \\
Dim 0 $q = p = 3$ & $1.0834 e{0}$ & $5.0134 e{-1}$ & $1.6655 e{0}$ & $1.7312 e{-1}$ & $5.3600 e{0}$ & $7.4000 e{-3}$ \\
Dim 0 $q = p = 20$ & $8.3403 e{-2}$ & $-4.1044 e{-3}$ & $1.7091 e{-1}$ & $2.4431 e{-2}$ & $4.3134 e{0}$ & $3.7840 e{-2}$ \\
Dim 0 $q = p = \infty$ & $5.2090 e{-1}$ & $2.0966 e{-1}$ & $8.3214 e{-1}$ & $5.4023 e{-2}$ & $1.0681 e{1}$ & $4.8000 e{-4}$ \\
Dim 1 $q = p = 1$ & $3.5372 e{-2}$ & $7.2709 e{-3}$ & $6.3473 e{-2}$ & $1.0270 e{-2}$ & $3.4113 e{0}$ & $4.5280 e{-2}$ \\
Dim 1 $q = p = 2$ & $-2.1050 e{0}$ & $-3.5764 e{0}$ & $-6.3351 e{-1}$ & $4.6105 e{-1}$ & $4.0441 e{0}$ & $3.9760 e{-2}$ \\
Dim 1 $q = p = \infty$ & $-1.3601 e{0}$ & $-2.5898 e{0}$ & $-1.3045 e{-1}$ & $3.1620 e{-1}$ & $5.8714 e{0}$ & $7.4800 e{-3}$ \\
Dim 2 $q = p = 11$ & $-1.9798 e{-1}$ & $-3.3485 e{-1}$ & $-6.1120 e{-2}$ & $3.0268 e{-2}$ & $6.4753 e{0}$ & $8.9200 e{-3}$ \\
Dim 2 $q = p = 12$ & $-1.5506 e{-1}$ & $-2.5847 e{-1}$ & $-5.1657 e{-2}$ & $1.9166 e{-2}$ & $8.0050 e{0}$ & $6.8000 e{-3}$ \\
Dim 2 $q = p = 16$ & $-7.5968 e{-2}$ & $-1.2718 e{-1}$ & $-2.4759 e{-2}$ & $1.3664 e{-2}$ & $5.5662 e{0}$ & $1.0080 e{-2}$ \\
Dim 2 $q = p = 20$ & $5.2582 e{-2}$ & $1.8756 e{-2}$ & $8.6409 e{-2}$ & $9.2219 e{-3}$ & $4.5193 e{0}$ & $3.6480 e{-2}$ \\

\bottomrule
\end{tabular}
\end{sc}
\end{small}
\end{center}
\vskip -0.1in
\end{table}

\def\columnWidthCoefResultsCoefsCI{1}
\def\columnWidthCoefResultsCoefsImportanceAndVariability{1}

\begin{figure}[H]
\vskip 0.2in
\begin{center}
\centerline{\includegraphics[width=\columnwidth*\columnWidthCoefResultsCoefsCI]{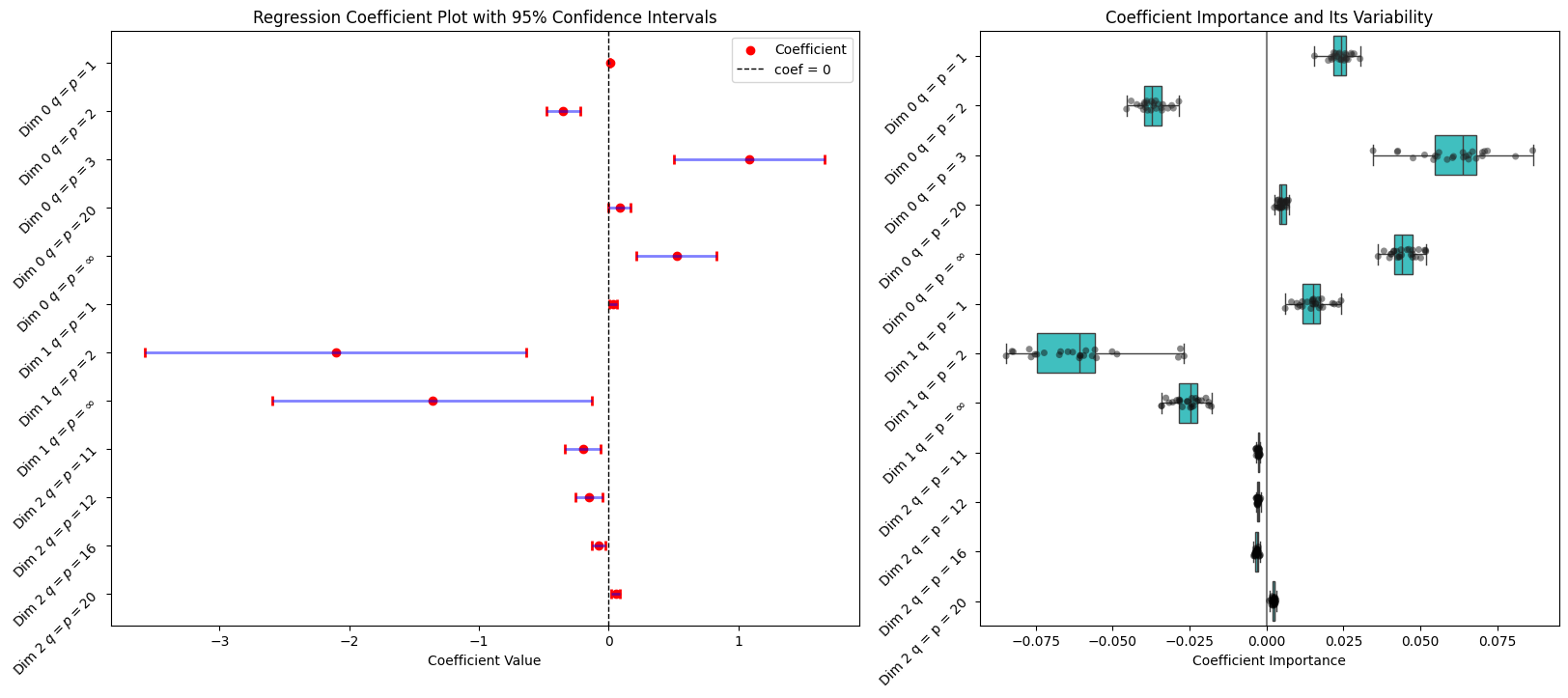}}
\caption{This figure illustrates the filtered-in features generated from the Second Pass (\cref{Second Pass}) for the ROI: Whole Brain Mask and the hemisphere: L(eft) where the training data contains "Only Trained" LLMs. On the left sub-figure, each feature's Weight is presented along with both the Lower and Upper Bounds of the $95\%$ Confidence Intervals, while the right sub-figure illustrates its Coefficient Importance and Variability.}
\label{only_trained_L_Whole_Brain_Mask_coefs_combined_20}
\end{center}
\vskip -0.2in
\end{figure}

\begin{table}[H]
\caption{This table summarizes the filtered-in features generated from the Second Pass (\cref{Second Pass}) for the ROI: Whole Brain Mask and the hemisphere: L(eft) where the training data contains both "Only Trained" and "Trained plus Untrained" LLMs.}
\label{with_untrained_L_Whole_Brain_Mask}
\vskip 0.15in
\begin{center}
\begin{small}
\begin{sc}
\begin{tabular}{ccccccc}
\toprule
Feature & Weight & $95\%$ CI & $95\%$ CI & SE & $|t|$ & $p$  \\
 &  & Lower & Upper &  & &  \\
\midrule
(Intercept) & $1.2627 e{-2}$ & $-3.8263 e{-3}$ & $2.9080 e{-2}$ & $5.2724 e{-3}$ & $3.5872 e{0}$ & $5.1160 e{-2}$ \\
Dim 0 $q = p = 1$ & $7.3745 e{-3}$ & $5.4757 e{-3}$ & $9.2733 e{-3}$ & $5.3823 e{-4}$ & $1.4333 e{1}$ & $0.0000 e{0}$ \\
Dim 0 $q = p = 2$ & $-3.1132 e{-1}$ & $-4.0518 e{-1}$ & $-2.1746 e{-1}$ & $2.6914 e{-2}$ & $1.1603 e{1}$ & $0.0000 e{0}$ \\
Dim 0 $q = p = 3$ & $9.6782 e{-1}$ & $5.5116 e{-1}$ & $1.3845 e{0}$ & $1.2708 e{-1}$ & $7.4729 e{0}$ & $8.0000 e{-5}$ \\
Dim 0 $q = p = 4$ & $-1.4886 e{0}$ & $-2.5840 e{0}$ & $-3.9314 e{-1}$ & $3.0016 e{-1}$ & $5.3065 e{0}$ & $1.2000 e{-2}$ \\
Dim 0 $q = p = 20$ & $8.0842 e{-2}$ & $1.7004 e{-2}$ & $1.4468 e{-1}$ & $1.7817 e{-2}$ & $4.3624 e{0}$ & $3.4440 e{-2}$ \\
Dim 0 $q = p = 30$ & $-7.9946 e{-2}$ & $-1.1570 e{-1}$ & $-4.4192 e{-2}$ & $1.0071 e{-2}$ & $8.6739 e{0}$ & $4.0000 e{-5}$ \\
Dim 1 $q = p = 1$ & $4.2343 e{-2}$ & $2.2814 e{-2}$ & $6.1872 e{-2}$ & $5.7902 e{-3}$ & $7.3898 e{0}$ & $3.6000 e{-4}$ \\
Dim 1 $q = p = 2$ & $-2.0656 e{0}$ & $-3.1950 e{0}$ & $-9.3612 e{-1}$ & $3.1271 e{-1}$ & $7.5218 e{0}$ & $2.8000 e{-4}$ \\
Dim 1 $q = p = 3$ & $7.2762 e{0}$ & $9.4592 e{-1}$ & $1.3607 e{1}$ & $1.8620 e{0}$ & $5.0218 e{0}$ & $1.0920 e{-2}$ \\
Dim 1 $q = p = 18$ & $-9.5618 e{-2}$ & $-1.5525 e{-1}$ & $-3.5987 e{-2}$ & $1.5532 e{-2}$ & $5.1992 e{0}$ & $1.6440 e{-2}$ \\
Dim 1 $q = p = \infty$ & $-1.1059 e{0}$ & $-1.9578 e{0}$ & $-2.5407 e{-1}$ & $2.6410 e{-1}$ & $4.5394 e{0}$ & $2.0000 e{-2}$ \\
Dim 2 $q = p = 12$ & $-1.5540 e{-1}$ & $-2.3276 e{-1}$ & $-7.8033 e{-2}$ & $2.3939 e{-2}$ & $6.3398 e{0}$ & $1.0400 e{-3}$ \\
Dim 2 $q = p = 16$ & $-3.2496 e{-2}$ & $-7.2192 e{-2}$ & $7.1994 e{-3}$ & $1.1286 e{-2}$ & $5.1708 e{0}$ & $1.1600 e{-2}$ \\
Dim 2 $q = p = 23$ & $5.2076 e{-2}$ & $2.0964 e{-2}$ & $8.3189 e{-2}$ & $6.5853 e{-3}$ & $6.8184 e{0}$ & $8.6000 e{-3}$ \\
Dim 2 $q = p = \infty$ & $-4.7243 e{0}$ & $-8.2617 e{0}$ & $-1.1869 e{0}$ & $8.5644 e{-1}$ & $5.6357 e{0}$ & $1.6480 e{-2}$ \\

\bottomrule
\end{tabular}
\end{sc}
\end{small}
\end{center}
\vskip -0.1in
\end{table}

\begin{figure}[H]
\vskip 0.2in
\begin{center}
\centerline{\includegraphics[width=\columnwidth*\columnWidthCoefResultsCoefsCI]{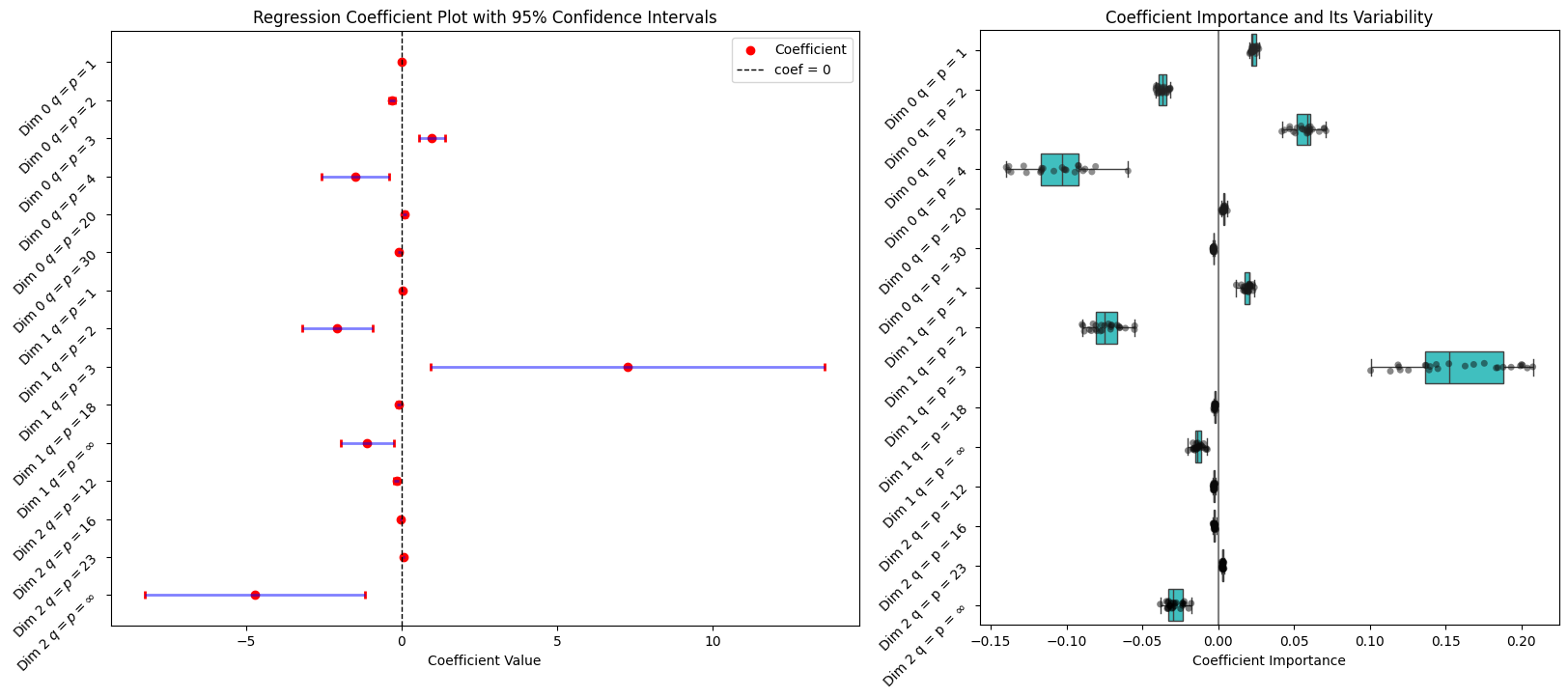}}
\caption{This figure illustrates the filtered-in features generated from the Second Pass (\cref{Second Pass}) for the ROI: Whole Brain Mask and the hemisphere: L(eft) where the training data contains both "Only Trained" and "Trained plus Untrained" LLMs. On the left sub-figure, each feature's Weight is presented along with both the Lower and Upper Bounds of the $95\%$ Confidence Intervals, while the right sub-figure illustrates its Coefficient Importance and Variability.}
\label{with_untrained_L_Whole_Brain_Mask_coefs_combined_30}
\end{center}
\vskip -0.2in
\end{figure}

\begin{table}[H]
\caption{This table summarizes the filtered-in features generated from the Second Pass (\cref{Second Pass}) for the ROI: Whole Brain Mask and the hemisphere: R(ight) where the training data contains "Only Trained" LLMs.}
\label{only_trained_R_Whole_Brain_Mask}
\vskip 0.15in
\begin{center}
\begin{small}
\begin{sc}
\begin{tabular}{ccccccc}
\toprule
Feature & Weight & $95\%$ CI & $95\%$ CI & SE & $|t|$ & $p$  \\
 &  & Lower & Upper &  & &  \\
\midrule
(Intercept) & $2.2858 e{-3}$ & $-2.1022 e{-2}$ & $2.5594 e{-2}$ & $6.4020 e{-3}$ & $1.7450 e{0}$ & $4.0984 e{-1}$ \\
Dim 0 $q = p = 1$ & $-3.5462 e{-3}$ & $-6.4684 e{-3}$ & $-6.2392 e{-4}$ & $7.9883 e{-4}$ & $4.7242 e{0}$ & $2.4320 e{-2}$ \\
Dim 0 $q = p = 11$ & $-3.8843 e{0}$ & $-5.9861 e{0}$ & $-1.7826 e{0}$ & $5.7617 e{-1}$ & $6.9440 e{0}$ & $2.4400 e{-3}$ \\
Dim 0 $q = p = 12$ & $6.4340 e{-1}$ & $-2.5351 e{-1}$ & $1.5403 e{0}$ & $2.0817 e{-1}$ & $4.9566 e{0}$ & $4.0440 e{-2}$ \\
Dim 0 $q = p = 14$ & $4.3350 e{-1}$ & $1.5303 e{-2}$ & $8.5169 e{-1}$ & $9.7953 e{-2}$ & $5.1920 e{0}$ & $2.9000 e{-2}$ \\
Dim 0 $q = p = 23$ & $8.1616 e{-2}$ & $2.5986 e{-3}$ & $1.6063 e{-1}$ & $1.8969 e{-2}$ & $4.5890 e{0}$ & $4.8920 e{-2}$ \\
Dim 0 $q = p = 26$ & $-9.9791 e{-2}$ & $-1.4632 e{-1}$ & $-5.3259 e{-2}$ & $1.5084 e{-2}$ & $5.6716 e{0}$ & $2.1600 e{-3}$ \\
Dim 1 $q = p = 11$ & $-3.6901 e{-1}$ & $-6.1385 e{-1}$ & $-1.2417 e{-1}$ & $6.3202 e{-2}$ & $5.3309 e{0}$ & $1.7080 e{-2}$ \\
Dim 1 $q = p = 26$ & $1.4624 e{-1}$ & $5.9673 e{-2}$ & $2.3281 e{-1}$ & $2.3238 e{-2}$ & $4.1770 e{0}$ & $4.8880 e{-2}$ \\
Dim 2 $q = p = 22$ & $7.4633 e{-2}$ & $2.8737 e{-2}$ & $1.2053 e{-1}$ & $1.2754 e{-2}$ & $4.3268 e{0}$ & $3.6320 e{-2}$ \\
Dim 2 $q = p = 26$ & $-6.6502 e{-2}$ & $-1.2655 e{-1}$ & $-6.4532 e{-3}$ & $1.2490 e{-2}$ & $5.9284 e{0}$ & $2.6280 e{-2}$ \\

\bottomrule
\end{tabular}
\end{sc}
\end{small}
\end{center}
\vskip -0.1in
\end{table}

\begin{figure}[H]
\vskip 0.2in
\begin{center}
\centerline{\includegraphics[width=\columnwidth*\columnWidthCoefResultsCoefsCI]{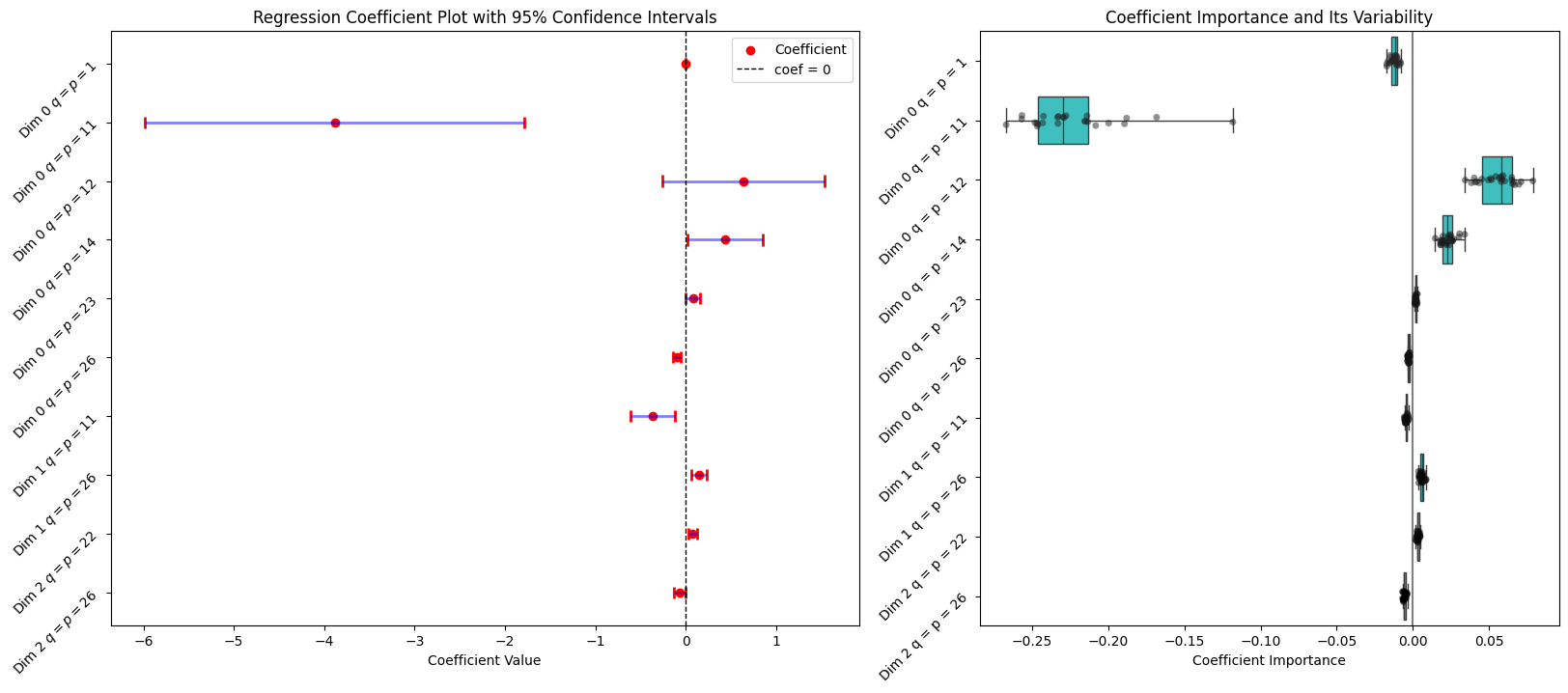}}
\caption{This figure illustrates the filtered-in features generated from the Second Pass (\cref{Second Pass}) for the ROI: Whole Brain Mask and the hemisphere: R(ight) where the training data contains "Only Trained" LLMs. On the left sub-figure, each feature's Weight is presented along with both the Lower and Upper Bounds of the $95\%$ Confidence Intervals, while the right sub-figure illustrates its Coefficient Importance and Variability.}
\label{only_trained_R_Whole_Brain_Mask_coefs_combined_26}
\end{center}
\vskip -0.2in
\end{figure}

\begin{table}[H]
\caption{This table summarizes the filtered-in features generated from the Second Pass (\cref{Second Pass}) for the ROI: Whole Brain Mask and the hemisphere: R(ight) where the training data contains both "Only Trained" and "Trained plus Untrained" LLMs.}
\label{with_untrained_R_Whole_Brain_Mask}
\vskip 0.15in
\begin{center}
\begin{small}
\begin{sc}
\begin{tabular}{ccccccc}
\toprule
Feature & Weight & $95\%$ CI & $95\%$ CI & SE & $|t|$ & $p$  \\
 &  & Lower & Upper &  & &  \\
\midrule
(Intercept) & $1.5559 e{-2}$ & $-2.6929 e{-3}$ & $3.3811 e{-2}$ & $4.7705 e{-3}$ & $2.2174 e{0}$ & $3.1344 e{-1}$ \\
Dim 0 $q = p = 2$ & $-1.8406 e{-1}$ & $-3.1568 e{-1}$ & $-5.2445 e{-2}$ & $3.8945 e{-2}$ & $3.8918 e{0}$ & $5.0000 e{-2}$ \\
Dim 0 $q = p = 3$ & $1.3628 e{0}$ & $5.3039 e{-1}$ & $2.1953 e{0}$ & $2.2354 e{-1}$ & $5.6215 e{0}$ & $1.0760 e{-2}$ \\
Dim 0 $q = p = 4$ & $-3.6328 e{0}$ & $-6.2187 e{0}$ & $-1.0470 e{0}$ & $6.7469 e{-1}$ & $5.0968 e{0}$ & $2.5160 e{-2}$ \\
Dim 0 $q = p = 20$ & $1.0482 e{-1}$ & $1.6801 e{-2}$ & $1.9284 e{-1}$ & $2.1982 e{-2}$ & $5.2974 e{0}$ & $1.7800 e{-2}$ \\
Dim 0 $q = p = 21$ & $1.0367 e{-1}$ & $3.1106 e{-2}$ & $1.7624 e{-1}$ & $1.4984 e{-2}$ & $6.1088 e{0}$ & $2.2720 e{-2}$ \\
Dim 0 $q = p = 26$ & $-5.8867 e{-2}$ & $-9.2287 e{-2}$ & $-2.5447 e{-2}$ & $8.2864 e{-3}$ & $7.5092 e{0}$ & $8.8000 e{-4}$ \\
Dim 1 $q = p = 1$ & $1.4654 e{-2}$ & $-1.9281 e{-3}$ & $3.1236 e{-2}$ & $4.3709 e{-3}$ & $5.3080 e{0}$ & $1.4680 e{-2}$ \\
Dim 1 $q = p = 11$ & $-3.1760 e{-1}$ & $-5.0292 e{-1}$ & $-1.3228 e{-1}$ & $3.6330 e{-2}$ & $8.6244 e{0}$ & $1.9200 e{-3}$ \\
Dim 1 $q = p = 13$ & $-9.4151 e{-2}$ & $-1.7372 e{-1}$ & $-1.4584 e{-2}$ & $2.0149 e{-2}$ & $5.3008 e{0}$ & $1.8200 e{-2}$ \\
Dim 2 $q = p = 13$ & $6.6123 e{-2}$ & $1.4182 e{-2}$ & $1.1807 e{-1}$ & $1.4526 e{-2}$ & $4.9253 e{0}$ & $1.8240 e{-2}$ \\
Dim 2 $q = p = 26$ & $-4.6102 e{-2}$ & $-8.5844 e{-2}$ & $-6.3609 e{-3}$ & $7.3413 e{-3}$ & $7.7983 e{0}$ & $7.7200 e{-3}$ \\

\bottomrule
\end{tabular}
\end{sc}
\end{small}
\end{center}
\vskip -0.1in
\end{table}

\begin{figure}[H]
\vskip 0.2in
\begin{center}
\centerline{\includegraphics[width=\columnwidth*\columnWidthCoefResultsCoefsCI]{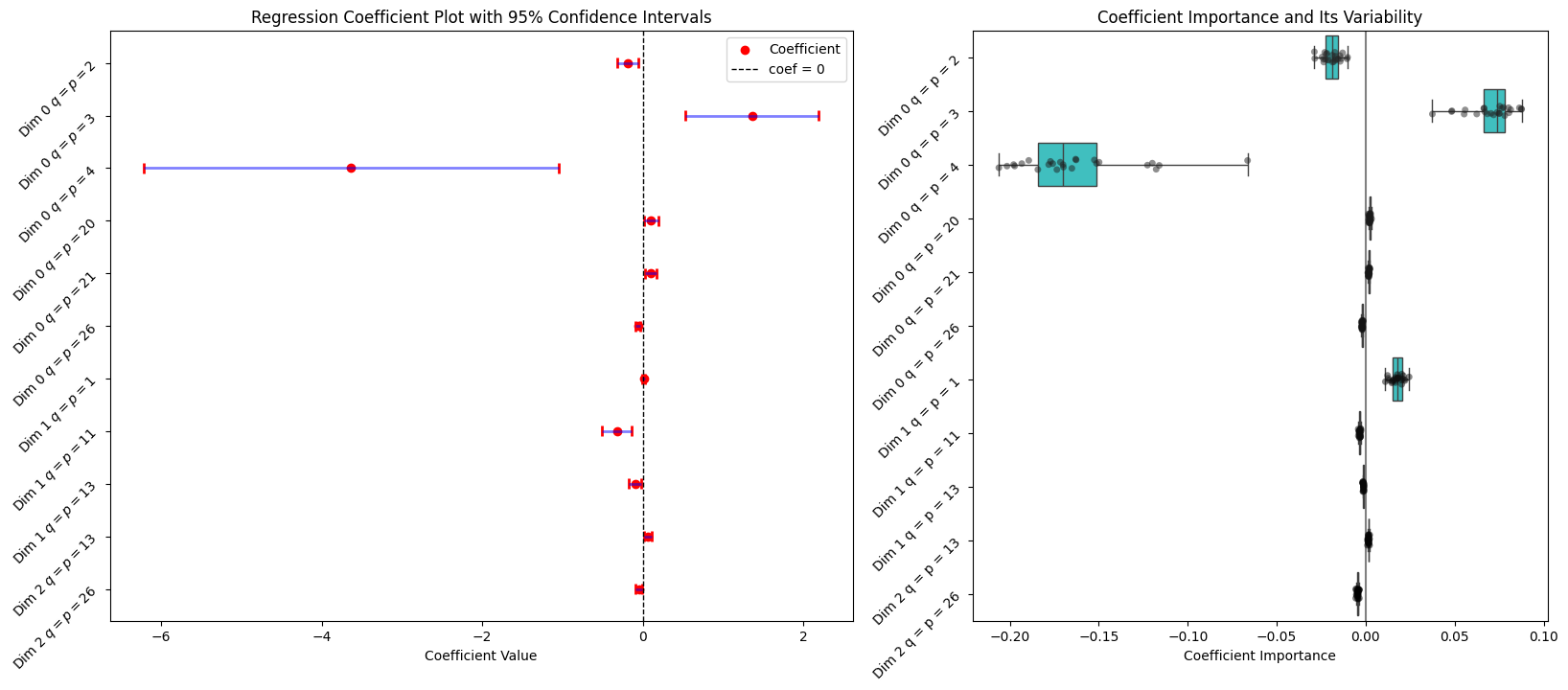}}
\caption{This figure illustrates the filtered-in features generated from the Second Pass (\cref{Second Pass}) for the ROI: Whole Brain Mask and the hemisphere: R(ight) where the training data contains both "Only Trained" and "Trained plus Untrained" LLMs. On the left sub-figure, each feature's Weight is presented along with both the Lower and Upper Bounds of the $95\%$ Confidence Intervals, while the right sub-figure illustrates its Coefficient Importance and Variability.}
\label{with_untrained_R_Whole_Brain_Mask_coefs_combined_26}
\end{center}
\vskip -0.2in
\end{figure}

\begin{table}[H]
\caption{This table summarizes the filtered-in features generated from the Second Pass (\cref{Second Pass}) for the ROI: AG and the hemisphere: L(eft) where the training data contains "Only Trained" LLMs.}
\label{only_trained_L_AG}
\vskip 0.15in
\begin{center}
\begin{small}
\begin{sc}
\begin{tabular}{ccccccc}
\toprule
Feature & Weight & $95\%$ CI & $95\%$ CI & SE & $|t|$ & $p$  \\
 &  & Lower & Upper &  & &  \\
\midrule
(Intercept) & $7.7569 e{-3}$ & $5.2319 e{-3}$ & $1.0282 e{-2}$ & $8.9245 e{-4}$ & $8.0471 e{0}$ & $0.0000 e{0}$ \\
Dim 0 $q = p = 1$ & $1.7418 e{-3}$ & $1.3126 e{-3}$ & $2.1710 e{-3}$ & $1.9325 e{-4}$ & $9.1925 e{0}$ & $0.0000 e{0}$ \\
Dim 0 $q = p = 2$ & $-8.6721 e{-2}$ & $-1.1524 e{-1}$ & $-5.8203 e{-2}$ & $1.9219 e{-2}$ & $4.9603 e{0}$ & $0.0000 e{0}$ \\
Dim 0 $q = p = 19$ & $1.7606 e{-2}$ & $2.5294 e{-3}$ & $3.2683 e{-2}$ & $4.8392 e{-3}$ & $5.4937 e{0}$ & $2.6400 e{-3}$ \\
Dim 1 $q = p = 1$ & $6.3380 e{-3}$ & $2.4140 e{-3}$ & $1.0262 e{-2}$ & $1.4778 e{-3}$ & $4.2500 e{0}$ & $1.0000 e{-2}$ \\
Dim 1 $q = p = 2$ & $-2.2954 e{-1}$ & $-4.2704 e{-1}$ & $-3.2041 e{-2}$ & $7.8173 e{-2}$ & $3.3021 e{0}$ & $3.3440 e{-2}$ \\
Dim 2 $q = p = 16$ & $4.3482 e{-3}$ & $-1.7606 e{-3}$ & $1.0457 e{-2}$ & $1.1942 e{-3}$ & $6.2958 e{0}$ & $2.5680 e{-2}$ \\
Dim 2 $q = p = 17$ & $-3.7191 e{-3}$ & $-9.1522 e{-3}$ & $1.7140 e{-3}$ & $1.2206 e{-3}$ & $4.9698 e{0}$ & $4.4200 e{-2}$ \\

\bottomrule
\end{tabular}
\end{sc}
\end{small}
\end{center}
\vskip -0.1in
\end{table}

\begin{figure}[H]
\vskip 0.2in
\begin{center}
\centerline{\includegraphics[width=\columnwidth*\columnWidthCoefResultsCoefsCI]{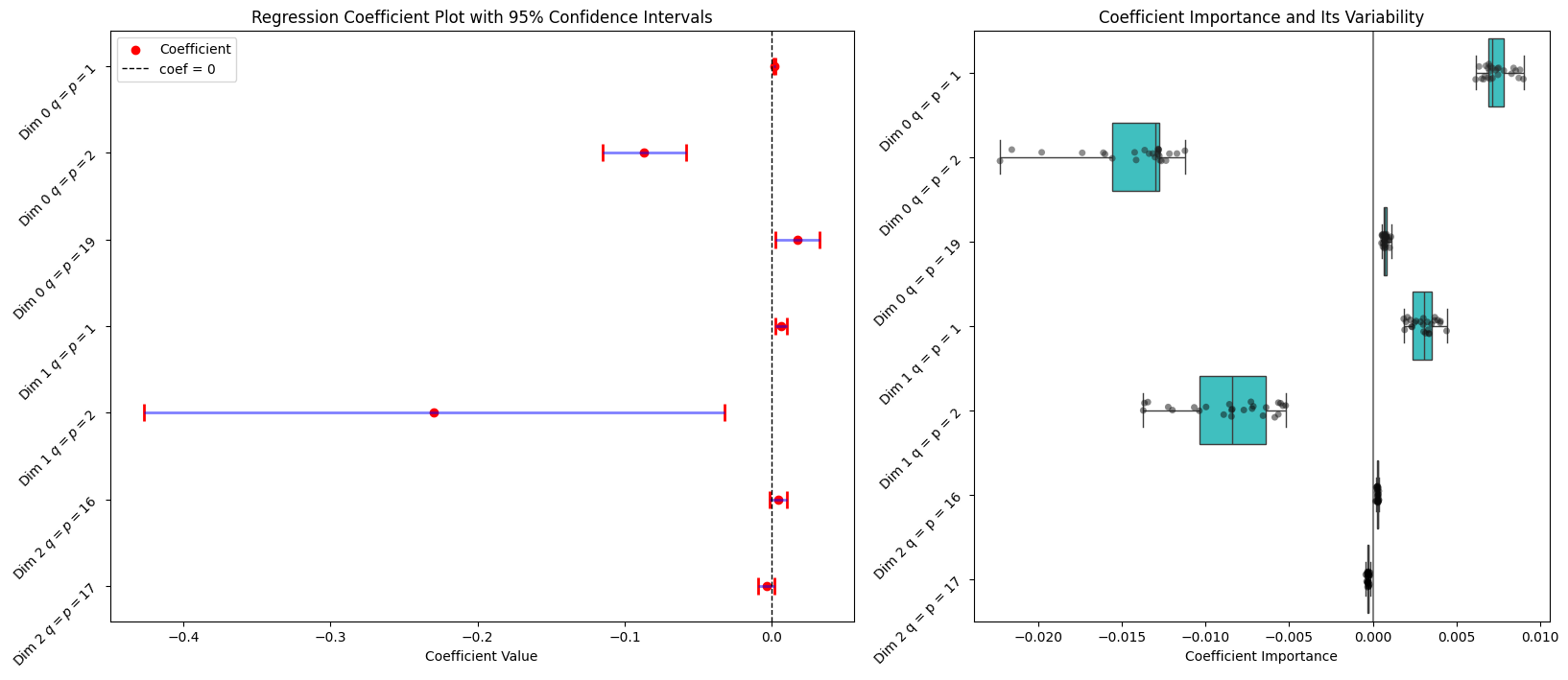}}
\caption{This figure illustrates the filtered-in features generated from the Second Pass (\cref{Second Pass}) for the ROI: AG and the hemisphere: L(eft) where the training data contains "Only Trained" LLMs. On the left sub-figure, each feature's Weight is presented along with both the Lower and Upper Bounds of the $95\%$ Confidence Intervals, while the right sub-figure illustrates its Coefficient Importance and Variability.}
\label{only_trained_L_AG_coefs_combined_19}
\end{center}
\vskip -0.2in
\end{figure}

\begin{table}[H]
\caption{This table summarizes the filtered-in features generated from the Second Pass (\cref{Second Pass}) for the ROI: AG and the hemisphere: L(eft) where the training data contains both "Only Trained" and "Trained plus Untrained" LLMs.}
\label{with_untrained_L_AG}
\vskip 0.15in
\begin{center}
\begin{small}
\begin{sc}
\begin{tabular}{ccccccc}
\toprule
Feature & Weight & $95\%$ CI & $95\%$ CI & SE & $|t|$ & $p$  \\
 &  & Lower & Upper &  & &  \\
\midrule
(Intercept) & $9.3740 e{-3}$ & $7.2599 e{-3}$ & $1.1488 e{-2}$ & $8.0076 e{-4}$ & $1.1223 e{1}$ & $0.0000 e{0}$ \\
Dim 0 $q = p = 1$ & $1.8345 e{-3}$ & $1.4837 e{-3}$ & $2.1853 e{-3}$ & $1.6255 e{-4}$ & $1.1828 e{1}$ & $0.0000 e{0}$ \\
Dim 0 $q = p = 2$ & $-8.7200 e{-2}$ & $-1.1267 e{-1}$ & $-6.1728 e{-2}$ & $2.3745 e{-2}$ & $4.3938 e{0}$ & $0.0000 e{0}$ \\
Dim 0 $q = p = 3$ & $1.2756 e{-1}$ & $-4.5414 e{-2}$ & $3.0054 e{-1}$ & $2.1260 e{-1}$ & $1.4209 e{0}$ & $2.6600 e{-2}$ \\
Dim 0 $q = p = 6$ & $5.7342 e{0}$ & $2.7947 e{0}$ & $8.6736 e{0}$ & $1.3476 e{0}$ & $3.4930 e{0}$ & $4.8840 e{-2}$ \\
Dim 0 $q = p = 7$ & $-4.8757 e{0}$ & $-7.6865 e{0}$ & $-2.0649 e{0}$ & $7.1900 e{-1}$ & $5.8813 e{0}$ & $1.2680 e{-2}$ \\
Dim 0 $q = p = 8$ & $1.9270 e{0}$ & $6.1853 e{-1}$ & $3.2355 e{0}$ & $2.8687 e{-1}$ & $5.6867 e{0}$ & $2.5040 e{-2}$ \\
Dim 0 $q = p = 16$ & $-1.8583 e{-2}$ & $-3.2215 e{-2}$ & $-4.9514 e{-3}$ & $3.7571 e{-3}$ & $4.5648 e{0}$ & $3.1000 e{-2}$ \\
Dim 0 $q = p = 18$ & $-1.7607 e{-2}$ & $-2.8040 e{-2}$ & $-7.1743 e{-3}$ & $2.9363 e{-3}$ & $6.5962 e{0}$ & $1.5200 e{-3}$ \\
Dim 0 $q = p = 19$ & $1.6683 e{-2}$ & $2.9630 e{-3}$ & $3.0403 e{-2}$ & $4.9227 e{-3}$ & $3.9165 e{0}$ & $2.1520 e{-2}$ \\
Dim 1 $q = p = 1$ & $6.1574 e{-3}$ & $3.2007 e{-3}$ & $9.1141 e{-3}$ & $8.8558 e{-4}$ & $6.6509 e{0}$ & $5.6000 e{-4}$ \\
Dim 1 $q = p = 2$ & $-3.3154 e{-1}$ & $-4.9844 e{-1}$ & $-1.6464 e{-1}$ & $6.6522 e{-2}$ & $4.6371 e{0}$ & $3.4400 e{-3}$ \\
Dim 1 $q = p = 10$ & $6.3321 e{-2}$ & $1.9917 e{-2}$ & $1.0672 e{-1}$ & $1.2613 e{-2}$ & $4.7153 e{0}$ & $1.9000 e{-2}$ \\

\bottomrule
\end{tabular}
\end{sc}
\end{small}
\end{center}
\vskip -0.1in
\end{table}

\begin{figure}[H]
\vskip 0.2in
\begin{center}
\centerline{\includegraphics[width=\columnwidth*\columnWidthCoefResultsCoefsCI]{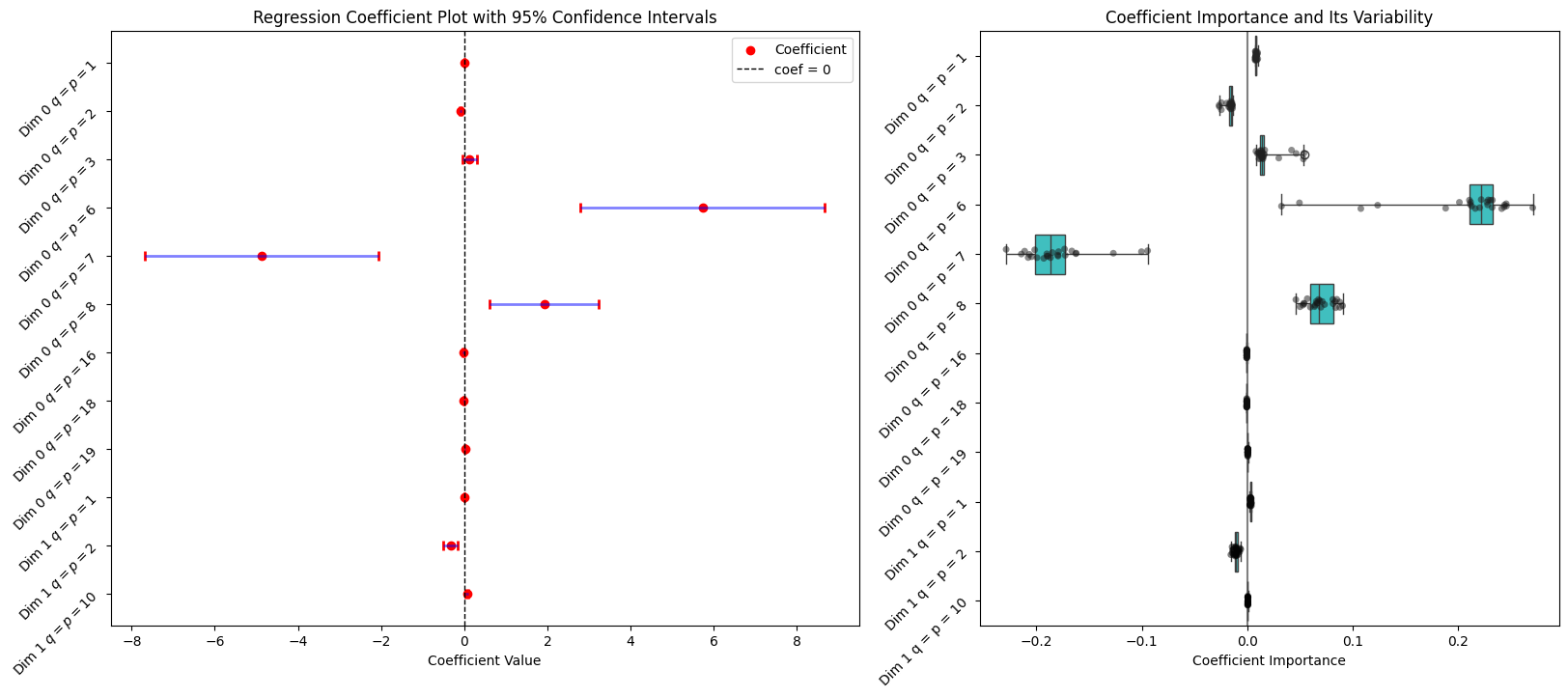}}
\caption{This figure illustrates the filtered-in features generated from the Second Pass (\cref{Second Pass}) for the ROI: AG and the hemisphere: L(eft) where the training data contains both "Only Trained" and "Trained plus Untrained" LLMs. On the left sub-figure, each feature's Weight is presented along with both the Lower and Upper Bounds of the $95\%$ Confidence Intervals, while the right sub-figure illustrates its Coefficient Importance and Variability.}
\label{with_untrained_L_AG_coefs_combined_27}
\end{center}
\vskip -0.2in
\end{figure}

\begin{table}[H]
\caption{This table summarizes the filtered-in features generated from the Second Pass (\cref{Second Pass}) for the ROI: AG and the hemisphere: R(ight) where the training data contains "Only Trained" LLMs.}
\label{only_trained_R_AG}
\vskip 0.15in
\begin{center}
\begin{small}
\begin{sc}
\begin{tabular}{ccccccc}
\toprule
Feature & Weight & $95\%$ CI & $95\%$ CI & SE & $|t|$ & $p$  \\
 &  & Lower & Upper &  & &  \\
\midrule
(Intercept) & $4.7500 e{-3}$ & $3.7984 e{-3}$ & $5.7016 e{-3}$ & $2.6542 e{-4}$ & $1.8713 e{1}$ & $0.0000 e{0}$ \\
Dim 0 $q = p = 1$ & $1.0109 e{-3}$ & $5.9847 e{-4}$ & $1.4233 e{-3}$ & $1.8252 e{-4}$ & $5.4804 e{0}$ & $1.2000 e{-4}$ \\
Dim 0 $q = p = 2$ & $-8.1555 e{-2}$ & $-1.0243 e{-1}$ & $-6.0683 e{-2}$ & $8.7707 e{-3}$ & $1.0170 e{1}$ & $0.0000 e{0}$ \\
Dim 0 $q = p = 3$ & $2.5801 e{-1}$ & $1.8472 e{-1}$ & $3.3130 e{-1}$ & $2.8475 e{-2}$ & $1.0503 e{1}$ & $0.0000 e{0}$ \\
Dim 0 $q = p = 4$ & $-1.8818 e{-1}$ & $-2.5087 e{-1}$ & $-1.2549 e{-1}$ & $2.1287 e{-2}$ & $1.0487 e{1}$ & $0.0000 e{0}$ \\
Dim 0 $q = p = \infty$ & $2.0329 e{-2}$ & $7.9035 e{-3}$ & $3.2755 e{-2}$ & $2.7765 e{-3}$ & $8.4606 e{0}$ & $4.8000 e{-4}$ \\
Dim 1 $q = p = 1$ & $1.5488 e{-2}$ & $1.1544 e{-2}$ & $1.9433 e{-2}$ & $1.1860 e{-3}$ & $1.2535 e{1}$ & $0.0000 e{0}$ \\
Dim 1 $q = p = 2$ & $-4.2374 e{-1}$ & $-5.8951 e{-1}$ & $-2.5796 e{-1}$ & $4.9491 e{-2}$ & $7.6941 e{0}$ & $8.0000 e{-5}$ \\
Dim 1 $q = p = 3$ & $1.0854 e{0}$ & $5.0985 e{-1}$ & $1.6610 e{0}$ & $1.7182 e{-1}$ & $5.3370 e{0}$ & $6.7200 e{-3}$ \\

\bottomrule
\end{tabular}
\end{sc}
\end{small}
\end{center}
\vskip -0.1in
\end{table}

\begin{figure}[H]
\vskip 0.2in
\begin{center}
\centerline{\includegraphics[width=\columnwidth*\columnWidthCoefResultsCoefsCI]{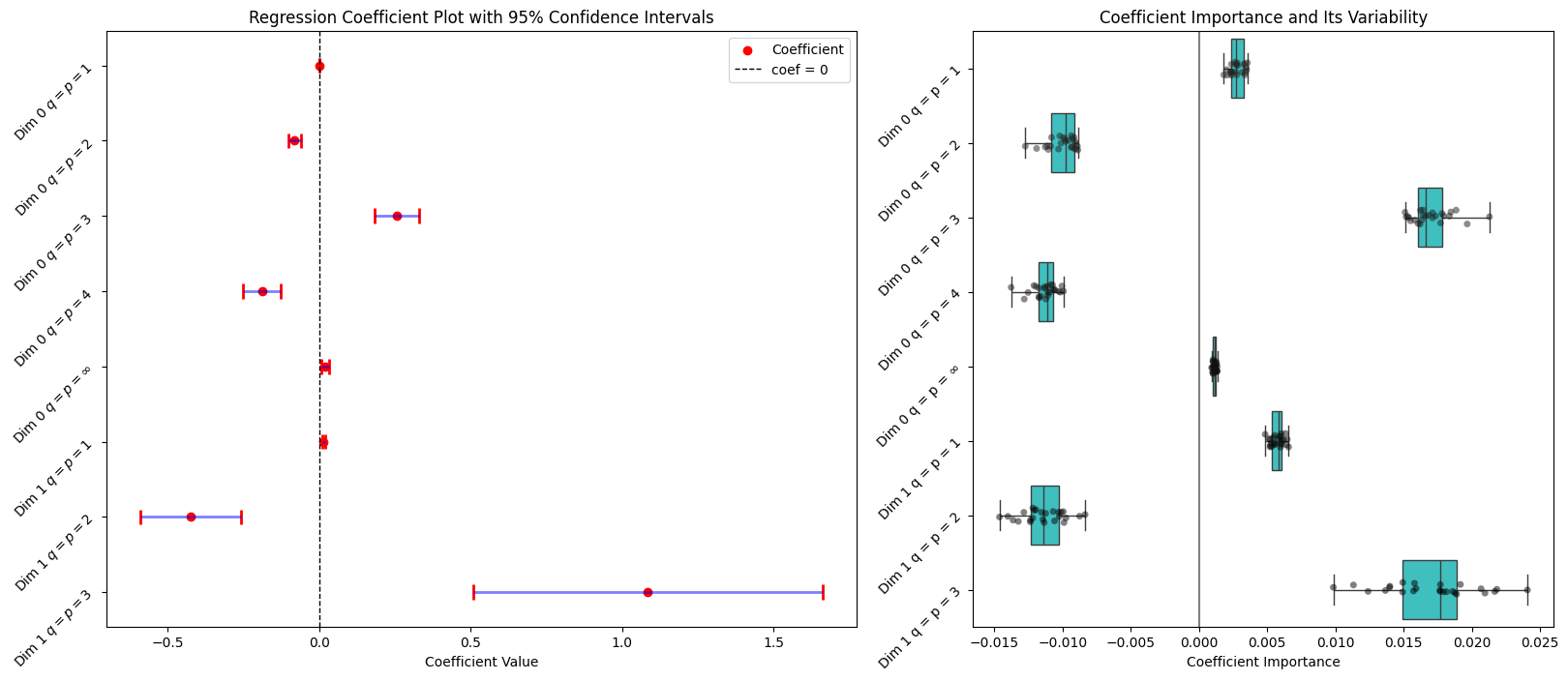}}
\caption{This figure illustrates the filtered-in features generated from the Second Pass (\cref{Second Pass}) for the ROI: AG and the hemisphere: R(ight) where the training data contains "Only Trained" LLMs. On the left sub-figure, each feature's Weight is presented along with both the Lower and Upper Bounds of the $95\%$ Confidence Intervals, while the right sub-figure illustrates its Coefficient Importance and Variability.}
\label{only_trained_R_AG_coefs_combined_4}
\end{center}
\vskip -0.2in
\end{figure}

\begin{table}[H]
\caption{This table summarizes the filtered-in features generated from the Second Pass (\cref{Second Pass}) for the ROI: AG and the hemisphere: R(ight) where the training data contains both "Only Trained" and "Trained plus Untrained" LLMs.}
\label{with_untrained_R_AG}
\vskip 0.15in
\begin{center}
\begin{small}
\begin{sc}
\begin{tabular}{ccccccc}
\toprule
Feature & Weight & $95\%$ CI & $95\%$ CI & SE & $|t|$ & $p$  \\
 &  & Lower & Upper &  & &  \\
\midrule
(Intercept) & $-8.2385 e{-4}$ & $-3.4681 e{-3}$ & $1.8204 e{-3}$ & $6.6444 e{-4}$ & $8.8795 e{-1}$ & $6.7984 e{-1}$ \\
Dim 0 $q = p = 1$ & $1.4035 e{-3}$ & $1.0627 e{-3}$ & $1.7444 e{-3}$ & $9.3760 e{-5}$ & $1.6129 e{1}$ & $0.0000 e{0}$ \\
Dim 0 $q = p = 2$ & $-9.0451 e{-2}$ & $-1.1485 e{-1}$ & $-6.6048 e{-2}$ & $7.8371 e{-3}$ & $1.1291 e{1}$ & $0.0000 e{0}$ \\
Dim 0 $q = p = 3$ & $3.5668 e{-1}$ & $1.7908 e{-1}$ & $5.3428 e{-1}$ & $5.1019 e{-2}$ & $5.9654 e{0}$ & $2.7200 e{-3}$ \\
Dim 1 $q = p = 1$ & $6.2163 e{-3}$ & $2.9964 e{-3}$ & $9.4362 e{-3}$ & $1.0081 e{-3}$ & $5.4609 e{0}$ & $4.0800 e{-3}$ \\
Dim 1 $q = p = 2$ & $-2.3875 e{-1}$ & $-4.1399 e{-1}$ & $-6.3505 e{-2}$ & $5.5314 e{-2}$ & $3.9595 e{0}$ & $3.4360 e{-2}$ \\
Dim 1 $q = p = 14$ & $1.5117 e{-2}$ & $2.9759 e{-3}$ & $2.7258 e{-2}$ & $3.9668 e{-3}$ & $3.9243 e{0}$ & $3.4200 e{-2}$ \\
Dim 2 $q = p = 10$ & $4.0770 e{-2}$ & $1.5110 e{-2}$ & $6.6430 e{-2}$ & $6.6048 e{-3}$ & $5.1724 e{0}$ & $1.8000 e{-2}$ \\
Dim 2 $q = p = \infty$ & $-7.4708 e{-1}$ & $-1.2996 e{0}$ & $-1.9459 e{-1}$ & $1.2706 e{-1}$ & $6.0437 e{0}$ & $1.2680 e{-2}$ \\

\bottomrule
\end{tabular}
\end{sc}
\end{small}
\end{center}
\vskip -0.1in
\end{table}

\begin{figure}[H]
\vskip 0.2in
\begin{center}
\centerline{\includegraphics[width=\columnwidth*\columnWidthCoefResultsCoefsCI]{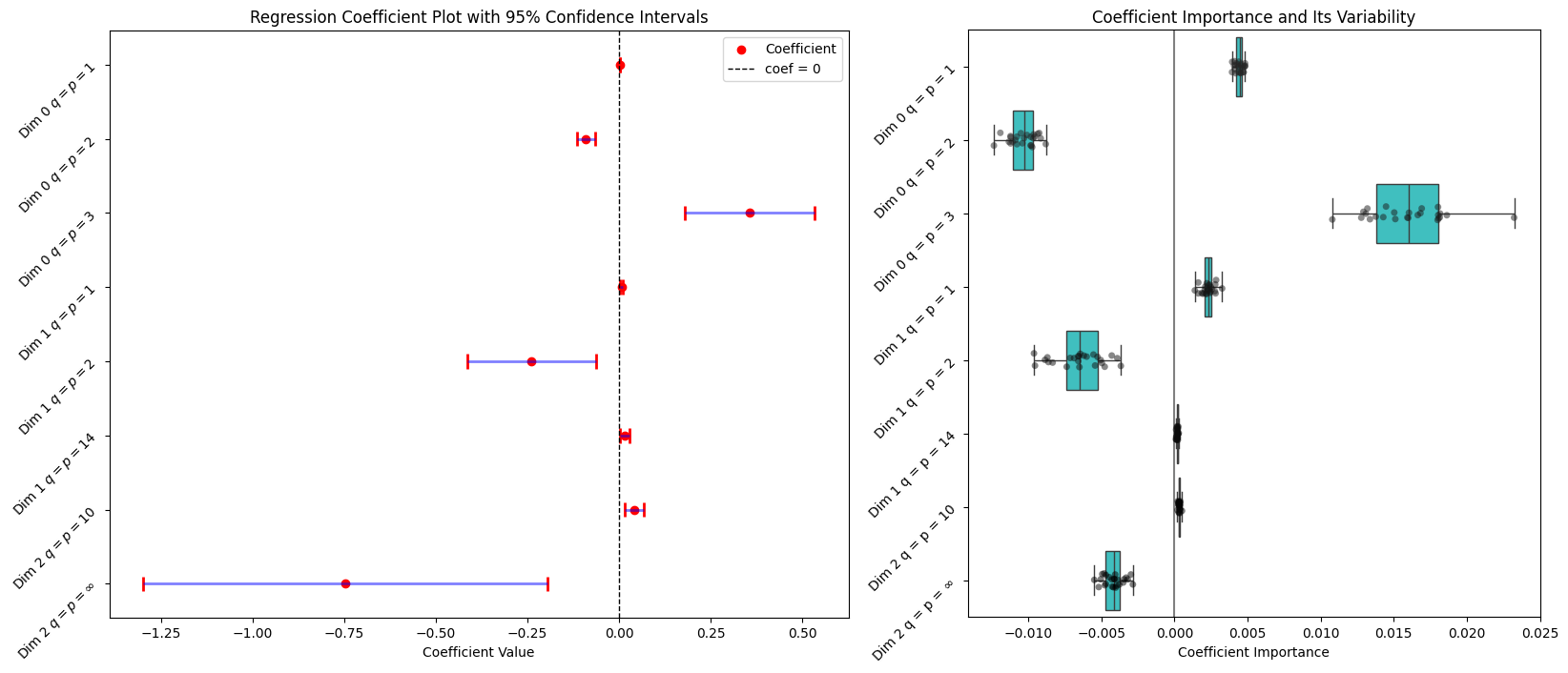}}
\caption{This figure illustrates the filtered-in features generated from the Second Pass (\cref{Second Pass}) for the ROI: AG and the hemisphere: R(ight) where the training data contains both "Only Trained" and "Trained plus Untrained" LLMs. On the left sub-figure, each feature's Weight is presented along with both the Lower and Upper Bounds of the $95\%$ Confidence Intervals, while the right sub-figure illustrates its Coefficient Importance and Variability.}
\label{with_untrained_R_AG_coefs_combined_26}
\end{center}
\vskip -0.2in
\end{figure}

\begin{table}[H]
\caption{This table summarizes the filtered-in features generated from the Second Pass (\cref{Second Pass}) for the ROI: ATL and the hemisphere: L(eft) where the training data contains "Only Trained" LLMs.}
\label{only_trained_L_ATL}
\vskip 0.15in
\begin{center}
\begin{small}
\begin{sc}
\begin{tabular}{ccccccc}
\toprule
Feature & Weight & $95\%$ CI & $95\%$ CI & SE & $|t|$ & $p$  \\
 &  & Lower & Upper &  & &  \\
\midrule
(Intercept) & $2.6347 e{-3}$ & $2.0549 e{-3}$ & $3.2144 e{-3}$ & $1.6393 e{-4}$ & $1.7366 e{1}$ & $0.0000 e{0}$ \\
Dim 0 $q = p = 1$ & $2.1014 e{-4}$ & $2.7715 e{-5}$ & $3.9257 e{-4}$ & $4.1501 e{-5}$ & $6.0691 e{0}$ & $1.3320 e{-2}$ \\
Dim 0 $q = p = 2$ & $-2.0689 e{-2}$ & $-2.9585 e{-2}$ & $-1.1793 e{-2}$ & $1.9951 e{-3}$ & $1.0714 e{1}$ & $0.0000 e{0}$ \\
Dim 0 $q = p = 3$ & $6.7637 e{-2}$ & $4.0634 e{-2}$ & $9.4640 e{-2}$ & $6.5712 e{-3}$ & $1.0587 e{1}$ & $0.0000 e{0}$ \\
Dim 0 $q = p = 4$ & $-4.1862 e{-2}$ & $-6.2291 e{-2}$ & $-2.1433 e{-2}$ & $5.6908 e{-3}$ & $7.5298 e{0}$ & $2.4000 e{-4}$ \\
Dim 0 $q = p = \infty$ & $-7.4256 e{-3}$ & $-1.1949 e{-2}$ & $-2.9021 e{-3}$ & $1.4971 e{-3}$ & $5.3081 e{0}$ & $3.3600 e{-3}$ \\

\bottomrule
\end{tabular}
\end{sc}
\end{small}
\end{center}
\vskip -0.1in
\end{table}

\begin{figure}[H]
\vskip 0.2in
\begin{center}
\centerline{\includegraphics[width=\columnwidth*\columnWidthCoefResultsCoefsCI]{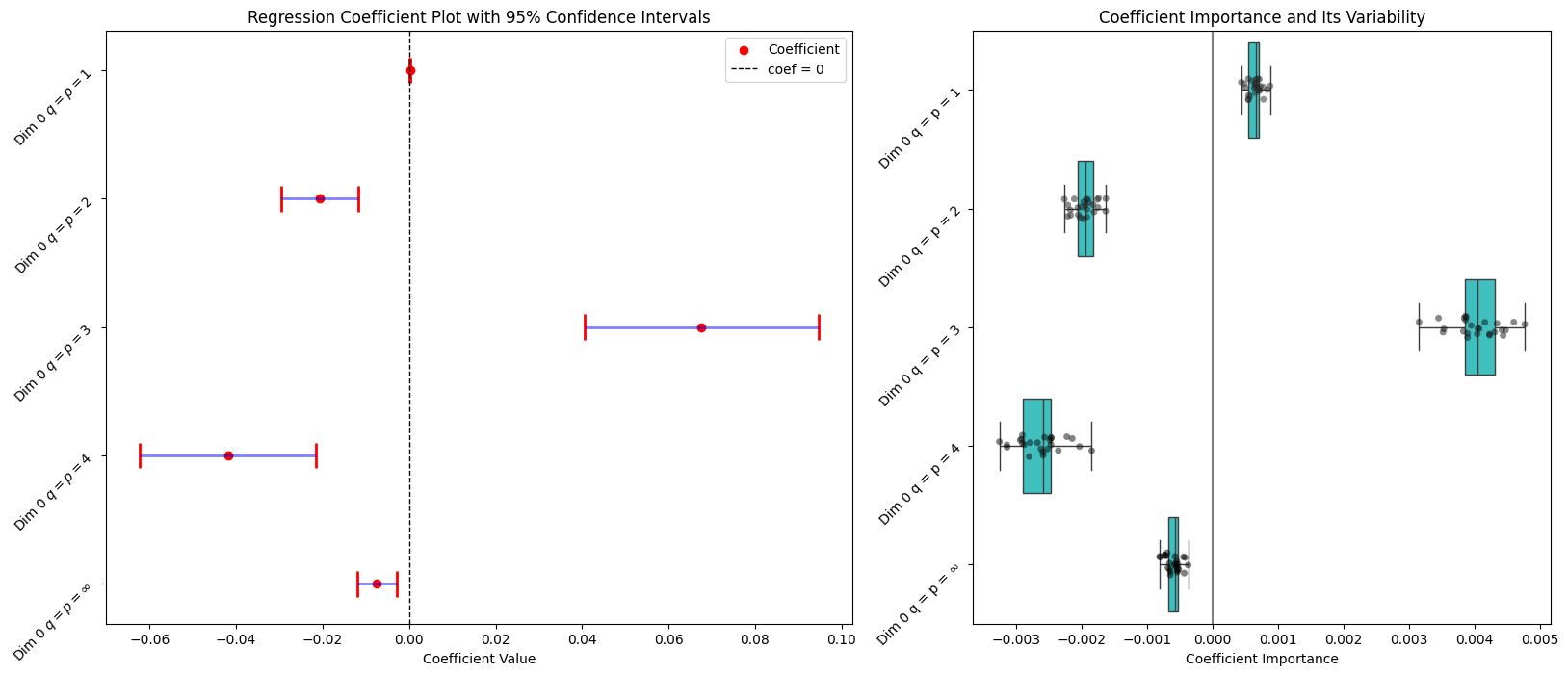}}
\caption{This figure illustrates the filtered-in features generated from the Second Pass (\cref{Second Pass}) for the ROI: ATL and the hemisphere: L(eft) where the training data contains "Only Trained" LLMs. On the left sub-figure, each feature's Weight is presented along with both the Lower and Upper Bounds of the $95\%$ Confidence Intervals, while the right sub-figure illustrates its Coefficient Importance and Variability.}
\label{only_trained_L_ATL_coefs_combined_4}
\end{center}
\vskip -0.2in
\end{figure}

\begin{table}[H]
\caption{This table summarizes the filtered-in features generated from the Second Pass (\cref{Second Pass}) for the ROI: ATL and the hemisphere: L(eft) where the training data contains both "Only Trained" and "Trained plus Untrained" LLMs.}
\label{with_untrained_L_ATL}
\vskip 0.15in
\begin{center}
\begin{small}
\begin{sc}
\begin{tabular}{ccccccc}
\toprule
Feature & Weight & $95\%$ CI & $95\%$ CI & SE & $|t|$ & $p$  \\
 &  & Lower & Upper &  & &  \\
\midrule
(Intercept) & $3.5153 e{-3}$ & $3.0754 e{-3}$ & $3.9552 e{-3}$ & $1.1921 e{-4}$ & $3.0574 e{1}$ & $0.0000 e{0}$ \\
Dim 0 $q = p = 1$ & $4.1029 e{-4}$ & $2.7661 e{-4}$ & $5.4397 e{-4}$ & $3.5847 e{-5}$ & $1.1788 e{1}$ & $0.0000 e{0}$ \\
Dim 0 $q = p = 2$ & $-2.6809 e{-2}$ & $-3.3389 e{-2}$ & $-2.0228 e{-2}$ & $2.0654 e{-3}$ & $1.2977 e{1}$ & $0.0000 e{0}$ \\
Dim 0 $q = p = 3$ & $7.7520 e{-2}$ & $5.7995 e{-2}$ & $9.7044 e{-2}$ & $5.5108 e{-3}$ & $1.3990 e{1}$ & $0.0000 e{0}$ \\
Dim 0 $q = p = 4$ & $-4.7567 e{-2}$ & $-6.2149 e{-2}$ & $-3.2984 e{-2}$ & $3.6455 e{-3}$ & $1.2877 e{1}$ & $0.0000 e{0}$ \\
Dim 0 $q = p = \infty$ & $-6.3964 e{-3}$ & $-9.5608 e{-3}$ & $-3.2319 e{-3}$ & $6.0558 e{-4}$ & $1.0680 e{1}$ & $1.2000 e{-4}$ \\

\bottomrule
\end{tabular}
\end{sc}
\end{small}
\end{center}
\vskip -0.1in
\end{table}

\begin{figure}[H]
\vskip 0.2in
\begin{center}
\centerline{\includegraphics[width=\columnwidth*\columnWidthCoefResultsCoefsCI]{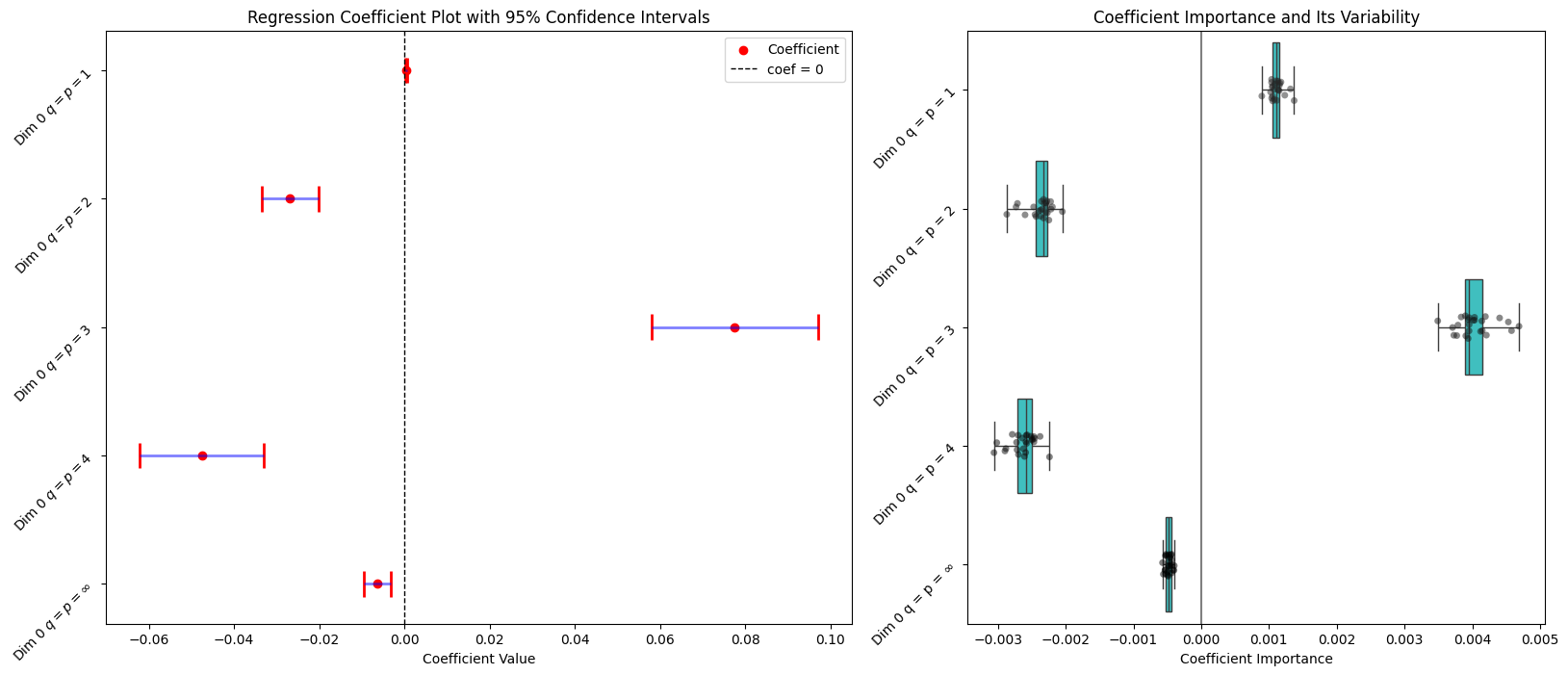}}
\caption{This figure illustrates the filtered-in features generated from the Second Pass (\cref{Second Pass}) for the ROI: ATL and the hemisphere: L(eft) where the training data contains both "Only Trained" and "Trained plus Untrained" LLMs. On the left sub-figure, each feature's Weight is presented along with both the Lower and Upper Bounds of the $95\%$ Confidence Intervals, while the right sub-figure illustrates its Coefficient Importance and Variability.}
\label{with_untrained_L_ATL_coefs_combined_4}
\end{center}
\vskip -0.2in
\end{figure}

\begin{table}[H]
\caption{This table summarizes the filtered-in features generated from the Second Pass (\cref{Second Pass}) for the ROI: ATL and the hemisphere: R(ight) where the training data contains "Only Trained" LLMs.}
\label{only_trained_R_ATL}
\vskip 0.15in
\begin{center}
\begin{small}
\begin{sc}
\begin{tabular}{ccccccc}
\toprule
Feature & Weight & $95\%$ CI & $95\%$ CI & SE & $|t|$ & $p$  \\
 &  & Lower & Upper &  & &  \\
\midrule
(Intercept) & $3.7035 e{-3}$ & $3.1671 e{-3}$ & $4.2399 e{-3}$ & $1.1697 e{-4}$ & $3.1591 e{1}$ & $0.0000 e{0}$ \\
Dim 0 $q = p = 1$ & $4.8066 e{-4}$ & $2.8894 e{-4}$ & $6.7239 e{-4}$ & $5.8894 e{-5}$ & $7.6347 e{0}$ & $4.0000 e{-5}$ \\
Dim 0 $q = p = 2$ & $-4.4705 e{-2}$ & $-5.9075 e{-2}$ & $-3.0335 e{-2}$ & $3.9365 e{-3}$ & $1.0551 e{1}$ & $0.0000 e{0}$ \\
Dim 0 $q = p = 3$ & $2.5911 e{-1}$ & $1.6485 e{-1}$ & $3.5338 e{-1}$ & $2.1661 e{-2}$ & $1.1428 e{1}$ & $0.0000 e{0}$ \\
Dim 0 $q = p = 4$ & $-3.8620 e{-1}$ & $-6.8443 e{-1}$ & $-8.7968 e{-2}$ & $6.1739 e{-2}$ & $6.7995 e{0}$ & $9.8400 e{-3}$ \\
Dim 1 $q = p = 2$ & $-7.7284 e{-2}$ & $-1.5978 e{-1}$ & $5.2087 e{-3}$ & $1.9573 e{-2}$ & $4.6817 e{0}$ & $4.7920 e{-2}$ \\
Dim 2 $q = p = 8$ & $-1.6718 e{-1}$ & $-2.4964 e{-1}$ & $-8.4720 e{-2}$ & $2.4815 e{-2}$ & $6.1501 e{0}$ & $1.4400 e{-3}$ \\

\bottomrule
\end{tabular}
\end{sc}
\end{small}
\end{center}
\vskip -0.1in
\end{table}

\begin{figure}[H]
\vskip 0.2in
\begin{center}
\centerline{\includegraphics[width=\columnwidth*\columnWidthCoefResultsCoefsCI]{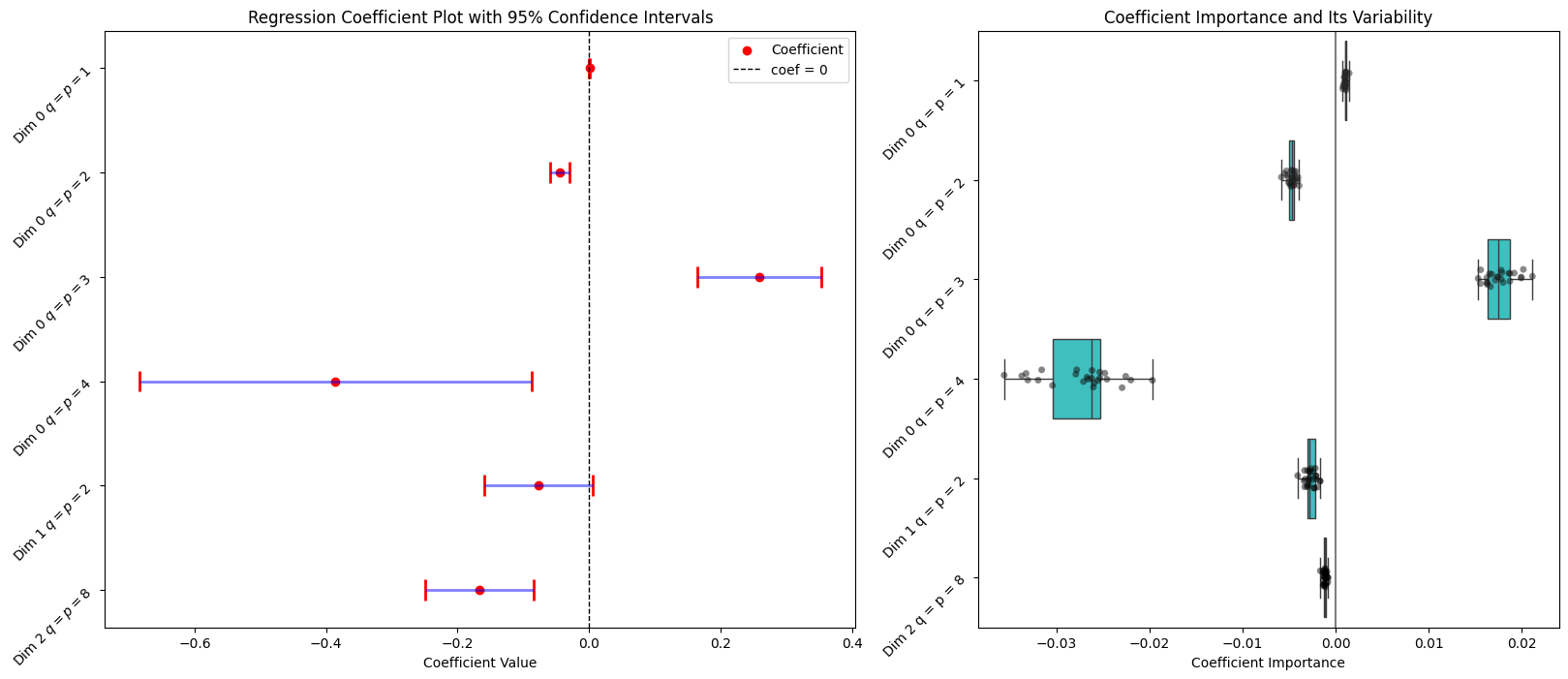}}
\caption{This figure illustrates the filtered-in features generated from the Second Pass (\cref{Second Pass}) for the ROI: ATL and the hemisphere: R(ight) where the training data contains "Only Trained" LLMs. On the left sub-figure, each feature's Weight is presented along with both the Lower and Upper Bounds of the $95\%$ Confidence Intervals, while the right sub-figure illustrates its Coefficient Importance and Variability.}
\label{only_trained_R_ATL_coefs_combined_9}
\end{center}
\vskip -0.2in
\end{figure}

\begin{table}[H]
\caption{This table summarizes the filtered-in features generated from the Second Pass (\cref{Second Pass}) for the ROI: ATL and the hemisphere: R(ight) where the training data contains both "Only Trained" and "Trained plus Untrained" LLMs.}
\label{with_untrained_R_ATL}
\vskip 0.15in
\begin{center}
\begin{small}
\begin{sc}
\begin{tabular}{ccccccc}
\toprule
Feature & Weight & $95\%$ CI & $95\%$ CI & SE & $|t|$ & $p$  \\
 &  & Lower & Upper &  & &  \\
\midrule
(Intercept) & $3.7193 e{-3}$ & $2.8529 e{-3}$ & $4.5856 e{-3}$ & $2.4718 e{-4}$ & $1.5594 e{1}$ & $0.0000 e{0}$ \\
Dim 0 $q = p = 1$ & $5.5864 e{-4}$ & $4.0255 e{-4}$ & $7.1474 e{-4}$ & $2.8772 e{-5}$ & $1.9914 e{1}$ & $0.0000 e{0}$ \\
Dim 0 $q = p = 2$ & $-4.3459 e{-2}$ & $-5.5079 e{-2}$ & $-3.1838 e{-2}$ & $2.3736 e{-3}$ & $1.8322 e{1}$ & $0.0000 e{0}$ \\
Dim 0 $q = p = 3$ & $2.2427 e{-1}$ & $1.5096 e{-1}$ & $2.9757 e{-1}$ & $1.8438 e{-2}$ & $1.2110 e{1}$ & $0.0000 e{0}$ \\
Dim 0 $q = p = 4$ & $-3.2287 e{-1}$ & $-5.5225 e{-1}$ & $-9.3487 e{-2}$ & $5.8378 e{-2}$ & $6.0121 e{0}$ & $8.6400 e{-3}$ \\
Dim 0 $q = p = 13$ & $-3.9876 e{-2}$ & $-6.4446 e{-2}$ & $-1.5307 e{-2}$ & $7.6197 e{-3}$ & $5.7064 e{0}$ & $2.6800 e{-3}$ \\
Dim 0 $q = p = 14$ & $2.7933 e{-2}$ & $4.5715 e{-3}$ & $5.1295 e{-2}$ & $7.6033 e{-3}$ & $4.0033 e{0}$ & $3.4480 e{-2}$ \\
Dim 1 $q = p = 17$ & $-3.4450 e{-3}$ & $-6.4170 e{-3}$ & $-4.7290 e{-4}$ & $6.1856 e{-4}$ & $6.1534 e{0}$ & $2.0560 e{-2}$ \\
Dim 2 $q = p = 8$ & $-8.6516 e{-2}$ & $-1.5178 e{-1}$ & $-2.1253 e{-2}$ & $1.7083 e{-2}$ & $7.1545 e{0}$ & $1.2000 e{-3}$ \\

\bottomrule
\end{tabular}
\end{sc}
\end{small}
\end{center}
\vskip -0.1in
\end{table}

\begin{figure}[H]
\vskip 0.2in
\begin{center}
\centerline{\includegraphics[width=\columnwidth*\columnWidthCoefResultsCoefsCI]{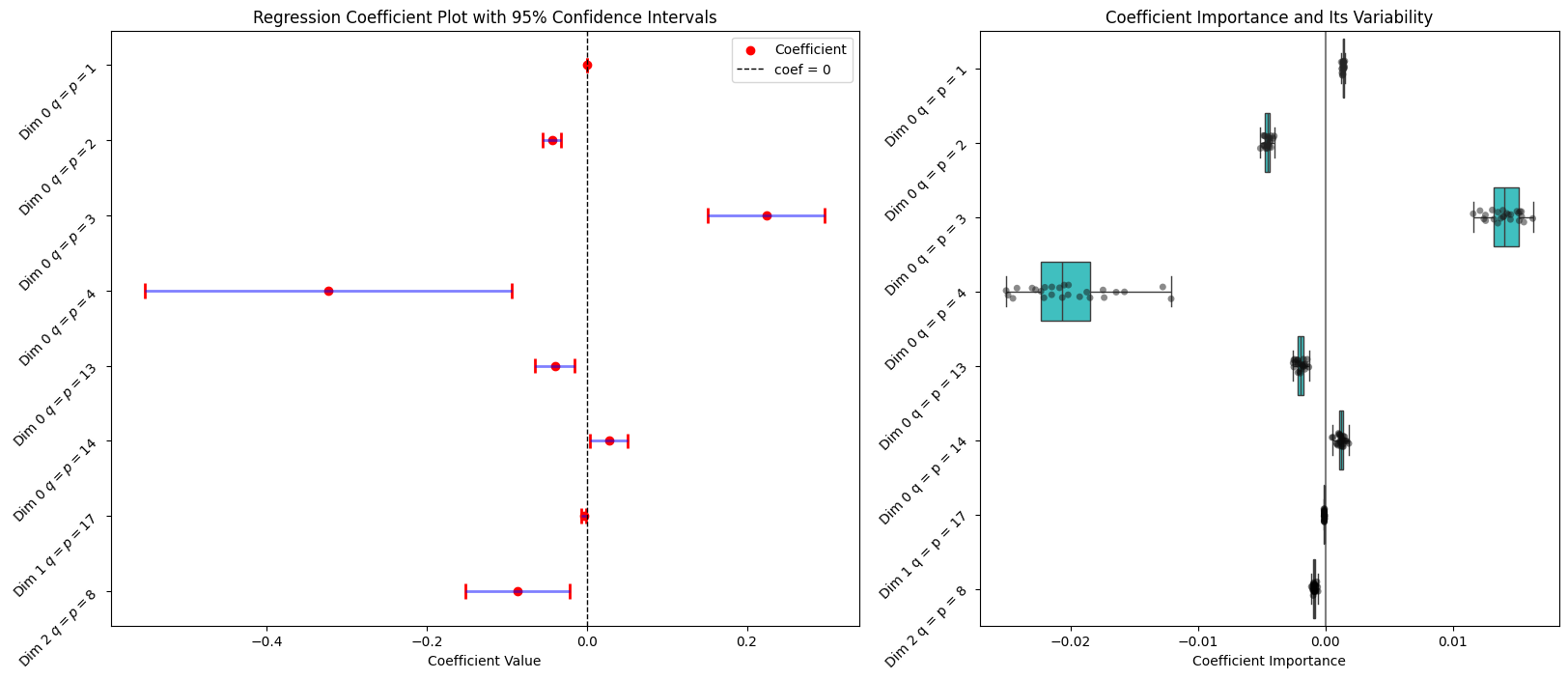}}
\caption{This figure illustrates the filtered-in features generated from the Second Pass (\cref{Second Pass}) for the ROI: ATL and the hemisphere: R(ight) where the training data contains both "Only Trained" and "Trained plus Untrained" LLMs. On the left sub-figure, each feature's Weight is presented along with both the Lower and Upper Bounds of the $95\%$ Confidence Intervals, while the right sub-figure illustrates its Coefficient Importance and Variability.}
\label{with_untrained_R_ATL_coefs_combined_17}
\end{center}
\vskip -0.2in
\end{figure}

\begin{table}[H]
\caption{This table summarizes the filtered-in features generated from the Second Pass (\cref{Second Pass}) for the ROI: PTL and the hemisphere: L(eft) where the training data contains "Only Trained" LLMs.}
\label{only_trained_L_PTL}
\vskip 0.15in
\begin{center}
\begin{small}
\begin{sc}
\begin{tabular}{ccccccc}
\toprule
Feature & Weight & $95\%$ CI & $95\%$ CI & SE & $|t|$ & $p$  \\
 &  & Lower & Upper &  & &  \\
\midrule
(Intercept) & $1.5585 e{-2}$ & $5.7167 e{-3}$ & $2.5453 e{-2}$ & $3.4389 e{-3}$ & $2.3243 e{0}$ & $1.8824 e{-1}$ \\
Dim 0 $q = p = 1$ & $7.5348 e{-4}$ & $-2.5819 e{-4}$ & $1.7652 e{-3}$ & $3.2716 e{-4}$ & $3.9641 e{0}$ & $3.3800 e{-2}$ \\
Dim 0 $q = p = 13$ & $2.6580 e{-1}$ & $3.7613 e{-3}$ & $5.2784 e{-1}$ & $5.8176 e{-2}$ & $5.0569 e{0}$ & $4.2080 e{-2}$ \\
Dim 0 $q = p = 18$ & $7.6117 e{-2}$ & $9.9656 e{-3}$ & $1.4227 e{-1}$ & $1.3776 e{-2}$ & $5.1739 e{0}$ & $4.8600 e{-2}$ \\
Dim 0 $q = p = 23$ & $3.5486 e{-2}$ & $1.0157 e{-2}$ & $6.0816 e{-2}$ & $6.3124 e{-3}$ & $5.5113 e{0}$ & $1.6000 e{-2}$ \\
Dim 1 $q = p = 9$ & $-5.2947 e{-1}$ & $-8.6968 e{-1}$ & $-1.8927 e{-1}$ & $5.6995 e{-2}$ & $8.8984 e{0}$ & $4.9200 e{-3}$ \\
Dim 1 $q = p = 10$ & $-1.2356 e{-1}$ & $-2.5348 e{-1}$ & $6.3622 e{-3}$ & $3.1852 e{-2}$ & $5.1894 e{0}$ & $2.4320 e{-2}$ \\
Dim 1 $q = p = 19$ & $-3.1893 e{-2}$ & $-6.8747 e{-2}$ & $4.9612 e{-3}$ & $9.3464 e{-3}$ & $4.4091 e{0}$ & $4.6160 e{-2}$ \\
Dim 1 $q = p = 33$ & $3.6272 e{-2}$ & $1.9693 e{-2}$ & $5.2852 e{-2}$ & $6.0548 e{-3}$ & $6.1350 e{0}$ & $5.2000 e{-4}$ \\
Dim 2 $q = p = 12$ & $-2.5913 e{-2}$ & $-5.9565 e{-2}$ & $7.7385 e{-3}$ & $7.6498 e{-3}$ & $4.9852 e{0}$ & $4.1120 e{-2}$ \\
Dim 2 $q = p = 22$ & $-2.9233 e{-2}$ & $-5.2112 e{-2}$ & $-6.3550 e{-3}$ & $5.8999 e{-3}$ & $5.5086 e{0}$ & $1.3320 e{-2}$ \\

\bottomrule
\end{tabular}
\end{sc}
\end{small}
\end{center}
\vskip -0.1in
\end{table}

\begin{figure}[H]
\vskip 0.2in
\begin{center}
\centerline{\includegraphics[width=\columnwidth*\columnWidthCoefResultsCoefsCI]{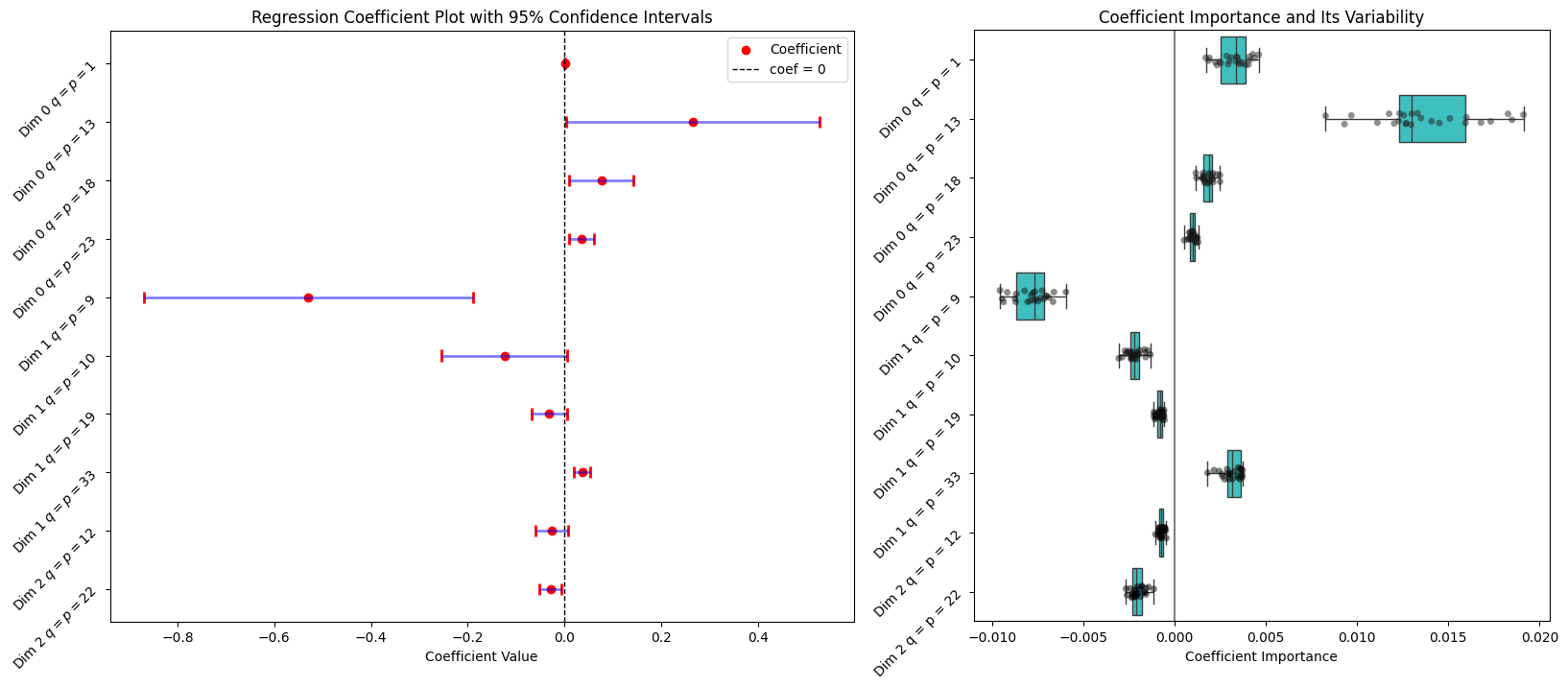}}
\caption{This figure illustrates the filtered-in features generated from the Second Pass (\cref{Second Pass}) for the ROI: PTL and the hemisphere: L(eft) where the training data contains "Only Trained" LLMs. On the left sub-figure, each feature's Weight is presented along with both the Lower and Upper Bounds of the $95\%$ Confidence Intervals, while the right sub-figure illustrates its Coefficient Importance and Variability.}
\label{only_trained_L_PTL_coefs_combined_33}
\end{center}
\vskip -0.2in
\end{figure}

\begin{table}[H]
\caption{This table summarizes the filtered-in features generated from the Second Pass (\cref{Second Pass}) for the ROI: PTL and the hemisphere: L(eft) where the training data contains both "Only Trained" and "Trained plus Untrained" LLMs.}
\label{with_untrained_L_PTL}
\vskip 0.15in
\begin{center}
\begin{small}
\begin{sc}
\begin{tabular}{ccccccc}
\toprule
Feature & Weight & $95\%$ CI & $95\%$ CI & SE & $|t|$ & $p$  \\
 &  & Lower & Upper &  & &  \\
\midrule
(Intercept) & $1.0268 e{-2}$ & $3.1473 e{-3}$ & $1.7389 e{-2}$ & $3.0379 e{-3}$ & $3.1600 e{0}$ & $3.5320 e{-2}$ \\
Dim 0 $q = p = 2$ & $8.0419 e{-2}$ & $3.5483 e{-2}$ & $1.2535 e{-1}$ & $1.7892 e{-2}$ & $4.4886 e{0}$ & $3.6000 e{-3}$ \\
Dim 0 $q = p = 3$ & $-3.5544 e{-1}$ & $-6.1130 e{-1}$ & $-9.9571 e{-2}$ & $9.5457 e{-2}$ & $3.6268 e{0}$ & $3.0440 e{-2}$ \\
Dim 0 $q = p = 9$ & $-3.2986 e{0}$ & $-5.4383 e{0}$ & $-1.1588 e{0}$ & $8.9503 e{-1}$ & $3.2245 e{0}$ & $4.0360 e{-2}$ \\
Dim 0 $q = p = 18$ & $7.0178 e{-2}$ & $2.0545 e{-2}$ & $1.1981 e{-1}$ & $1.2005 e{-2}$ & $5.4401 e{0}$ & $2.0560 e{-2}$ \\
Dim 0 $q = p = 23$ & $2.3662 e{-2}$ & $7.3796 e{-3}$ & $3.9945 e{-2}$ & $3.6630 e{-3}$ & $5.5682 e{0}$ & $2.6040 e{-2}$ \\
Dim 0 $q = p = 32$ & $-2.2148 e{-2}$ & $-3.6412 e{-2}$ & $-7.8849 e{-3}$ & $3.2943 e{-3}$ & $7.5597 e{0}$ & $1.6000 e{-3}$ \\
Dim 0 $q = p = 35$ & $1.9227 e{-2}$ & $5.1714 e{-4}$ & $3.7937 e{-2}$ & $5.8812 e{-3}$ & $3.7248 e{0}$ & $4.5240 e{-2}$ \\
Dim 1 $q = p = 9$ & $-4.6281 e{-1}$ & $-7.0142 e{-1}$ & $-2.2420 e{-1}$ & $6.0491 e{-2}$ & $7.0598 e{0}$ & $1.6400 e{-3}$ \\
Dim 1 $q = p = 10$ & $-1.3839 e{-1}$ & $-2.2630 e{-1}$ & $-5.0483 e{-2}$ & $1.6225 e{-2}$ & $7.7840 e{0}$ & $7.5200 e{-3}$ \\
Dim 1 $q = p = 19$ & $-3.0829 e{-2}$ & $-5.6030 e{-2}$ & $-5.6277 e{-3}$ & $7.2475 e{-3}$ & $4.3476 e{0}$ & $3.5880 e{-2}$ \\
Dim 1 $q = p = 35$ & $3.5057 e{-2}$ & $1.8972 e{-2}$ & $5.1142 e{-2}$ & $6.1734 e{-3}$ & $4.1389 e{0}$ & $1.1440 e{-2}$ \\
Dim 2 $q = p = 6$ & $-1.4861 e{1}$ & $-2.8733 e{1}$ & $-9.8855 e{-1}$ & $3.8387 e{0}$ & $4.5248 e{0}$ & $2.9120 e{-2}$ \\
Dim 2 $q = p = 18$ & $2.3164 e{-2}$ & $7.8150 e{-3}$ & $3.8512 e{-2}$ & $3.5853 e{-3}$ & $6.0314 e{0}$ & $1.2400 e{-2}$ \\
Dim 2 $q = p = 19$ & $-2.4316 e{-2}$ & $-3.8713 e{-2}$ & $-9.9189 e{-3}$ & $4.6657 e{-3}$ & $3.6749 e{0}$ & $4.6480 e{-2}$ \\
Dim 2 $q = p = 20$ & $2.9925 e{-2}$ & $1.4701 e{-2}$ & $4.5150 e{-2}$ & $4.5436 e{-3}$ & $4.8935 e{0}$ & $1.1920 e{-2}$ \\
Dim 2 $q = p = 22$ & $-2.2814 e{-2}$ & $-3.9557 e{-2}$ & $-6.0712 e{-3}$ & $3.8954 e{-3}$ & $5.9441 e{0}$ & $1.4160 e{-2}$ \\

\bottomrule
\end{tabular}
\end{sc}
\end{small}
\end{center}
\vskip -0.1in
\end{table}

\begin{figure}[H]
\vskip 0.2in
\begin{center}
\centerline{\includegraphics[width=\columnwidth*\columnWidthCoefResultsCoefsCI]{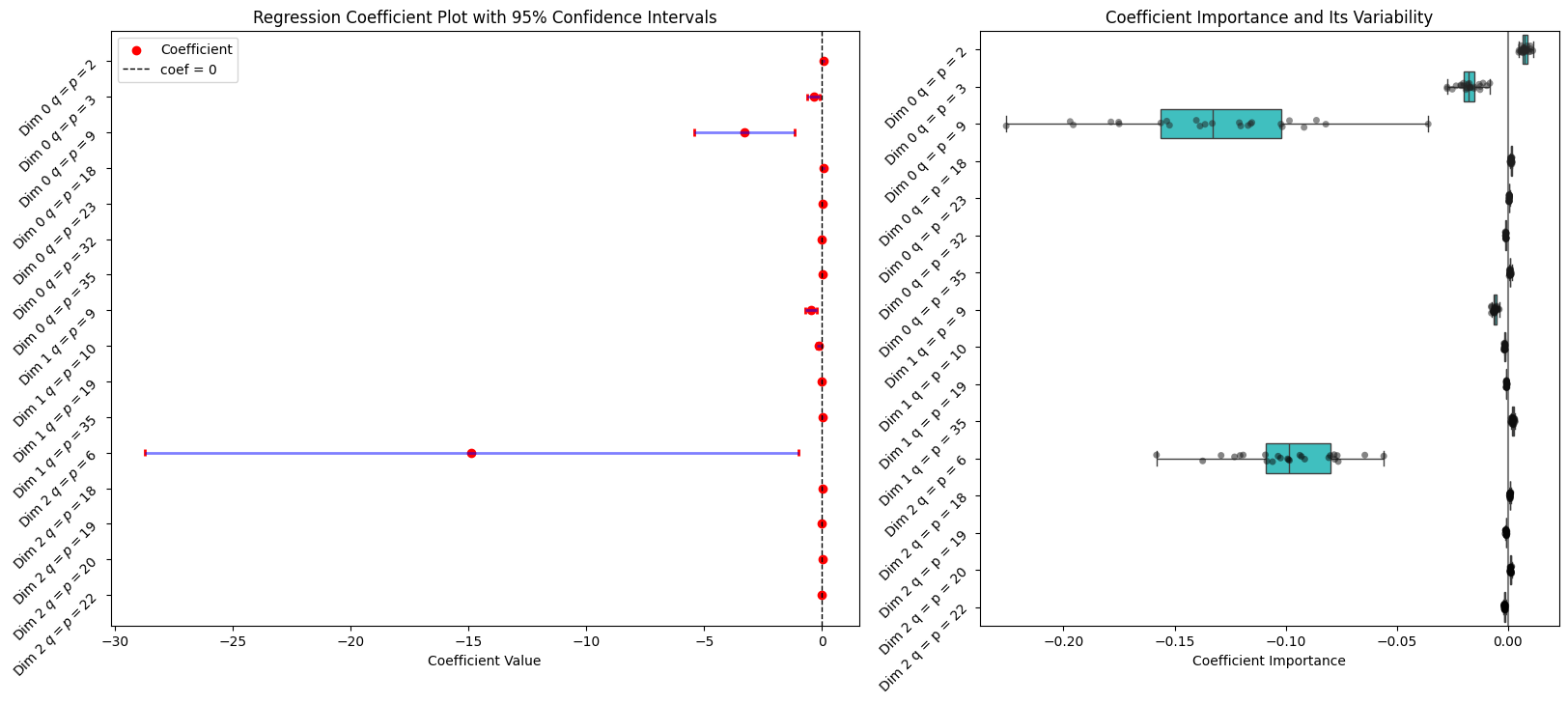}}
\caption{This figure illustrates the filtered-in features generated from the Second Pass (\cref{Second Pass}) for the ROI: PTL and the hemisphere: L(eft) where the training data contains both "Only Trained" and "Trained plus Untrained" LLMs. On the left sub-figure, each feature's Weight is presented along with both the Lower and Upper Bounds of the $95\%$ Confidence Intervals, while the right sub-figure illustrates its Coefficient Importance and Variability.}
\label{with_untrained_L_PTL_coefs_combined_35}
\end{center}
\vskip -0.2in
\end{figure}

\begin{table}[H]
\caption{This table summarizes the filtered-in features generated from the Second Pass (\cref{Second Pass}) for the ROI: PTL and the hemisphere: R(ight) where the training data contains "Only Trained" LLMs.}
\label{only_trained_R_PTL}
\vskip 0.15in
\begin{center}
\begin{small}
\begin{sc}
\begin{tabular}{ccccccc}
\toprule
Feature & Weight & $95\%$ CI & $95\%$ CI & SE & $|t|$ & $p$  \\
 &  & Lower & Upper &  & &  \\
\midrule
(Intercept) & $-5.2923 e{-3}$ & $-1.6630 e{-2}$ & $6.0458 e{-3}$ & $4.4800 e{-3}$ & $1.7909 e{0}$ & $2.5344 e{-1}$ \\
Dim 0 $q = p = 11$ & $-5.8523 e{-1}$ & $-1.1147 e{0}$ & $-5.5797 e{-2}$ & $1.7857 e{-1}$ & $4.2272 e{0}$ & $1.6520 e{-2}$ \\
Dim 0 $q = p = 19$ & $-1.1286 e{-1}$ & $-1.7756 e{-1}$ & $-4.8149 e{-2}$ & $1.4506 e{-2}$ & $6.2654 e{0}$ & $1.0680 e{-2}$ \\
Dim 0 $q = p = \infty$ & $1.7962 e{-1}$ & $5.1970 e{-2}$ & $3.0728 e{-1}$ & $2.6219 e{-2}$ & $6.6191 e{0}$ & $1.3560 e{-2}$ \\
Dim 1 $q = p = 1$ & $-1.7499 e{-2}$ & $-2.7279 e{-2}$ & $-7.7197 e{-3}$ & $3.3008 e{-3}$ & $5.3990 e{0}$ & $2.4400 e{-3}$ \\
Dim 1 $q = p = 2$ & $7.4784 e{-1}$ & $2.5412 e{-1}$ & $1.2416 e{0}$ & $1.5426 e{-1}$ & $4.4264 e{0}$ & $2.0920 e{-2}$ \\
Dim 1 $q = p = 10$ & $1.4712 e{-1}$ & $1.4032 e{-2}$ & $2.8020 e{-1}$ & $3.1971 e{-2}$ & $6.4546 e{0}$ & $6.2000 e{-3}$ \\
Dim 1 $q = p = 12$ & $5.3914 e{-2}$ & $1.4478 e{-2}$ & $9.3351 e{-2}$ & $1.2628 e{-2}$ & $5.1767 e{0}$ & $5.6400 e{-3}$ \\
Dim 2 $q = p = 1$ & $2.5956 e{-1}$ & $1.6803 e{-1}$ & $3.5110 e{-1}$ & $3.3323 e{-2}$ & $5.9471 e{0}$ & $3.6000 e{-4}$ \\
Dim 2 $q = p = 2$ & $-9.3916 e{0}$ & $-1.3255 e{1}$ & $-5.5281 e{0}$ & $1.6880 e{0}$ & $3.9068 e{0}$ & $1.3040 e{-2}$ \\
Dim 2 $q = p = 14$ & $-3.9924 e{-2}$ & $-6.8999 e{-2}$ & $-1.0850 e{-2}$ & $7.5473 e{-3}$ & $4.3309 e{0}$ & $4.9200 e{-2}$ \\
Dim 2 $q = p = 28$ & $3.9050 e{-2}$ & $1.4120 e{-2}$ & $6.3979 e{-2}$ & $9.7509 e{-3}$ & $4.3271 e{0}$ & $1.5960 e{-2}$ \\

\bottomrule
\end{tabular}
\end{sc}
\end{small}
\end{center}
\vskip -0.1in
\end{table}

\begin{figure}[H]
\vskip 0.2in
\begin{center}
\centerline{\includegraphics[width=\columnwidth*\columnWidthCoefResultsCoefsCI]{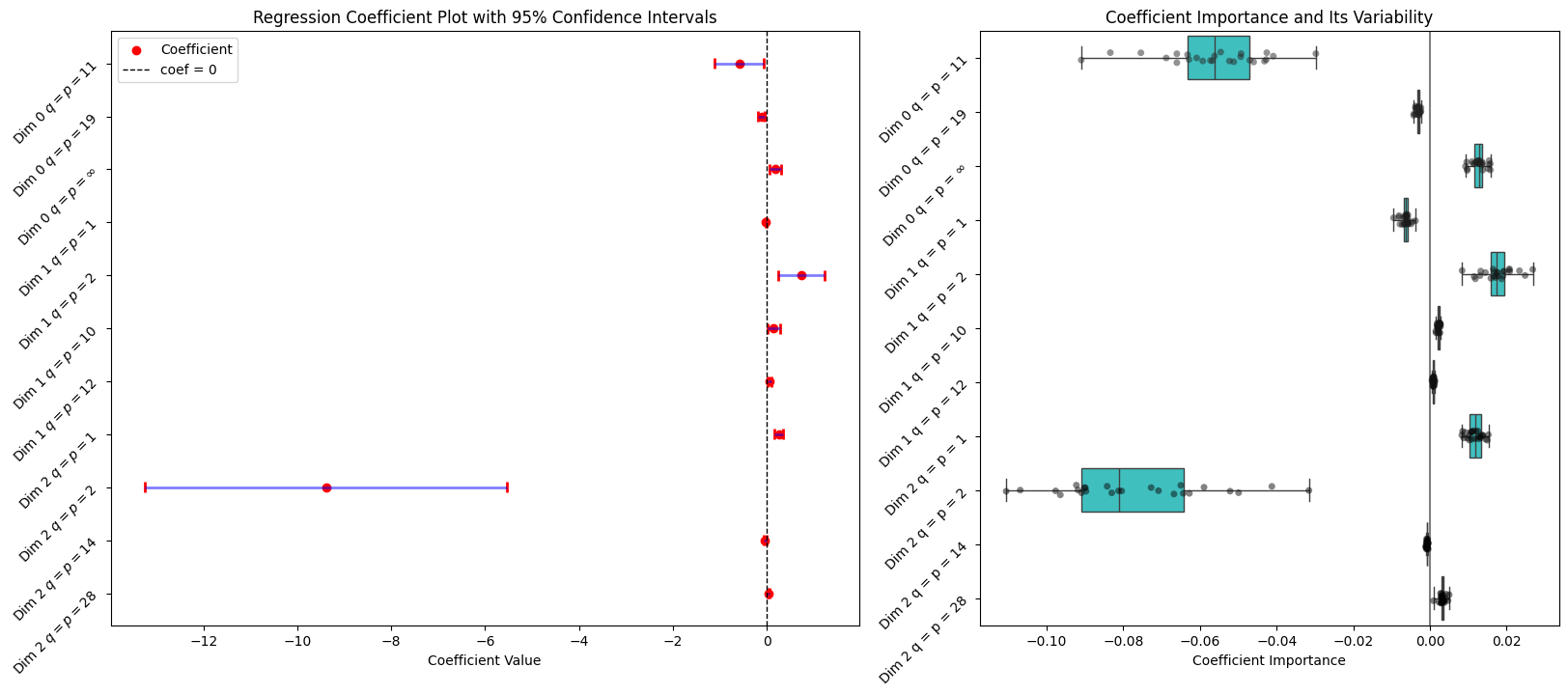}}
\caption{This figure illustrates the filtered-in features generated from the Second Pass (\cref{Second Pass}) for the ROI: PTL and the hemisphere: R(ight) where the training data contains "Only Trained" LLMs. On the left sub-figure, each feature's Weight is presented along with both the Lower and Upper Bounds of the $95\%$ Confidence Intervals, while the right sub-figure illustrates its Coefficient Importance and Variability.}
\label{only_trained_R_PTL_coefs_combined_28}
\end{center}
\vskip -0.2in
\end{figure}

\begin{table}[H]
\caption{This table summarizes the filtered-in features generated from the Second Pass (\cref{Second Pass}) for the ROI: PTL and the hemisphere: R(ight) where the training data contains both "Only Trained" and "Trained plus Untrained" LLMs.}
\label{with_untrained_R_PTL}
\vskip 0.15in
\begin{center}
\begin{small}
\begin{sc}
\begin{tabular}{ccccccc}
\toprule
Feature & Weight & $95\%$ CI & $95\%$ CI & SE & $|t|$ & $p$  \\
 &  & Lower & Upper &  & &  \\
\midrule
(Intercept) & $-2.0458 e{-2}$ & $-2.9290 e{-2}$ & $-1.1626 e{-2}$ & $4.6079 e{-3}$ & $1.3974 e{0}$ & $2.8840 e{-1}$ \\
Dim 0 $q = p = 19$ & $-4.4577 e{-2}$ & $-8.7760 e{-2}$ & $-1.3936 e{-3}$ & $1.1679 e{-2}$ & $4.4383 e{0}$ & $3.8120 e{-2}$ \\
Dim 0 $q = p = 29$ & $2.6851 e{-2}$ & $1.1790 e{-2}$ & $4.1912 e{-2}$ & $2.7699 e{-3}$ & $8.8938 e{0}$ & $2.2000 e{-3}$ \\
Dim 0 $q = p = \infty$ & $1.5250 e{-1}$ & $6.4918 e{-2}$ & $2.4008 e{-1}$ & $1.8435 e{-2}$ & $8.5399 e{0}$ & $9.6000 e{-4}$ \\
Dim 1 $q = p = 1$ & $-7.2707 e{-3}$ & $-1.4486 e{-2}$ & $-5.5603 e{-5}$ & $2.6363 e{-3}$ & $3.8352 e{0}$ & $1.9560 e{-2}$ \\
Dim 1 $q = p = 10$ & $2.1890 e{-1}$ & $1.2080 e{-1}$ & $3.1700 e{-1}$ & $1.5871 e{-2}$ & $1.1342 e{1}$ & $4.4000 e{-4}$ \\
Dim 1 $q = p = 12$ & $3.6330 e{-2}$ & $8.1226 e{-3}$ & $6.4537 e{-2}$ & $7.3995 e{-3}$ & $5.1413 e{0}$ & $1.8200 e{-2}$ \\
Dim 2 $q = p = 1$ & $1.5831 e{-1}$ & $1.0277 e{-1}$ & $2.1384 e{-1}$ & $1.6827 e{-2}$ & $9.5096 e{0}$ & $0.0000 e{0}$ \\
Dim 2 $q = p = 2$ & $-5.0261 e{0}$ & $-7.0774 e{0}$ & $-2.9747 e{0}$ & $6.3579 e{-1}$ & $7.5334 e{0}$ & $4.0000 e{-5}$ \\
Dim 2 $q = p = 14$ & $-2.9303 e{-2}$ & $-5.1254 e{-2}$ & $-7.3525 e{-3}$ & $5.5759 e{-3}$ & $5.7891 e{0}$ & $7.7200 e{-3}$ \\
Dim 2 $q = p = 15$ & $-2.9817 e{-2}$ & $-4.9779 e{-2}$ & $-9.8541 e{-3}$ & $4.9745 e{-3}$ & $5.6993 e{0}$ & $1.1400 e{-2}$ \\
Dim 2 $q = p = 16$ & $-2.8393 e{-2}$ & $-4.6181 e{-2}$ & $-1.0604 e{-2}$ & $6.3687 e{-3}$ & $4.2164 e{0}$ & $1.1880 e{-2}$ \\
Dim 2 $q = p = 20$ & $1.3214 e{-2}$ & $2.7608 e{-3}$ & $2.3667 e{-2}$ & $2.1354 e{-3}$ & $6.4423 e{0}$ & $1.5160 e{-2}$ \\
Dim 2 $q = p = 29$ & $7.5011 e{-2}$ & $4.7473 e{-2}$ & $1.0255 e{-1}$ & $6.6481 e{-3}$ & $1.1495 e{1}$ & $0.0000 e{0}$ \\

\bottomrule
\end{tabular}
\end{sc}
\end{small}
\end{center}
\vskip -0.1in
\end{table}

\begin{figure}[H]
\vskip 0.2in
\begin{center}
\centerline{\includegraphics[width=\columnwidth*\columnWidthCoefResultsCoefsCI]{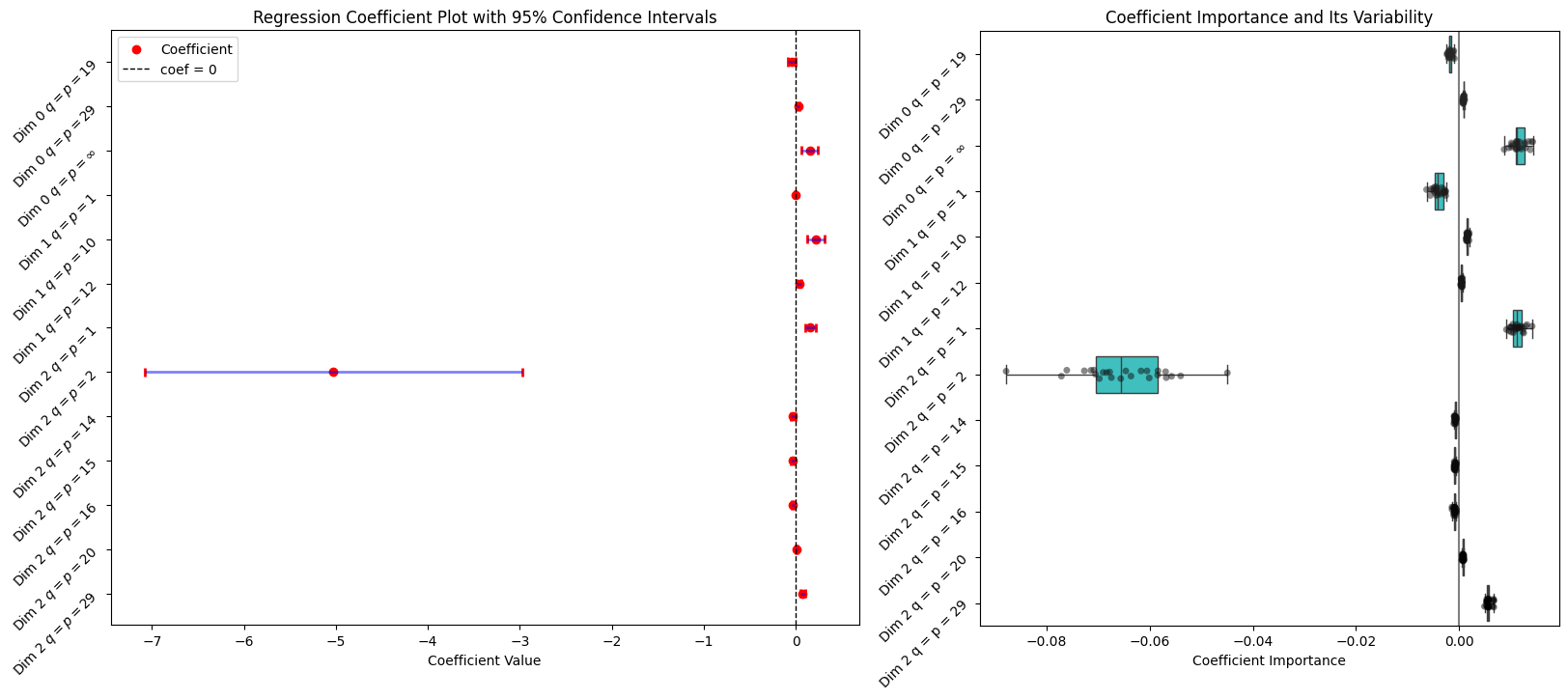}}
\caption{This figure illustrates the filtered-in features generated from the Second Pass (\cref{Second Pass}) for the ROI: PTL and the hemisphere: R(ight) where the training data contains both "Only Trained" and "Trained plus Untrained" LLMs. On the left sub-figure, each feature's Weight is presented along with both the Lower and Upper Bounds of the $95\%$ Confidence Intervals, while the right sub-figure illustrates its Coefficient Importance and Variability.}
\label{with_untrained_R_PTL_coefs_combined_29}
\end{center}
\vskip -0.2in
\end{figure}

\begin{table}[H]
\caption{This table summarizes the filtered-in features generated from the Second Pass (\cref{Second Pass}) for the ROI: IFG and the hemisphere: L(eft) where the training data contains "Only Trained" LLMs.}
\label{only_trained_L_IFG}
\vskip 0.15in
\begin{center}
\begin{small}
\begin{sc}
\begin{tabular}{ccccccc}
\toprule
Feature & Weight & $95\%$ CI & $95\%$ CI & SE & $|t|$ & $p$  \\
 &  & Lower & Upper &  & &  \\
\midrule
(Intercept) & $1.2152 e{-2}$ & $1.0494 e{-2}$ & $1.3810 e{-2}$ & $5.8618 e{-4}$ & $1.8943 e{1}$ & $0.0000 e{0}$ \\
Dim 0 $q = p = 1$ & $3.4086 e{-3}$ & $2.8282 e{-3}$ & $3.9890 e{-3}$ & $1.8962 e{-4}$ & $1.8268 e{1}$ & $0.0000 e{0}$ \\
Dim 0 $q = p = 2$ & $-1.8577 e{-1}$ & $-2.1542 e{-1}$ & $-1.5612 e{-1}$ & $1.0606 e{-2}$ & $1.7234 e{1}$ & $0.0000 e{0}$ \\
Dim 0 $q = p = 3$ & $5.1693 e{-1}$ & $4.2193 e{-1}$ & $6.1194 e{-1}$ & $3.3823 e{-2}$ & $1.4441 e{1}$ & $0.0000 e{0}$ \\
Dim 0 $q = p = 4$ & $-3.4284 e{-1}$ & $-4.2229 e{-1}$ & $-2.6339 e{-1}$ & $2.6157 e{-2}$ & $1.1914 e{1}$ & $0.0000 e{0}$ \\
Dim 0 $q = p = \infty$ & $3.6600 e{-2}$ & $1.9244 e{-2}$ & $5.3956 e{-2}$ & $4.9681 e{-3}$ & $6.0574 e{0}$ & $2.6400 e{-3}$ \\
Dim 1 $q = p = 1$ & $8.0467 e{-3}$ & $2.9246 e{-3}$ & $1.3169 e{-2}$ & $1.4108 e{-3}$ & $6.2215 e{0}$ & $2.4800 e{-3}$ \\
Dim 1 $q = p = 2$ & $-3.4178 e{-1}$ & $-5.4512 e{-1}$ & $-1.3844 e{-1}$ & $6.3094 e{-2}$ & $6.8980 e{0}$ & $1.2000 e{-4}$ \\
Dim 1 $q = p = 3$ & $1.0227 e{0}$ & $3.1841 e{-1}$ & $1.7270 e{0}$ & $2.2583 e{-1}$ & $6.2329 e{0}$ & $6.8000 e{-4}$ \\
Dim 1 $q = p = 4$ & $-7.0560 e{-1}$ & $-1.3062 e{0}$ & $-1.0494 e{-1}$ & $1.7930 e{-1}$ & $5.4286 e{0}$ & $5.9200 e{-3}$ \\

\bottomrule
\end{tabular}
\end{sc}
\end{small}
\end{center}
\vskip -0.1in
\end{table}

\begin{figure}[H]
\vskip 0.2in
\begin{center}
\centerline{\includegraphics[width=\columnwidth*\columnWidthCoefResultsCoefsCI]{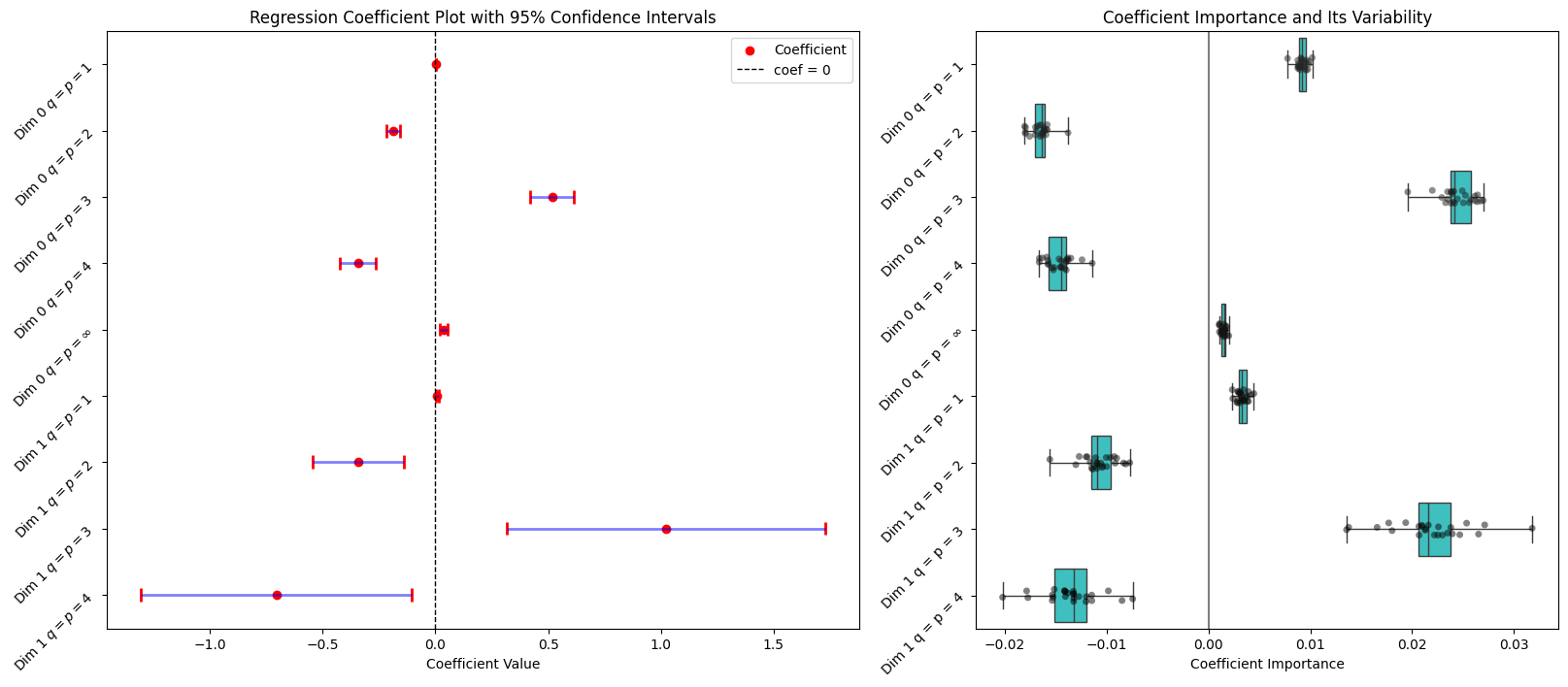}}
\caption{This figure illustrates the filtered-in features generated from the Second Pass (\cref{Second Pass}) for the ROI: IFG and the hemisphere: L(eft) where the training data contains "Only Trained" LLMs. On the left sub-figure, each feature's Weight is presented along with both the Lower and Upper Bounds of the $95\%$ Confidence Intervals, while the right sub-figure illustrates its Coefficient Importance and Variability.}
\label{only_trained_L_IFG_coefs_combined_4}
\end{center}
\vskip -0.2in
\end{figure}

\begin{table}[H]
\caption{This table summarizes the filtered-in features generated from the Second Pass (\cref{Second Pass}) for the ROI: IFG and the hemisphere: L(eft) where the training data contains both "Only Trained" and "Trained plus Untrained" LLMs.}
\label{with_untrained_L_IFG}
\vskip 0.15in
\begin{center}
\begin{small}
\begin{sc}
\begin{tabular}{ccccccc}
\toprule
Feature & Weight & $95\%$ CI & $95\%$ CI & SE & $|t|$ & $p$  \\
 &  & Lower & Upper &  & &  \\
\midrule
(Intercept) & $1.4132 e{-2}$ & $1.2836 e{-2}$ & $1.5428 e{-2}$ & $3.5541 e{-4}$ & $4.0213 e{1}$ & $0.0000 e{0}$ \\
Dim 0 $q = p = 1$ & $3.5718 e{-3}$ & $3.1534 e{-3}$ & $3.9901 e{-3}$ & $1.3535 e{-4}$ & $2.7680 e{1}$ & $0.0000 e{0}$ \\
Dim 0 $q = p = 2$ & $-1.8855 e{-1}$ & $-2.1799 e{-1}$ & $-1.5912 e{-1}$ & $1.0323 e{-2}$ & $2.0283 e{1}$ & $0.0000 e{0}$ \\
Dim 0 $q = p = 3$ & $5.9426 e{-1}$ & $4.3111 e{-1}$ & $7.5740 e{-1}$ & $4.8883 e{-2}$ & $1.3972 e{1}$ & $0.0000 e{0}$ \\
Dim 0 $q = p = 4$ & $-6.2831 e{-1}$ & $-9.1999 e{-1}$ & $-3.3663 e{-1}$ & $8.0982 e{-2}$ & $9.2786 e{0}$ & $0.0000 e{0}$ \\
Dim 0 $q = p = 5$ & $2.3541 e{-1}$ & $6.2997 e{-2}$ & $4.0782 e{-1}$ & $4.6127 e{-2}$ & $6.3883 e{0}$ & $2.5200 e{-3}$ \\
Dim 1 $q = p = 1$ & $9.1191 e{-3}$ & $5.2884 e{-3}$ & $1.2950 e{-2}$ & $9.6548 e{-4}$ & $8.2422 e{0}$ & $1.2000 e{-4}$ \\
Dim 1 $q = p = 2$ & $-4.5065 e{-1}$ & $-6.4678 e{-1}$ & $-2.5452 e{-1}$ & $4.8220 e{-2}$ & $8.0120 e{0}$ & $4.0000 e{-4}$ \\

\bottomrule
\end{tabular}
\end{sc}
\end{small}
\end{center}
\vskip -0.1in
\end{table}

\begin{figure}[H]
\vskip 0.2in
\begin{center}
\centerline{\includegraphics[width=\columnwidth*\columnWidthCoefResultsCoefsCI]{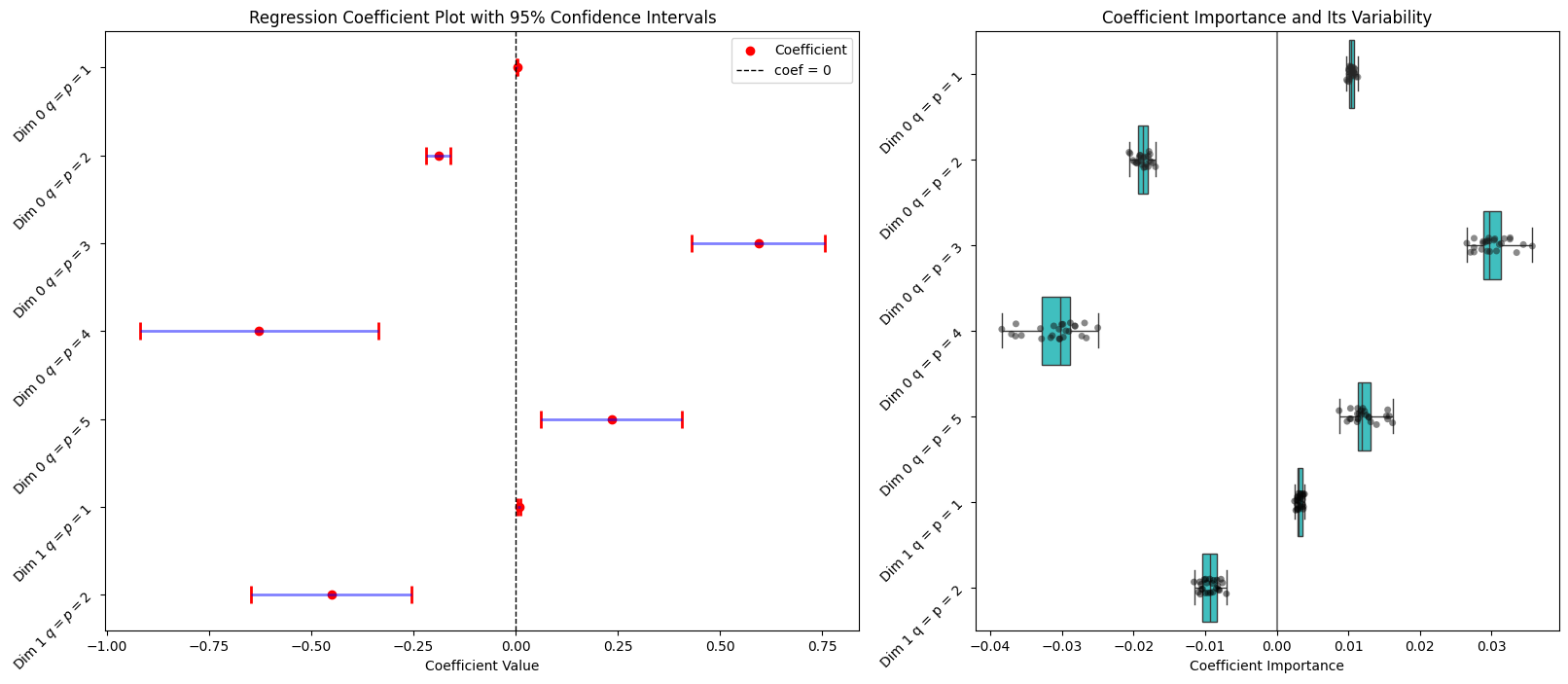}}
\caption{This figure illustrates the filtered-in features generated from the Second Pass (\cref{Second Pass}) for the ROI: IFG and the hemisphere: L(eft) where the training data contains both "Only Trained" and "Trained plus Untrained" LLMs. On the left sub-figure, each feature's Weight is presented along with both the Lower and Upper Bounds of the $95\%$ Confidence Intervals, while the right sub-figure illustrates its Coefficient Importance and Variability.}
\label{with_untrained_L_IFG_coefs_combined_5}
\end{center}
\vskip -0.2in
\end{figure}

\begin{table}[H]
\caption{This table summarizes the filtered-in features generated from the Second Pass (\cref{Second Pass}) for the ROI: IFG and the hemisphere: R(ight) where the training data contains "Only Trained" LLMs.}
\label{only_trained_R_IFG}
\vskip 0.15in
\begin{center}
\begin{small}
\begin{sc}
\begin{tabular}{ccccccc}
\toprule
Feature & Weight & $95\%$ CI & $95\%$ CI & SE & $|t|$ & $p$  \\
 &  & Lower & Upper &  & &  \\
\midrule
(Intercept) & $7.6982 e{-3}$ & $5.2758 e{-4}$ & $1.4869 e{-2}$ & $2.0497 e{-3}$ & $3.9723 e{0}$ & $4.7920 e{-2}$ \\
Dim 0 $q = p = 1$ & $2.4968 e{-3}$ & $1.7600 e{-3}$ & $3.2336 e{-3}$ & $2.4391 e{-4}$ & $1.0014 e{1}$ & $0.0000 e{0}$ \\
Dim 0 $q = p = 2$ & $-1.3040 e{-1}$ & $-1.7178 e{-1}$ & $-8.9025 e{-2}$ & $1.9467 e{-2}$ & $5.5568 e{0}$ & $0.0000 e{0}$ \\
Dim 0 $q = p = 11$ & $-1.6410 e{-1}$ & $-3.0071 e{-1}$ & $-2.7502 e{-2}$ & $2.7423 e{-2}$ & $5.6992 e{0}$ & $3.6080 e{-2}$ \\
Dim 1 $q = p = 1$ & $1.1696 e{-2}$ & $5.6391 e{-3}$ & $1.7753 e{-2}$ & $2.2258 e{-3}$ & $5.5314 e{0}$ & $6.4000 e{-4}$ \\
Dim 1 $q = p = 2$ & $-3.5225 e{-1}$ & $-6.5589 e{-1}$ & $-4.8604 e{-2}$ & $1.1324 e{-1}$ & $3.7184 e{0}$ & $2.0640 e{-2}$ \\
Dim 1 $q = p = 3$ & $1.7138 e{0}$ & $2.2874 e{-1}$ & $3.1988 e{0}$ & $5.6054 e{-1}$ & $3.7436 e{0}$ & $1.5360 e{-2}$ \\
Dim 1 $q = p = 4$ & $-4.9525 e{0}$ & $-8.6806 e{0}$ & $-1.2243 e{0}$ & $1.2788 e{0}$ & $4.1077 e{0}$ & $1.6600 e{-2}$ \\
Dim 1 $q = p = 11$ & $-5.7324 e{-2}$ & $-9.7330 e{-2}$ & $-1.7318 e{-2}$ & $7.0890 e{-3}$ & $6.9779 e{0}$ & $2.2200 e{-2}$ \\
Dim 1 $q = p = 17$ & $2.2443 e{-2}$ & $2.4178 e{-4}$ & $4.4645 e{-2}$ & $5.8184 e{-3}$ & $5.1042 e{0}$ & $2.1840 e{-2}$ \\
Dim 2 $q = p = 19$ & $-1.5601 e{-2}$ & $-2.8013 e{-2}$ & $-3.1898 e{-3}$ & $3.3055 e{-3}$ & $5.6017 e{0}$ & $8.7600 e{-3}$ \\

\bottomrule
\end{tabular}
\end{sc}
\end{small}
\end{center}
\vskip -0.1in
\end{table}

\begin{figure}[H]
\vskip 0.2in
\begin{center}
\centerline{\includegraphics[width=\columnwidth*\columnWidthCoefResultsCoefsCI]{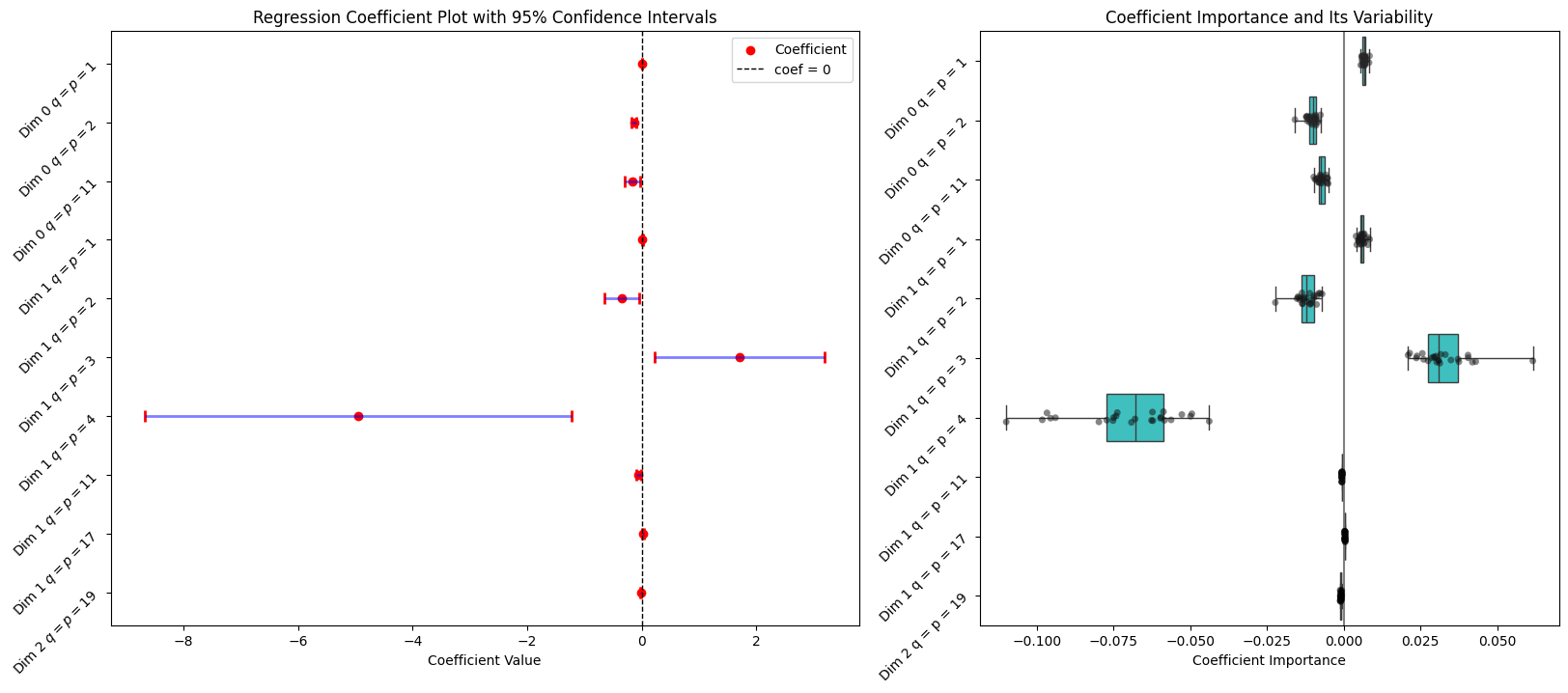}}
\caption{This figure illustrates the filtered-in features generated from the Second Pass (\cref{Second Pass}) for the ROI: IFG and the hemisphere: R(ight) where the training data contains "Only Trained" LLMs. On the left sub-figure, each feature's Weight is presented along with both the Lower and Upper Bounds of the $95\%$ Confidence Intervals, while the right sub-figure illustrates its Coefficient Importance and Variability.}
\label{only_trained_R_IFG_coefs_combined_19}
\end{center}
\vskip -0.2in
\end{figure}

\begin{table}[H]
\caption{This table summarizes the filtered-in features generated from the Second Pass (\cref{Second Pass}) for the ROI: IFG and the hemisphere: R(ight) where the training data contains both "Only Trained" and "Trained plus Untrained" LLMs.}
\label{with_untrained_R_IFG}
\vskip 0.15in
\begin{center}
\begin{small}
\begin{sc}
\begin{tabular}{ccccccc}
\toprule
Feature & Weight & $95\%$ CI & $95\%$ CI & SE & $|t|$ & $p$  \\
 &  & Lower & Upper &  & &  \\
\midrule
(Intercept) & $1.0846 e{-2}$ & $5.5497 e{-3}$ & $1.6142 e{-2}$ & $1.9059 e{-3}$ & $3.9423 e{0}$ & $2.5320 e{-2}$ \\
Dim 0 $q = p = 1$ & $2.9811 e{-3}$ & $2.4451 e{-3}$ & $3.5171 e{-3}$ & $1.8159 e{-4}$ & $1.6916 e{1}$ & $0.0000 e{0}$ \\
Dim 0 $q = p = 2$ & $-1.4613 e{-1}$ & $-1.8164 e{-1}$ & $-1.1062 e{-1}$ & $1.8017 e{-2}$ & $8.5486 e{0}$ & $0.0000 e{0}$ \\
Dim 0 $q = p = 3$ & $3.9672 e{-1}$ & $1.5808 e{-1}$ & $6.3536 e{-1}$ & $1.4258 e{-1}$ & $2.8312 e{0}$ & $3.8480 e{-2}$ \\
Dim 1 $q = p = 1$ & $1.5140 e{-2}$ & $1.0645 e{-2}$ & $1.9635 e{-2}$ & $1.1797 e{-3}$ & $1.0358 e{1}$ & $0.0000 e{0}$ \\
Dim 1 $q = p = 2$ & $-5.5080 e{-1}$ & $-7.9807 e{-1}$ & $-3.0353 e{-1}$ & $7.3269 e{-2}$ & $5.9266 e{0}$ & $3.4400 e{-3}$ \\
Dim 1 $q = p = 3$ & $1.9133 e{0}$ & $6.3249 e{-1}$ & $3.1941 e{0}$ & $3.7926 e{-1}$ & $4.7817 e{0}$ & $1.8600 e{-2}$ \\
Dim 1 $q = p = 9$ & $4.5235 e{-1}$ & $1.7541 e{-1}$ & $7.2928 e{-1}$ & $3.8963 e{-2}$ & $1.0325 e{1}$ & $6.4400 e{-3}$ \\
Dim 1 $q = p = 11$ & $-3.2621 e{-2}$ & $-6.3679 e{-2}$ & $-1.5635 e{-3}$ & $8.0543 e{-3}$ & $5.2390 e{0}$ & $1.8800 e{-2}$ \\
Dim 1 $q = p = 13$ & $-1.5051 e{-2}$ & $-3.1761 e{-2}$ & $1.6596 e{-3}$ & $4.7727 e{-3}$ & $4.0961 e{0}$ & $4.6600 e{-2}$ \\
Dim 1 $q = p = 19$ & $1.4403 e{-2}$ & $2.4424 e{-3}$ & $2.6364 e{-2}$ & $2.9250 e{-3}$ & $5.4581 e{0}$ & $1.8040 e{-2}$ \\
Dim 2 $q = p = 19$ & $-2.5808 e{-2}$ & $-3.7367 e{-2}$ & $-1.4250 e{-2}$ & $3.2017 e{-3}$ & $6.4056 e{0}$ & $2.3200 e{-3}$ \\

\bottomrule
\end{tabular}
\end{sc}
\end{small}
\end{center}
\vskip -0.1in
\end{table}

\begin{figure}[H]
\vskip 0.2in
\begin{center}
\centerline{\includegraphics[width=\columnwidth*\columnWidthCoefResultsCoefsCI]{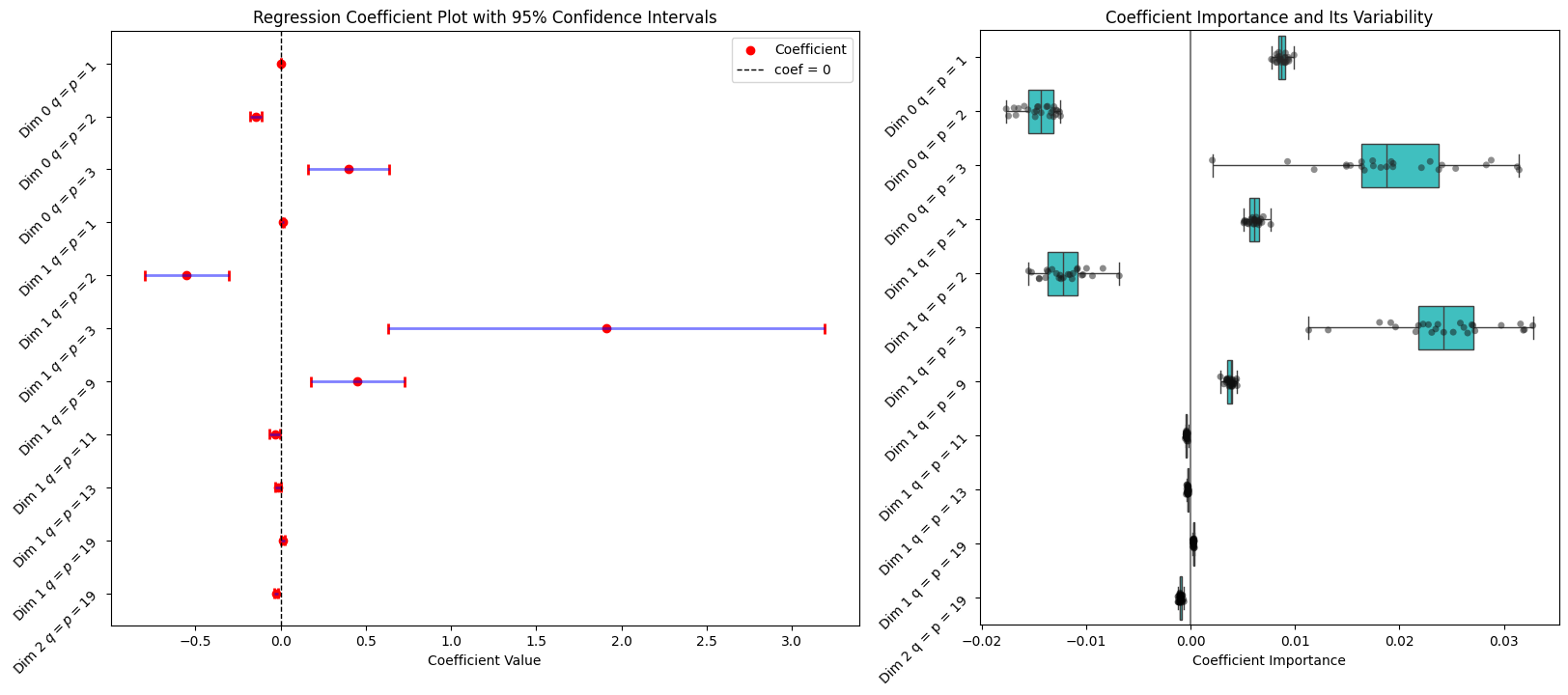}}
\caption{This figure illustrates the filtered-in features generated from the Second Pass (\cref{Second Pass}) for the ROI: IFG and the hemisphere: R(ight) where the training data contains both "Only Trained" and "Trained plus Untrained" LLMs. On the left sub-figure, each feature's Weight is presented along with both the Lower and Upper Bounds of the $95\%$ Confidence Intervals, while the right sub-figure illustrates its Coefficient Importance and Variability.}
\label{with_untrained_R_IFG_coefs_combined_19}
\end{center}
\vskip -0.2in
\end{figure}

\begin{table}[H]
\caption{This table summarizes the filtered-in features generated from the Second Pass (\cref{Second Pass}) for the ROI: MFG and the hemisphere: L(eft) where the training data contains "Only Trained" LLMs.}
\label{only_trained_L_MFG}
\vskip 0.15in
\begin{center}
\begin{small}
\begin{sc}
\begin{tabular}{ccccccc}
\toprule
Feature & Weight & $95\%$ CI & $95\%$ CI & SE & $|t|$ & $p$  \\
 &  & Lower & Upper &  & &  \\
\midrule
(Intercept) & $1.3996 e{-3}$ & $-1.7717 e{-5}$ & $2.8168 e{-3}$ & $3.2696 e{-4}$ & $4.8814 e{0}$ & $4.2120 e{-2}$ \\
Dim 0 $q = p = \infty$ & $-2.6933 e{-2}$ & $-4.0844 e{-2}$ & $-1.3022 e{-2}$ & $3.2240 e{-3}$ & $6.7520 e{0}$ & $4.6800 e{-3}$ \\
Dim 1 $q = p = 1$ & $1.0460 e{-2}$ & $6.3590 e{-3}$ & $1.4562 e{-2}$ & $1.1523 e{-3}$ & $8.4706 e{0}$ & $4.0000 e{-5}$ \\
Dim 1 $q = p = 2$ & $-3.5318 e{-1}$ & $-5.5446 e{-1}$ & $-1.5191 e{-1}$ & $6.1924 e{-2}$ & $5.2441 e{0}$ & $9.3600 e{-3}$ \\
Dim 2 $q = p = 1$ & $8.2001 e{-2}$ & $5.0725 e{-2}$ & $1.1328 e{-1}$ & $1.0095 e{-2}$ & $8.2064 e{0}$ & $0.0000 e{0}$ \\
Dim 2 $q = p = 2$ & $-2.2388 e{0}$ & $-3.1626 e{0}$ & $-1.3149 e{0}$ & $3.4067 e{-1}$ & $6.9419 e{0}$ & $4.0000 e{-5}$ \\
Dim 2 $q = p = 3$ & $5.7072 e{0}$ & $2.5076 e{0}$ & $8.9068 e{0}$ & $1.1902 e{0}$ & $5.1960 e{0}$ & $3.1600 e{-3}$ \\
Dim 2 $q = p = 4$ & $-3.9269 e{0}$ & $-6.6403 e{0}$ & $-1.2135 e{0}$ & $9.9198 e{-1}$ & $4.2785 e{0}$ & $1.7200 e{-2}$ \\

\bottomrule
\end{tabular}
\end{sc}
\end{small}
\end{center}
\vskip -0.1in
\end{table}

\begin{figure}[H]
\vskip 0.2in
\begin{center}
\centerline{\includegraphics[width=\columnwidth*\columnWidthCoefResultsCoefsCI]{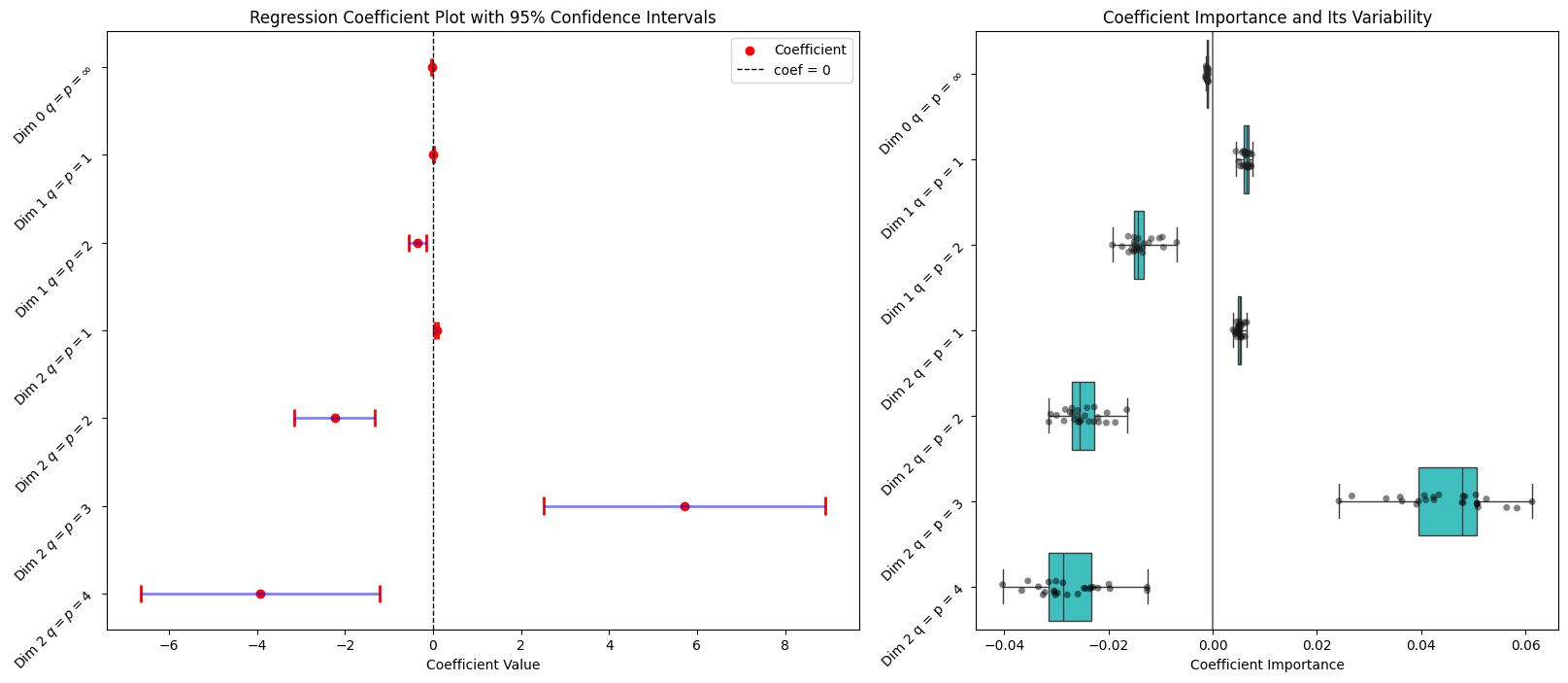}}
\caption{This figure illustrates the filtered-in features generated from the Second Pass (\cref{Second Pass}) for the ROI: MFG and the hemisphere: L(eft) where the training data contains "Only Trained" LLMs. On the left sub-figure, each feature's Weight is presented along with both the Lower and Upper Bounds of the $95\%$ Confidence Intervals, while the right sub-figure illustrates its Coefficient Importance and Variability.}
\label{only_trained_L_MFG_coefs_combined_4}
\end{center}
\vskip -0.2in
\end{figure}

\begin{table}[H]
\caption{This table summarizes the filtered-in features generated from the Second Pass (\cref{Second Pass}) for the ROI: MFG and the hemisphere: L(eft) where the training data contains both "Only Trained" and "Trained plus Untrained" LLMs.}
\label{with_untrained_L_MFG}
\vskip 0.15in
\begin{center}
\begin{small}
\begin{sc}
\begin{tabular}{ccccccc}
\toprule
Feature & Weight & $95\%$ CI & $95\%$ CI & SE & $|t|$ & $p$  \\
 &  & Lower & Upper &  & &  \\
\midrule
(Intercept) & $-4.7110 e{-3}$ & $-8.5453 e{-3}$ & $-8.7671 e{-4}$ & $1.2991 e{-3}$ & $3.4611 e{0}$ & $5.7360 e{-2}$ \\
Dim 0 $q = p = 1$ & $5.7421 e{-4}$ & $1.6202 e{-4}$ & $9.8641 e{-4}$ & $1.2700 e{-4}$ & $5.2930 e{0}$ & $9.2800 e{-3}$ \\
Dim 0 $q = p = 25$ & $-2.9137 e{-2}$ & $-5.2704 e{-2}$ & $-5.5700 e{-3}$ & $8.2875 e{-3}$ & $3.6270 e{0}$ & $3.2360 e{-2}$ \\
Dim 0 $q = p = 27$ & $3.1774 e{-2}$ & $2.9214 e{-3}$ & $6.0626 e{-2}$ & $6.5506 e{-3}$ & $4.9152 e{0}$ & $4.5000 e{-2}$ \\
Dim 1 $q = p = 1$ & $6.1024 e{-3}$ & $2.7949 e{-3}$ & $9.4098 e{-3}$ & $8.7892 e{-4}$ & $7.1884 e{0}$ & $8.4000 e{-4}$ \\
Dim 1 $q = p = 2$ & $-2.7618 e{-1}$ & $-4.5964 e{-1}$ & $-9.2723 e{-2}$ & $5.1548 e{-2}$ & $5.9465 e{0}$ & $3.5200 e{-3}$ \\
Dim 1 $q = p = 9$ & $3.1466 e{-1}$ & $5.9412 e{-2}$ & $5.6991 e{-1}$ & $5.2077 e{-2}$ & $5.3063 e{0}$ & $4.9400 e{-2}$ \\
Dim 1 $q = p = 14$ & $-2.4714 e{-2}$ & $-3.6881 e{-2}$ & $-1.2547 e{-2}$ & $4.0324 e{-3}$ & $4.2467 e{0}$ & $1.9320 e{-2}$ \\
Dim 1 $q = p = 27$ & $2.8023 e{-2}$ & $4.7186 e{-3}$ & $5.1327 e{-2}$ & $8.8894 e{-3}$ & $3.2704 e{0}$ & $4.7680 e{-2}$ \\
Dim 2 $q = p = 8$ & $-7.9962 e{-1}$ & $-1.3330 e{0}$ & $-2.6619 e{-1}$ & $9.0912 e{-2}$ & $7.4300 e{0}$ & $1.8280 e{-2}$ \\
Dim 2 $q = p = 16$ & $6.8196 e{-3}$ & $7.1987 e{-4}$ & $1.2919 e{-2}$ & $1.6549 e{-3}$ & $4.3284 e{0}$ & $4.2920 e{-2}$ \\

\bottomrule
\end{tabular}
\end{sc}
\end{small}
\end{center}
\vskip -0.1in
\end{table}

\begin{figure}[H]
\vskip 0.2in
\begin{center}
\centerline{\includegraphics[width=\columnwidth*\columnWidthCoefResultsCoefsCI]{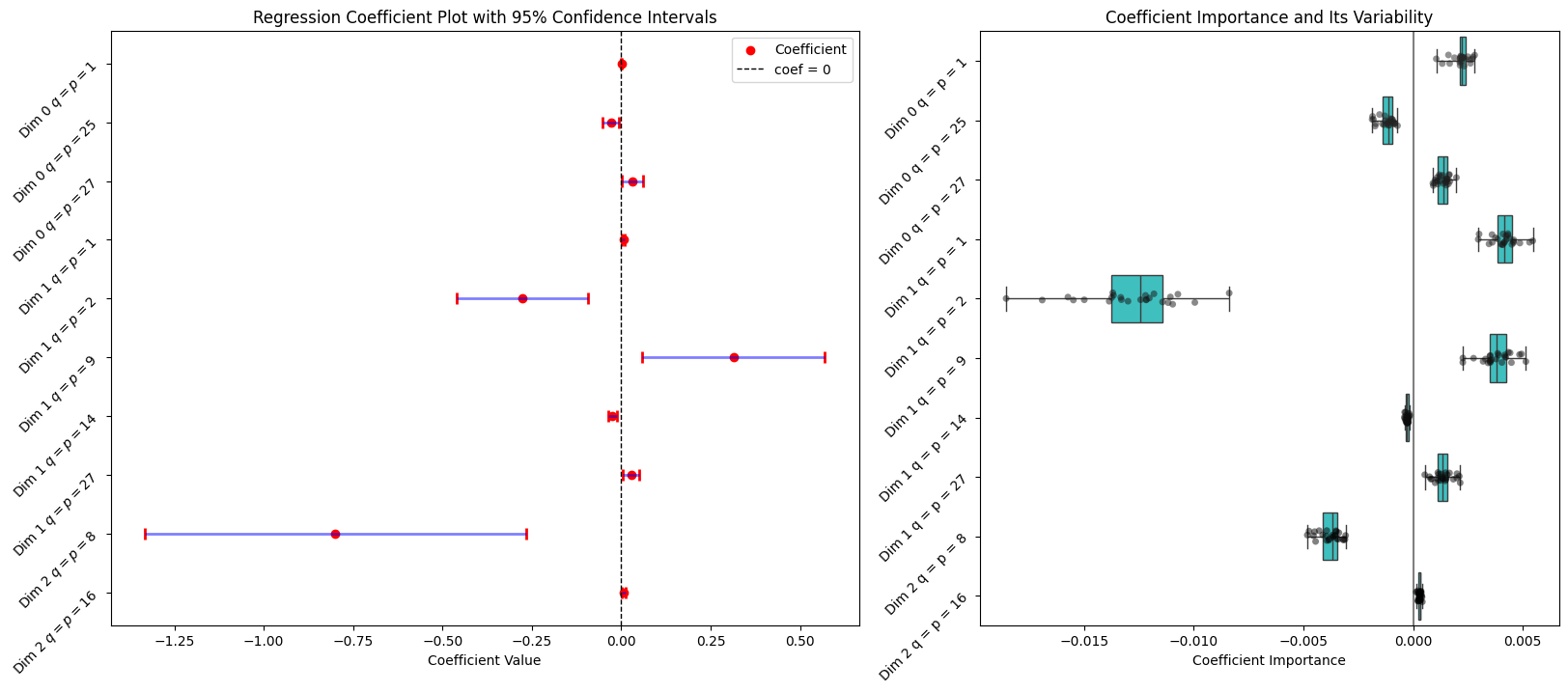}}
\caption{This figure illustrates the filtered-in features generated from the Second Pass (\cref{Second Pass}) for the ROI: MFG and the hemisphere: L(eft) where the training data contains both "Only Trained" and "Trained plus Untrained" LLMs. On the left sub-figure, each feature's Weight is presented along with both the Lower and Upper Bounds of the $95\%$ Confidence Intervals, while the right sub-figure illustrates its Coefficient Importance and Variability.}
\label{with_untrained_L_MFG_coefs_combined_27}
\end{center}
\vskip -0.2in
\end{figure}

\begin{table}[H]
\caption{This table summarizes the filtered-in features generated from the Second Pass (\cref{Second Pass}) for the ROI: MFG and the hemisphere: R(ight) where the training data contains "Only Trained" LLMs.}
\label{only_trained_R_MFG}
\vskip 0.15in
\begin{center}
\begin{small}
\begin{sc}
\begin{tabular}{ccccccc}
\toprule
Feature & Weight & $95\%$ CI & $95\%$ CI & SE & $|t|$ & $p$  \\
 &  & Lower & Upper &  & &  \\
\midrule
(Intercept) & $-5.1684 e{-3}$ & $-1.0312 e{-2}$ & $-2.4684 e{-5}$ & $9.4711 e{-4}$ & $4.6363 e{0}$ & $1.1100 e{-1}$ \\
Dim 0 $q = p = 1$ & $5.2862 e{-4}$ & $-8.4094 e{-5}$ & $1.1413 e{-3}$ & $1.8096 e{-4}$ & $4.5573 e{0}$ & $2.1240 e{-2}$ \\
Dim 1 $q = p = 1$ & $9.6453 e{-3}$ & $4.9470 e{-3}$ & $1.4344 e{-2}$ & $1.1250 e{-3}$ & $7.1302 e{0}$ & $2.4000 e{-3}$ \\

\bottomrule
\end{tabular}
\end{sc}
\end{small}
\end{center}
\vskip -0.1in
\end{table}

\begin{figure}[H]
\vskip 0.2in
\begin{center}
\centerline{\includegraphics[width=\columnwidth*\columnWidthCoefResultsCoefsCI]{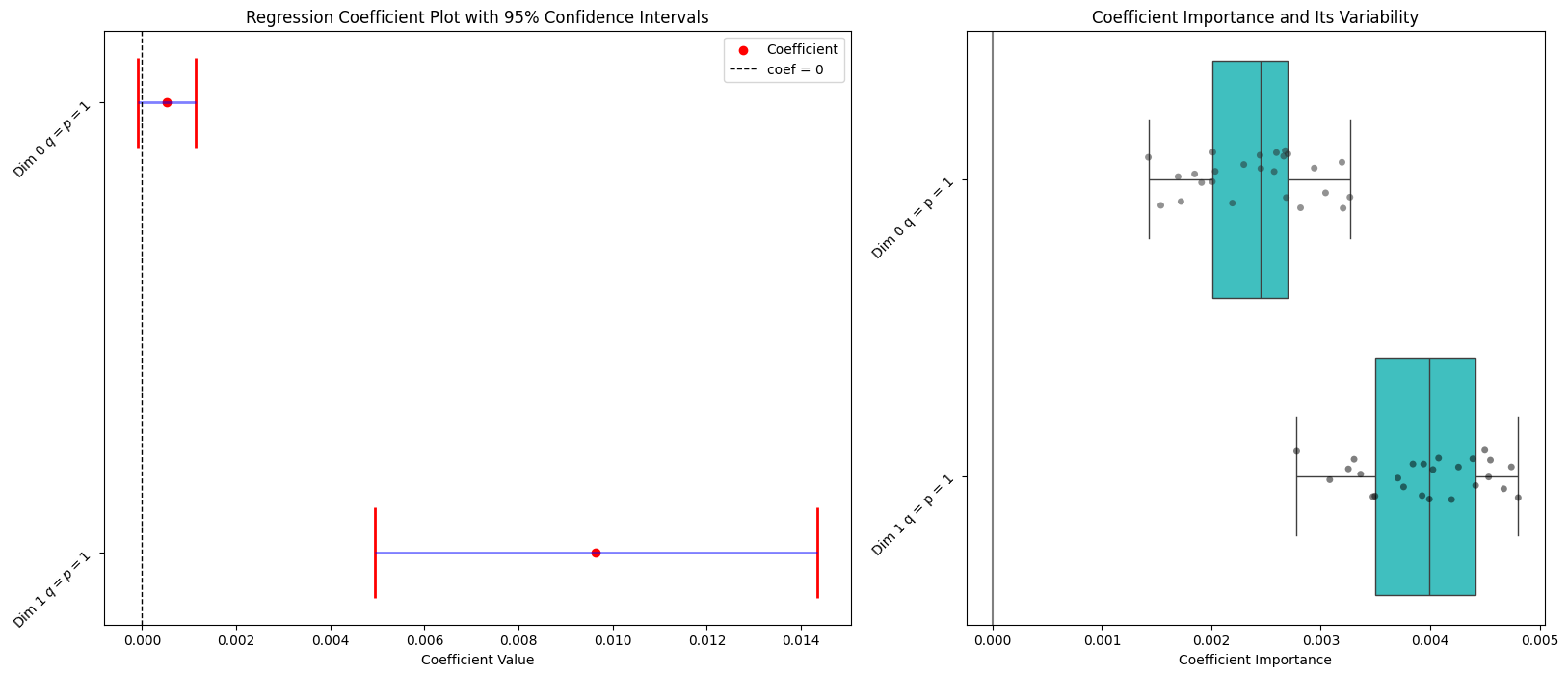}}
\caption{This figure illustrates the filtered-in features generated from the Second Pass (\cref{Second Pass}) for the ROI: MFG and the hemisphere: R(ight) where the training data contains "Only Trained" LLMs. On the left sub-figure, each feature's Weight is presented along with both the Lower and Upper Bounds of the $95\%$ Confidence Intervals, while the right sub-figure illustrates its Coefficient Importance and Variability.}
\label{only_trained_R_MFG_coefs_combined_21}
\end{center}
\vskip -0.2in
\end{figure}

\begin{table}[H]
\caption{This table summarizes the filtered-in features generated from the Second Pass (\cref{Second Pass}) for the ROI: MFG and the hemisphere: R(ight) where the training data contains both "Only Trained" and "Trained plus Untrained" LLMs.}
\label{with_untrained_R_MFG}
\vskip 0.15in
\begin{center}
\begin{small}
\begin{sc}
\begin{tabular}{ccccccc}
\toprule
Feature & Weight & $95\%$ CI & $95\%$ CI & SE & $|t|$ & $p$  \\
 &  & Lower & Upper &  & &  \\
\midrule
(Intercept) & $-6.1642 e{-3}$ & $-9.9033 e{-3}$ & $-2.4250 e{-3}$ & $1.1508 e{-3}$ & $4.8619 e{0}$ & $9.4400 e{-3}$ \\
Dim 0 $q = p = 1$ & $1.5097 e{-3}$ & $1.0501 e{-3}$ & $1.9693 e{-3}$ & $1.1655 e{-4}$ & $1.0806 e{1}$ & $0.0000 e{0}$ \\
Dim 0 $q = p = 2$ & $-1.2065 e{-1}$ & $-1.5796 e{-1}$ & $-8.3346 e{-2}$ & $1.2607 e{-2}$ & $7.2642 e{0}$ & $0.0000 e{0}$ \\
Dim 0 $q = p = 3$ & $6.5482 e{-1}$ & $4.1103 e{-1}$ & $8.9862 e{-1}$ & $8.2447 e{-2}$ & $5.8368 e{0}$ & $5.2000 e{-4}$ \\
Dim 0 $q = p = 4$ & $-1.3070 e{0}$ & $-2.0279 e{0}$ & $-5.8613 e{-1}$ & $2.3516 e{-1}$ & $3.7931 e{0}$ & $3.6440 e{-2}$ \\
Dim 0 $q = p = 16$ & $3.6950 e{-2}$ & $1.0030 e{-2}$ & $6.3870 e{-2}$ & $6.5351 e{-3}$ & $5.1724 e{0}$ & $2.7240 e{-2}$ \\
Dim 1 $q = p = 1$ & $6.5595 e{-3}$ & $3.0950 e{-3}$ & $1.0024 e{-2}$ & $9.5399 e{-4}$ & $7.6541 e{0}$ & $2.8000 e{-4}$ \\
Dim 1 $q = p = 2$ & $-1.8015 e{-1}$ & $-3.6976 e{-1}$ & $9.4571 e{-3}$ & $6.1966 e{-2}$ & $3.6304 e{0}$ & $4.8040 e{-2}$ \\
Dim 1 $q = p = 17$ & $1.6031 e{-2}$ & $1.5339 e{-3}$ & $3.0529 e{-2}$ & $3.4386 e{-3}$ & $5.0093 e{0}$ & $3.4280 e{-2}$ \\
Dim 2 $q = p = \infty$ & $-5.8378 e{-1}$ & $-1.3998 e{0}$ & $2.3229 e{-1}$ & $2.0274 e{-1}$ & $4.5419 e{0}$ & $4.4680 e{-2}$ \\

\bottomrule
\end{tabular}
\end{sc}
\end{small}
\end{center}
\vskip -0.1in
\end{table}

\begin{figure}[H]
\vskip 0.2in
\begin{center}
\centerline{\includegraphics[width=\columnwidth*\columnWidthCoefResultsCoefsCI]{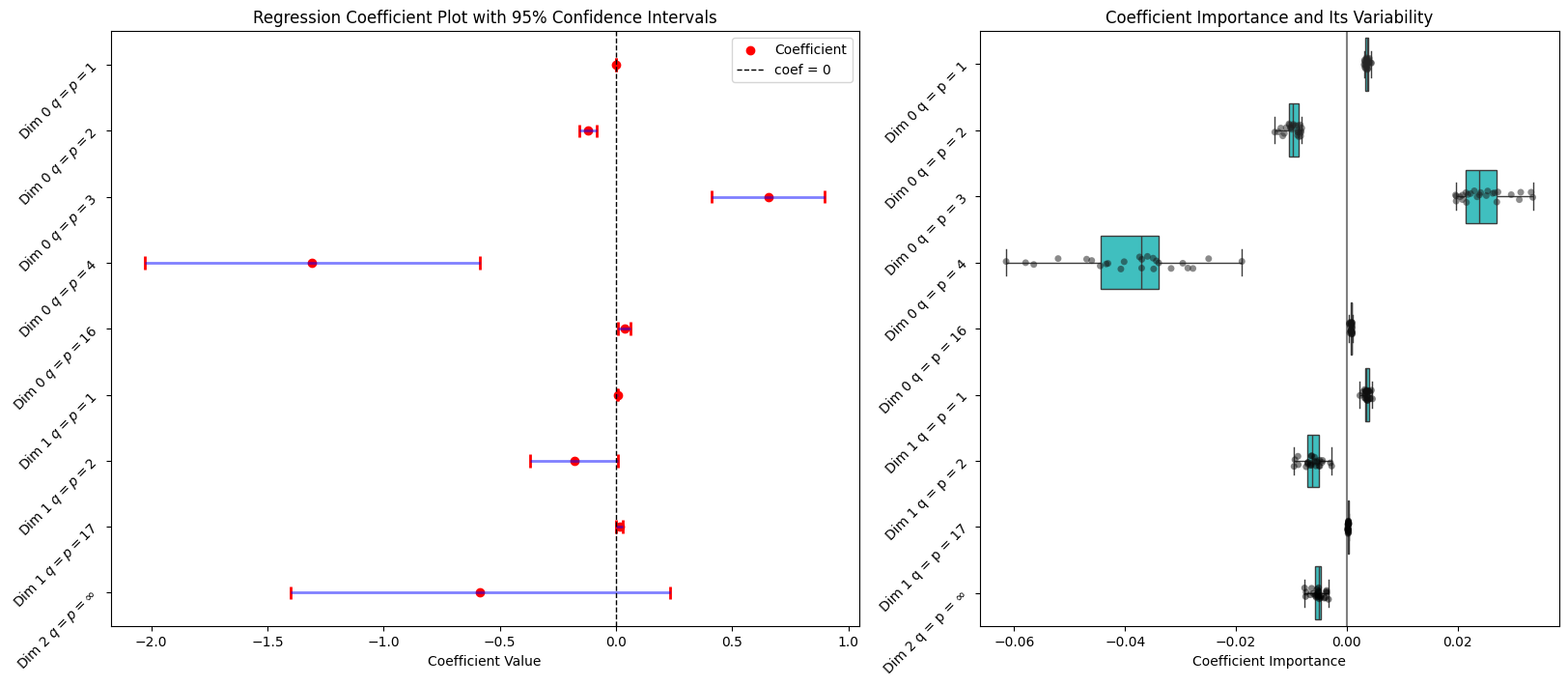}}
\caption{This figure illustrates the filtered-in features generated from the Second Pass (\cref{Second Pass}) for the ROI: MFG and the hemisphere: R(ight) where the training data contains both "Only Trained" and "Trained plus Untrained" LLMs. On the left sub-figure, each feature's Weight is presented along with both the Lower and Upper Bounds of the $95\%$ Confidence Intervals, while the right sub-figure illustrates its Coefficient Importance and Variability.}
\label{with_untrained_R_MFG_coefs_combined_22}
\end{center}
\vskip -0.2in
\end{figure}

\begin{table}[H]
\caption{This table summarizes the filtered-in features generated from the Second Pass (\cref{Second Pass}) for the ROI: IFGorb and the hemisphere: L(eft) where the training data contains "Only Trained" LLMs.}
\label{only_trained_L_IFGorb}
\vskip 0.15in
\begin{center}
\begin{small}
\begin{sc}
\begin{tabular}{ccccccc}
\toprule
Feature & Weight & $95\%$ CI & $95\%$ CI & SE & $|t|$ & $p$  \\
 &  & Lower & Upper &  & &  \\
\midrule
(Intercept) & $4.0872 e{-3}$ & $3.4022 e{-3}$ & $4.7722 e{-3}$ & $2.6092 e{-4}$ & $1.4420 e{1}$ & $0.0000 e{0}$ \\
Dim 0 $q = p = 1$ & $1.3223 e{-3}$ & $1.1003 e{-3}$ & $1.5443 e{-3}$ & $1.0719 e{-4}$ & $1.2588 e{1}$ & $0.0000 e{0}$ \\
Dim 0 $q = p = 2$ & $-8.1721 e{-2}$ & $-9.4914 e{-2}$ & $-6.8527 e{-2}$ & $6.6871 e{-3}$ & $1.1971 e{1}$ & $0.0000 e{0}$ \\
Dim 0 $q = p = 3$ & $2.4845 e{-1}$ & $2.0322 e{-1}$ & $2.9368 e{-1}$ & $2.2562 e{-2}$ & $1.0374 e{1}$ & $0.0000 e{0}$ \\
Dim 0 $q = p = 4$ & $-1.7140 e{-1}$ & $-2.0797 e{-1}$ & $-1.3484 e{-1}$ & $1.7459 e{-2}$ & $8.8373 e{0}$ & $0.0000 e{0}$ \\
Dim 0 $q = p = \infty$ & $1.2560 e{-2}$ & $6.2300 e{-3}$ & $1.8889 e{-2}$ & $2.0932 e{-3}$ & $3.9695 e{0}$ & $2.6040 e{-2}$ \\
Dim 1 $q = p = 1$ & $5.6828 e{-3}$ & $3.7792 e{-3}$ & $7.5864 e{-3}$ & $6.4386 e{-4}$ & $8.0676 e{0}$ & $0.0000 e{0}$ \\
Dim 1 $q = p = 2$ & $-2.3237 e{-1}$ & $-3.1542 e{-1}$ & $-1.4931 e{-1}$ & $2.6543 e{-2}$ & $8.2521 e{0}$ & $0.0000 e{0}$ \\
Dim 1 $q = p = 3$ & $5.8867 e{-1}$ & $3.0317 e{-1}$ & $8.7416 e{-1}$ & $9.0539 e{-2}$ & $6.3086 e{0}$ & $4.0000 e{-4}$ \\
Dim 1 $q = p = 4$ & $-3.2586 e{-1}$ & $-5.6097 e{-1}$ & $-9.0763 e{-2}$ & $7.0794 e{-2}$ & $4.6249 e{0}$ & $1.5760 e{-2}$ \\

\bottomrule
\end{tabular}
\end{sc}
\end{small}
\end{center}
\vskip -0.1in
\end{table}

\begin{figure}[H]
\vskip 0.2in
\begin{center}
\centerline{\includegraphics[width=\columnwidth*\columnWidthCoefResultsCoefsCI]{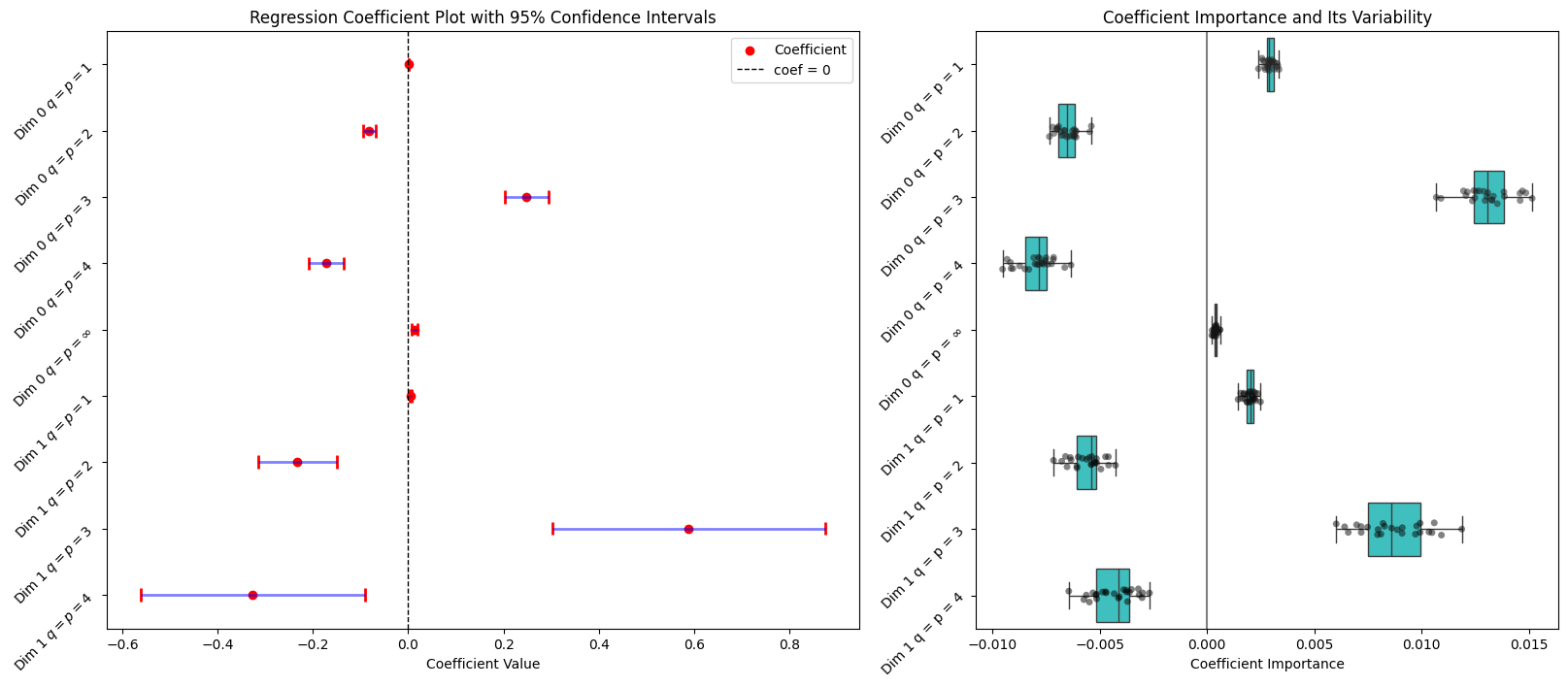}}
\caption{This figure illustrates the filtered-in features generated from the Second Pass (\cref{Second Pass}) for the ROI: IFGorb and the hemisphere: L(eft) where the training data contains "Only Trained" LLMs. On the left sub-figure, each feature's Weight is presented along with both the Lower and Upper Bounds of the $95\%$ Confidence Intervals, while the right sub-figure illustrates its Coefficient Importance and Variability.}
\label{only_trained_L_IFGorb_coefs_combined_4}
\end{center}
\vskip -0.2in
\end{figure}

\begin{table}[H]
\caption{This table summarizes the filtered-in features generated from the Second Pass (\cref{Second Pass}) for the ROI: IFGorb and the hemisphere: L(eft) where the training data contains both "Only Trained" and "Trained plus Untrained" LLMs.}
\label{with_untrained_L_IFGorb}
\vskip 0.15in
\begin{center}
\begin{small}
\begin{sc}
\begin{tabular}{ccccccc}
\toprule
Feature & Weight & $95\%$ CI & $95\%$ CI & SE & $|t|$ & $p$  \\
 &  & Lower & Upper &  & &  \\
\midrule
(Intercept) & $4.8182 e{-3}$ & $4.2913 e{-3}$ & $5.3451 e{-3}$ & $2.1064 e{-4}$ & $2.2569 e{1}$ & $0.0000 e{0}$ \\
Dim 0 $q = p = 1$ & $1.5360 e{-3}$ & $1.3878 e{-3}$ & $1.6842 e{-3}$ & $4.0320 e{-5}$ & $3.6555 e{1}$ & $0.0000 e{0}$ \\
Dim 0 $q = p = 2$ & $-9.3796 e{-2}$ & $-1.0299 e{-1}$ & $-8.4602 e{-2}$ & $3.5408 e{-3}$ & $2.6042 e{1}$ & $0.0000 e{0}$ \\
Dim 0 $q = p = 3$ & $2.7084 e{-1}$ & $2.3770 e{-1}$ & $3.0397 e{-1}$ & $1.5124 e{-2}$ & $1.8135 e{1}$ & $0.0000 e{0}$ \\
Dim 0 $q = p = 4$ & $-1.7918 e{-1}$ & $-2.0667 e{-1}$ & $-1.5169 e{-1}$ & $1.3121 e{-2}$ & $1.4126 e{1}$ & $0.0000 e{0}$ \\
Dim 0 $q = p = \infty$ & $1.0916 e{-2}$ & $6.3192 e{-3}$ & $1.5514 e{-2}$ & $1.5222 e{-3}$ & $7.7552 e{0}$ & $0.0000 e{0}$ \\
Dim 1 $q = p = 1$ & $4.7381 e{-3}$ & $3.3740 e{-3}$ & $6.1021 e{-3}$ & $4.7664 e{-4}$ & $1.1084 e{1}$ & $0.0000 e{0}$ \\
Dim 1 $q = p = 2$ & $-2.1427 e{-1}$ & $-2.7891 e{-1}$ & $-1.4964 e{-1}$ & $2.5617 e{-2}$ & $8.5284 e{0}$ & $0.0000 e{0}$ \\
Dim 1 $q = p = 3$ & $6.1890 e{-1}$ & $3.8935 e{-1}$ & $8.4846 e{-1}$ & $9.3479 e{-2}$ & $6.2885 e{0}$ & $4.0000 e{-5}$ \\
Dim 1 $q = p = 4$ & $-4.0466 e{-1}$ & $-5.9288 e{-1}$ & $-2.1644 e{-1}$ & $7.1281 e{-2}$ & $5.1461 e{0}$ & $1.6800 e{-3}$ \\
Dim 2 $q = p = \infty$ & $-2.3388 e{-1}$ & $-3.7978 e{-1}$ & $-8.7981 e{-2}$ & $3.8588 e{-2}$ & $5.6401 e{0}$ & $9.5600 e{-3}$ \\

\bottomrule
\end{tabular}
\end{sc}
\end{small}
\end{center}
\vskip -0.1in
\end{table}

\begin{figure}[H]
\vskip 0.2in
\begin{center}
\centerline{\includegraphics[width=\columnwidth*\columnWidthCoefResultsCoefsCI]{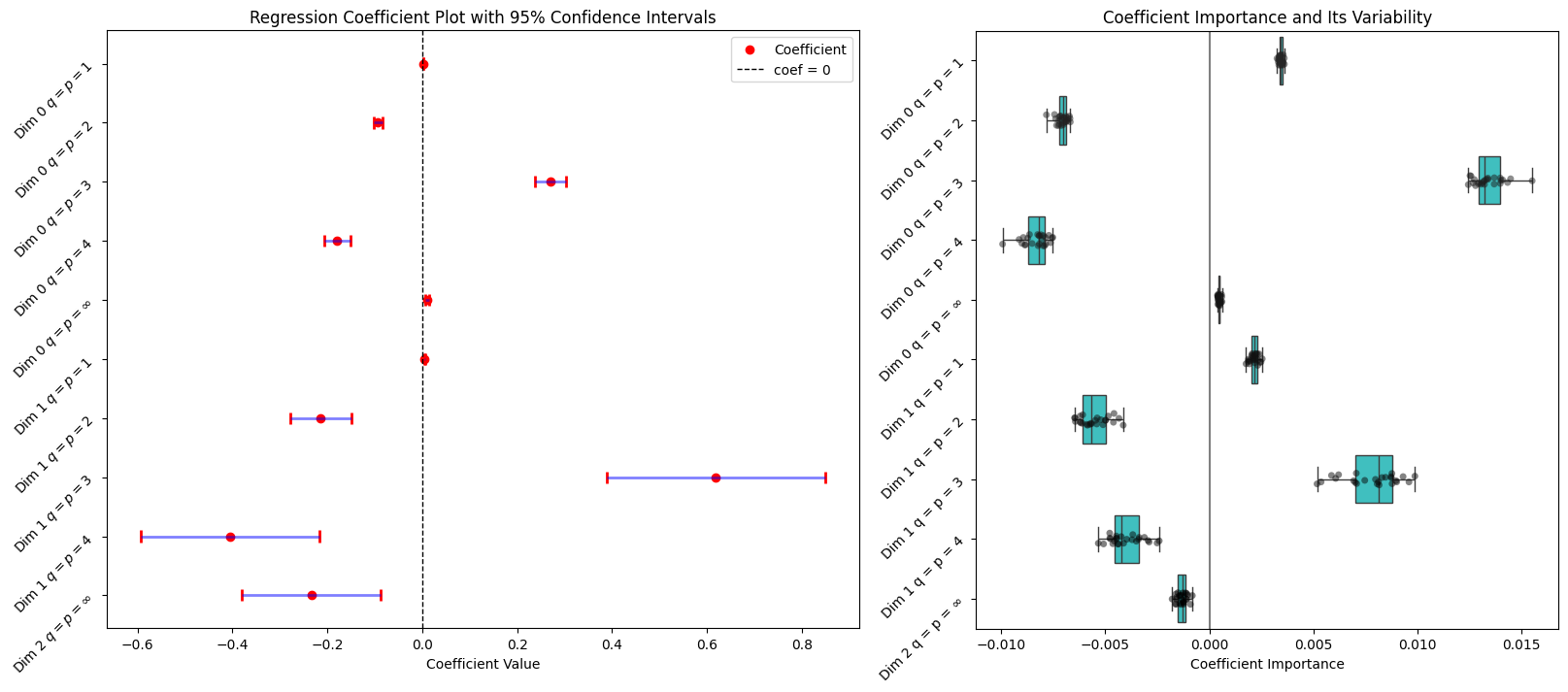}}
\caption{This figure illustrates the filtered-in features generated from the Second Pass (\cref{Second Pass}) for the ROI: IFGorb and the hemisphere: L(eft) where the training data contains both "Only Trained" and "Trained plus Untrained" LLMs. On the left sub-figure, each feature's Weight is presented along with both the Lower and Upper Bounds of the $95\%$ Confidence Intervals, while the right sub-figure illustrates its Coefficient Importance and Variability.}
\label{with_untrained_L_IFGorb_coefs_combined_4}
\end{center}
\vskip -0.2in
\end{figure}

\begin{table}[H]
\caption{This table summarizes the filtered-in features generated from the Second Pass (\cref{Second Pass}) for the ROI: IFGorb and the hemisphere: R(ight) where the training data contains "Only Trained" LLMs.}
\label{only_trained_R_IFGorb}
\vskip 0.15in
\begin{center}
\begin{small}
\begin{sc}
\begin{tabular}{ccccccc}
\toprule
Feature & Weight & $95\%$ CI & $95\%$ CI & SE & $|t|$ & $p$  \\
 &  & Lower & Upper &  & &  \\
\midrule
(Intercept) & $4.1390 e{-3}$ & $3.4394 e{-3}$ & $4.8387 e{-3}$ & $2.2546 e{-4}$ & $1.7859 e{1}$ & $0.0000 e{0}$ \\
Dim 0 $q = p = 1$ & $1.4564 e{-3}$ & $1.2480 e{-3}$ & $1.6647 e{-3}$ & $7.8326 e{-5}$ & $1.8145 e{1}$ & $0.0000 e{0}$ \\
Dim 0 $q = p = 2$ & $-7.5929 e{-2}$ & $-8.8622 e{-2}$ & $-6.3237 e{-2}$ & $5.2309 e{-3}$ & $1.3765 e{1}$ & $0.0000 e{0}$ \\
Dim 0 $q = p = 3$ & $1.9063 e{-1}$ & $1.4891 e{-1}$ & $2.3236 e{-1}$ & $1.6993 e{-2}$ & $1.0269 e{1}$ & $0.0000 e{0}$ \\
Dim 0 $q = p = 4$ & $-1.0865 e{-1}$ & $-1.4263 e{-1}$ & $-7.4677 e{-2}$ & $1.3197 e{-2}$ & $7.0897 e{0}$ & $0.0000 e{0}$ \\
Dim 1 $q = p = 1$ & $2.2672 e{-3}$ & $2.7942 e{-4}$ & $4.2551 e{-3}$ & $5.0943 e{-4}$ & $4.4725 e{0}$ & $4.3880 e{-2}$ \\
Dim 2 $q = p = 1$ & $2.5250 e{-2}$ & $1.0881 e{-2}$ & $3.9619 e{-2}$ & $4.4580 e{-3}$ & $5.7732 e{0}$ & $3.4800 e{-3}$ \\
Dim 2 $q = p = 2$ & $-5.6038 e{-1}$ & $-9.4559 e{-1}$ & $-1.7516 e{-1}$ & $1.2595 e{-1}$ & $4.4972 e{0}$ & $1.7400 e{-2}$ \\

\bottomrule
\end{tabular}
\end{sc}
\end{small}
\end{center}
\vskip -0.1in
\end{table}

\begin{figure}[H]
\vskip 0.2in
\begin{center}
\centerline{\includegraphics[width=\columnwidth*\columnWidthCoefResultsCoefsCI]{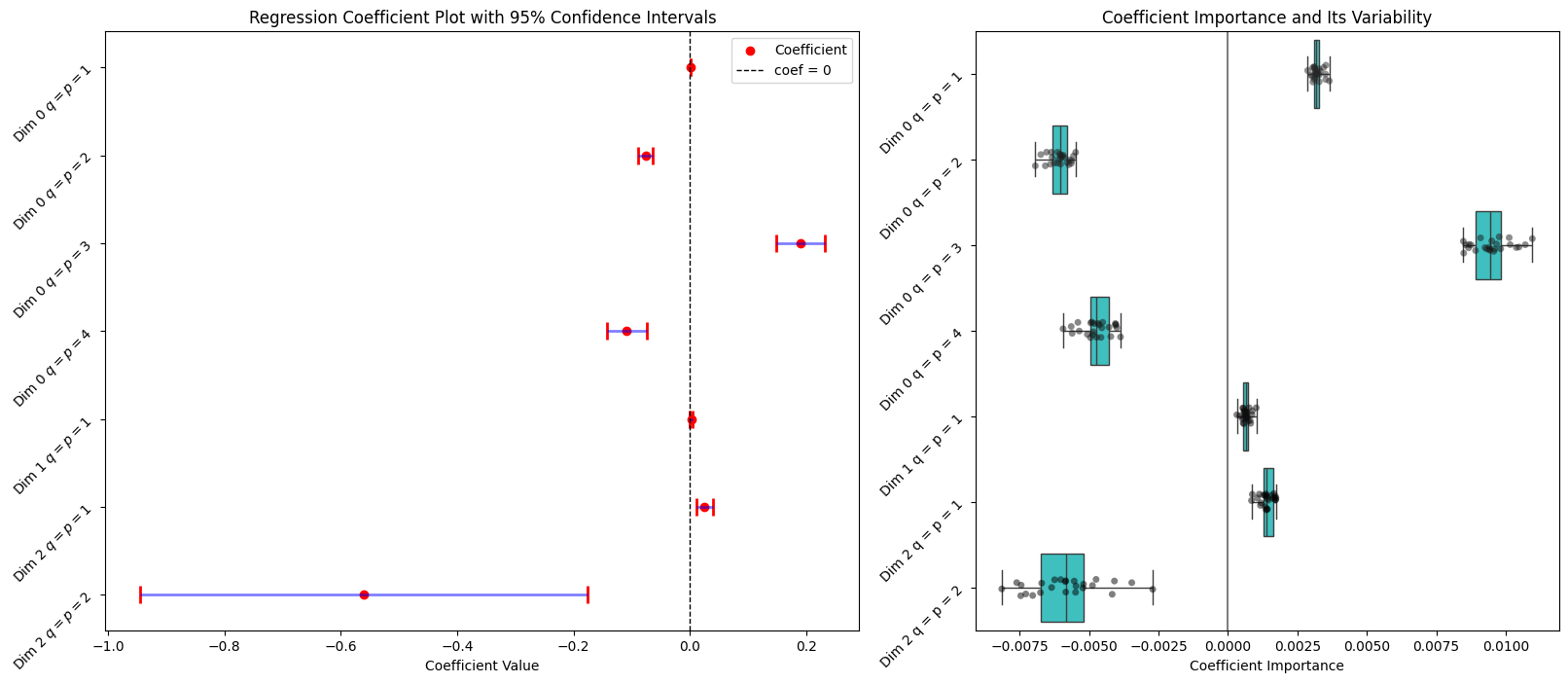}}
\caption{This figure illustrates the filtered-in features generated from the Second Pass (\cref{Second Pass}) for the ROI: IFGorb and the hemisphere: R(ight) where the training data contains "Only Trained" LLMs. On the left sub-figure, each feature's Weight is presented along with both the Lower and Upper Bounds of the $95\%$ Confidence Intervals, while the right sub-figure illustrates its Coefficient Importance and Variability.}
\label{only_trained_R_IFGorb_coefs_combined_4}
\end{center}
\vskip -0.2in
\end{figure}

\begin{table}[H]
\caption{This table summarizes the filtered-in features generated from the Second Pass (\cref{Second Pass}) for the ROI: IFGorb and the hemisphere: R(ight) where the training data contains both "Only Trained" and "Trained plus Untrained" LLMs.}
\label{with_untrained_R_IFGorb}
\vskip 0.15in
\begin{center}
\begin{small}
\begin{sc}
\begin{tabular}{ccccccc}
\toprule
Feature & Weight & $95\%$ CI & $95\%$ CI & SE & $|t|$ & $p$  \\
 &  & Lower & Upper &  & &  \\
\midrule
(Intercept) & $4.8847 e{-3}$ & $4.3702 e{-3}$ & $5.3992 e{-3}$ & $1.4309 e{-4}$ & $3.3710 e{1}$ & $0.0000 e{0}$ \\
Dim 0 $q = p = 1$ & $1.4334 e{-3}$ & $1.2884 e{-3}$ & $1.5784 e{-3}$ & $5.8138 e{-5}$ & $2.4453 e{1}$ & $0.0000 e{0}$ \\
Dim 0 $q = p = 2$ & $-7.1814 e{-2}$ & $-8.1013 e{-2}$ & $-6.2615 e{-2}$ & $4.0488 e{-3}$ & $1.8175 e{1}$ & $0.0000 e{0}$ \\
Dim 0 $q = p = 3$ & $1.7307 e{-1}$ & $1.4182 e{-1}$ & $2.0432 e{-1}$ & $1.3839 e{-2}$ & $1.2966 e{1}$ & $0.0000 e{0}$ \\
Dim 0 $q = p = 4$ & $-9.4650 e{-2}$ & $-1.2056 e{-1}$ & $-6.8738 e{-2}$ & $1.0656 e{-2}$ & $9.2866 e{0}$ & $0.0000 e{0}$ \\
Dim 1 $q = p = 1$ & $2.7981 e{-3}$ & $1.3820 e{-3}$ & $4.2143 e{-3}$ & $4.5536 e{-4}$ & $6.4843 e{0}$ & $4.8000 e{-4}$ \\
Dim 1 $q = p = 2$ & $-1.5449 e{-1}$ & $-2.2333 e{-1}$ & $-8.5637 e{-2}$ & $2.7692 e{-2}$ & $4.8937 e{0}$ & $2.4000 e{-3}$ \\
Dim 1 $q = p = 3$ & $4.6717 e{-1}$ & $2.2176 e{-1}$ & $7.1258 e{-1}$ & $1.0415 e{-1}$ & $3.6902 e{0}$ & $2.5640 e{-2}$ \\

\bottomrule
\end{tabular}
\end{sc}
\end{small}
\end{center}
\vskip -0.1in
\end{table}

\begin{figure}[H]
\vskip 0.2in
\begin{center}
\centerline{\includegraphics[width=\columnwidth*\columnWidthCoefResultsCoefsCI]{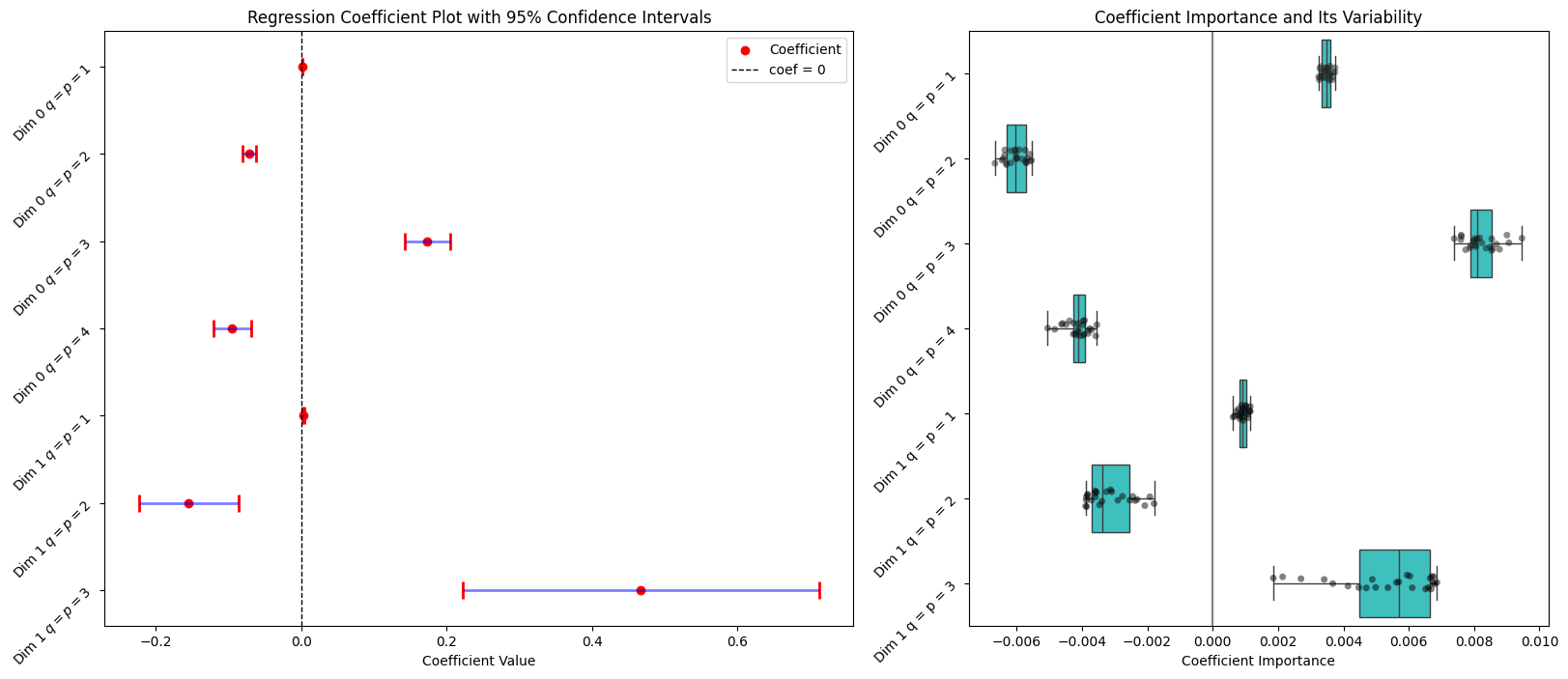}}
\caption{This figure illustrates the filtered-in features generated from the Second Pass (\cref{Second Pass}) for the ROI: IFGorb and the hemisphere: R(ight) where the training data contains both "Only Trained" and "Trained plus Untrained" LLMs. On the left sub-figure, each feature's Weight is presented along with both the Lower and Upper Bounds of the $95\%$ Confidence Intervals, while the right sub-figure illustrates its Coefficient Importance and Variability.}
\label{with_untrained_R_IFGorb_coefs_combined_4}
\end{center}
\vskip -0.2in
\end{figure}

\begin{table}[H]
\caption{This table summarizes the filtered-in features generated from the Second Pass (\cref{Second Pass}) for the ROI: PCC and the hemisphere: L(eft) where the training data contains "Only Trained" LLMs.}
\label{only_trained_L_PCC}
\vskip 0.15in
\begin{center}
\begin{small}
\begin{sc}
\begin{tabular}{ccccccc}
\toprule
Feature & Weight & $95\%$ CI & $95\%$ CI & SE & $|t|$ & $p$  \\
 &  & Lower & Upper &  & &  \\
\midrule
(Intercept) & $2.8611 e{-3}$ & $2.1406 e{-3}$ & $3.5817 e{-3}$ & $2.1186 e{-4}$ & $1.4518 e{1}$ & $0.0000 e{0}$ \\
Dim 0 $q = p = 1$ & $9.0675 e{-4}$ & $7.0720 e{-4}$ & $1.1063 e{-3}$ & $6.4641 e{-5}$ & $1.2677 e{1}$ & $0.0000 e{0}$ \\
Dim 0 $q = p = 2$ & $-6.5188 e{-2}$ & $-7.7791 e{-2}$ & $-5.2586 e{-2}$ & $4.8102 e{-3}$ & $1.2467 e{1}$ & $0.0000 e{0}$ \\
Dim 0 $q = p = 3$ & $1.9371 e{-1}$ & $1.4809 e{-1}$ & $2.3934 e{-1}$ & $1.6057 e{-2}$ & $1.1113 e{1}$ & $0.0000 e{0}$ \\
Dim 0 $q = p = 4$ & $-1.2030 e{-1}$ & $-1.5858 e{-1}$ & $-8.2019 e{-2}$ & $1.2365 e{-2}$ & $9.2920 e{0}$ & $0.0000 e{0}$ \\
Dim 1 $q = p = 1$ & $4.7096 e{-3}$ & $2.7731 e{-3}$ & $6.6460 e{-3}$ & $5.4751 e{-4}$ & $9.0836 e{0}$ & $0.0000 e{0}$ \\
Dim 1 $q = p = 2$ & $-1.5204 e{-1}$ & $-2.2845 e{-1}$ & $-7.5625 e{-2}$ & $2.4904 e{-2}$ & $6.2067 e{0}$ & $6.8000 e{-4}$ \\
Dim 1 $q = p = 3$ & $4.5030 e{-1}$ & $1.7614 e{-1}$ & $7.2447 e{-1}$ & $8.9790 e{-2}$ & $5.0927 e{0}$ & $5.4800 e{-3}$ \\
Dim 1 $q = p = 4$ & $-3.2848 e{-1}$ & $-5.6898 e{-1}$ & $-8.7975 e{-2}$ & $7.6992 e{-2}$ & $4.2219 e{0}$ & $2.2800 e{-2}$ \\
Dim 2 $q = p = 1$ & $1.4710 e{-2}$ & $1.6354 e{-4}$ & $2.9256 e{-2}$ & $4.7872 e{-3}$ & $3.6213 e{0}$ & $4.3920 e{-2}$ \\

\bottomrule
\end{tabular}
\end{sc}
\end{small}
\end{center}
\vskip -0.1in
\end{table}

\begin{figure}[H]
\vskip 0.2in
\begin{center}
\centerline{\includegraphics[width=\columnwidth*\columnWidthCoefResultsCoefsCI]{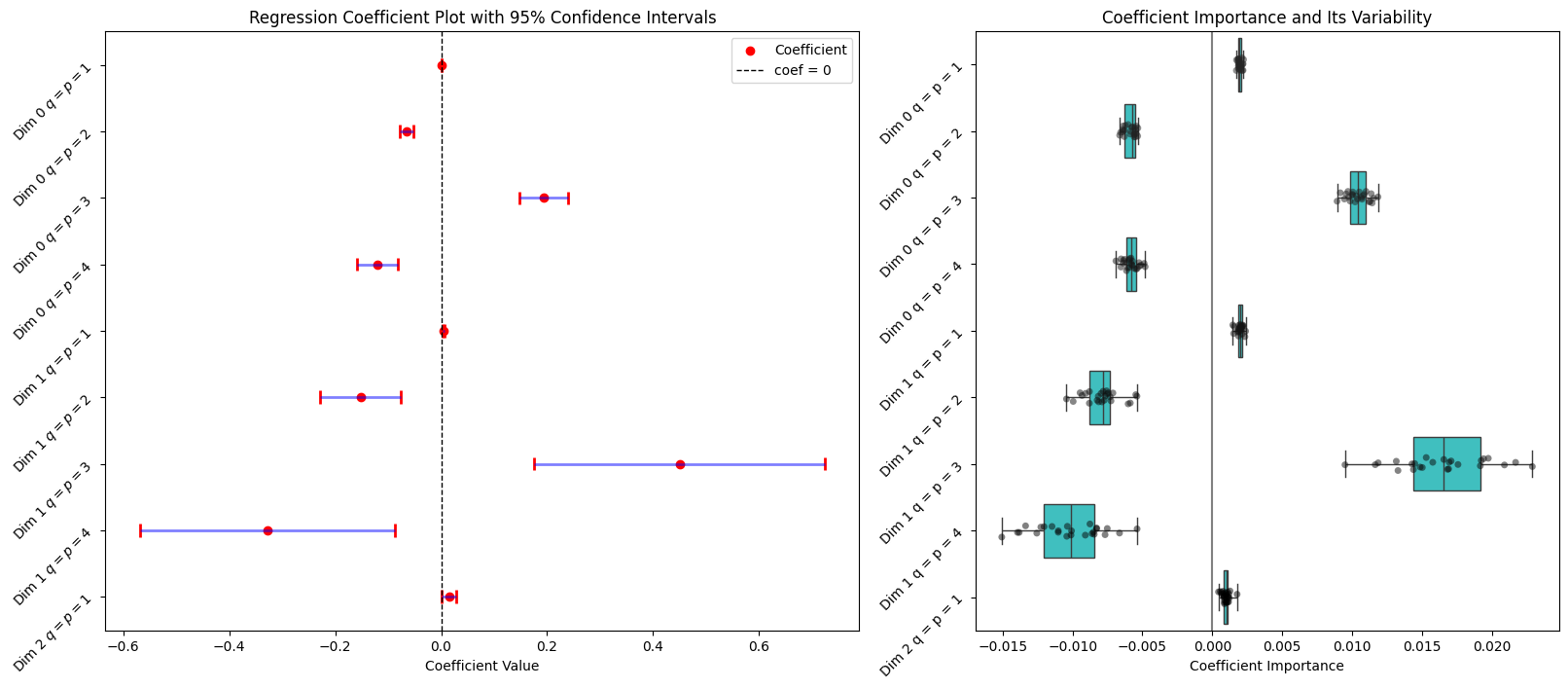}}
\caption{This figure illustrates the filtered-in features generated from the Second Pass (\cref{Second Pass}) for the ROI: PCC and the hemisphere: L(eft) where the training data contains "Only Trained" LLMs. On the left sub-figure, each feature's Weight is presented along with both the Lower and Upper Bounds of the $95\%$ Confidence Intervals, while the right sub-figure illustrates its Coefficient Importance and Variability.}
\label{only_trained_L_PCC_coefs_combined_4}
\end{center}
\vskip -0.2in
\end{figure}

\begin{table}[H]
\caption{This table summarizes the filtered-in features generated from the Second Pass (\cref{Second Pass}) for the ROI: PCC and the hemisphere: L(eft) where the training data contains both "Only Trained" and "Trained plus Untrained" LLMs.}
\label{with_untrained_L_PCC}
\vskip 0.15in
\begin{center}
\begin{small}
\begin{sc}
\begin{tabular}{ccccccc}
\toprule
Feature & Weight & $95\%$ CI & $95\%$ CI & SE & $|t|$ & $p$  \\
 &  & Lower & Upper &  & &  \\
\midrule
(Intercept) & $3.7603 e{-3}$ & $3.1880 e{-3}$ & $4.3325 e{-3}$ & $1.8512 e{-4}$ & $1.9837 e{1}$ & $0.0000 e{0}$ \\
Dim 0 $q = p = 1$ & $8.0481 e{-4}$ & $6.6375 e{-4}$ & $9.4586 e{-4}$ & $4.2085 e{-5}$ & $2.0047 e{1}$ & $0.0000 e{0}$ \\
Dim 0 $q = p = 2$ & $-4.8646 e{-2}$ & $-5.7792 e{-2}$ & $-3.9500 e{-2}$ & $2.4564 e{-3}$ & $2.0622 e{1}$ & $0.0000 e{0}$ \\
Dim 0 $q = p = 3$ & $1.3175 e{-1}$ & $9.6681 e{-2}$ & $1.6683 e{-1}$ & $8.7956 e{-3}$ & $1.6049 e{1}$ & $0.0000 e{0}$ \\
Dim 0 $q = p = 4$ & $-8.1977 e{-2}$ & $-1.1222 e{-1}$ & $-5.1737 e{-2}$ & $7.0822 e{-3}$ & $1.2585 e{1}$ & $0.0000 e{0}$ \\
Dim 1 $q = p = 1$ & $5.7958 e{-3}$ & $4.4040 e{-3}$ & $7.1875 e{-3}$ & $4.2205 e{-4}$ & $1.2288 e{1}$ & $0.0000 e{0}$ \\
Dim 1 $q = p = 2$ & $-2.3960 e{-1}$ & $-2.9739 e{-1}$ & $-1.8181 e{-1}$ & $1.7591 e{-2}$ & $1.1567 e{1}$ & $0.0000 e{0}$ \\
Dim 1 $q = p = 3$ & $7.0155 e{-1}$ & $4.8696 e{-1}$ & $9.1615 e{-1}$ & $6.3730 e{-2}$ & $8.8358 e{0}$ & $0.0000 e{0}$ \\
Dim 1 $q = p = 4$ & $-4.6938 e{-1}$ & $-6.5634 e{-1}$ & $-2.8241 e{-1}$ & $5.3688 e{-2}$ & $6.8522 e{0}$ & $7.2000 e{-4}$ \\

\bottomrule
\end{tabular}
\end{sc}
\end{small}
\end{center}
\vskip -0.1in
\end{table}

\begin{figure}[H]
\vskip 0.2in
\begin{center}
\centerline{\includegraphics[width=\columnwidth*\columnWidthCoefResultsCoefsCI]{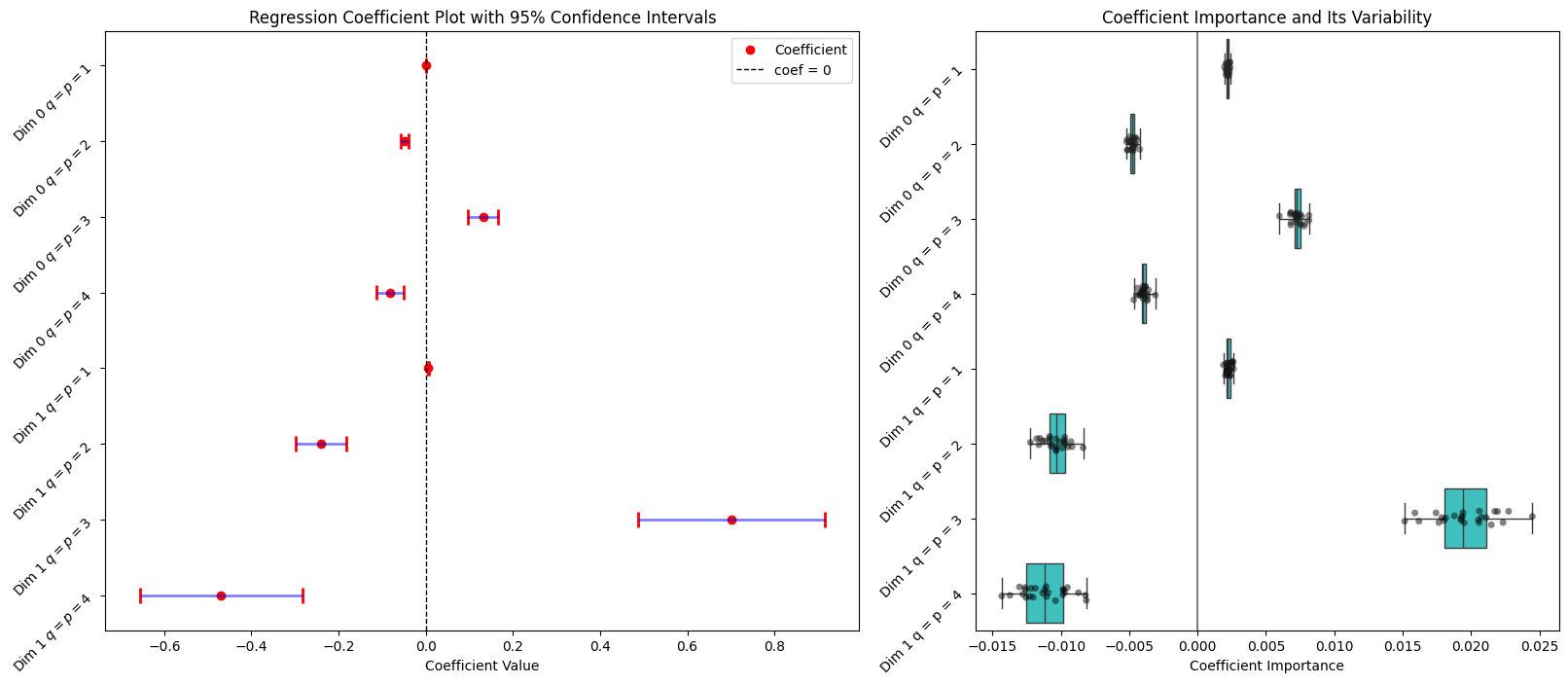}}
\caption{This figure illustrates the filtered-in features generated from the Second Pass (\cref{Second Pass}) for the ROI: PCC and the hemisphere: L(eft) where the training data contains both "Only Trained" and "Trained plus Untrained" LLMs. On the left sub-figure, each feature's Weight is presented along with both the Lower and Upper Bounds of the $95\%$ Confidence Intervals, while the right sub-figure illustrates its Coefficient Importance and Variability.}
\label{with_untrained_L_PCC_coefs_combined_4}
\end{center}
\vskip -0.2in
\end{figure}

\begin{table}[H]
\caption{This table summarizes the filtered-in features generated from the Second Pass (\cref{Second Pass}) for the ROI: PCC and the hemisphere: R(ight) where the training data contains "Only Trained" LLMs.}
\label{only_trained_R_PCC}
\vskip 0.15in
\begin{center}
\begin{small}
\begin{sc}
\begin{tabular}{ccccccc}
\toprule
Feature & Weight & $95\%$ CI & $95\%$ CI & SE & $|t|$ & $p$  \\
 &  & Lower & Upper &  & &  \\
\midrule
(Intercept) & $1.7410 e{-3}$ & $1.1174 e{-3}$ & $2.3645 e{-3}$ & $1.7364 e{-4}$ & $1.0852 e{1}$ & $0.0000 e{0}$ \\
Dim 0 $q = p = 1$ & $4.4869 e{-4}$ & $2.0435 e{-4}$ & $6.9302 e{-4}$ & $6.0081 e{-5}$ & $7.5765 e{0}$ & $8.8000 e{-4}$ \\
Dim 0 $q = p = 2$ & $-2.5439 e{-2}$ & $-3.6694 e{-2}$ & $-1.4184 e{-2}$ & $2.6778 e{-3}$ & $1.0059 e{1}$ & $0.0000 e{0}$ \\
Dim 0 $q = p = 3$ & $7.2124 e{-2}$ & $3.6451 e{-2}$ & $1.0780 e{-1}$ & $8.5342 e{-3}$ & $9.5325 e{0}$ & $0.0000 e{0}$ \\
Dim 0 $q = p = 4$ & $-3.9465 e{-2}$ & $-6.8156 e{-2}$ & $-1.0774 e{-2}$ & $6.5266 e{-3}$ & $7.4546 e{0}$ & $2.4400 e{-3}$ \\
Dim 1 $q = p = 1$ & $5.8645 e{-3}$ & $3.7476 e{-3}$ & $7.9814 e{-3}$ & $5.7891 e{-4}$ & $1.0361 e{1}$ & $0.0000 e{0}$ \\
Dim 1 $q = p = 2$ & $-1.8806 e{-1}$ & $-2.8471 e{-1}$ & $-9.1417 e{-2}$ & $2.5273 e{-2}$ & $6.9312 e{0}$ & $1.3600 e{-3}$ \\
Dim 1 $q = p = 3$ & $5.1224 e{-1}$ & $1.6175 e{-1}$ & $8.6273 e{-1}$ & $9.3985 e{-2}$ & $4.3896 e{0}$ & $4.0080 e{-2}$ \\

\bottomrule
\end{tabular}
\end{sc}
\end{small}
\end{center}
\vskip -0.1in
\end{table}

\begin{figure}[H]
\vskip 0.2in
\begin{center}
\centerline{\includegraphics[width=\columnwidth*\columnWidthCoefResultsCoefsCI]{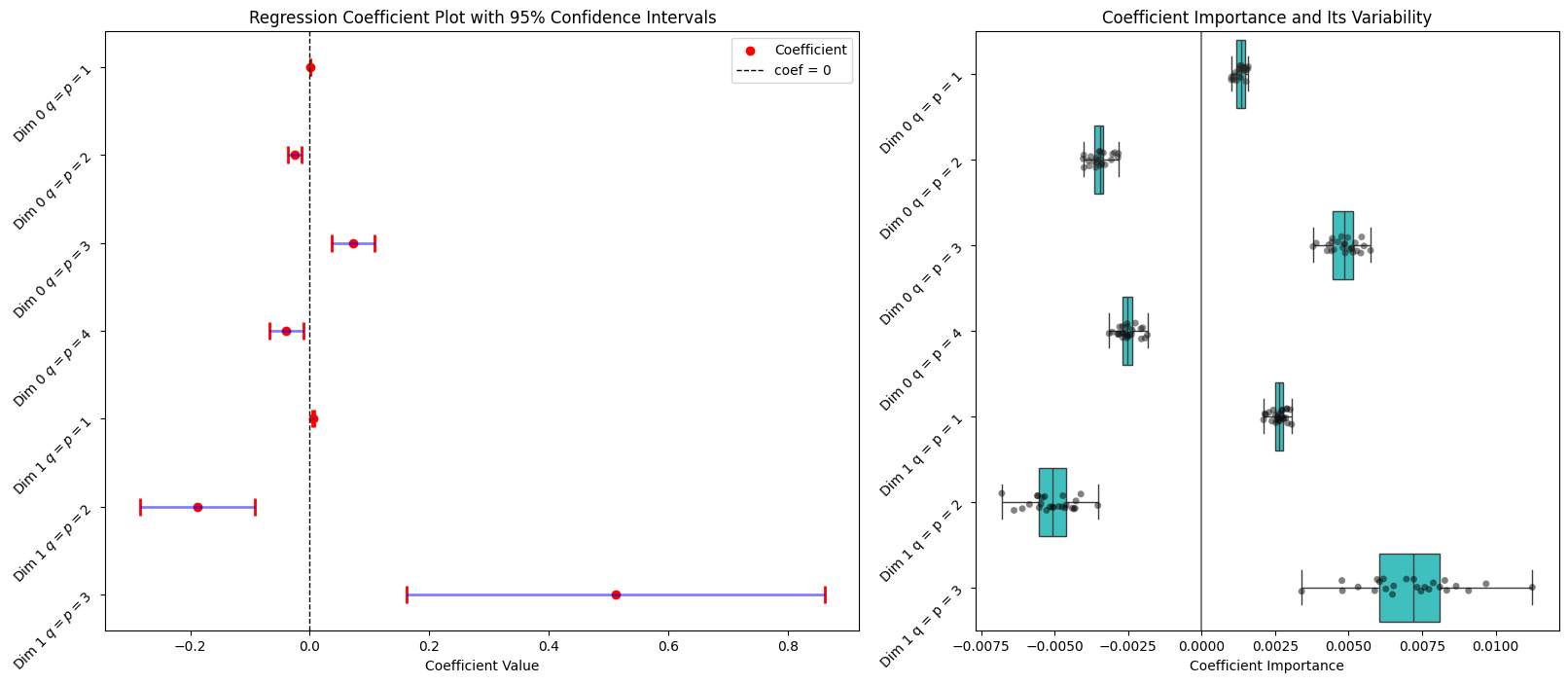}}
\caption{This figure illustrates the filtered-in features generated from the Second Pass (\cref{Second Pass}) for the ROI: PCC and the hemisphere: R(ight) where the training data contains "Only Trained" LLMs. On the left sub-figure, each feature's Weight is presented along with both the Lower and Upper Bounds of the $95\%$ Confidence Intervals, while the right sub-figure illustrates its Coefficient Importance and Variability.}
\label{only_trained_R_PCC_coefs_combined_4}
\end{center}
\vskip -0.2in
\end{figure}

\begin{table}[H]
\caption{This table summarizes the filtered-in features generated from the Second Pass (\cref{Second Pass}) for the ROI: PCC and the hemisphere: R(ight) where the training data contains both "Only Trained" and "Trained plus Untrained" LLMs.}
\label{with_untrained_R_PCC}
\vskip 0.15in
\begin{center}
\begin{small}
\begin{sc}
\begin{tabular}{ccccccc}
\toprule
Feature & Weight & $95\%$ CI & $95\%$ CI & SE & $|t|$ & $p$  \\
 &  & Lower & Upper &  & &  \\
\midrule
(Intercept) & $2.9357 e{-3}$ & $2.4457 e{-3}$ & $3.4258 e{-3}$ & $9.3103 e{-5}$ & $2.9678 e{1}$ & $0.0000 e{0}$ \\
Dim 0 $q = p = 1$ & $7.9441 e{-4}$ & $5.9643 e{-4}$ & $9.9239 e{-4}$ & $5.0970 e{-5}$ & $1.5234 e{1}$ & $0.0000 e{0}$ \\
Dim 0 $q = p = 2$ & $-3.9411 e{-2}$ & $-5.0555 e{-2}$ & $-2.8267 e{-2}$ & $3.0967 e{-3}$ & $1.1528 e{1}$ & $0.0000 e{0}$ \\
Dim 0 $q = p = 3$ & $1.4536 e{-1}$ & $8.6666 e{-2}$ & $2.0406 e{-1}$ & $1.5243 e{-2}$ & $7.7238 e{0}$ & $3.2000 e{-4}$ \\
Dim 0 $q = p = 4$ & $-2.6746 e{-1}$ & $-4.1629 e{-1}$ & $-1.1862 e{-1}$ & $3.4332 e{-2}$ & $5.9503 e{0}$ & $1.3400 e{-2}$ \\
Dim 0 $q = p = 5$ & $3.6745 e{-1}$ & $1.6384 e{-1}$ & $5.7106 e{-1}$ & $4.4056 e{-2}$ & $6.7737 e{0}$ & $7.4400 e{-3}$ \\
Dim 0 $q = p = 6$ & $-2.1043 e{-1}$ & $-3.2185 e{-1}$ & $-9.9004 e{-2}$ & $2.3175 e{-2}$ & $7.8935 e{0}$ & $2.4800 e{-3}$ \\
Dim 0 $q = p = \infty$ & $1.0477 e{-2}$ & $-3.4399 e{-4}$ & $2.1299 e{-2}$ & $2.2325 e{-3}$ & $5.2951 e{0}$ & $4.6320 e{-2}$ \\
Dim 1 $q = p = 1$ & $5.0501 e{-3}$ & $3.3402 e{-3}$ & $6.7601 e{-3}$ & $5.6453 e{-4}$ & $1.0015 e{1}$ & $0.0000 e{0}$ \\
Dim 1 $q = p = 2$ & $-2.0141 e{-1}$ & $-3.0629 e{-1}$ & $-9.6519 e{-2}$ & $4.0548 e{-2}$ & $6.6267 e{0}$ & $0.0000 e{0}$ \\
Dim 1 $q = p = 3$ & $5.6797 e{-1}$ & $-1.1902 e{-1}$ & $1.2550 e{0}$ & $2.5080 e{-1}$ & $3.9494 e{0}$ & $1.9440 e{-2}$ \\

\bottomrule
\end{tabular}
\end{sc}
\end{small}
\end{center}
\vskip -0.1in
\end{table}

\begin{figure}[H]
\vskip 0.2in
\begin{center}
\centerline{\includegraphics[width=\columnwidth*\columnWidthCoefResultsCoefsCI]{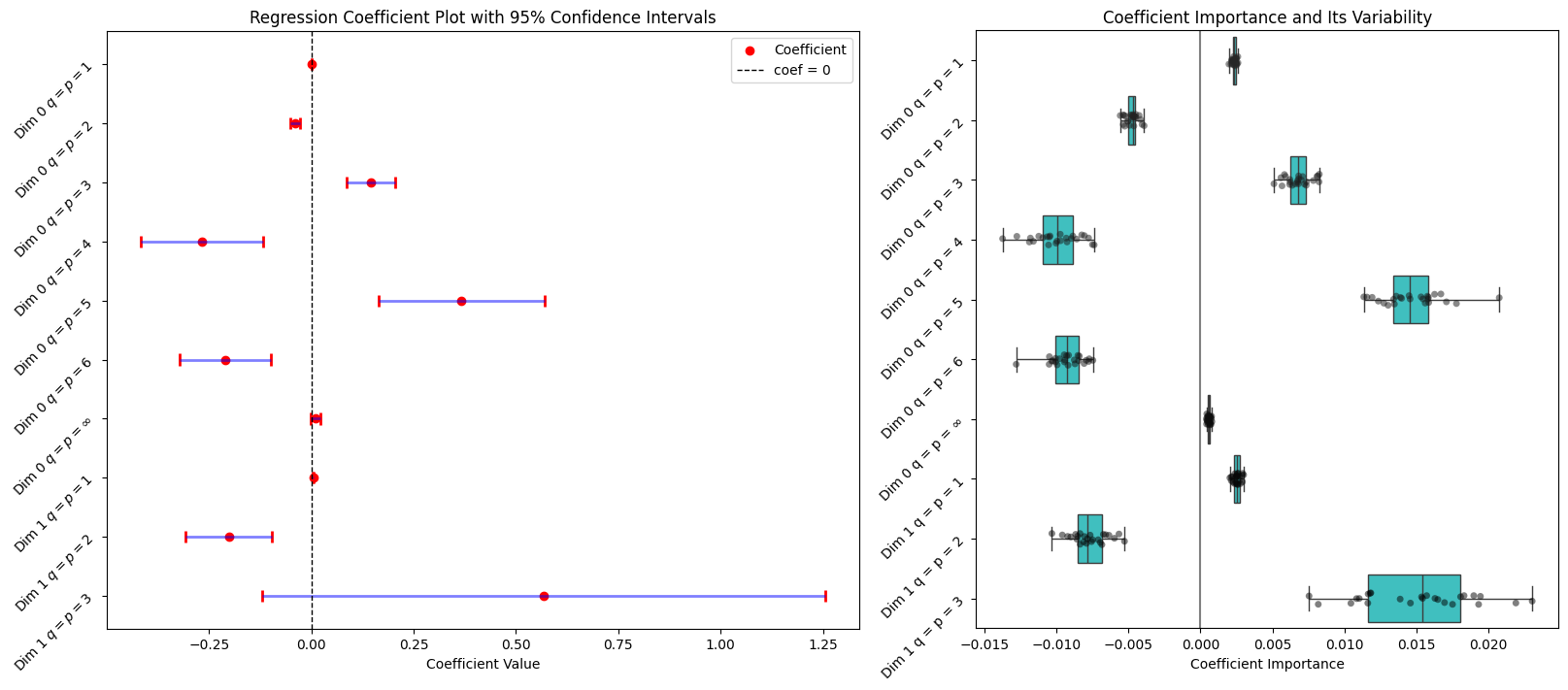}}
\caption{This figure illustrates the filtered-in features generated from the Second Pass (\cref{Second Pass}) for the ROI: PCC and the hemisphere: R(ight) where the training data contains both "Only Trained" and "Trained plus Untrained" LLMs. On the left sub-figure, each feature's Weight is presented along with both the Lower and Upper Bounds of the $95\%$ Confidence Intervals, while the right sub-figure illustrates its Coefficient Importance and Variability.}
\label{with_untrained_R_PCC_coefs_combined_6}
\end{center}
\vskip -0.2in
\end{figure}

\begin{table}[H]
\caption{This table summarizes the filtered-in features generated from the Second Pass (\cref{Second Pass}) for the ROI: dmPFC and the hemisphere: L(eft) where the training data contains "Only Trained" LLMs.}
\label{only_trained_L_dmPFC}
\vskip 0.15in
\begin{center}
\begin{small}
\begin{sc}
\begin{tabular}{ccccccc}
\toprule
Feature & Weight & $95\%$ CI & $95\%$ CI & SE & $|t|$ & $p$  \\
 &  & Lower & Upper &  & &  \\
\midrule
(Intercept) & $1.8114 e{-3}$ & $4.4714 e{-4}$ & $3.1757 e{-3}$ & $3.5342 e{-4}$ & $5.2203 e{0}$ & $2.0280 e{-2}$ \\
Dim 0 $q = p = 1$ & $5.6317 e{-4}$ & $3.5552 e{-4}$ & $7.7083 e{-4}$ & $5.1113 e{-5}$ & $1.1222 e{1}$ & $0.0000 e{0}$ \\
Dim 0 $q = p = 2$ & $-3.3020 e{-2}$ & $-4.5732 e{-2}$ & $-2.0308 e{-2}$ & $3.5458 e{-3}$ & $9.0300 e{0}$ & $0.0000 e{0}$ \\
Dim 0 $q = p = 3$ & $8.8753 e{-2}$ & $2.0050 e{-2}$ & $1.5746 e{-1}$ & $2.3931 e{-2}$ & $3.7293 e{0}$ & $3.4800 e{-2}$ \\
Dim 0 $q = p = 11$ & $6.7070 e{-2}$ & $2.5333 e{-2}$ & $1.0881 e{-1}$ & $1.2946 e{-2}$ & $5.1823 e{0}$ & $7.5200 e{-3}$ \\
Dim 1 $q = p = 1$ & $2.1597 e{-3}$ & $7.1246 e{-4}$ & $3.6069 e{-3}$ & $2.9764 e{-4}$ & $6.4576 e{0}$ & $1.4760 e{-2}$ \\
Dim 1 $q = p = 2$ & $-9.8263 e{-2}$ & $-1.6514 e{-1}$ & $-3.1383 e{-2}$ & $1.3596 e{-2}$ & $5.4787 e{0}$ & $4.2560 e{-2}$ \\
Dim 1 $q = p = 14$ & $-1.1407 e{-2}$ & $-1.7035 e{-2}$ & $-5.7794 e{-3}$ & $1.5985 e{-3}$ & $6.5236 e{0}$ & $1.2400 e{-3}$ \\
Dim 1 $q = p = \infty$ & $-8.3111 e{-2}$ & $-1.4977 e{-1}$ & $-1.6452 e{-2}$ & $1.5753 e{-2}$ & $4.7382 e{0}$ & $4.5160 e{-2}$ \\
Dim 2 $q = p = 16$ & $3.0546 e{-3}$ & $6.0739 e{-4}$ & $5.5017 e{-3}$ & $6.9384 e{-4}$ & $4.0559 e{0}$ & $4.7080 e{-2}$ \\
Dim 2 $q = p = 17$ & $-2.5786 e{-3}$ & $-4.3843 e{-3}$ & $-7.7285 e{-4}$ & $4.2269 e{-4}$ & $7.1505 e{0}$ & $2.2800 e{-3}$ \\

\bottomrule
\end{tabular}
\end{sc}
\end{small}
\end{center}
\vskip -0.1in
\end{table}

\begin{figure}[H]
\vskip 0.2in
\begin{center}
\centerline{\includegraphics[width=\columnwidth*\columnWidthCoefResultsCoefsCI]{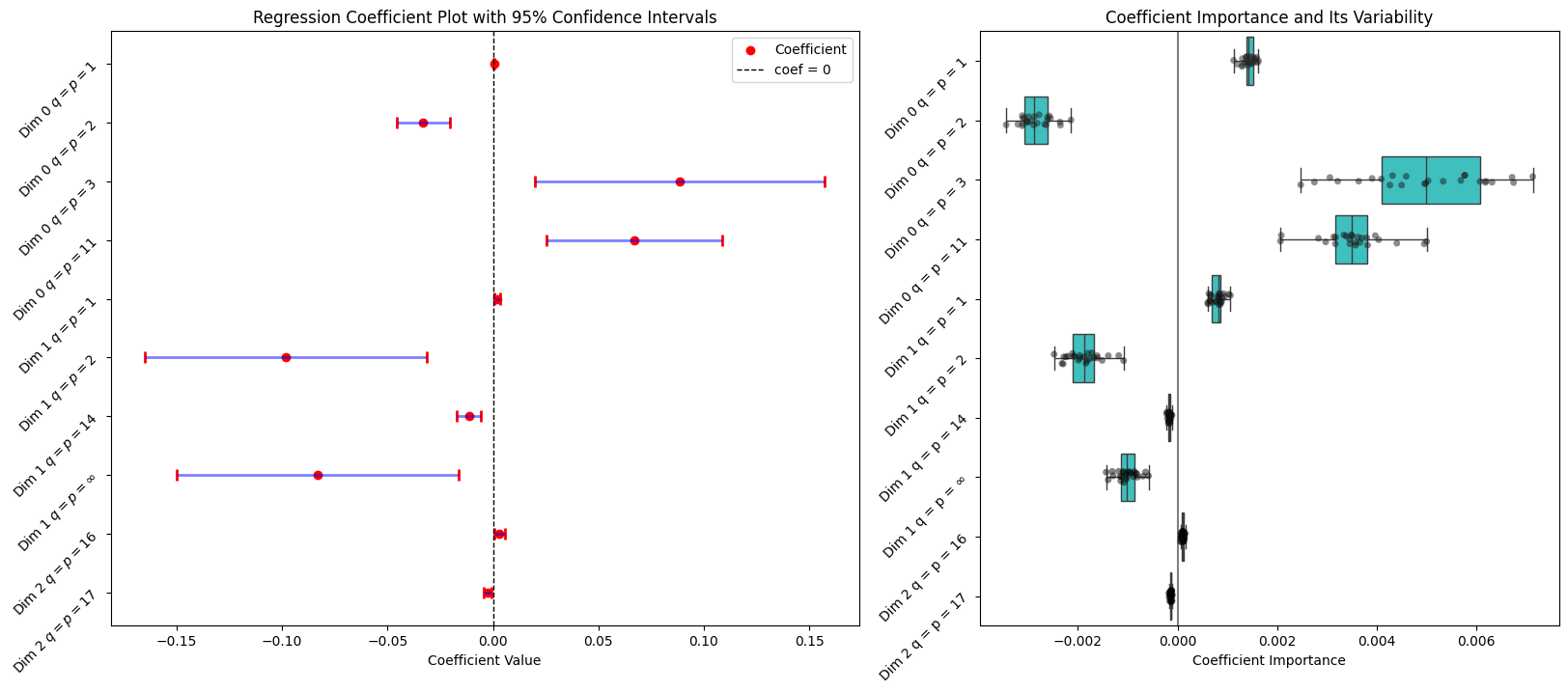}}
\caption{This figure illustrates the filtered-in features generated from the Second Pass (\cref{Second Pass}) for the ROI: dmPFC and the hemisphere: L(eft) where the training data contains "Only Trained" LLMs. On the left sub-figure, each feature's Weight is presented along with both the Lower and Upper Bounds of the $95\%$ Confidence Intervals, while the right sub-figure illustrates its Coefficient Importance and Variability.}
\label{only_trained_L_dmPFC_coefs_combined_17}
\end{center}
\vskip -0.2in
\end{figure}

\begin{table}[H]
\caption{This table summarizes the filtered-in features generated from the Second Pass (\cref{Second Pass}) for the ROI: dmPFC and the hemisphere: L(eft) where the training data contains both "Only Trained" and "Trained plus Untrained" LLMs.}
\label{with_untrained_L_dmPFC}
\vskip 0.15in
\begin{center}
\begin{small}
\begin{sc}
\begin{tabular}{ccccccc}
\toprule
Feature & Weight & $95\%$ CI & $95\%$ CI & SE & $|t|$ & $p$  \\
 &  & Lower & Upper &  & &  \\
\midrule
(Intercept) & $3.1380 e{-4}$ & $-8.5901 e{-4}$ & $1.4866 e{-3}$ & $3.1654 e{-4}$ & $1.3174 e{0}$ & $5.2556 e{-1}$ \\
Dim 0 $q = p = 1$ & $5.7039 e{-4}$ & $4.2105 e{-4}$ & $7.1974 e{-4}$ & $5.4346 e{-5}$ & $1.1831 e{1}$ & $0.0000 e{0}$ \\
Dim 0 $q = p = 2$ & $-2.9279 e{-2}$ & $-3.9455 e{-2}$ & $-1.9103 e{-2}$ & $5.5756 e{-3}$ & $6.4241 e{0}$ & $0.0000 e{0}$ \\
Dim 0 $q = p = 3$ & $1.0095 e{-1}$ & $3.9324 e{-2}$ & $1.6258 e{-1}$ & $4.3705 e{-2}$ & $3.0269 e{0}$ & $8.4000 e{-4}$ \\
Dim 0 $q = p = 11$ & $6.7116 e{-2}$ & $3.3675 e{-2}$ & $1.0056 e{-1}$ & $1.0482 e{-2}$ & $4.5881 e{0}$ & $1.4600 e{-2}$ \\
Dim 1 $q = p = 1$ & $1.4933 e{-3}$ & $4.5474 e{-4}$ & $2.5318 e{-3}$ & $2.9036 e{-4}$ & $5.8492 e{0}$ & $4.1200 e{-3}$ \\
Dim 1 $q = p = 2$ & $-5.6171 e{-2}$ & $-1.0607 e{-1}$ & $-6.2675 e{-3}$ & $1.3891 e{-2}$ & $4.8249 e{0}$ & $2.0200 e{-2}$ \\
Dim 1 $q = p = 22$ & $-6.8932 e{-3}$ & $-1.0936 e{-2}$ & $-2.8502 e{-3}$ & $7.2532 e{-4}$ & $9.1046 e{0}$ & $2.3200 e{-3}$ \\
Dim 1 $q = p = 23$ & $4.5706 e{-3}$ & $1.2615 e{-3}$ & $7.8797 e{-3}$ & $8.2250 e{-4}$ & $4.4876 e{0}$ & $4.8360 e{-2}$ \\
Dim 1 $q = p = \infty$ & $-4.9737 e{-2}$ & $-9.6498 e{-2}$ & $-2.9763 e{-3}$ & $1.0266 e{-2}$ & $6.1257 e{0}$ & $1.5040 e{-2}$ \\

\bottomrule
\end{tabular}
\end{sc}
\end{small}
\end{center}
\vskip -0.1in
\end{table}

\begin{figure}[H]
\vskip 0.2in
\begin{center}
\centerline{\includegraphics[width=\columnwidth*\columnWidthCoefResultsCoefsCI]{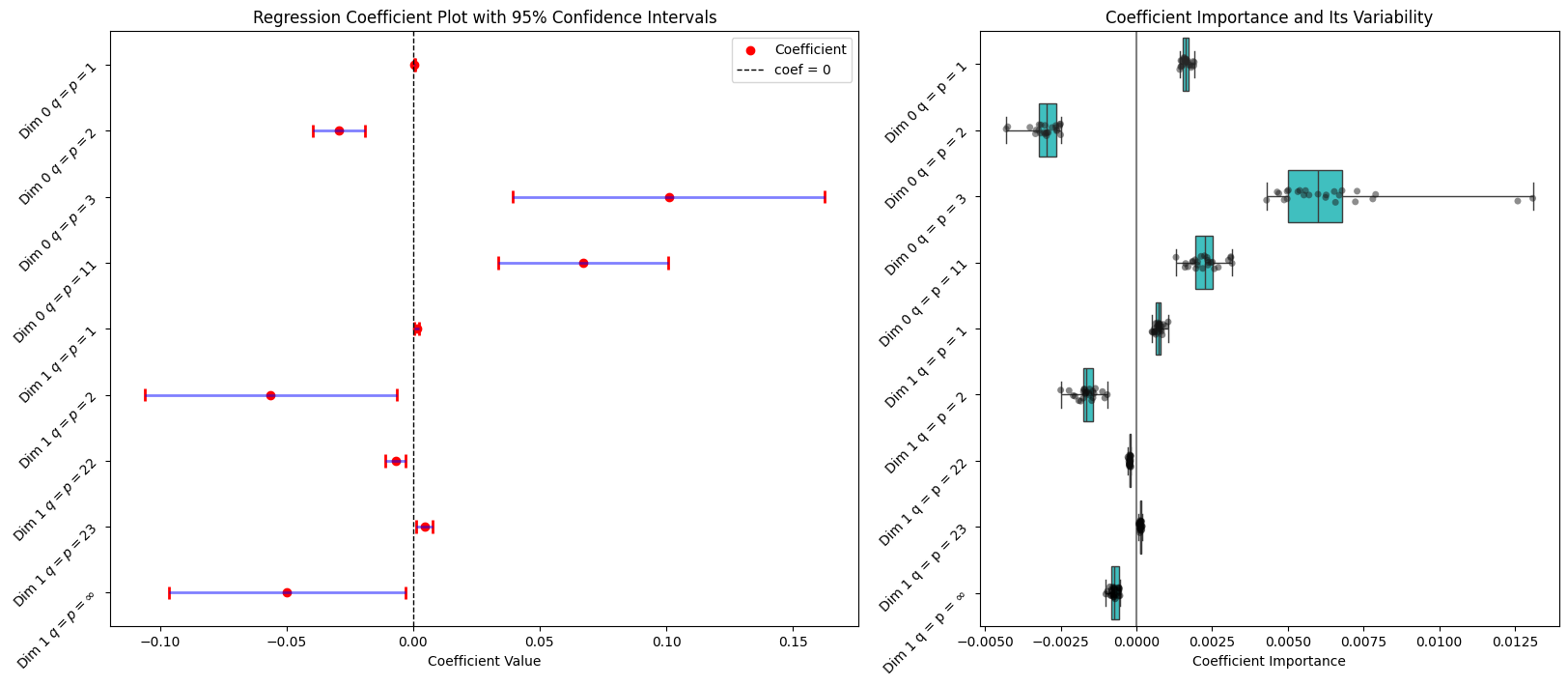}}
\caption{This figure illustrates the filtered-in features generated from the Second Pass (\cref{Second Pass}) for the ROI: dmPFC and the hemisphere: L(eft) where the training data contains both "Only Trained" and "Trained plus Untrained" LLMs. On the left sub-figure, each feature's Weight is presented along with both the Lower and Upper Bounds of the $95\%$ Confidence Intervals, while the right sub-figure illustrates its Coefficient Importance and Variability.}
\label{with_untrained_L_dmPFC_coefs_combined_23}
\end{center}
\vskip -0.2in
\end{figure}

\begin{table}[H]
\caption{This table summarizes the filtered-in features generated from the Second Pass (\cref{Second Pass}) for the ROI: dmPFC and the hemisphere: R(ight) where the training data contains "Only Trained" LLMs.}
\label{only_trained_R_dmPFC}
\vskip 0.15in
\begin{center}
\begin{small}
\begin{sc}
\begin{tabular}{ccccccc}
\toprule
Feature & Weight & $95\%$ CI & $95\%$ CI & SE & $|t|$ & $p$  \\
 &  & Lower & Upper &  & &  \\
\midrule
(Intercept) & $4.8202 e{-4}$ & $-7.8001 e{-4}$ & $1.7440 e{-3}$ & $2.9272 e{-4}$ & $1.9039 e{0}$ & $4.3344 e{-1}$ \\
Dim 0 $q = p = 1$ & $4.0236 e{-4}$ & $2.1314 e{-4}$ & $5.9157 e{-4}$ & $7.8672 e{-5}$ & $5.4965 e{0}$ & $8.0000 e{-5}$ \\
Dim 0 $q = p = 2$ & $-3.3138 e{-2}$ & $-4.5556 e{-2}$ & $-2.0720 e{-2}$ & $6.0554 e{-3}$ & $5.5138 e{0}$ & $0.0000 e{0}$ \\
Dim 0 $q = p = 3$ & $1.3683 e{-1}$ & $6.3979 e{-2}$ & $2.0968 e{-1}$ & $3.5150 e{-2}$ & $4.1794 e{0}$ & $1.0400 e{-3}$ \\
Dim 1 $q = p = 1$ & $3.8741 e{-3}$ & $2.3882 e{-3}$ & $5.3600 e{-3}$ & $4.1502 e{-4}$ & $8.7766 e{0}$ & $0.0000 e{0}$ \\
Dim 1 $q = p = 2$ & $-1.1151 e{-1}$ & $-1.8843 e{-1}$ & $-3.4600 e{-2}$ & $1.9431 e{-2}$ & $5.1319 e{0}$ & $2.2000 e{-2}$ \\
Dim 1 $q = p = 14$ & $-1.0395 e{-2}$ & $-1.6193 e{-2}$ & $-4.5961 e{-3}$ & $1.6721 e{-3}$ & $7.1680 e{0}$ & $2.4000 e{-4}$ \\
Dim 1 $q = p = 15$ & $7.3163 e{-3}$ & $1.3269 e{-3}$ & $1.3306 e{-2}$ & $1.5463 e{-3}$ & $4.8113 e{0}$ & $3.2680 e{-2}$ \\
Dim 1 $q = p = 16$ & $6.5807 e{-3}$ & $1.2895 e{-3}$ & $1.1872 e{-2}$ & $1.0768 e{-3}$ & $6.0820 e{0}$ & $2.4000 e{-2}$ \\

\bottomrule
\end{tabular}
\end{sc}
\end{small}
\end{center}
\vskip -0.1in
\end{table}

\begin{figure}[H]
\vskip 0.2in
\begin{center}
\centerline{\includegraphics[width=\columnwidth*\columnWidthCoefResultsCoefsCI]{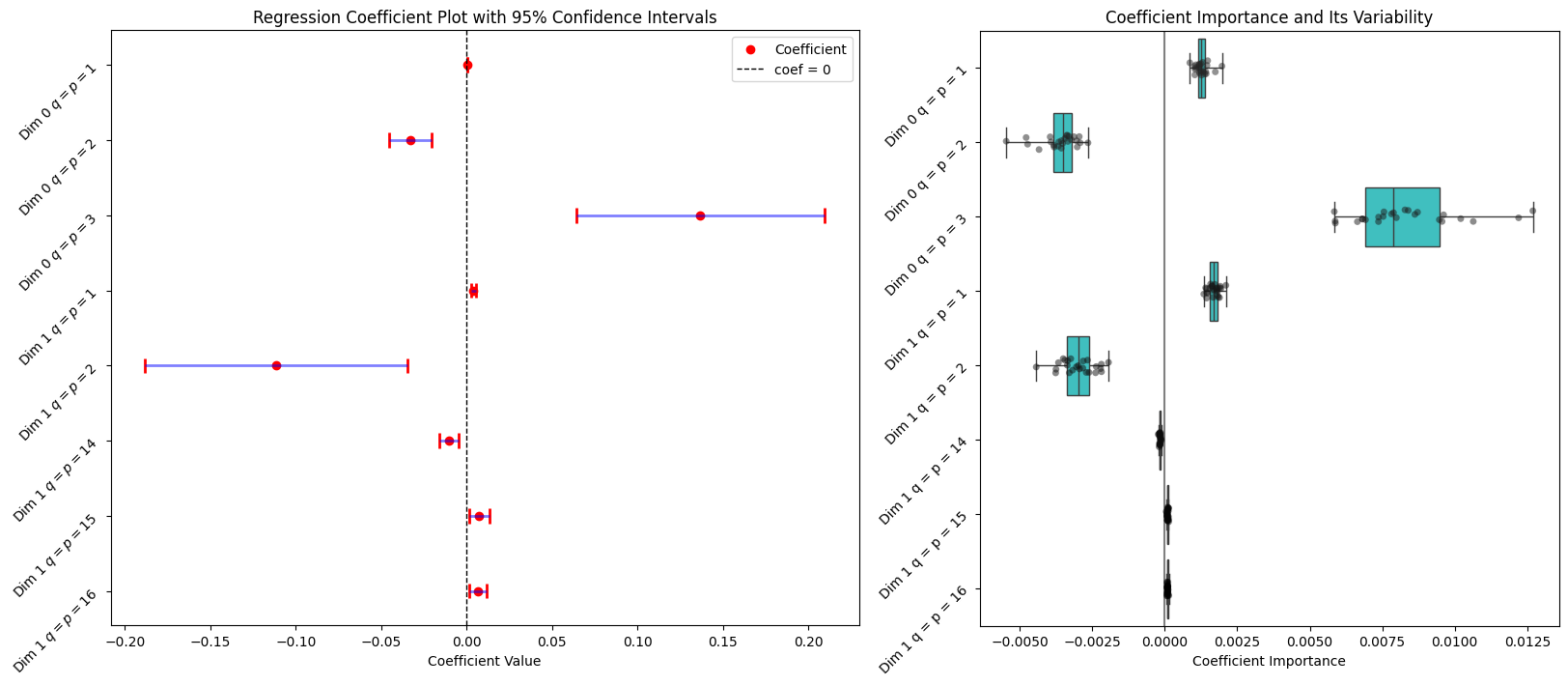}}
\caption{This figure illustrates the filtered-in features generated from the Second Pass (\cref{Second Pass}) for the ROI: dmPFC and the hemisphere: R(ight) where the training data contains "Only Trained" LLMs. On the left sub-figure, each feature's Weight is presented along with both the Lower and Upper Bounds of the $95\%$ Confidence Intervals, while the right sub-figure illustrates its Coefficient Importance and Variability.}
\label{only_trained_R_dmPFC_coefs_combined_16}
\end{center}
\vskip -0.2in
\end{figure}

\begin{table}[H]
\caption{This table summarizes the filtered-in features generated from the Second Pass (\cref{Second Pass}) for the ROI: dmPFC and the hemisphere: R(ight) where the training data contains both "Only Trained" and "Trained plus Untrained" LLMs.}
\label{with_untrained_R_dmPFC}
\vskip 0.15in
\begin{center}
\begin{small}
\begin{sc}
\begin{tabular}{ccccccc}
\toprule
Feature & Weight & $95\%$ CI & $95\%$ CI & SE & $|t|$ & $p$  \\
 &  & Lower & Upper &  & &  \\
\midrule
(Intercept) & $1.9088 e{-4}$ & $-9.4506 e{-4}$ & $1.3268 e{-3}$ & $2.2561 e{-4}$ & $9.9811 e{-1}$ & $7.0760 e{-1}$ \\
Dim 0 $q = p = 1$ & $6.6875 e{-4}$ & $5.2661 e{-4}$ & $8.1088 e{-4}$ & $5.6867 e{-5}$ & $1.0986 e{1}$ & $0.0000 e{0}$ \\
Dim 0 $q = p = 2$ & $-4.0338 e{-2}$ & $-5.0386 e{-2}$ & $-3.0290 e{-2}$ & $3.3882 e{-3}$ & $1.2002 e{1}$ & $0.0000 e{0}$ \\
Dim 0 $q = p = 3$ & $1.7443 e{-1}$ & $1.1521 e{-1}$ & $2.3,366 e{-1}$ & $1.8315 e{-2}$ & $9.8858 e{0}$ & $0.0000 e{0}$ \\
Dim 0 $q = p = 4$ & $-3.2980 e{-1}$ & $-5.0822 e{-1}$ & $-1.5138 e{-1}$ & $5.6767 e{-2}$ & $5.9246 e{0}$ & $1.1600 e{-3}$ \\
Dim 1 $q = p = 1$ & $2.5034 e{-3}$ & $1.3587 e{-3}$ & $3.6481 e{-3}$ & $4.2438 e{-4}$ & $6.8876 e{0}$ & $0.0000 e{0}$ \\
Dim 1 $q = p = 2$ & $-9.7239 e{-2}$ & $-1.5799 e{-1}$ & $-3.6484 e{-2}$ & $2.1368 e{-2}$ & $4.2936 e{0}$ & $1.1640 e{-2}$ \\
Dim 1 $q = p = 13$ & $-9.8151 e{-3}$ & $-1.4318 e{-2}$ & $-5.3121 e{-3}$ & $1.5374 e{-3}$ & $4.0777 e{0}$ & $2.7280 e{-2}$ \\
Dim 1 $q = p = 14$ & $-6.9471 e{-3}$ & $-1.1090 e{-2}$ & $-2.8043 e{-3}$ & $9.1087 e{-4}$ & $9.0745 e{0}$ & $2.0000 e{-4}$ \\
Dim 1 $q = p = 22$ & $3.4595 e{-3}$ & $1.0818 e{-3}$ & $5.8371 e{-3}$ & $4.9270 e{-4}$ & $6.2763 e{0}$ & $1.7400 e{-2}$ \\
Dim 2 $q = p = 9$ & $-1.6268 e{-2}$ & $-3.2134 e{-2}$ & $-4.0227 e{-4}$ & $3.8915 e{-3}$ & $4.5707 e{0}$ & $4.7760 e{-2}$ \\

\bottomrule
\end{tabular}
\end{sc}
\end{small}
\end{center}
\vskip -0.1in
\end{table}

\begin{figure}[H]
\vskip 0.2in
\begin{center}
\centerline{\includegraphics[width=\columnwidth*\columnWidthCoefResultsCoefsCI]{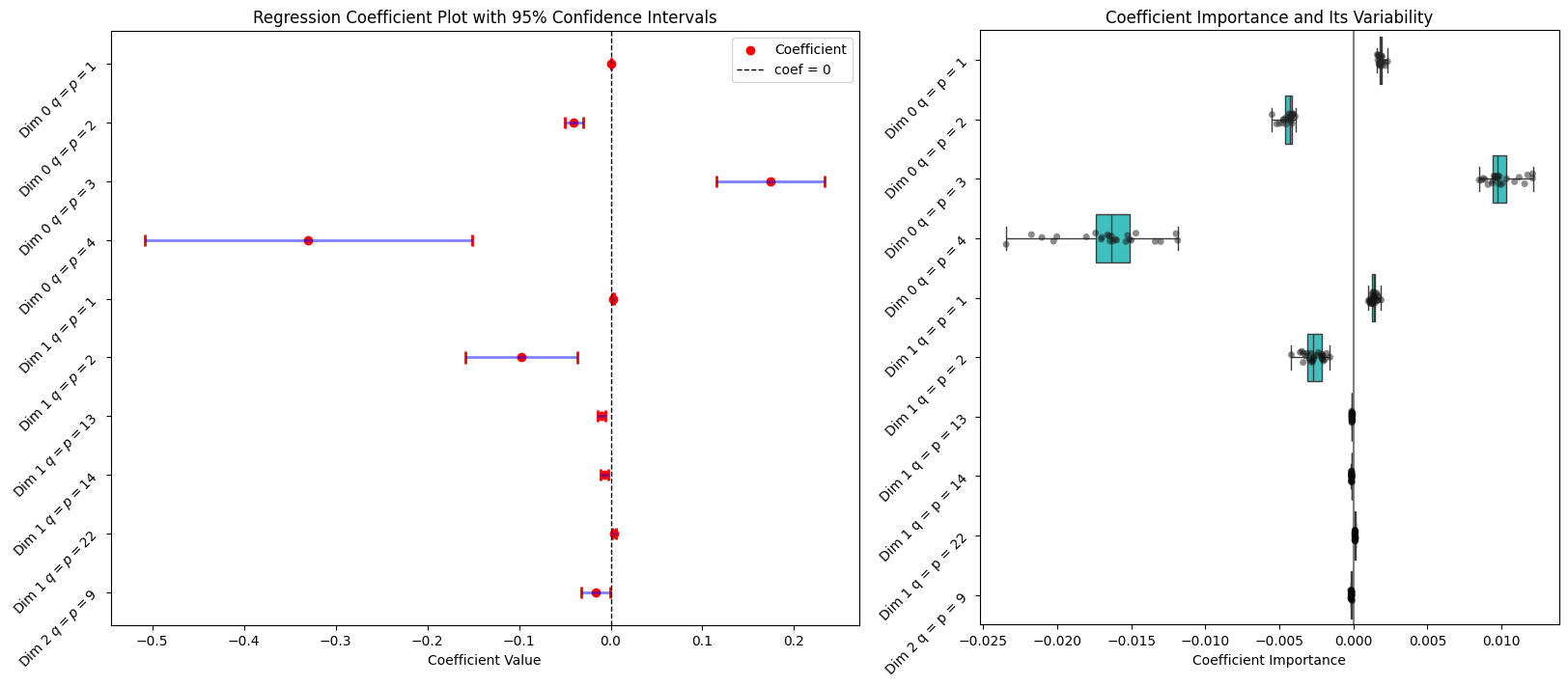}}
\caption{This figure illustrates the filtered-in features generated from the Second Pass (\cref{Second Pass}) for the ROI: dmPFC and the hemisphere: R(ight) where the training data contains both "Only Trained" and "Trained plus Untrained" LLMs. On the left sub-figure, each feature's Weight is presented along with both the Lower and Upper Bounds of the $95\%$ Confidence Intervals, while the right sub-figure illustrates its Coefficient Importance and Variability.}
\label{with_untrained_R_dmPFC_coefs_combined_22}
\end{center}
\vskip -0.2in
\end{figure}

\section{Illustrations for Brainscores in Discussions and Limitations}
\label{Illustrations for Brainscores in Discussions and Limitations}

This section presents various representations of brainscores corresponding to \cref{Discussions and Limitations}. 

Additionally, we augment our illustrations with statistical analyses by examining correlations between the computed brainscores and the number of parameters \footnote{Refer to \cref{LLMs-refs} in \cref{LLMs Data Representation Stats}.} in each LLM (\cref{Comparisons of the Brainscores between LLMs}), the proportion of LLMs where trained ones outperform their untrained counterparts w.r.t their brainscores (\cref{Does the Training Help Increase the LLM's Brainscore?}), and whether the original LLM achieves a higher brainscore compared to its quantized version (\cref{Future Direction in Discussions and Limitations}), as outlined in \cref{Discussions and Limitations}.

\begin{figure}[H]
\vskip 0.2in
\begin{center}
\centerline{\includegraphics[width=\columnwidth]{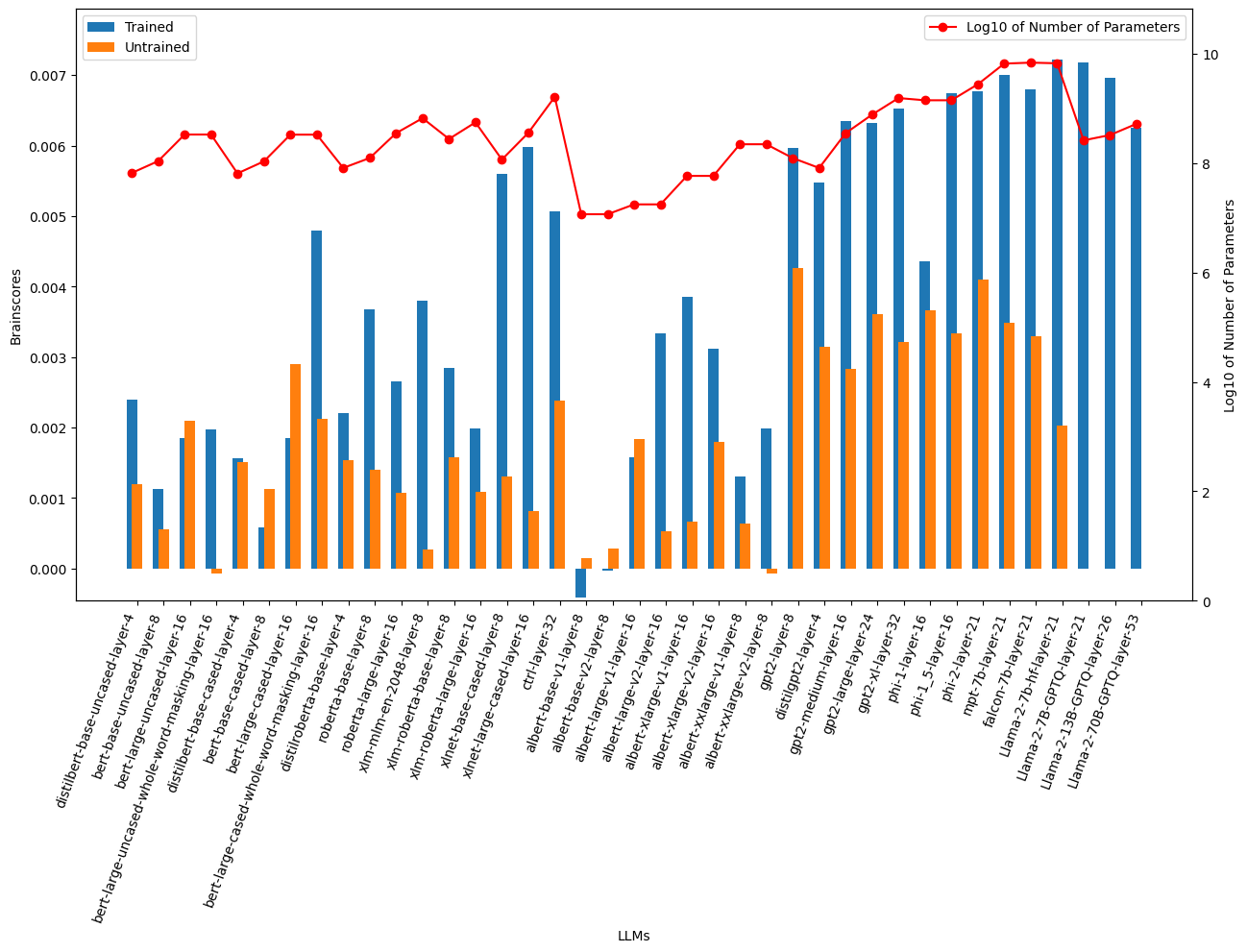}}
\caption{This figure illustrates the averaged brainscores computed for the intermediate-to-deep layer ($l = \frac{2}{3} n_{\text{layers}}$) according to \citet{caucheteux2023evidence} across all 39 LLMs, along with their untrained counterparts (excluding the last 3 LLMs). The brainscores are evaluated against the ROI: PCC and the (L)eft hemisphere, within the context of the task: "It's Not the Fall That Gets You". In this specific instance, $83.33\%$ of the LLMs demonstrate that trained models outperform their untrained counterparts concerning their brainscores. Additionally, the Llama-2-7b-hf (trained) exhibits a higher brainscore compared to its quantized version, Llama-2-7B-GPTQ (trained). The figure also displays the logarithmic (with base 10) transformation of each LLM's number of parameters (refer to \cref{LLMs-refs} in \cref{LLMs Data Representation Stats}).}
\label{notthefallintact-two_thirds_layer-L-PCC}
\end{center}
\vskip -0.2in
\end{figure}

\begin{table}[H]
\caption{This table encapsulates the correlations, for the instance illustrated in \cref{notthefallintact-two_thirds_layer-L-PCC}, across three metrics: Pearson $r$, Spearman $\rho$, and Kendall $\tau$, supplemented with corresponding $p$ values, obtained from correlating the computed brainscores with the number of parameters in each LLM (trained only). Each correlation is assessed within four distinct groups, determined by whether each LLM's number of parameters are represented in their raw value or transformed logarithmically (with base 10), and whether data points on quantized LLMs are incorporated. The decision to exclude quantized LLMs stems from their substantial reduction in the number of parameters, while maintaining comparable brainscores with their full-resolution configurations.}
\label{notthefallintact-two_thirds_layer-L-PCC_correlations}
\vskip 0.15in
\begin{center}
\begin{small}
\begin{sc}
\begin{tabular}{cccc}
\toprule
Group & Pearson $r$ & Spearman $\rho$ & kendall $\tau$ \\
\midrule
Corr All Trained & $50.87\%$  & $63.62\%$  & $47.46\%$  \\
$p$ & $9.44e{-4}$  & $1.35e{-5}$  & $2.27e{-5}$  \\
Corr All Trained Log & $66.67\%$  & $63.62\%$  & $47.46\%$  \\
$p$ & $3.59e{-6}$  & $1.35e{-5}$  & $2.27e{-5}$ \\
Corr No Quantized & $58.37\%$  & $67.80\%$  & $51.56\%$ \\
$p$ & $1.87e{-4}$  & $5.53e{-6}$  & $1.07e{-5}$ \\
Corr No Quantized Log & $69.92\%$  & $67.80\%$  & $51.56\%$ \\
$p$ & $2.10e{-6}$  & $5.53e{-6}$  & $1.07e{-5}$ \\
\bottomrule
\end{tabular}
\end{sc}
\end{small}
\end{center}
\vskip -0.1in
\end{table}

\begin{figure}[H]
\vskip 0.2in
\begin{center}
\centerline{\includegraphics[width=\columnwidth]{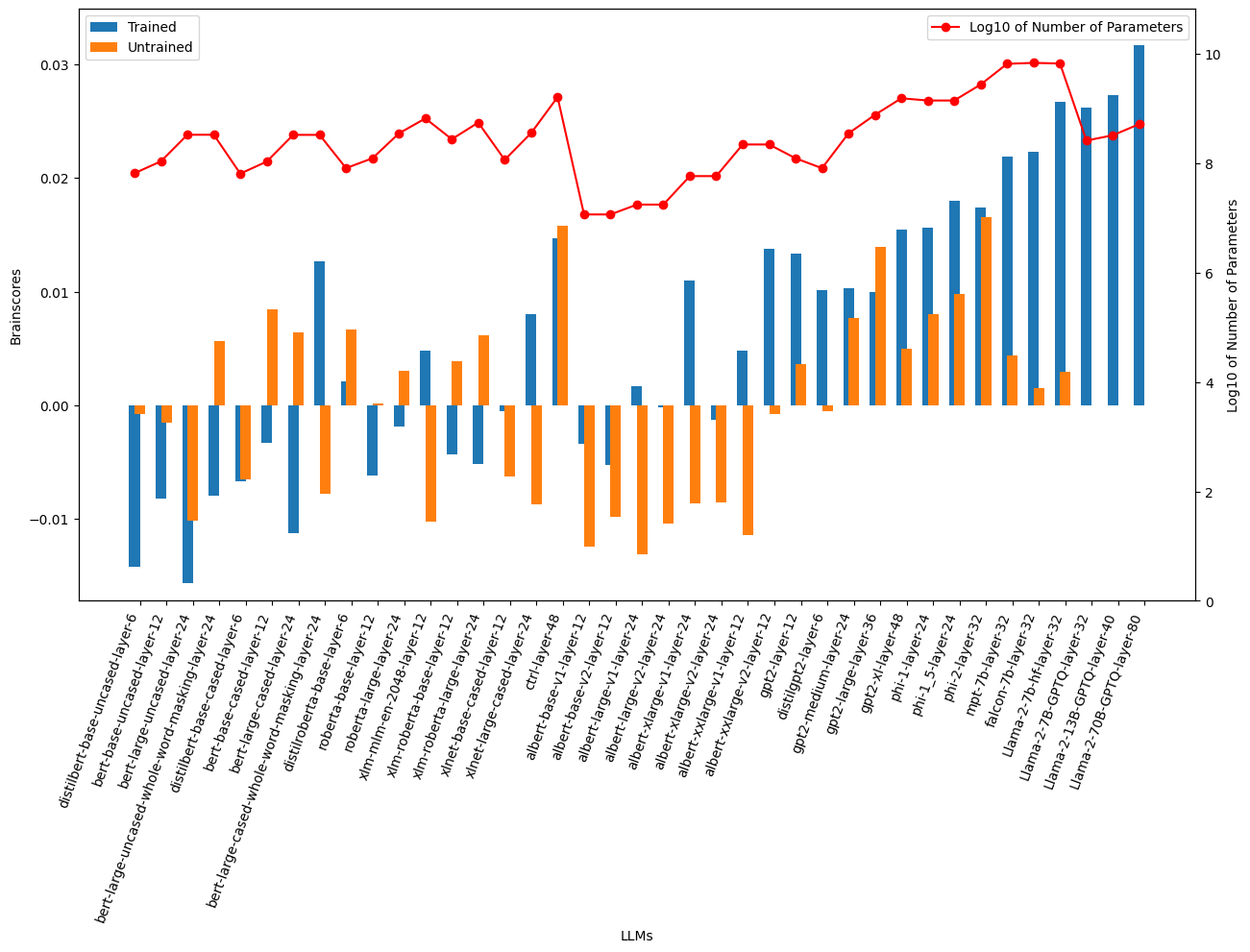}}
\caption{This figure illustrates the averaged brainscores computed for the last layer across all 39 LLMs, along with their untrained counterparts (excluding the last 3 LLMs). The brainscores are evaluated against the ROI: Whole Brain Mask and the (L)eft hemisphere, within the context of the task: "Pie Man". In this specific instance, $61.11\%$ of the LLMs demonstrate that trained models outperform their untrained counterparts concerning their brainscores. Additionally, the Llama-2-7b-hf (trained) exhibits a higher brainscore compared to its quantized version, Llama-2-7B-GPTQ (trained). The figure also displays the logarithmic (with base 10) transformation of each LLM's number of parameters (refer to \cref{LLMs-refs} in \cref{LLMs Data Representation Stats}).}
\label{pieman-linspace_layer_8-L-Evidence}
\end{center}
\vskip -0.2in
\end{figure}

\begin{table}[H]
\caption{This table encapsulates the correlations, for the instance illustrated in \cref{pieman-linspace_layer_8-L-Evidence}, across three metrics: Pearson $r$, Spearman $\rho$, and Kendall $\tau$, supplemented with corresponding $p$ values, obtained from correlating the computed brainscores with the number of parameters in each LLM (trained only). Each correlation is assessed within four distinct groups, determined by whether each LLM's number of parameters are represented in their raw value or transformed logarithmically (with base 10), and whether data points on quantized LLMs are incorporated. The decision to exclude quantized LLMs stems from their substantial reduction in the number of parameters, while maintaining comparable brainscores with their full-resolution configurations.}
\label{pieman-linspace_layer_8-L-Evidence_correlations}
\vskip 0.15in
\begin{center}
\begin{small}
\begin{sc}
\begin{tabular}{cccc}
\toprule
Group & Pearson $r$ & Spearman $\rho$ & kendall $\tau$ \\
\midrule
Corr All Trained & $51.33\%$  & $53.10\%$  & $37.70\%$  \\
$p$ & $8.33e{-4}$  & $5.06e{-4}$  & $7.66e{-4}$  \\
Corr All Trained Log & $57.30\%$  & $53.10\%$  & $37.70\%$  \\
$p$ & $1.38e{-4}$  & $5.06e{-4}$  & $7.66e{-4}$ \\
Corr No Quantized & $65.98\%$  & $57.61\%$  & $41.66\%$ \\
$p$ & $1.20e{-5}$  & $2.36e{-4}$  & $3.75e{-4}$ \\
Corr No Quantized Log & $63.93\%$  & $57.61\%$  & $41.66\%$ \\
$p$ & $2.70e{-5}$  & $2.36e{-4}$  & $3.75e{-4}$ \\
\bottomrule
\end{tabular}
\end{sc}
\end{small}
\end{center}
\vskip -0.1in
\end{table}

\begin{figure}[H]
\vskip 0.2in
\begin{center}
\centerline{\includegraphics[width=\columnwidth]{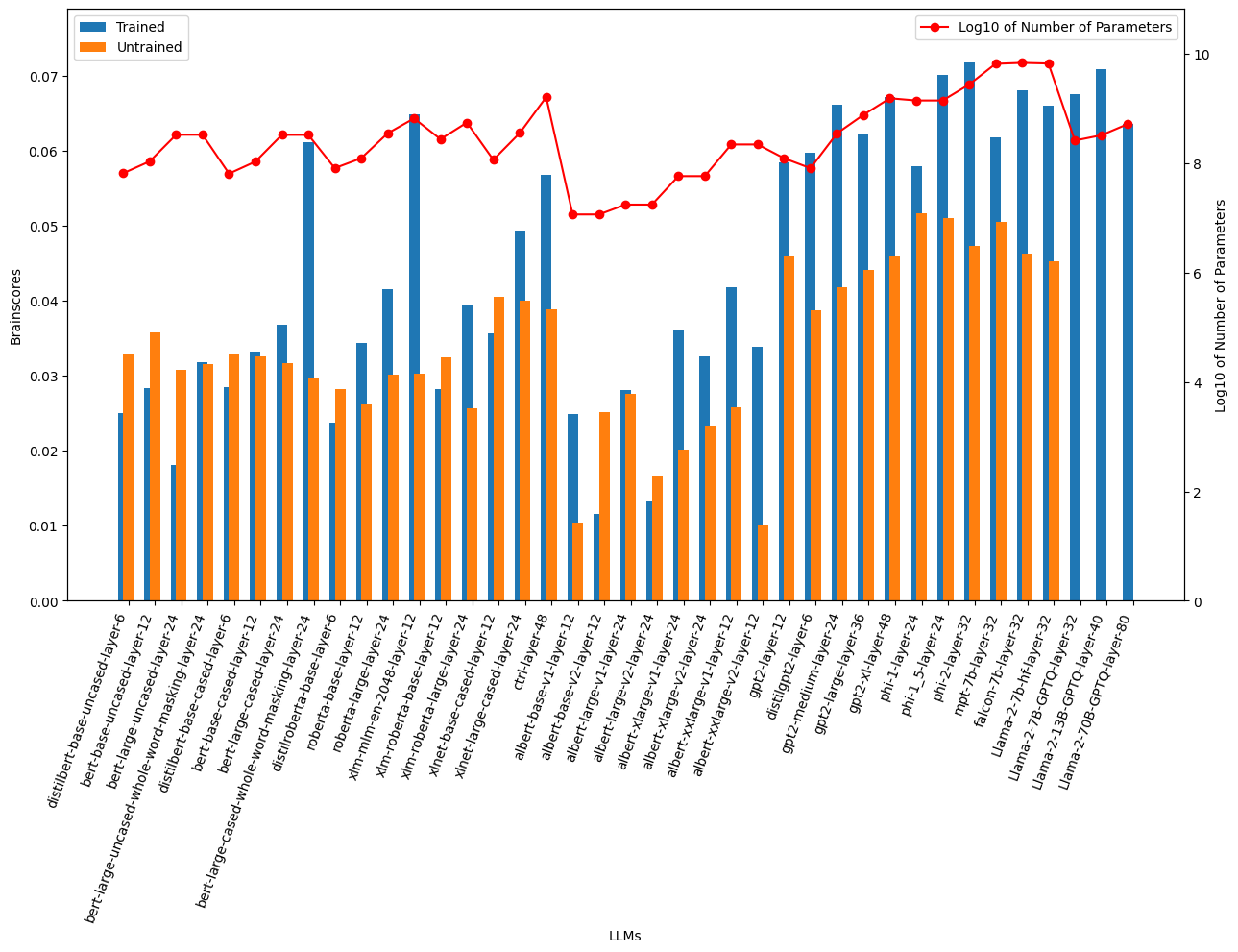}}
\caption{This figure illustrates the averaged brainscores computed for the last layer across all 39 LLMs, along with their untrained counterparts (excluding the last 3 LLMs). The brainscores are evaluated against the ROI: Whole Brain Mask and the (L)eft hemisphere, within the context of the task: "It's Not the Fall That Gets You". In this specific instance, $75\%$ of the LLMs demonstrate that trained models outperform their untrained counterparts concerning their brainscores. But the Llama-2-7b-hf (trained) fails to exhibit a higher brainscore compared to its quantized version, Llama-2-7B-GPTQ (trained). The figure also displays the logarithmic (with base 10) transformation of each LLM's number of parameters (refer to \cref{LLMs-refs} in \cref{LLMs Data Representation Stats}).}
\label{notthefallintact-linspace_layer_8-L-Evidence}
\end{center}
\vskip -0.2in
\end{figure}

\begin{table}[H]
\caption{This table encapsulates the correlations, for the instance illustrated in \cref{notthefallintact-linspace_layer_8-L-Evidence}, across three metrics: Pearson $r$, Spearman $\rho$, and Kendall $\tau$, supplemented with corresponding $p$ values, obtained from correlating the computed brainscores with the number of parameters in each LLM (trained only). Each correlation is assessed within four distinct groups, determined by whether each LLM's number of parameters are represented in their raw value or transformed logarithmically (with base 10), and whether data points on quantized LLMs are incorporated. The decision to exclude quantized LLMs stems from their substantial reduction in the number of parameters, while maintaining comparable brainscores with their full-resolution configurations.}
\label{notthefallintact-linspace_layer_8-L-Evidence_correlations}
\vskip 0.15in
\begin{center}
\begin{small}
\begin{sc}
\begin{tabular}{cccc}
\toprule
Group & Pearson $r$ & Spearman $\rho$ & kendall $\tau$ \\
\midrule
Corr All Trained & $48.09\%$  & $72.99\%$  & $54.51\%$  \\
$p$ & $1.94e{-3}$  & $1.34e{-7}$  & $1.14e{-6}$  \\
Corr All Trained Log & $74.67\%$  & $72.99\%$  & $54.51\%$  \\
$p$ & $4.79e{-8}$  & $1.34e{-7}$  & $1.14e{-6}$ \\
Corr No Quantized & $55.05\%$  & $78.17\%$  & $59.54\%$ \\
$p$ & $5.04e{-4}$  & $1.82e{-8}$  & $3.70e{-7}$ \\
Corr No Quantized Log & $78.19\%$  & $78.17\%$  & $59.54\%$ \\
$p$ & $1.80e{-8}$  & $1.82e{-8}$  & $3.70e{-7}$ \\
\bottomrule
\end{tabular}
\end{sc}
\end{small}
\end{center}
\vskip -0.1in
\end{table}

\begin{figure}[H]
\vskip 0.2in
\begin{center}
\centerline{\includegraphics[width=\columnwidth]{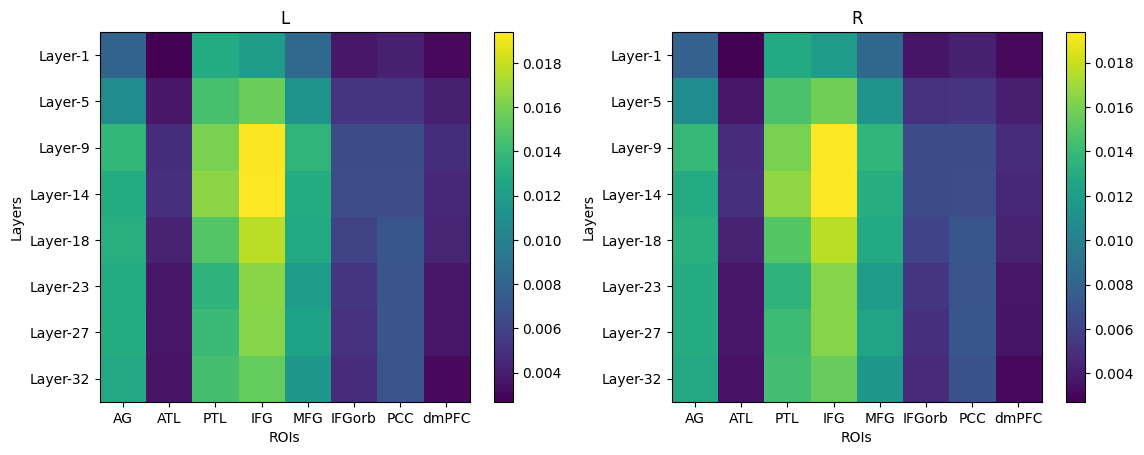}}
\caption{This figure depicts the averaged brainscores across the uniformly sampled eight layers, spanning from the first to the last layer (inclusive at both ends), of the LLM: Llama-2-7b-hf. These scores are computed concerning all eight language-related brain ROIs, within the context of the task: "It's Not the Fall That Gets You". The heatmap on the left represents the outcomes for the left hemisphere, while the one on the right illustrates those for the right hemisphere.}
\label{notthefallintact-Llama-2-7b-hf-L-and-R}
\end{center}
\vskip -0.2in
\end{figure}

\begin{figure}[H]
\vskip 0.2in
\begin{center}
\centerline{\includegraphics[width=\columnwidth]{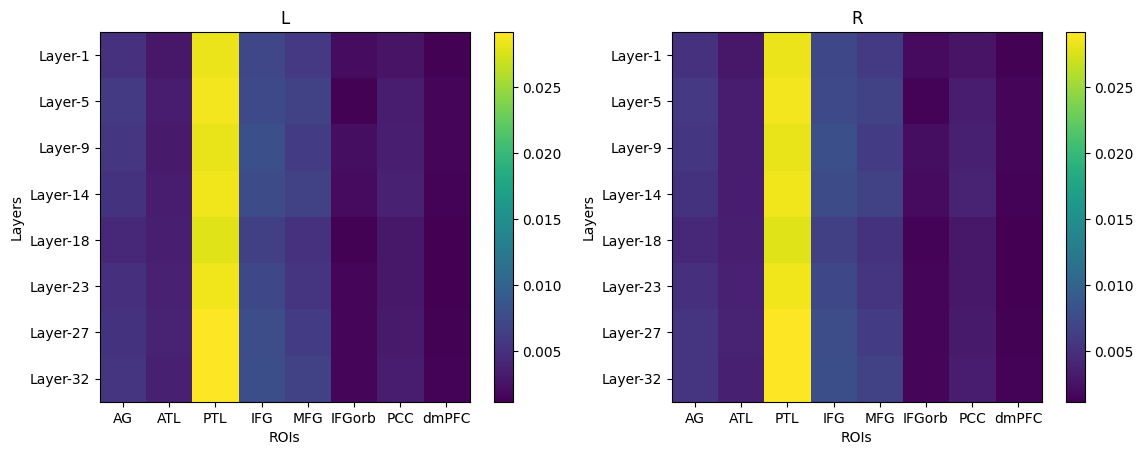}}
\caption{This figure depicts the averaged brainscores across the uniformly sampled eight layers, spanning from the first to the last layer (inclusive at both ends), of the LLM: Llama-2-7b-hf. These scores are computed concerning all eight language-related brain ROIs, within the context of the task: "Shapes". The heatmap on the left represents the outcomes for the left hemisphere, while the one on the right illustrates those for the right hemisphere.}
\label{shapessocial-Llama-2-7b-hf-L-and-R}
\end{center}
\vskip -0.2in
\end{figure}

\begin{figure}[H]
\vskip 0.2in
\begin{center}
\centerline{\includegraphics[width=\columnwidth]{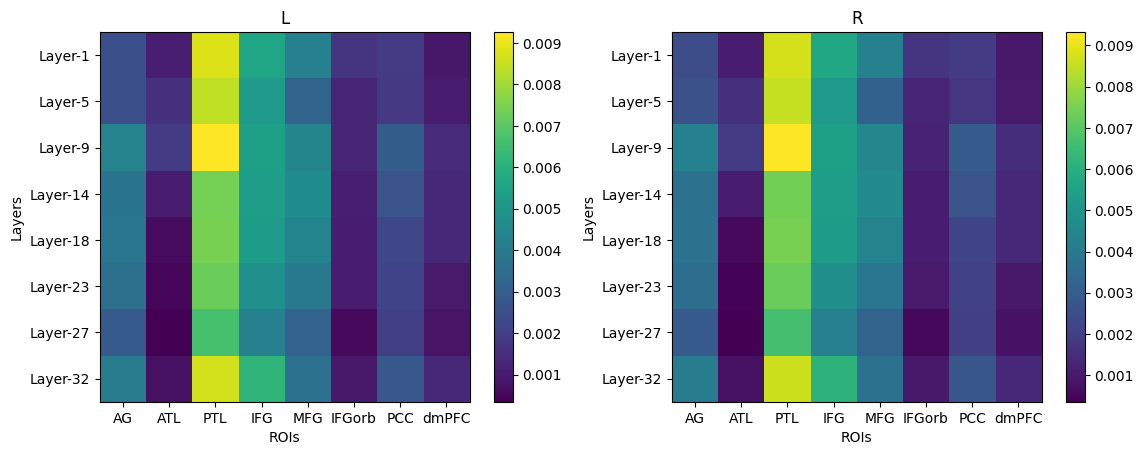}}
\caption{This figure depicts the averaged brainscores across the uniformly sampled eight layers, spanning from the first to the last layer (inclusive at both ends), of the LLM: Llama-2-7b-hf. These scores are computed concerning all eight language-related brain ROIs, within the context of the task: "Pie Man". The heatmap on the left represents the outcomes for the left hemisphere, while the one on the right illustrates those for the right hemisphere.}
\label{pieman-Llama-2-7b-hf-L-and-R}
\end{center}
\vskip -0.2in
\end{figure}

\end{document}